\documentclass[12pt,preprint]{aastex}

\shorttitle{13 Powerful Classical Double Radio Galaxies}
\shortauthors{Kharb et al.}

\begin{document}

\title{A Study of 13 Powerful Classical Double Radio Galaxies}
\author{P. Kharb\altaffilmark{~}, C. P. O'Dea\altaffilmark{~}, 
S. A. Baum\altaffilmark{~}}
\affil{Rochester Institute of Technology, 54 Lomb Memorial Drive, 
Rochester, NY 14623}
\author{R. A. Daly\altaffilmark{~}, M. P. Mory}
\affil{Department of Physics, Penn State University, Berks Campus, P. O. 
Box 7009, Reading, PA 19610}
\author{M. Donahue\altaffilmark{~}}
\affil{Department of Physics and Astronomy, BPS Building, Michigan State 
University, East Lansing, MI 48824} 
\and
\author{E. J. Guerra\altaffilmark{~}}
\affil{Department of Physics and Astronomy, 
Rowan University, 201 Mullica Hill Rd. Glassboro, NJ 08028}

\begin{abstract}
We have carried out an extensive study of a sample of 13 large, powerful 
Fanaroff-Riley type II radio galaxies with the Very Large Array in multiple
configurations at 330~MHz, 1.4, 5 and 8~GHz. We present the total intensity, 
polarization, spectral index, and rotation measure maps of the sources. 
On the whole the 13 FRII sources have symmetric structures with arm-length
ratios close to unity, small misalignment angles and low values of radio
core prominence, suggesting that these radio galaxies lie close to the plane 
of the sky. We have revisited some well known radio galaxy correlations using a 
large combined dataset comprising our radio galaxies and others from the 
literature. We confirm that the hotspot size correlates with the core-hotspot 
distance. The hotspot spectral index is correlated with, and flatter than the 
lobe spectral index, consistent with the assumptions of spectral aging
models. Both the hotspot and lobe spectral index are correlated with redshift.  
The depolarization asymmetry in the lobes is not correlated with the
radio core prominence or misalignment angle, which are statistical 
indicators of orientation. The `Liu-Pooley' correlation of lobe depolarization 
with the lobe spectral index is significant in our radio galaxy sample. 
Further, the lobe with the steeper spectral index and greater depolarization 
is shorter and fatter. The arm-length ratio seems to 
be correlated with the misalignment angle between the two sides of the radio 
source and strongly anti-correlated with the axial ratio, consistent with
environmental effects and/or a change in the outflow direction. 
In this sample, asymmetries in the local environments and/or motion of the 
outflow axis are likely to be more important than relativistic beaming effects.

\end{abstract}


\section{INTRODUCTION}

Classical double radio galaxies are active galactic nuclei with regions of
synchrotron-emitting plasma that can extend to thousands of kiloparsecs. These
Fanaroff-Riley type II radio galaxies \citep{FanaroffRiley74} are thought to
be powered by narrow collimated jets, terminating in high 
surface brightness regions called `hotspots'. Regions of low surface 
brightness emission lying between the galaxy center and the hotspots, 
called radio bridges, are believed to be a result of the accumulation of 
relativistic particles accelerated at the hotspots over the lifetime of 
an FRII source that form a low-density cocoon around the jet
\citep[see][for a theoretical review]{Begelman84}.  
The ultimate motivation of our project is to study the radio bridges in 
FRII radio galaxies and do a spectral aging analysis. 
In this paper we present the first results of our study and describe
the global properties of our FRII radio galaxy sample.

In order to comprehend the phenomenology of FRII galaxies, large samples
of radio galaxies have hitherto been observed, yielding an extensive 
database of information. Results from arcsecond-scale radio observations 
of FRII radio galaxies have been presented by many authors, for example,
\citet{Laing81,LeahyWilliams84,AlexanderLeahy87,LeahyMuxlow89,
Pedelty89,GarringtonConway91,Liu92,Hardcastle98,Ishwara-Chandra01,
Goodlet04,Gilbert04} and \citet{Mullin06}, and others. 
Previous studies of the radio bridges have included relatively 
low-resolution observations ($\sim3\arcsec-4\arcsec$) at 151~MHz and 1.4~GHz 
of the full bridge region \citep[e.g.,][]{LeahyMuxlow89}, while the higher 
resolution observations ($\sim1\arcsec$) at 1.4, 5 and 15~GHz have sampled 
the bridge emission partially \citep[e.g.,][]{Liu92}. 
However, to empirically address and 
understand the different physical processes
that are important at different radio frequencies, physical locations, and
stages of evolution of a source, both low-frequency radio data 
and high-resolution radio data are required. 

To this end, we observed 13 FRII radio galaxies with the Very Large 
Array (VLA) in multiple configurations at 330~MHz, 1.4~GHz, 5~GHz, and 8~GHz.
Here we present images of the total and polarized radio intensity,
spectral index between 1.4 and 5 GHz, 0.3 and 1.4 GHz, and rotation measure 
between 1.4 and 5 GHz\footnote{Spectral index and 
RM maps are available only in the electronic version}. Further,
we probe the relationship between different global characteristics of the  
radio galaxies by augmenting our data with additional galaxy data
gleaned from the literature. We will subsequently refer to this extended,
eclectic sample as the ``combined'' dataset, while the 13 FRII galaxies
will be referred to as such.

The paper is arranged as follows: the radio galaxy sample is described
in Sect.~\ref{sample} while the observations and data reduction are
discussed in Sect.~\ref{datareduction}; the source properties are presented in 
Sect.~\ref{results}; the correlations are discussed in Sect.~\ref{corr};
and the summary and conclusions follow in Sect.~\ref{conclusions}.
This paper is the first in a series of four papers. In Paper~II, we will 
present the results of the 
spectral aging analysis, lobe propagation velocities, pressures, beam 
powers, and ambient gas densities. In Paper~III, we will describe the use 
of these sources for cosmological studies and present the results for a large
sample of powerful FRII galaxies. Finally, in Paper IV, we will present a 
detailed analysis of the radio bridge structure of the sources.

Throughout the paper, we have adopted the cosmology in which 
$H_0$=71 km s$^{-1}$ Mpc$^{-1}$, $\Omega_m$=0.27 and $\Omega_{\Lambda}$=0.73.
The spectral indices $\alpha$ are defined such that the flux density 
$S_\nu$ at frequency $\nu$, is $S_\nu \propto \nu^{-\alpha}$.

\section{THE SAMPLE}
\label{sample}

The FRII radio galaxies considered for the present study are part of the
3CR sample of radio sources \citep{Bennett62}.
The 13 sources which are a focus of this 
paper were those for which we were alloctated observing time, from a larger 
sample of powerful FRII sources selected for an extensive study of their radio 
bridges. These sources satisfy the following criteria: their radio power at 
178~MHz is greater than $10^{28}$~W~Hz$^{-1}$; their angular sizes are larger 
than $27\arcsec$, they span the redshift range of $z\simeq$~0.4--1.65, and 
are classified as narrow-line FRII radio galaxies 
\citep{Hes96,JacksonRawlings97}. 
The largest source in our sample, 3C172, has an angular 
extent of $\geqslant100\arcsec$, which corresponds to a linear size of 
$\simeq680$ kpc. The large angular sizes ensure that the radio lobes span 
several beam widths, crucial for the spectral aging study, which will be 
presented in Paper~II. A compilation of the basic parameters for each source 
is given in Table~\ref{tabsample}. 
\begin{deluxetable}{llllccllllllllllllllllll}
\tabletypesize{\small}
\tablecaption{The sample of FRII radio galaxies}
\tablewidth{0pt}
\tablehead{
\colhead{Source}        & \colhead{IAU name}  & \colhead{z} & 
\colhead{S$_{178}$}     & \colhead{log(P$_{178}$)} &
\colhead{Scale}         & \colhead{$\Theta$, PA} &\colhead{LAS} &\colhead{LAS} & \colhead{Ref} \\
\colhead{}              &\colhead{}           &\colhead{}   &
\colhead{(Jy)}          &\colhead{(W~Hz$^{-1}$)}& 
\colhead{(kpc/$\arcsec$)}&\colhead{($\arcsec$, deg)}
&\colhead{($\arcsec$)} &\colhead{(kpc)} & \colhead{}\\
\colhead{(1)}         & \colhead{(2)}    & \colhead{(3)} &
\colhead{(4)}         & \colhead{(5)}    & \colhead{(6)} &
\colhead{(7)}         & \colhead{(8)}    & \colhead{(9)} & \colhead{(10)}}     
\startdata
\object{3C6.1}  & 0013+790 & 0.8404 & 14.9 & 28.68& 7.6 & 26.0, 27.3  & 31.5  &241 &  9, 12  \\
\object{3C13}   & 0031+391 & 1.3510 & 13.1 & 29.14& 8.5 & 28.6, 145.1 & 32.5  &276 &  4, 8 \\
\object{3C34}   & 0107+315 & 0.6900 & 13.0 & 28.41& 7.1 & 45.1, 82.9  & 48.7  &346 &  9, 12 \\
\object{3C41}   & 0123+329 & 0.7940 & 11.6 & 28.51& 7.5 & 23.8, 145.1 & 27.2  &204 &  9, 12 \\
\object{3C44}   & 0128+061 & 0.6600 & 10.0 & 28.24& 6.9 & 64.8, 12.8  & 72.0  &502 &  7 \\
\object{3C54}   & 0152+435 & 0.8274 & 10.5 & 28.51& 7.6 & 51.1, 23.6  & 60.2  &458 & 1, 2  \\
\object{3C114}  & 0417+177 & 0.8150 & 8.2  & 28.38& 7.6 & 53.2, 44.3  & 59.6  &451 & 1, 6\\
\object{3C142.1}& 0528+064 & 0.4061 & 18.6 & 28.00& 5.4 & 52.4, 130.1 & 55.5  &299 & 7 \\
\object{3C169.1}& 0647+452 & 0.6330 & 6.6  & 28.02& 6.8 & 45.9$^\ast$, 135.7$^\ast$ & 53.6  &367 & 9 \\
\object{3C172}  & 0659+253 & 0.5191 & 16.5 & 28.21& 6.2 & 101.2, 16.6 & 109.0 &676 &  5, 11 \\
\object{3C441}  & 2203+292 & 0.7070 & 13.7 & 28.45& 7.2 & 33.3, 149.8 & 41.7  &299 &  9, 12 \\
\object{3C469.1}& 2352+796 & 1.3360 & 12.1 & 29.09& 8.5 & 75.4, 171.1 & 84.1  &712 &  2 \\
\object{3C470}  & 2356+437 & 1.6530 & 11.0 & 29.28& 8.6 & 24.9, 37.9  & 29.8  &255 & 3, 10 \\
\enddata
\tablecomments{Cols.1 \& 2: Common and IAU names of the FRII radio galaxies.
Col.3: Redshifts were obtained from the NASA/IPAC Extragalactic
Database (NED). Col.4 : Total flux density at 178 MHz in Jy on the 
\citet{Baars77} scale taken from http://www.3crr.dyndns.org/cgi/database
$-$ for 5 sources not listed in \citet{LaingRiley83}, we used the
4C flux densities from \citet{Gower67} multiplied by a factor of
1.09 to convert them to Baars et al. scale following \citet{Roger73}.
Col.5: Logarithm of the total luminosity at 178 MHz.
Col.6: Spatial scale in source corresponging to 1$\arcsec$. 
Col.7: Angular extent and position angle of the source, measured from the 
VLA A-array 8 GHz maps using the brightest hotspot. The position angle is 
defined as 
counter-clockwise from North. $\ast$ For 3C169.1, the northern hotspot was not 
detected in the 8 GHz image -- the 2.5$\arcsec$ 5~GHz image was used 
instead to obtain the extent.
Col.8: Largest angular size measured from the $\sim 2\arcsec$ image at 1.4 GHz.
The AIPS task TVDIST was used to measure the entire radio extent (from
hotspot-to-hotspot) of the source.
Col.9: Largest projected linear size of source estimated using values in 
Col.s 7 and 9.
Col.10: References for previous observations (this is not an exhaustive list):
1 - \citet{MacDonald68},
2 - \citet{Longair75},
3 - \citet{RileyPooley75},
4 - \citet{Schilizzi82},
5 - \citet{Strom85},
6 - \citet{Strom90},
7 - \citet{Bogers94},
8 - \citet{LawGreen95},
9 - \citet{Neff95},
10 - \citet{Best97},
11 - \citet{Gilbert04},
12 - \citet{Mullin06}.}
\label{tabsample}
\end{deluxetable}
\section{OBSERVATIONS AND DATA ANALYSIS}
\label{datareduction}

The 13 FRII radio galaxies were observed with the A, B, C and 
D-array configurations of the VLA \citep{Napier83} between November, 2002 
and November, 2003 at 330~MHz, 1.4, 4.8 and 8.4~GHz. Table~\ref{tabsummary} 
provides a summary of the VLA observations. The sources were observed in 
a $\sim$10 min snapshot mode. In order to be able to extract possible radio 
frequency interference, which can become significant at low frequencies, 
the 330~MHz data were observed in the spectral-channel mode. 
While making the radio maps at 330~MHz,
only the central 80$\%$ of the spectral channels were averaged. Scheduling 
constraints resulted in some sources not being observed at some frequencies. 
For a few sources, VLA archival data existed at one or more frequencies and 
these were not observed (see Table~\ref{tabarchive}).  
We used the VLA archival data for these sources, the program IDs of which
are listed in Table~\ref{tabarchive}.

The data were reduced using the standard calibration and reduction procedures 
in the Astronomical Image Processing System (AIPS). 3C48  
was used as the primary flux density calibrator for the experiment. 
The instrumental polarization and electric vector position angles
were calibrated using 0217+738 and 3C138, respectively.
The polarization data at 1.4~GHz were corrected for 
ionospheric Faraday rotation using the total electron content maps 
available at CDDIS data archive\footnote{ftp cddisa.gsfc.nasa.gov} 
in the AIPS task TECOR. 

After the initial amplitude and phase calibration using the calibrators,
the source data were phase and amplitude self-calibrated 
\citep{Schwab80} and imaged in an iterative manner, using the tasks CALIB 
and IMAGR. In order to improve the UV coverage, we combined the 
self-calibrated datasets from the various configurations at a given 
frequency using the AIPS task DBCON, and performed additional rounds of phase 
and amplitude self-calibration on the combined datasets. 
Table~\ref{tabsummary} lists the largest angular scale structure which can
be imaged well in single snapshot observations with the different VLA
configurations. A quick comparison with the angular sizes of our 13 FRII
galaxies listed in Table~\ref{tabsample} demonstrates that our sources
were imaged reasonably well when the data from A, B and C-arrays were combined
at the L-band, B, C and D-arrays were combined at the C-band, and A and 
C-arrays were combined at the P-band. 

The polarization maps were made by combining the Stokes $Q$ and $U$ images 
using the AIPS task COMB. We split our L-band data into its constituent
Intermediate frequencies (IFs, see Table~\ref{tabsummary}) and used the lower 
frequency, $ie.$, 1345 MHz and the
averaged C-band data at 4860 MHz, to create the spectral index maps using 
the AIPS task COMB.
For the rotation measure maps, we used the three frequencies, 1345, 1665
and 4860 MHz, in the AIPS task RM. 
When archival data were used, we could
still combine the respective L-band IFs, which sometimes differed from our
frequencies, to the closest frequency in our data.
For 3C142.1 we noticed RM flips across the southern lobe, which were a result 
of the n$\pi$ ambiguity in the polarization angles. We therefore used a 
guess (integrated) RM value of +84 rad~m$^{-2}$ \citep{SimardNormandin81} in 
the AIPS task RM for the rotation measure map of 3C142.1. 

We typically recover 95\% to 100\% of the integrated flux density listed
in \citet{Kuehr96} at the L and C-bands. For 3C441 and 3C469.1, between
85\% and 90\% of the integrated flux density is recovered in our maps.
The registration of images for the spectral index and rotation measure maps
is better than $\sim0.1\arcsec$.
By blanking values in the task COMB, the polarization intensity, 
polarization angle and fractional polarization maps were restricted to 
have S/N$>3$, output errors less than $\approx$10 degrees and 10\%, 
respectively. No Ricean bias correction \citep[e.g.,][]{Wardle74} was made 
to the polarized
intensity. The mean rotation measure and the dispersion/spread in 
rotation measure, $\sigma_{RM}$ was obtained by using the AIPS task IMEAN.
\begin{deluxetable}{lcccccccccc}
\tabletypesize{\small}
\tablecaption{Summary of VLA Observations}
\tablewidth{0pt}
\tablehead{
\colhead{Obs. Date} & \colhead{Config.}&\colhead{Band} & \colhead{Frequency}&
\colhead{Bandwidth} & \colhead{Int. Time} & \colhead{$\theta_{LAS}$} \\
\colhead{}& \colhead{}& \colhead{} & \colhead{(MHz)} & \colhead{(MHz)} & 
\colhead{(mins)} & \colhead{($\arcsec$)}\\
\colhead{(1)}       & \colhead{(2)}    & \colhead{(3)} & \colhead{(4)} &
\colhead{(5)}       & \colhead{(6)}   & \colhead{(7)} }
\startdata
June 17, 2003 & A   &    P   & 327.5, 321.5&  3.1   &       8 & 85.0   \\
&    &    L   & 1344.9, 1664.9&  25    &      11              & 19.0   \\
&    &    X   & 8435.1, 8485.1&  25    &      11              & 3.5    \\
Nov. 29, 2003 &B   &    L   & 1344.9, 1664.9&  25    &       9& 60.0   \\
&    &    C   & 4885.1, 4835.1&  50    &       9              & 18.0   \\
Nov. 4, 2002& C   &    P   & 327.5, 321.5 &  3.1   &       6  & 2100.0 \\
&    &    L   & 1344.9, 1664.9&  25    &       9              & 450.0  \\
&    &    C   & 4885.1, 4835.1&  50    &       9              & 150.0  \\
Apr. 11, 2003&D   &    C   & 4885.1, 4835.1&  50    &       8 & 150.0  \\
\enddata
\vspace{-0.8cm}
\tablecomments
{Col.1: Observing date, Col.2: VLA Array configuration, 
Col.3: Observing frequency band, Col.4 \& 5: Central frequency 
for the 2 constituent IFs in each frequency band, and bandwidth respectively,
Col.6: Integration time in minutes, Col.7: The largest angular scale 
structure which can be imaged reasonably well in single snapshot observations
(see The VLA Observational Status Summary). 
The P-band data were observed in the spectral-line mode. }
\label{tabsummary}
\end{deluxetable}

\begin{deluxetable}{lcccccccccc}
\tabletypesize{\small}
\tablecaption{Observation Summary of the Sample}
\tablewidth{0pt}
\tablehead{
\colhead{Source} & \colhead{A/P} & \colhead{A/L} & \colhead{A/X} & 
\colhead{B/L} & \colhead{B/C} & \colhead{C/P} & \colhead{C/L} &
\colhead{C/C} & \colhead{D/C} }
\startdata
3C6.1    &   ...&AH291&AP380&  ... &  N & ...& ...&AJ111& ...\\
3C13     &    N &AA150&AL322&   N  &  N & ...& ...& ... &VAH54\\
3C34     &    N &AG247&AL322&   N  &  N&  N &  N &  N  &AP380\\
3C41     &   ...&AH291&AP380&  ... &AG220& ...& ...& ... & ...\\
3C44     &    N &   N &  N&   N  &  N &  N &  N &  N  & ...\\
3C54     &    N &   N &  N &   N  &  N &  N &  N &  N  &  N\\
3C114    &    N &   N &  N &   N &  N &  N &  N &  N  &  N\\
3C142.1  &    N &AH480&  N &   N  &  N &  N &  N &  N  &  N\\
3C169.1  &    N &   N &  N &   N  &  N &  N &  N &  N  &  N\\
3C172    &    N &   N &AP361&   N  &  N &  N &  N &  N  &  N\\
3C441    &    N &   N&  N &   N  &  N&  N &  N &  N  &  N\\
3C469.1  &    N &   N&  N &   N  &  N&  N &  N &  N &  N\\
3C470    & AB917&   N &AL322&   N &  N & ...& ...&AH343& ...\\
\enddata
\vspace{-0.8cm}
\tablecomments{N: New observations made by us (Program ID - AO170). 
AH291, AA150, etc.,: Program IDs for Archival observations.
... : No observations due to scheduling constraints.}
\label{tabarchive}
\end{deluxetable}
\section{RESULTS}
\label{results}

The final maps at 1.4~GHz and 5~GHz, restored with a 2.0$\arcsec$ or 
2.5$\arcsec$ circular Gaussian beam, the 330~MHz maps restored with 
a 5.0$\arcsec$ circular Gaussian beam, and the 8~GHz maps with insets
for the hotspots, are presented in Figs.~\ref{fig3c6.1} to 49. The total 
intensity contours are superimposed with the fractional polarization vectors. 
1.4~GHz polarization maps could not be made for 3C6.1 and 3C41 because
their respective (archival) datasets did not include scans of a 
polarization angle calibrator.
The dynamic range in the final images (defined as peak brightness/rms 
noise) varies from 7400 to 400, being typically greater than 3000. The 
1.4$-$5 GHz spectral index and RM maps were made at a resolution of 
2.0$\arcsec$ or 2.5$\arcsec$, while the 0.3$-$1.4 GHz spectral index maps 
were made at a resolution of 5.0$\arcsec$.

The high resolution A-array 8 GHz images reveal the details of the hotspot 
structure. Most sources show the presence of multiple hotspots. These could be
due to the plasma beam moving about along the cavity wall as in the 
`dentist's drill' model \citep{Scheuer82}. Alternatively, multiple hotspots 
could be interpreted as `splatter-spots' formed by a jet with a variable 
direction \citep{WilliamsGull85}. Table~\ref{tabspots} presents selected 
characteristics of the hotspots in our 13 FRII galaxies. Our high-resolution 
8 GHz images reveal that the brightest hotspots at either end, which probably 
are the current acceleration sites of particles, lie roughly along a line 
connecting the hotspots and the core for each of the 13 sources.

Radio bridges are detected in all but two sample radio galaxies $viz.,$
3C13 and 3C470. These sources are the most luminous and the most distant in 
our sample, and are likely to be strongly affected by (1+$z$)$^4$ surface 
brightness dimming. Therefore, more sensitive observations are required to 
detect the radio bridges in these sources. The spectral index 
maps show a gradual steepening of the radio spectra away from the hotspots 
with some spectral index variations superimposed, in agreement with 
previous studies \citep[e.g.,][]{MyersSpangler85,Alexander87,AlexanderLeahy87}. 
Some quantitative estimates of the radio lobe properties are tabulated in
Table~\ref{tablobes}. 

The majority of the sources exhibit a regular bridge structure, with some 
distortions becoming apparent at very low surface brightness.
We have classified distortions using the radio morphology classification 
of \citet{LeahyWilliams84}: Type~1 (hereafter LW1) morphology describes 
a bridge with no distortions but a marked decrease in the bridge 
surface-brightness towards the centre; Type~2 (hereafter LW2) describes a 
bridge which bends away from the central galaxy only on one side; Type~3 
(LW3) describes a bridge which bends away from the galaxy on both sides in a 
cross-shape; Type~4 (LW4) is a radio bridge which bends on both sides in the 
same direction away from the host galaxy; and Type~5 (LW5) describes a bridge
that is continuous across the galaxy which is displaced towards the steepest
edge of the radio source. The bridge types are presented in
Table~\ref{tablobes}. 

The radio core prominence ($R_c$) is the ratio of the flux density from
the radio core ($S_{core}$) to the extended radio lobe emission ($S_{ext}$).
In the restframe of the source, 
$R_c=\frac{S_{core}}{S_{ext}}(1+z)^{\alpha_{core}-\alpha_{ext}}$, where
$z$ is the redshift of the source, and $\alpha$ is
the spectral index, assumed to be 0 and 1 for the core and extended emission, 
respectively. Due to Doppler beaming effects in the radio core (which is
the unresolved base of the relativistic jet), $R_c$ can be used as a 
statistical indicator of beaming and therefore orientation
\citep{KapahiSaikia82,OrrBrowne82}. We have derived $R_c$ using the 
radio core flux densities from the high-resolution 8~GHz maps (where the core
is clearly delineated) and lobe flux densities from the 1.4~GHz maps 
(see Table~\ref{tabcomps}). The lobe flux density is primarily 
all the extended radio emission minus the radio core, and therefore includes 
the emission from the hotspots and jets.
We find that the core prominence in our 13 FRII radio galaxies is small --
$R_c$ varies from 4.7$\times10^{-5}$ to 0.0029. This is consistent with
the picture of these radio galaxies oriented close to the plane of the sky.

In Table~\ref{tablobes} we have tabulated the mean values of the rotation
measure ($RM$) and the corresponding dispersion ($\sigma_{RM}$) in the 
radio lobes of our 13 FRII galaxies. 
The mean $RM$ was 
averaged over the two lobes. We have estimated the lobe depolarization
parameter, $DP$, as $DP=\frac{m_l}{m_h}$, where
$m_h$, $m_l$ are the fractional polarizations at high and low frequencies, 
respectively. A higher value of $DP$ implies lower depolarization.
We created $DP$ maps using the lower frequency L-band (IF1) and the C-band 
fractional polarization maps in AIPS using the task COMB. 
We used these maps to obtain a mean lobe $DP$ 
by putting a box around each lobe (using AIPS verbs TVWIN and IMSTAT).

Note that a few sources have $DP>1$. While the degree of polarization will
occasionally increase with wavelength by chance due to Faraday effects, a close
examination of our maps revealed that the high $DP$ values were mostly 
obtained for those sources which showed polarization at 1.4~GHz only in the 
hotspots, where the systematic errors are expected to be the highest 
(e.g., 3C13, 3C54-North, 3C469.1-North, 3C470). Residual instrumental 
polarization errors and errors in the position-angle zero-point contribute 
to the total error in $DP$, especially since data from multiple 
configurations has been combined to get the polarization maps. 
The $DP$ errors quoted in Table~\ref{tablobes} were estimated by
taking the difference in the $DP$ obtained using the L-band IF1 or IF2, and
the C-band, respectively. In our discussions of lobe depolarization we have 
therefore excluded all values of $DP$ that are not estimated over the radio 
lobes. 

The spectral index for 
the lobes in Table~\ref{tablobes} was obtained by putting a box 
around the lobe in the spectral index map and getting a mean value.
The hotspot spectral index was taken as the value in this map corresponding to
the total intensity peak position at the hotspot.
The arm-length ratio ($Q$) is the ratio of the angular extent of the larger 
radio lobe (core to hotspot) to the smaller. Another fundamental structural 
parameter is the axial ratio ($AR$), which is the ratio of the length to 
width of the radio lobe. The axial ratio was estimated for our radio galaxies 
following 
\citet{LeahyWilliams84} - the width was estimated of a section of the radio
lobe roughly half-way between the hotspot and the core and
defined as $2/\sqrt{3}$ times the FWHM of the Gaussian function fitted to the 
lobe surface brightness. We used the 1.4 GHz maps to get the widths and the
8 GHz maps to get accurate lengths.
Table~\ref{tablobes} presents the arm-length and axial ratios for the sample 
FRII radio galaxies.

\begin{deluxetable}{lccllllll}
\tabletypesize{\scriptsize}
\tablecaption{Properties of the Radio Components}
\tablewidth{0pt}
\tablehead{
\colhead{Source} & \colhead{Freq} & \colhead{$S_\nu$ (N lobe)}
& \colhead{$S_\nu$ (S lobe)} & \colhead{$S_\nu$ (Core)} 
& \colhead{$S_8$ (Core)} & \colhead{Log $R_c$} & \colhead{Lobe} &\colhead{Jet-side}\\
\colhead{} & \colhead{(GHz)} & \colhead{(Jy)} & \colhead{(Jy)} 
&\colhead{(mJy)} & \colhead{(mJy)}& \colhead{} & \colhead{Type} &\colhead{N, S} \\
\colhead{(1)}    & \colhead{(2)}  & \colhead{(3)}  & \colhead{(4)}  &
\colhead{(5)}    & \colhead{(6)}  & \colhead{(7)} & \colhead{(8)} & \colhead{(9)}}
\startdata
3C6.1   & 5.0  & 0.58 & 0.51 & ...   &7.52 & -2.90 & LW1 & ...,...\\
	& 1.4  & 1.75 & 1.49 & 10.14 &&& \\
	& 0.33 & ...  & ...  & ...   &&&\\
3C13    & 5.0  & 0.29 & 0.11 & ...   &$<0.29^\dagger$ & -4.19 & ...& ...,...\\
	& 1.4  & 1.25 & 0.61 & ...   &&&\\
	& 0.33 & 4.56 & 3.16 & ...   &&&\\
3C34    & 5.0  & 0.22 & 0.16 & 1.91  &0.69 & -3.47 & LW1 & ...,Jet\\
	& 1.4  & 0.69 & 0.54 & 3.23  &&&\\
	& 0.33 & 2.79 & 2.38 & ...   &&&\\
3C41    & 5.0  & 0.44 & 0.86 & ...   &2.38 & -3.40 & LW3 & ...,Jet$^a$\\
	& 1.4  & 1.29 & 2.05 & ...   &&&\\
	& 0.33 & ...  & ...  & ...   &&&\\
3C44    & 5.0  & 0.12 & 0.21 & 0.21  &0.25 & -3.94 & LW1 & ...,...\\
	& 1.4  & 0.48 & 0.79 & ...   &&&\\
	& 0.33 & 1.58 & 3.11 & ...   &&&\\
3C54    & 5.0  & 0.21 & 0.36 & ...   &0.36 & -3.96 & LW1 & ...,Jet \\
	& 1.4  & 0.65 & 1.11 & ...   &&&\\
	& 0.33 & 2.03 & 3.66 & ...   &&&\\
3C114   & 5.0  & 0.13 & 0.14 & 12.38 &5.41 & -2.53 & LW1 & ...,Jet?\\
	& 1.4  & 0.47 & 0.52 & 13.85 &&&\\
	& 0.33 & 1.81 & 1.96 & ...   &&&\\
3C142.1 & 5.0  & 0.51 & 0.38 & 4.97  &1.17 & -3.57 & LW1 & ...,...\\
	& 1.4  & 1.88 & 1.28 & 7.95  &&&\\
	& 0.33 & 6.89 & 3.97 & ...   &&&\\
3C169.1 & 5.0  & 0.14 & 0.21 & 0.81  &0.75 & -3.40 & LW1  & ...,...\\   
	& 1.4  & 0.44 & 0.68 & 5.13  &&&\\
	& 0.33 & 1.42 & 2.12 & ...   &&&\\
3C172   & 5.0  & 0.39 & 0.51 & ...   &0.29 & -4.18 & LW5  & ...,...\\
	& 1.4  & 1.29 & 1.55 & ...   &&&\\
	& 0.33 & 3.79 & 4.14 & ...   &&&\\
3C441   & 5.0  & 0.55 & 0.23 & ...   &$<0.19^\dagger$ & -4.32 & LW3 & Jet,...\\
	& 1.4  & 1.65 & 0.79 & ...   &&&\\
	& 0.33 & 5.60 & 3.25 & ...   &&&\\
3C469.1 & 5.0  & 0.19 & 0.22 & 4.08  &1.58 & -3.40 & LW1 & ...,Jet?\\
	& 1.4  & 0.85 & 0.84 & 4.77  &&&\\
	& 0.33 & 3.69 & 2.81 & ...   &&&\\
3C470   & 5.0  & 0.07 & 0.47 & 1.85  &0.55 & -3.97 & ... & ...,...\\
	& 1.4  & 0.17 & 1.75 & ...   &&&\\
	& 0.33 & 0.42 & 6.43 & ...   &&&\\
\enddata
\vspace{-0.6cm}

\tablecomments{
Col.1 \& 2: Source name and the observing frequency in GHz, respectively.
Col.3 \& 4: Radio flux density in the northern and
southern radio lobe, respectively, obtained by putting a box around the lobe 
region and using the AIPS verb IMSTAT. 
Col.5: Radio core flux density in mJy. Col.6: 8 GHz core flux density in 
mJy. $\dagger$ For sources where core was not detected, an upper limit
is derived assuming the core flux density to be below
five times the rms noise level. Col.7: Logarithm of the K-corrected radio 
core prominence parameter, 
defined as the ratio of (8~GHz) core to (1.4~GHz) lobe flux density 
(see Sect.~\ref{results}). Col.8: Lobe type (see Sect.~\ref{results}). 
Col.9: The side of the source on which a jet-like feature was observed.
These tend to be faint and only visible in the grey-scale images 
(see Fig.~\ref{figgrey3c44}). Tentative features are marked with ``?''.
$^a$ The jet in 3C41 was noted by 
\citet{Mullin06}. Note that all estimates for 3C34 are for the eastern and 
western lobe, respectively.}
\label{tabcomps}
\end{deluxetable}
\thispagestyle{empty}
\begin{deluxetable}{llccrrrrrrrcr}
\tabletypesize{\scriptsize}
\rotate
\tablecaption{Properties of the Radio Lobes}
\tablewidth{0pt}
\tablehead{
\colhead{Source}&\colhead{$\Theta_{N,~S}$}&\colhead{Q}&\colhead{$\zeta$}
&\colhead{$\langle RM \rangle$} 
& \colhead{$\sigma_{RM}$}&\colhead{$DP$}&\colhead{Error}&
\colhead{$\alpha^{1.4}_{5}$} & \colhead{$\alpha^{0.3}_{1.4}$} & 
\colhead{AR} & \colhead{log($l_{j}$/$l_{cj}$)}\\
\colhead{}&\colhead{($\arcsec$,~$\arcsec$)}&\colhead{}&\colhead{(deg)}
&\colhead{(rad~m$^{-2}$)} &\colhead{(rad~m$^{-2}$)}& \colhead{$N,~S$}&
\colhead{in $DP$} & \colhead{$N,~S$} 
&\colhead{$N,~S$} & \colhead{$N,~S$} & \colhead{}\\
\colhead{(1)} & \colhead{(2)} & \colhead{(3)} & \colhead{(4)} & \colhead{(5)} &
\colhead{(6)} & \colhead{(7)} & \colhead{(8)} & \colhead{(9)} & 
\colhead{(10)} & \colhead{(11)} & \colhead{(12)} } 
\startdata
3C6.1          & 13.89,12.09 &1.15 &2.4 & ...,...       & ...,...   &...,...   &...,...  &1.121,1.123&...,...& 2.74,2.65 & ...\\ 
3C13$^\ast$    & 16.46,12.11 &1.36&3.4 &60.49,...      &28.04,...  &0.73,1.43 &0.37,0.5 &1.278,1.539&0.983,1.121&...,...& ...\\
3C34$^\ddagger$& 23.98,21.10 &1.14 &3.0 &--62.30,--63.16&19.13,7.73 &0.91,0.75 &0.07,0.15&1.281,1.311&0.955,0.942&4.79,3.81 &--0.056\\
3C41           & 12.53,11.32 &1.11 &1.5 & ...,...       & ..., ...   & ...,...  &...,...  &0.996,1.095&...,...&3.21,2.68 & --0.044\\
3C44           & 40.15,24.69 &1.63 &5.9 &4.01,14.60     &11.93,31.23&1.79,1.00 &0.25,0.19&1.294,1.451&0.876,1.172&7.55,5.82 & ...\\
3C54           & 26.59,24.71 &1.07&8.9 &--78.95,--70.15&3.90,14.11 &1.27,1.23 &0.36,0.40&1.093,1.161&0.919,1.030&3.69,4.08& --0.032\\ 
3C114          & 25.43,27.73 &1.09 &0.1 &--2.10,--6.50  &12.70,17.80&0.39,0.59 &0.27,0.01&1.267,1.247&1.031,1.205&4.32,4.68 & 0.037\\
3C142.1        & 18.30,34.14 &1.87 &3.4 &83.02,85.01    &19.03,12.16&0.62,0.94 &0.13,0.17&1.179,1.018&0.886,0.752&1.98,5.20 & ...\\
3C169.1        & 19.19,27.69 &1.44 &14.6&9.22,11.29     &9.63,15.40 &0.99,0.89 &0.07,0.08&1.057,1.105 &0.888,0.876&2.05,3.39 & ...\\
3C172          & 47.33,53.49 &1.13 &7.8 &25.87,1.72     &13.68,34.55&1.09,1.54 &0.07,0.50&0.859,0.772&0.753,0.770&4.29,5.14 & ...\\
3C441$^\ast$   & 9.67,23.66 &2.45&4.8 &75.05,68.23     &9.97,16.23 &0.44,0.76 &0.02,0.35&1.129,1.176&1.002,1.024&1.78,4.41& --0.389\\
3C469.1        & 36.44,38.99 &1.07 &2.1 &19.60,8.99     &14.70,16.34&1.26,1.09 &0.12,0.07&1.280,1.262&1.277,0.994&7.30,6.41 & 0.028\\
3C470          & 14.36,10.54 &1.36 &3.7 &...,--7.68     &...,32.10  &...,1.23  &...,0.5&0.708,1.118  &0.670,0.920&...,... & ...\\
\enddata
\tablecomments{Col.1: Source names. 
Col.2: Angular extent of source from the hotspot to the radio core, 
for the northern and the southern side, respectively, obtained using the high
resolution 8 GHz maps. Sources with an $\ast$ in Col.1 did not show a
radio core in the 8 GHz image - host galaxy positions for 3C13 and 3C441
were obtained from \citet{McCarthy97} and \citet{Fernini97}, respectively.
Col.3: Arm-length ratio, 
derived by taking the ratio of the two angular extents listed in Col.2. Col.4: 
Misalignment angle in degrees, derived by taking the difference between PAs 
obtained for both sides of the source. Col.5: Mean value of rotation measure
($RM$) for the northern and the southern side of the source, respectively. 
Col.6: Dispersion in $RM$ for the radio lobes. 
Col.7: Depolarization parameter for the northern and southern lobes,
respectively. A higher value of $DP$ implies lower lobe depolarization.
Col.8: Error in Depolarization estimated by taking a difference in the
DP obtained by the C-band and the two IFs in the L-band, respectively.
Col.9: Average spectral index between 1.4 and 5~GHz, 
for the northern and southern lobes, respectively. 
Col.10: Average lobe spectral index between 330 MHz and 1.4 GHz.
Col.11: Axial-ratio for 
both the radio lobes. $N,~S$ stand for the northern and southern side of 
source. 
Col.12: Logarithm of the arm-length ratio of the jet to the counterjet
side.
$\ddagger$ For 3C34, all the estimates are for the eastern and western lobe, 
respectively.}
\label{tablobes}
\end{deluxetable}

\begin{deluxetable}{lllllllllllllccccccccccc}
\tabletypesize{\scriptsize}
\tablecaption{Properties of Hotspots}
\tablewidth{0pt}
\tablehead{\colhead{Source}&\colhead{Morph}&\colhead{Comp}&\colhead{Size} 
&\colhead{I$_{peak}^8$} & \colhead{I$_{peak}^{1.4}$}& \colhead{I$_{peak}^{5}$} &
\colhead{$\alpha^5_{1.4}$} & \colhead{$\alpha^{0.3}_{1.4}$}\\
\colhead{}& \colhead{} & \colhead{} 
&\colhead{($\theta_{maj}\arcsec$,$\theta_{min}\arcsec$,PA$\degr$)}
&\colhead{(mJy/bm)}& \colhead{(mJy/bm)}& \colhead{(mJy/bm)} &
\colhead{(2$\arcsec$)}& \colhead{(5$\arcsec$)} \\
\colhead{(1)} & \colhead{(2)} & \colhead{(3)} & \colhead{(4)} & \colhead{(5)} &
\colhead{(6)} & \colhead{(7)} & \colhead{(8)} & \colhead{(9)} }
\startdata
3C6.1 & Multiple  &NH1 &0.39,~0.25,~134.69 & 46.7 & 1217.0 &435.1  &0.85&...\\
               &  &NH2 &0.33,~0.24,~99.64  & 26.4 & ...    &...    &... & ...\\
               &  &SH  &0.49,~0.44,~77.25  & 17.1 & 891.5  &324.9  &0.83&...\\
3C13 & Single     &NH  &0.23,~0.09,~130.63 & 57.2 & 1066.0 &270.1  &1.13&0.86\\
               &  &SH  &0.22,~0.17,~161.55 & 17.8 & 499.3  &91.6   &1.41&1.09\\
3C34 & Multiple   &EH1 &0.56,~0.47,~169.23 & 2.7  & 145.7  &57.8   &0.82&0.74\\
               &  &EH2 &1.24,~0.87,~9.55   & 1.8  & ...    &...    &... &...\\
               &  &WH  &0.39,~0.26,~179.44 & 1.5  & 38.3   &15.0   &0.86&0.70\\
3C41 & Multiple   &NH  &0.46,~0.22,~20.53  & 8.0  & 446.9  &178.8  &0.73&...\\
               &  &SH1 &0.42,~0.32,~120.46 & 33.2 & 981.8  &502.1  &0.55&...\\
               &  &SH2 &0.33,~0.23,~76.19  & 33.3 & ...    &...    &... &...\\
3C44 & Multiple   &NH1 &0.30,~0.22,~68.96  & 4.3  & 198.5  &63.2   &0.89$\ast$&0.91\\
               &  &NH2 &0.45,~0.28,~153.92 & 2.7  & ...    &...    &... & ... \\
               &  &SH  &0.26,~0.17,~44.66  & 22.7 & 429.9  &137.1  &0.89$\ast$&0.81\\
3C54 & Single     &NH  &0.19,~0.16,~31.99  & 7.5  & 289.6  &105.7  &0.79$\ast$&0.74\\
               &  &SH  &0.18,~0.10,~28.75  & 68.9 & 592.6  &228.9  &0.73$\ast$&0.71\\
3C114 & Single    &NH  &0.33,~0.23,~104.83 & 2.2  & 108.5  &37.5   &0.82$\ast$&0.79\\
               &  &SH  &0.30,~0.15,~142.78 & 6.7  & 206.5  &69.7   &0.83$\ast$&0.81\\
3C142.1 & Single  &NH  &0.42,~0.17,~41.45  & 1.8  & 376.8  &123.3  &0.89&0.83\\
               &  &SH  &0.46,~0.25,~67.12  & 1.6  & 183.3  &66.3   &0.77&0.82\\
3C169.1 & Single&NH&..........................& ...& 49.6  &18.2   &0.80$\ast$&0.82\\
        &         &SH  &0.41,~0.23,~50.37  & 1.6  & 153.8  &57.5   &0.77$\ast$&0.77\\
3C172 & Multiple  &NH1 &0.99,~0.76,~174.07 & 3.4  & 394.5  &119.8  &0.93$\ast$&0.73\\ 
               &  &NH2 &0.72,~0.49,~42.54  & 1.9  & ...    &...    &... & ...\\
               &  &NH3 &1.22,~0.42,~173.03 & 1.9  & ...    &...    &... & ...\\
               &  &NH4 &0.79,~0.58,~66.43  & 1.64 & ...    &...    &... & ...\\
               &  &SH  &0.35,~0.23,~124.89 & 11.1 & 486.2  &153.9  &0.89$\ast$&0.71\\
3C441 & Single    &NH  &0.29,~0.18,~98.68  & 31.9 & 424.8  &186.1  &0.64$\ast$&0.84\\
               &  &SH  &0.30,~0.17,~110.01 & 1.6  & 226.8  &68.6   &0.95$\ast$&0.98\\
3C469.1 & Multiple&NH1 &0.44,~0.30,~1.12   & 2.6  & 549.7  &131.5  &1.13$\ast$&0.97\\
        &         &NH2 &1.16,~0.77,~67.51  & 1.4  & ...    &...    &... & ...\\  
        &         &NH3 &0.69,~0.26,~27.40  & 1.8  & ...    &...    &... & ... \\ 
               &  &SH1 &0.80,~0.53,~60.79  & 4.2  & 395.6  &114.8  &0.96$\ast$&0.83\\
        &         &SH2 &0.42,~0.33,~28.95  & 2.2  & ...    &...    &... &... \\
3C470 & Multiple  &NH  &0.30,~0.13,~15.80  & 20.1 & 156.5  &62.5   &0.71&0.59\\
               &  &SH1 &0.22,~0.18,~57.18  & 38.6 & 1201.0 &318.7  &1.02&0.91\\
      &           &SH2 &0.91,~0.43,~33.06  & 17.0 & ...    &...    &... & ... \\
\enddata
\tablecomments
{Hotspot properties derived from the high-resolution 8 GHz images. 
Col.2: Morphology of the hotspot region -- whether single or multiple hotspots.
Col.3: NH, SH, EH and WH denote northern, southern, eastern and 
western hotspots respectively; NH2 is the second brightest hotspot and so on.
Col.4: Major and minor axes of the deconvolved hotspot sizes, estimated using AIPS task JMFIT.
Col.5: Peak surface brightness of hotspot at 8 GHz derived by JMFIT.
Col.6 \& 7: Peak surface brightness of hotspot at 1.4 and 5.0 GHz,
respectively, from the 2.0$\arcsec$ (or 2.5$\arcsec$) total intensity map
(data from both IFs were averaged). 
Col.8, \& 9: Hotspot spectral index between 1.4 and 5 GHz, \& between
$\sim$330 MHz and 1.4 GHz, respectively. $\ast$= Maps at 2.5$\arcsec$ were used.}
\label{tabspots}
\end{deluxetable}
\section{Radio Galaxy Properties and Correlations}
\label{corr}

In the following sections we describe some global properties of the 13 FRII
galaxies. Further, we have revisited some well-known correlations 
for FRII radio galaxies using the ``combined'' dataset comprising our sample 
galaxies and other FRII radio galaxies taken from the literature.

\subsection{Data from the Literature}
In order to understand the properties of FRII radio galaxies better,
we augmented our data with similar radio galaxy data from the
literature. The additional data comes primarily from 
\citet{LeahyWilliams84,LeahyMuxlow89,Pedelty89,GarringtonConway91,
LiuPooleyA91,Hardcastle98} and \citet{Goodlet04} (see Table~\ref{tablitt}).
We emphasize here that the data from the literature is inhomogeneous and chosen
on the basis of the availabilty of the relevant physical parameters
that we wished to examine in samples of (primarily 3C) FRII radio galaxies 
observed at a resolution of a few arcseconds. We have excluded
radio-loud quasars from our analysis, and restricted our selection to 
powerful FRII radio galaxies at a redshift-range matching our sample.
The combined dataset includes a few broad-line radio galaxies.

\begin{deluxetable}{lll|lll}
\tabletypesize{\scriptsize}
\tablecaption{Data from the Literature}
\tablewidth{0pt}
\tablehead{\colhead{Source}&\colhead{z}&\colhead{Ref}&\colhead{Source}&
\colhead{z}&\colhead{Ref}}
\startdata
3C16 & 0.405 &\citet{Goodlet04}	&	                3C268.1 & 0.970 &\citet{Goodlet04}	\\    			      
3C20   & 0.174 & \citet{Hardcastle98}	        &       3C274.1 & 0.422 & \citet{LeahyWilliams84}\\      		  	  
3C27 & 0.184 & \citet{LeahyMuxlow89}	&               3C277.2 & 0.766 &\citet{Pedelty89}	\\    			  
3C42 & 0.395 &\citet{Goodlet04}	& 	                3C280 & 0.996 & \citet{Goodlet04}\\	    			    	  
3C46 & 0.437 &\citet{Goodlet04}	&	                3C284 & 0.239 & \citet{LeahyWilliams84}\\	     	  
3C52 & 0.285 & \citet{LeahyWilliams84} &	     	3C285 & 0.079 & \citet{LeahyWilliams84}\\	     		  	  
3C55 & 0.734 &\citet{LeahyMuxlow89}	&   		3C288  & 0.246 &\citet{LiuPooleyA91}\\	    			  	  
3C65 & 1.176 &\citet{Goodlet04}	&	    		3C289 & 0.967 &\citet{LiuPooleyA91}\\	    			  	  
3C68.2 & 1.575 &\citet{LeahyMuxlow89}	&   		3C294 & 1.779 &\citet{LiuPooleyA91}\\	    			  
3C79   & 0.2559& \citet{Hardcastle98}	  &  		3C299 & 0.367 &\citet{Goodlet04}\\	    			  	  
3C98   & 0.0306& \citet{Hardcastle98}	  &3C300 & 0.270 & \citet{LeahyWilliams84} \\		  	  
3C103 & 0.330 & \citet{LeahyWilliams84} &	     	3C322 & 1.681 &\citet{LeahyMuxlow89}	 \\  			  
3C105  & 0.089 & \citet{Hardcastle98}	  &  		3C324 & 1.206 &\citet{Goodlet04}	\\    			  	  
3C111 & 0.048 & \citet{LeahyWilliams84} &	     	3C327  & 0.1039& \citet{Hardcastle98}	\\   			  	  
3C123  & 0.2177& \citet{Hardcastle98}	  &  		3C330 & 0.550 &\citet{LeahyMuxlow89}	 \\  			  
3C132  & 0.214 & \citet{Hardcastle98}	  &  		3C337 & 0.635 &\citet{Pedelty89}\\	    			  	  
3C135  & 0.1273& \citet{Hardcastle98}	  &  		3C341 & 0.448 &	\citet{Goodlet04} \\    	    			  	  
3C136.1 & 0.064 & \citet{LeahyWilliams84} &3C349  & 0.205 & \citet{Hardcastle98}	  \\ 			  	   
3C139.2 & ... & \citet{LeahyWilliams84} &	     	3C352 & 0.806 & \citet{GarringtonConway91} \\			  	  
3C153  & 0.2771& \citet{Hardcastle98}	   & 		3C353  & 0.0304& \citet{Hardcastle98}	  \\ 			  
3C165 & 0.295 & \citet{LeahyWilliams84} &	     	3C356 & 1.079 &\citet{Pedelty89}\\	    			  
3C166 & 0.244 & \citet{LeahyWilliams84} &	     	3C368 & 1.131 &\citet{Pedelty89}\\	    			  	  
3C171  & 0.2384& \citet{Hardcastle98}	   & 		3C401  & 0.201 & \citet{Hardcastle98}	  \\ 			  	  
3C173.1& 0.292 & \citet{Hardcastle98}	   & 		3C403  & 0.059 & \citet{Hardcastle98}	  \\ 			  
3C184.1& 0.1182& \citet{Hardcastle98}	   & 		3C405 & 0.056 & \citet{LeahyWilliams84}\\	     		  	  
3C192  & 0.0598& \citet{Hardcastle98}	   & 		3C424  & 0.127 & \citet{Hardcastle98}	  \\ 			  	  
3C197.1& 0.1301& \citet{Hardcastle98}	   & 		3C436  & 0.2145& \citet{Hardcastle98}	  \\ 			  	  
3C200 & 0.458 &\citet{GarringtonConway91}&  		3C438	& 0.290 & \citet{Hardcastle98}  \\   			  	  
3C217 & 0.897 & \citet{Pedelty89}&	    		3C452  & 0.0811& \citet{Hardcastle98}     \\ 			  
3C223  & 0.1368& \citet{Hardcastle98}	   & 		3C457 & 0.428 &\citet{Goodlet04}	  \\  			  	  
3C223.1& 0.1075& \citet{Hardcastle98}	   & 		4C14.11 & 0.206 & \citet{Hardcastle98}	  \\ 			  	  
3C234 & 0.184 & \citet{LeahyWilliams84} &	4C14.27 & 0.392 &\citet{Goodlet04}	\\    			  	  
3C239 & 1.781 & \citet{LiuPooleyA91}&	    		4C53.16 & 0.064 & \citet{GarringtonConway91}\\			  	  
3C244 & ... &  \citet{LeahyWilliams84}	 &   		4C74.16 & 0.81 &\citet{GarringtonConway91} \\			  	  
3C247 & 0.748 &\citet{LiuPooleyA91}&	    		6C0943+39 & 1.040 & \citet{Goodlet04}	\\    			  	  
3C252 & 1.100 &\citet{Goodlet04}&	    		6C1011+36 & 1.040 & \citet{Goodlet04}   \\   			  	  
3C263.1 & 0.824 &\citet{LiuPooleyA91}&	    		6C1129+37 & 1.060 & \citet{Goodlet04}	\\    			  	  
3C265 & 0.811 &\citet{Goodlet04}&	    		6C1256+36 & 1.127 & \citet{Goodlet04}	\\    			      
3C266 & 1.275 &\citet{LiuPooleyA91}&	    		6C1257+36 & 1.000 &\citet{Goodlet04} \\                            
3C267 & 1.140 &\citet{Goodlet04}&	   & & \\ 
\enddata
\tablecomments
{Col.1. Source name. Col.2. Redshift. Col.3. Reference.}
\label{tablitt}
\end{deluxetable}

When not listed explicitly in the above-mentioned references, the arm-length 
ratios and misalignment angles were directly estimated (using a ruler and
protractor) from the maps presented in these papers. Therefore, the angle 
estimates have errors of the order of a few degrees, while the error in 
the arm-length ratios is expected to be less than 10\%.
When no radio core was observed in the maps, the position of the centre of
the host galaxy was used for the core position. The host galaxy positions
were procured either from the papers directly or through the NASA/IPAC 
Extragalactic Database (NED)\footnote{http://nedwww.ipac.caltech.edu/}. 
For the 13 FRIIs, radio core prominence was derived 
using the 8~GHz core and 1.4~GHz lobe flux densities. If similar data were
not available in the literature, we converted flux densities from other
frequencies to 1.4~GHz for lobe and 8~GHz for core emission by assuming a
spectral index of 1 and 0 respectively.

\subsection{Hotspot Sizes}
\begin{figure}[h]
\centering{
\includegraphics[width=8.1cm]{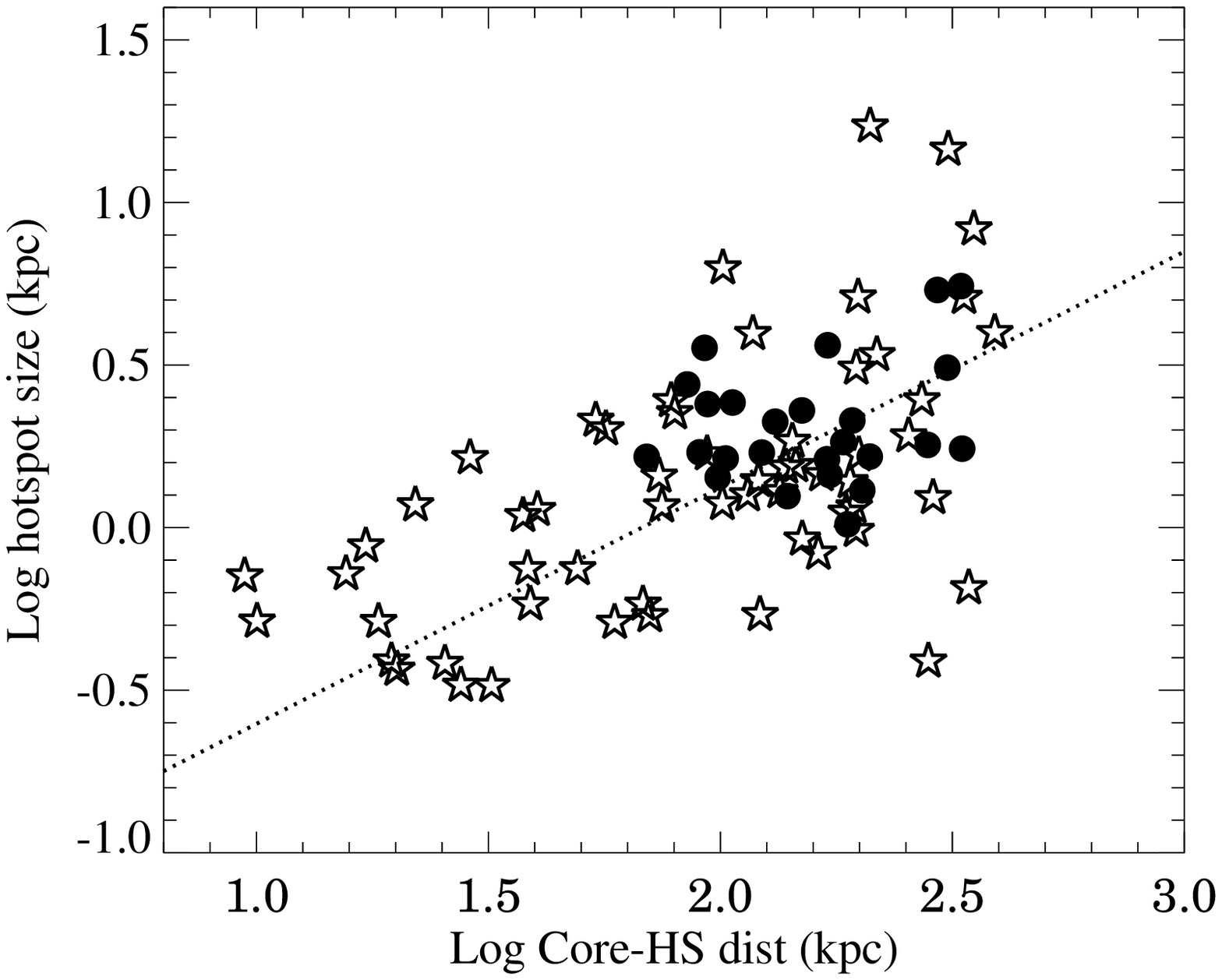}
\includegraphics[width=8.1cm]{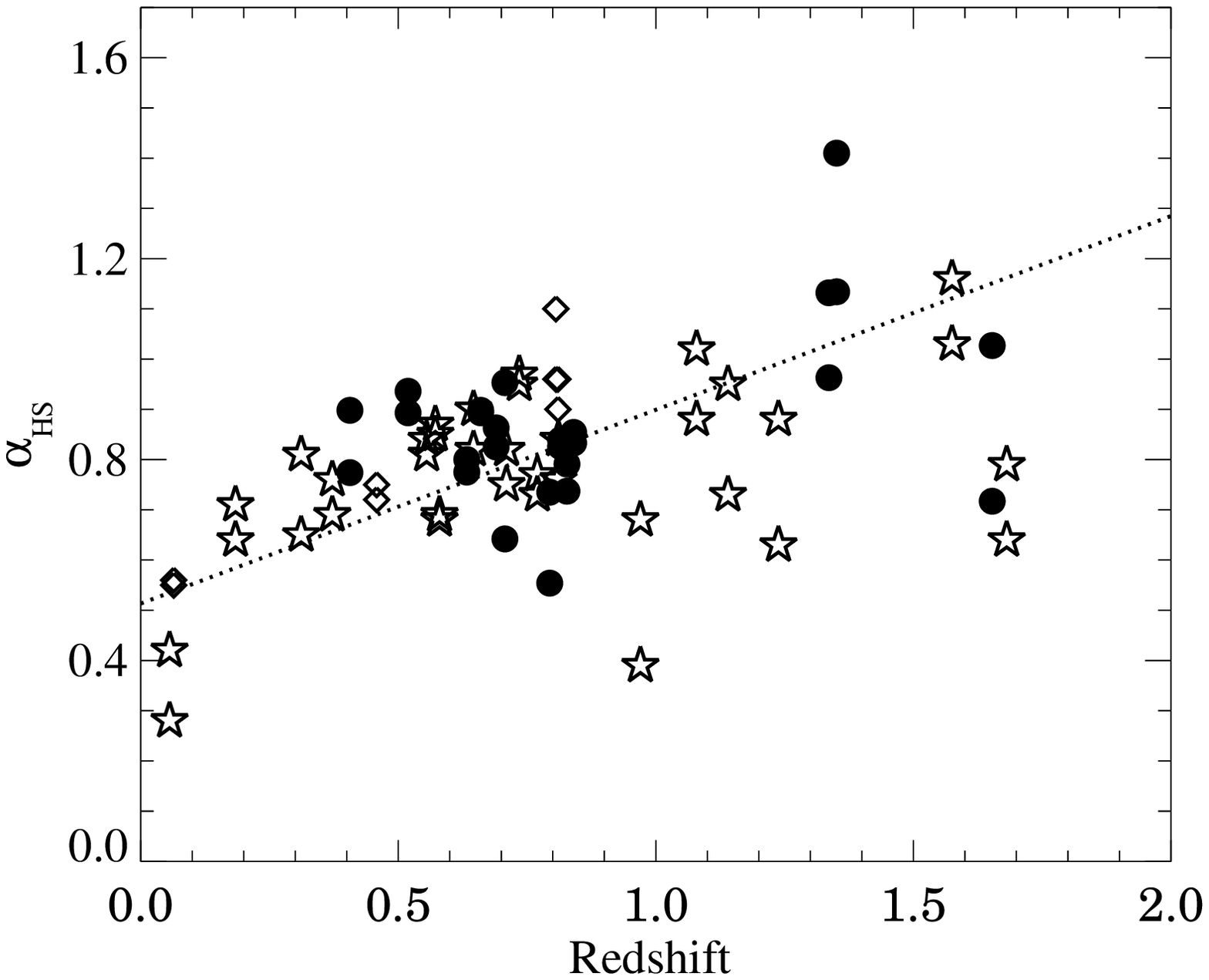}}
\caption{(Left) Log-log plot of the hotspot size versus the core-hotspot
distance of the sources. Filled circles denote our sample FRII
sources; open stars denote radio galaxies from \citet{Hardcastle98}.
The slope of the correlation is $\sim$0.7.
(Right) The hotspot spectral index versus
redshift. The filled circles denote the hotspots for our 13
FRII sources, open stars are hotspot values from \citet{LeahyMuxlow89} while
diamonds are from \citet{GarringtonConway91}. The slope of the
correlation is $\sim$0.4.}
\label{fighssize}
\end{figure}

A strong correlation has previously been noticed between linear sizes of 
FRII radio galaxies and their hotspot sizes \citep[e.g.,][]{Hardcastle98}. 
We find that our 13 FRII radio galaxies follow the same trend. 
In Fig.~\ref{fighssize} we have plotted the hotspot size (of the brighest
hotspot at each end) against the core-hotspot distance. Here we have 
considered only the 8~GHz measurements for the hotspot sizes from 
\citet{Hardcastle98} and excluded the broad-line radio galaxies from the 
analysis. The size of an individual hotspot was defined as the geometric mean 
of the largest and smallest angular sizes (see Table~\ref{tabspots}). 

Using the least absolute deviation method (implemented in the LADFIT routine 
in IDL), we fitted a linear model in the log-log space to the combined 
dataset and obtained a slope of 0.72$\pm$0.06. Note that by fitting a linear 
model to the combined dataset using chi-square minimization (implemented in 
the LINFIT routine in IDL), we obtain a slope of 0.52$\pm$0.08 
(\citet{Perucho03} have found the slope of the radio galaxy correlation to 
be 0.40$\pm$0.11). However since the use of the chi-square error statistic can 
result in a poor fit due to an undesired sensitivity to outlying data,
we consider the slope of the correlation to be close to 0.7. 
Thus the following relation holds for the hotspot-size $r_h$ and core-hotspot
distance $l$ of the radio galaxy: $r_h\propto l^{0.7}$.

This correlation provides support for suggestions that the hotspot sizes 
scale with source linear size \citep[see][]{Hardcastle98,Laing89,Bridle94}. 
This is consistent with the hotspot maintaining ram pressure balance as
the source propagates through a medium with declining ambient density
\citep{Carvalho02b}. \citet{Carvalho02a,Carvalho02b} discuss the results of 
two-dimensional axisymmetric numerical simulations of light, supersonic jets 
propagating into different density atmospheres and compare them with the 
predictions of three types of self-similar models. Of the self-similar models 
discussed, type I corresponds to the jet moving with a constant speed
in a constant density atmosphere \citep[e.g.,][]{BegelmanCioffi89,Daly90}; 
the type II model corresponds to jet motion in a medium where the ambient 
density falls off as $d^{-2}$ \citep[e.g.,][]{Daly90}, where $d$ is the
distance from the source; while the type III model assumes a power-law density 
distribution with an exponent $\delta$ \citep[e.g.,][]{KaiserAlexander97}.

Following the relations between hotspot size, expansion time and the 
core-hotspot distance
for type III models in \citet{Carvalho02a}, $i.e.,$
$r_h^2\propto t^{(\delta+4)/(5-\delta)}$ and $l\propto t^{3/(5-\delta)}$,
the derived exponent of 0.7 for the $r_h - l$ relation
translates to a $\delta$ of 0.2. This implies that the jet propagates
in a nearly constant ambient density medium ($\rho_a\propto d^{-0.2}$). 
This inference is consistent with the sources being located in the 
cores of protoclusters or clusters of galaxies, which have a roughly 
constant ambient gas density. Indeed, a couple of our 13 FRII radio 
galaxies seem to reside in such environments (see notes on
individual sources in the Appendix). This is supported by the previous studies 
of Cygnus~A \citep{Carilli91}, 3C295 \citep{PerleyTaylor91}, and the 
study of \citet{WellmanDaly97}. Finally, with a $\delta$ of 0.2, 
the hotspot advance speed is proportional to $t^{-0.4}$, thereby implying a 
decelerating source.

\subsection{Spectral index vs. Redshift for Hotspots and Lobes}

The hotspot spectral index ($\alpha_{HS}$) is found to strongly correlate with 
redshift ($z$) -- the spectra become steeper at higher redshifts 
(Fig.~\ref{fighssize}). This is consistent with the findings of
\citet{WellmanDaly97,Dennett-Thorpe99,Blundell99} and \citet{Ishwara-Chandra00}.
We find that the following approximate relation holds 
-- $\alpha_{HS} \propto z^{0.4}$, with the absolute deviation in the 
exponent being $\sim$0.03. Note that the $\alpha_{HS}$ for our sources
and from \citet{GarringtonConway91} have been estimated between
1.4 and 5 GHz, while the spectral index from \citet{LeahyMuxlow89} were derived 
between 151 MHz and 1.4 GHz. Excluding the data from \citet{LeahyMuxlow89}
yields a slope of 0.43$\pm$0.04. The chi-square minimization method 
(LINFIT in IDL) results in a slope of 0.22$\pm$0.05 for all the data,
and 0.26$\pm$0.07 when the data from \citet{LeahyMuxlow89} is excluded. 

\begin{figure}[h]
\centering{
\includegraphics[width=8.0cm]{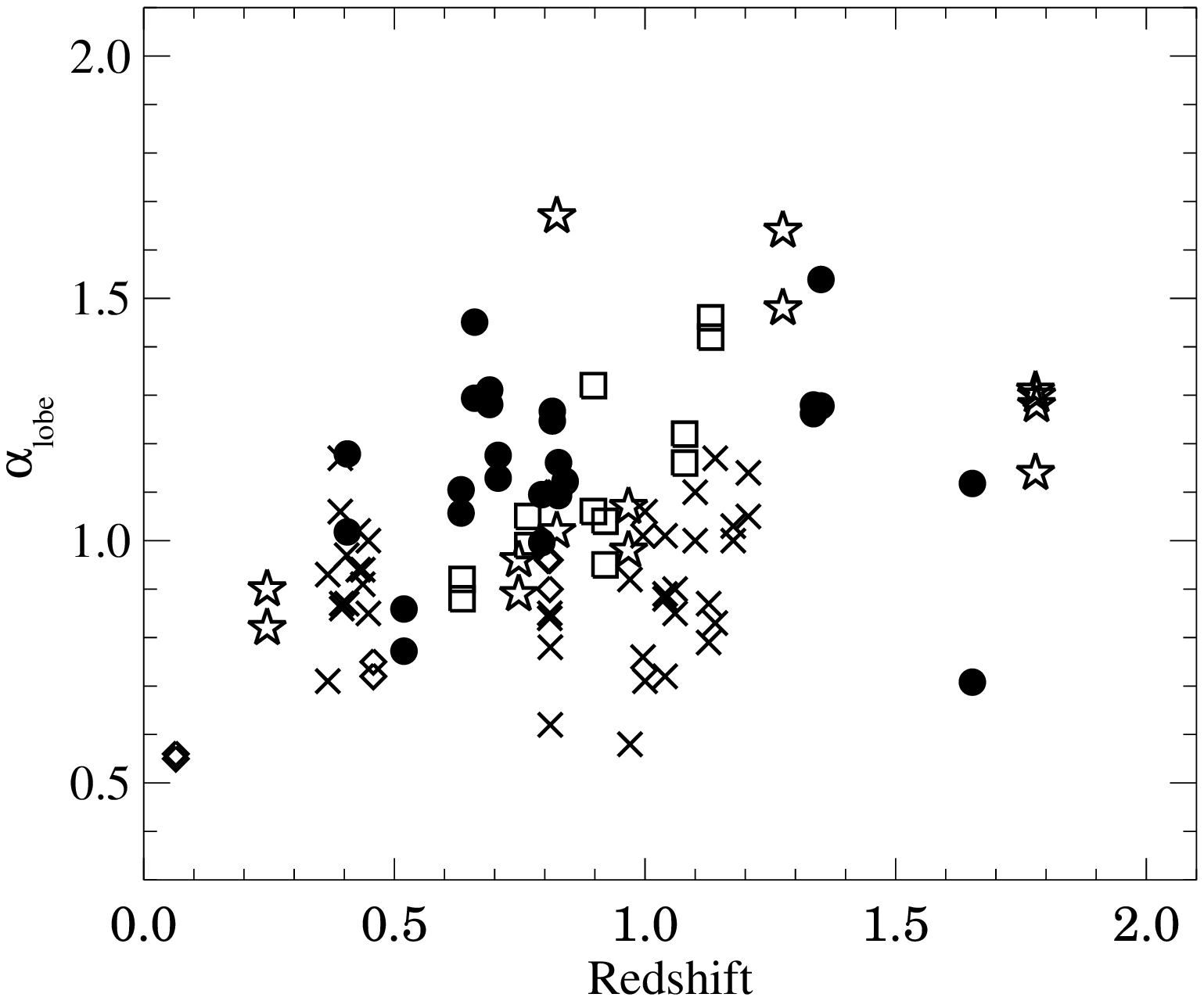}
\includegraphics[width=8.0cm]{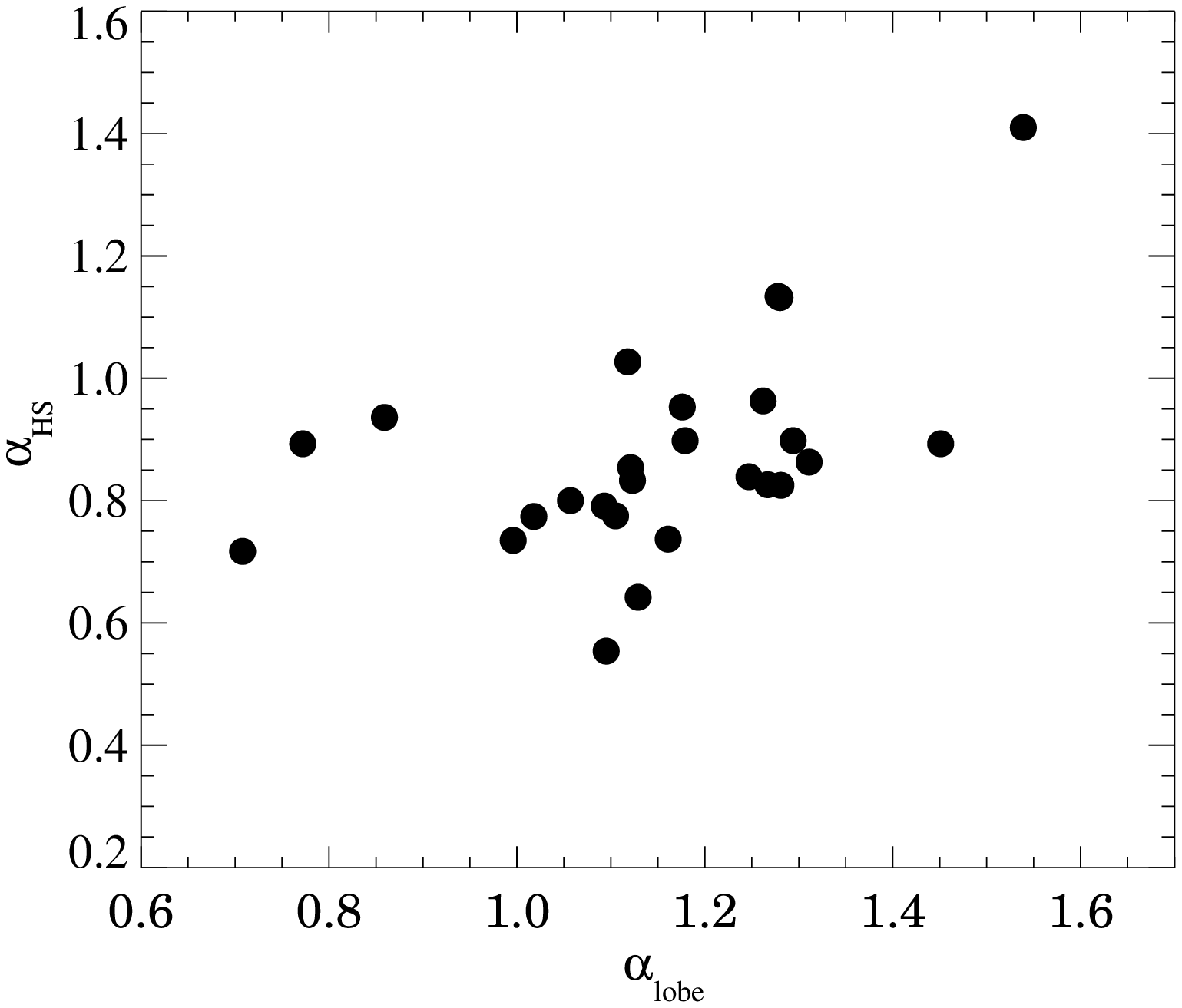}}
\caption{(Left) The lobe spectral index versus redshift. 
The filled circles denote our 13 radio galaxies,
stars are galaxies from \citet{LiuPooleyA91}, squares from \citet{Pedelty89},
crosses from \citet{Goodlet04}, diamonds from \citet{GarringtonConway91}.
(Right) Hotspot 
spectral index versus lobe spectral index for our 13 FRII radio galaxies.}
\label{figspvssp}
\end{figure}

A correlation between spectral index and redshift is already well 
established for the integrated extended emission in FRII 
galaxies \citep[e.g.,][]{LaingPeacock80,Chambers90,Athreya99,DeBreuck00}. 
In Fig.~\ref{figspvssp}, we have plotted the $z-\alpha$ correlation
for our combined radio galaxy dataset. We do not find a correlation between 
the spectral index {\it difference} between the lobes and redshift,
consistent with \citet{Goodlet05}.

It has been pointed out that although a simple redshifting of the 
curved spectrum contributes to the $z-\alpha$ relation 
\citep[e.g.,][]{Gopal-Krishna88,Lacy93}, it cannot account 
for the whole effect \citep[e.g.,][]{Lacy93,Blundell99}.
It has been proposed that the hotspot spectral
index is affected by inverse-Compton cooling owing to scattering  with
microwave background photons, which is a strong function of redshift
\citep[e.g.,][]{ReesSetti68,KrolikChen91,WellmanDaly97}. It has also been 
suggested that enhanced synchrotron losses in more powerful 
hotspots could also produce a spectral steepening at higher luminosity
\citep[e.g.,][]{Blundell99,Dennett-Thorpe99}.
\citet{Blundell99} had examined three complete samples with successively
fainter flux-limits, in order to break the luminosity-redshift degeneracy.
They reached the conclusion that the
rest-frame spectral indices have a stronger dependence on luminosity 
than on redshift, except at GHz frequencies.
Based on the finding that the extremely steep-spectrum radio galaxies
in the local universe inevitably reside at the centres of rich galaxy clusters
\citep[e.g.,][]{BaldwinScott73}, \citet{Klamer06} have postulated that 
$z-\alpha$ correlation could be a result of a higher fraction of radio galaxies
being located, as a function of redshift,
in environments with densities similar to nearby rich clusters.

The hotspot spectral index
correlates with the lobe spectral index (Fig.~\ref{figspvssp} Right). 
Further, the lobe spectral index is systematically steeper than the
hotspot spectral index. These two results provide support to the spectral 
aging model in the radio lobes. 
Electrons are injected into the lobe with the spectral index of the hotspots. 
They subsequently age, making the spectra steeper.

\subsection{The Rotation Measures and Dispersion}

The mean rotation measure ($RM$) across the lobes of 11 of our 13 FRIIs 
varies from $\sim10-80$~rad~m$^{-2}$.
On comparison with integrated $RM$ measurements of \citet{SimardNormandin81}, 
we find that our observed $RM$ values are comparable to that expected from 
our Galaxy. The error in the rotation measure is typically less than 
5~rad~m$^{-2}$. The rotation measure dispersion ($\sigma_{RM}$) is typically
less than $20$~rad~m$^{-2}$ for our sources. 
For two of the sources with $\sigma_{RM}$ greater than $30$~rad~m$^{-2}$,
$viz.,$ 3C13 and 3C470, large systematic errors were probably present in 
their bright hotspots.

\begin{figure}[h]
\centering{
\includegraphics[width=8.0cm]{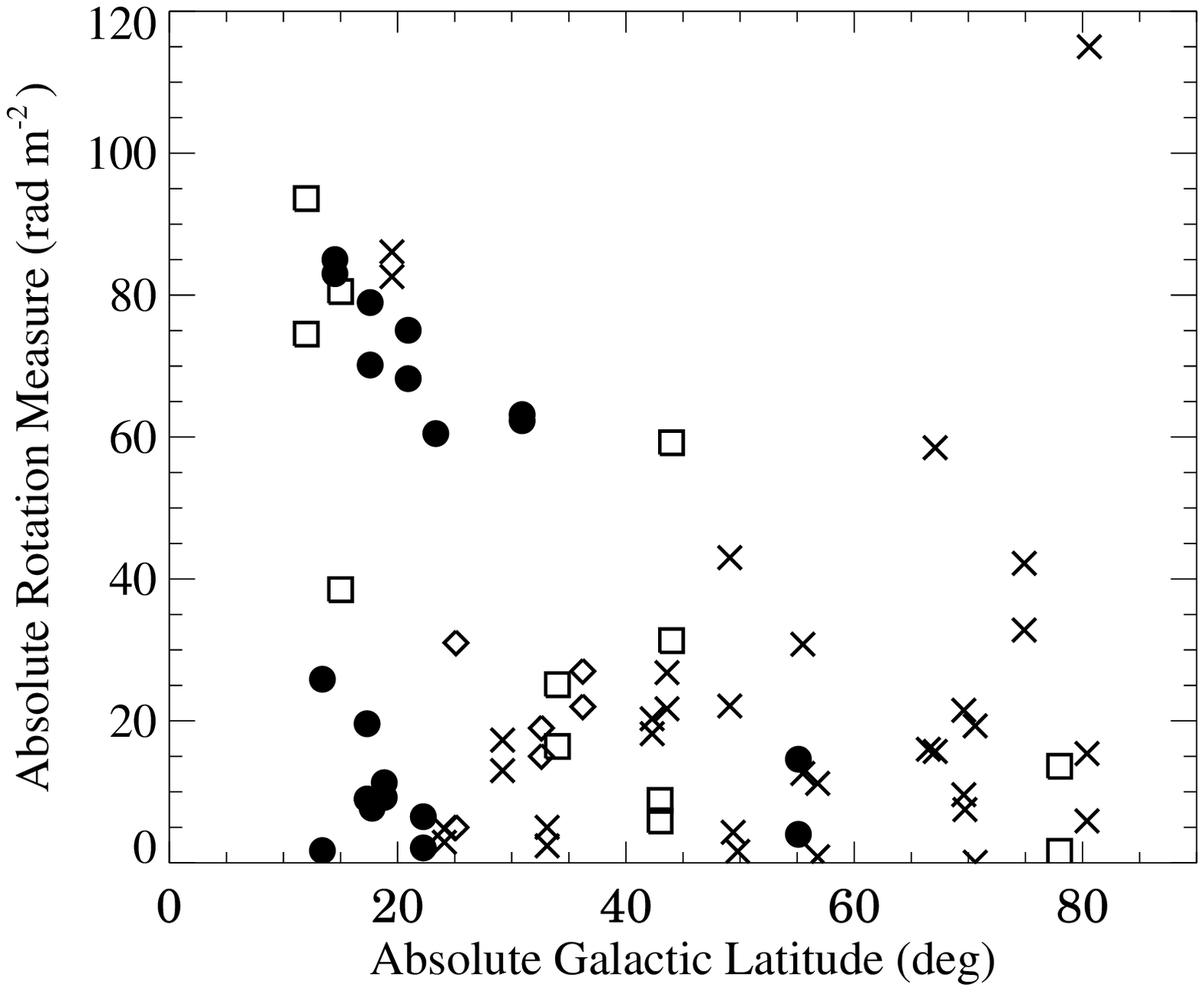}
\includegraphics[width=8.0cm]{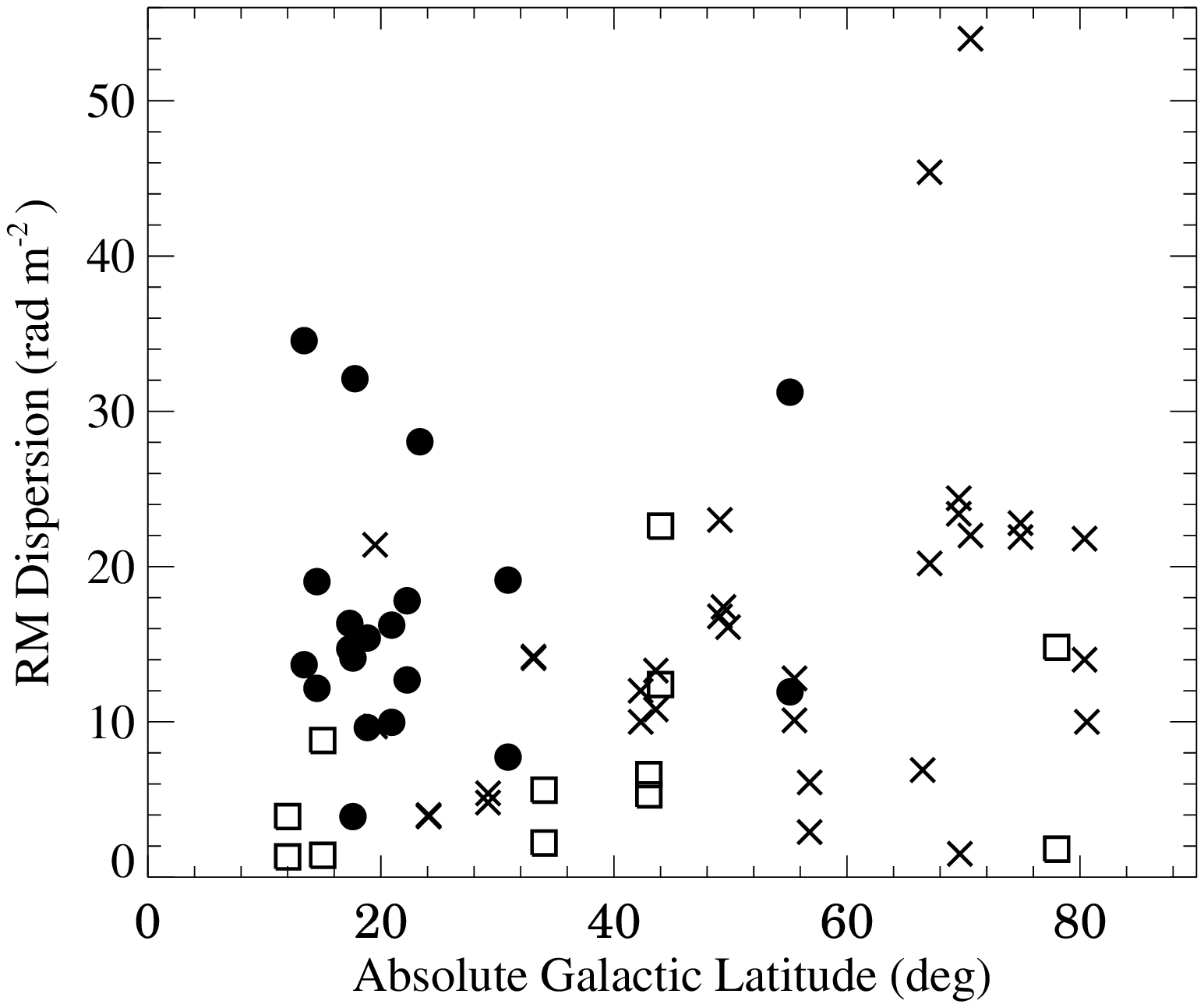}}
\caption{(Left) The absolute mean rotation measure for each of the two radio
lobes versus the absolute Galactic latitude. (Right) The rotation measure
dispersion in the two lobes versus the absolute Galactic latitude.
Filled circles denote our sample FRII radio galaxies, stars and crosses
denote radio galaxies from \citet{LiuPooleyA91} and \citet{Goodlet04},
respectively, while squares and diamonds are galaxies from
\citet{Pedelty89} and \citet{GarringtonConway91}, respectively.}
\label{figrmgal}
\end{figure}

\citet{Pedelty89} had found a weak correlation between the rotation 
measures across the radio lobes and source Galactic latitudes, with the 
higher values of rotation measure occuring at lower Galactic latitudes. 
We find that such a correlation is significant at the 99\% level
(Spearman-Rank and Kendall's-tau correlation tests)
for the combined radio galaxy dataset (Fig.~\ref{figrmgal}). 
The rotation measure dispersion and Galactic latitude are
not correlated (see Table~\ref{tabcorrel}).
These results suggest that the large-scale rotation measure is likely
dominated by the Galactic foreground, whereas the small-scale structure
in rotation measure ($\sigma_{RM}$) is produced in the source and/or its 
environment for these relatively high latitude sources
\citep[eg.,][]{Kronberg72,Simonetti84,Leahy87}.

\begin{figure}[h]
\centering{
\includegraphics[width=8.7cm]{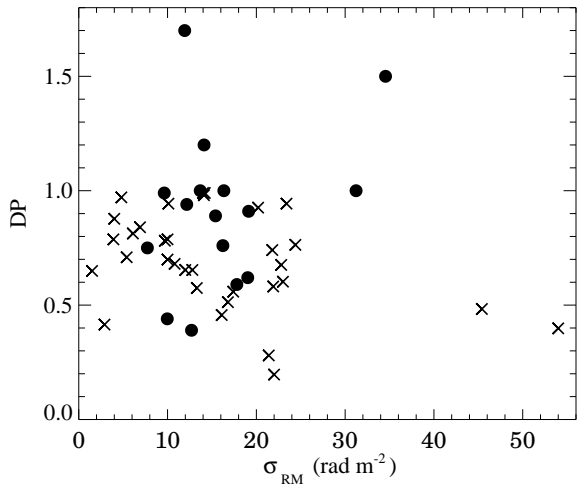}}
\caption{The two depolarization parameters $DP$ and $\sigma_{RM}$
plotted against each other. Filled circles denote our sample FRII radio 
galaxies while crosses denote radio galaxies from \citet{Goodlet04}.}
\label{figdpmsig}
\end{figure}

Figure~\ref{figdpmsig} shows the rotation measure dispersion plotted against 
the depolarization measure. We find no correlation between these two 
parameters (see Table~\ref{tabcorrel}). 
The lack of a correlation is consistent with the possibility that both the 
RM dispersion and depolarization result from a partially resolved 
foreground screen, associated with the radio source but in front of it
\citep{Tribble91,Johnson95,Goodlet05}. However this cannot be 
demonstrated unequivocally for these sources.

\subsection{The Depolarization Asymmetry in Radio Lobes}
Several strong correlations have been observed with lobe
depolarization asymmetries in FRII radio galaxies. In sources with strong
one-sided jets, the jet side tends to depolarize less rapidly with increasing
wavelength, also referred to as the 
`Laing-Garrington' effect \citep{Laing88,Garrington88}. Further, the more depolarized lobe has a steeper 
spectrum (the `Liu-Pooley' effect) and smaller angular distance between the 
core and the hotspot \citep{Laing88,LiuPooleyB91,LiuPooleyA91}. 
Also, an asymmetry exists in the distribution of the emission-line gas in 
radio galaxies, with more emission-line gas lying on that side of the radio 
source for which the depolarization is stronger \citep{McCarthy89}. 

Most of the 13 FRII radio galaxies have an arm-length ratio close to unity 
(Table~\ref{tablobes}). The misalignment angles are typically 
less than 8 degrees. Further, the radio core prominence in these 13 FRIIs is 
small, $-4.3\leqslant$ Log~$R_c$~$\leqslant-2.5$ (see Table~\ref{tabcomps}). 
The small misalignment angles and radio core prominence suggest that our 
13 FRII galaxies are largely plane-of-sky objects. This is consistent with the 
absence of bright radio jets in these radio 
galaxies. This makes them useful for a spectral aging study (Paper II). 

\subsubsection{The Laing-Garrington effect}

\begin{figure}[h]
\centering{
\includegraphics[width=8.0cm]{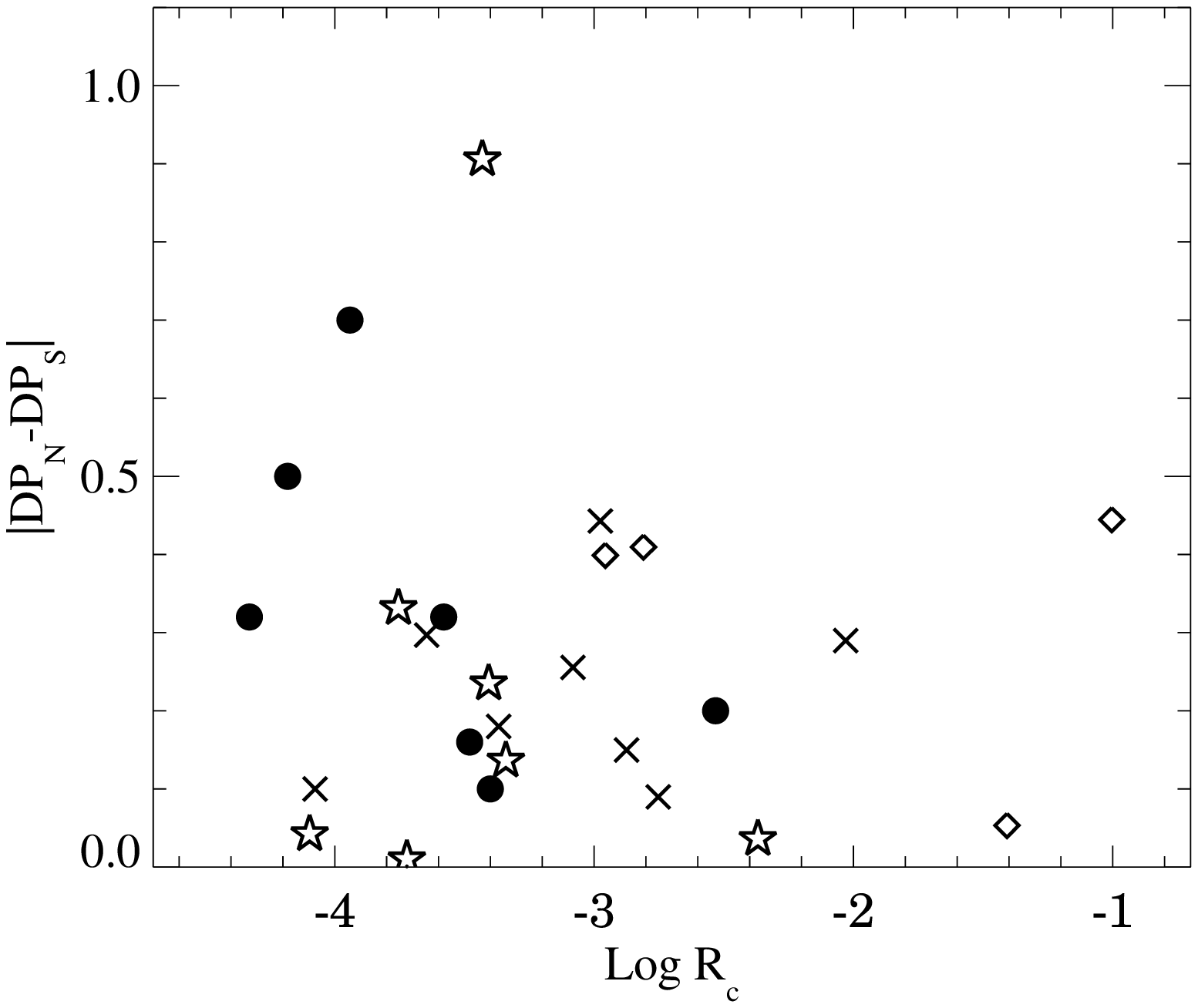}
\includegraphics[width=8.0cm]{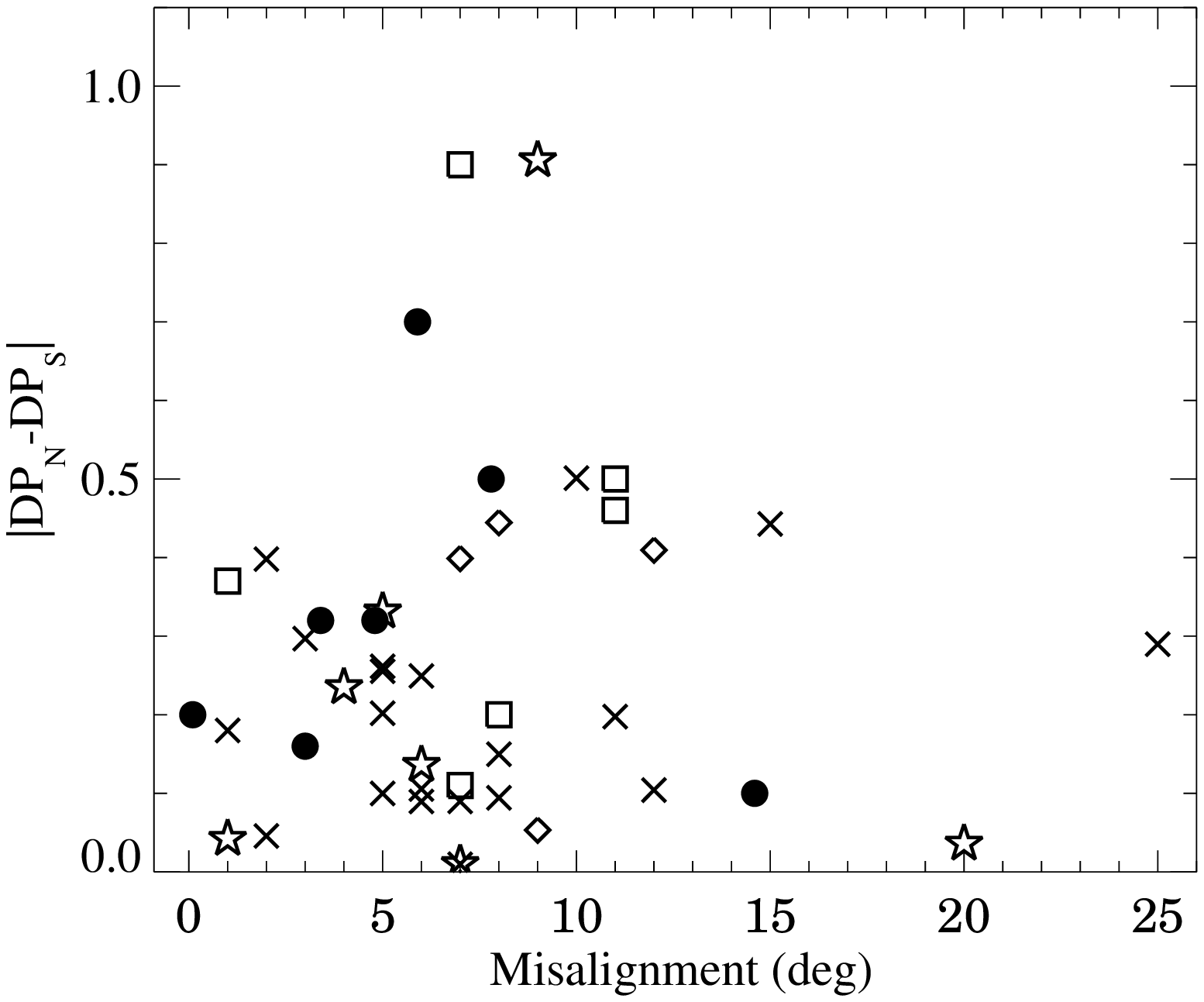}}
\caption{The absolute North-to-South lobe difference in depolarization 
versus the (Left) radio core prominence and (Right) misalignment angle.
Filled circles denote our sample FRII radio galaxies, stars and crosses 
denote radio galaxies from \citet{LiuPooleyA91} and \citet{Goodlet04},
respectively, while squares and diamonds are galaxies from
\citet{Pedelty89} and \citet{GarringtonConway91}, respectively.}
\label{figdpmmis}
\end{figure}

The jet-sidedness $-$ depolarization relation cannot be unambiguously examined 
for our 6 FRII radio galaxies with mostly tentative jet detections.
As the Laing-Garrington effect is suggested to be a consequence of 
Doppler-beaming, we examined the difference in lobe depolarization with 
respect to the (statistical) orientation-indicators -- the radio core 
prominence and the misalignment angle, for the larger combined dataset
(Fig.~\ref{figdpmmis}). We find that the lobe-to-lobe differences in 
depolarization are not correlated with either the radio core prominence
or the misalignment angle. The lack of a correlation is consistent with the 
picture of these radio galaxies lying close to the plane of the sky. 
However it must be noted that \citet{GarringtonConway91} failed to see a
correlation between depolarization asymmetry ratio and core strength, even in 
quasars. This is probably due to the large scatter in core prominence
relation and the fact that this test is intrinsically less sensitive than
one which compares quantities pertaining to two lobes. The lack of any 
correlation between the sidedness of faint jets with depolarization in a 
radio galaxy sample with weak jets has been noted by \citet{Laing93}.

\subsubsection{The Liu-Pooley effect}
\begin{figure}[h]
\centering{
\includegraphics[width=8.0cm]{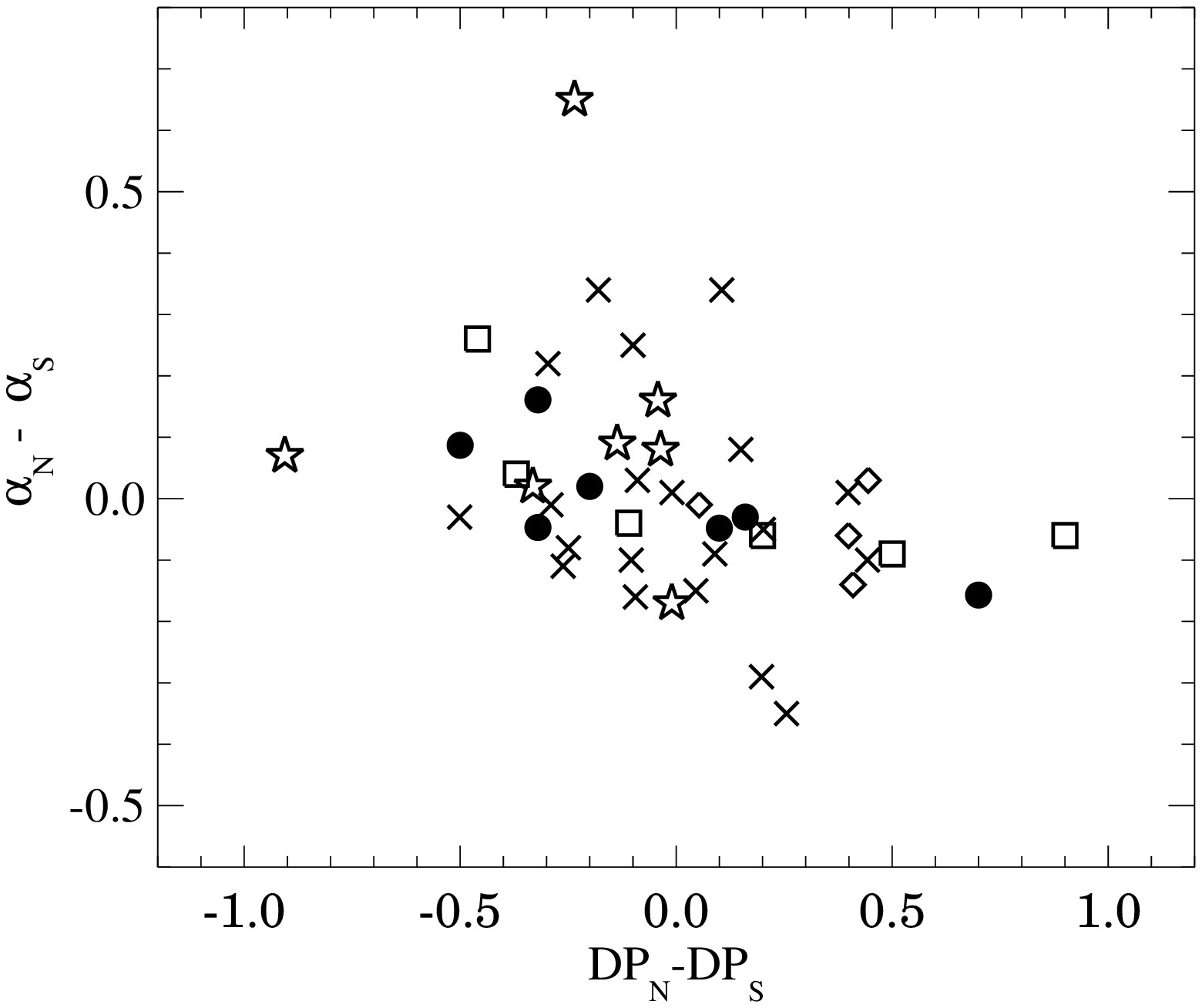}
\includegraphics[width=8.0cm]{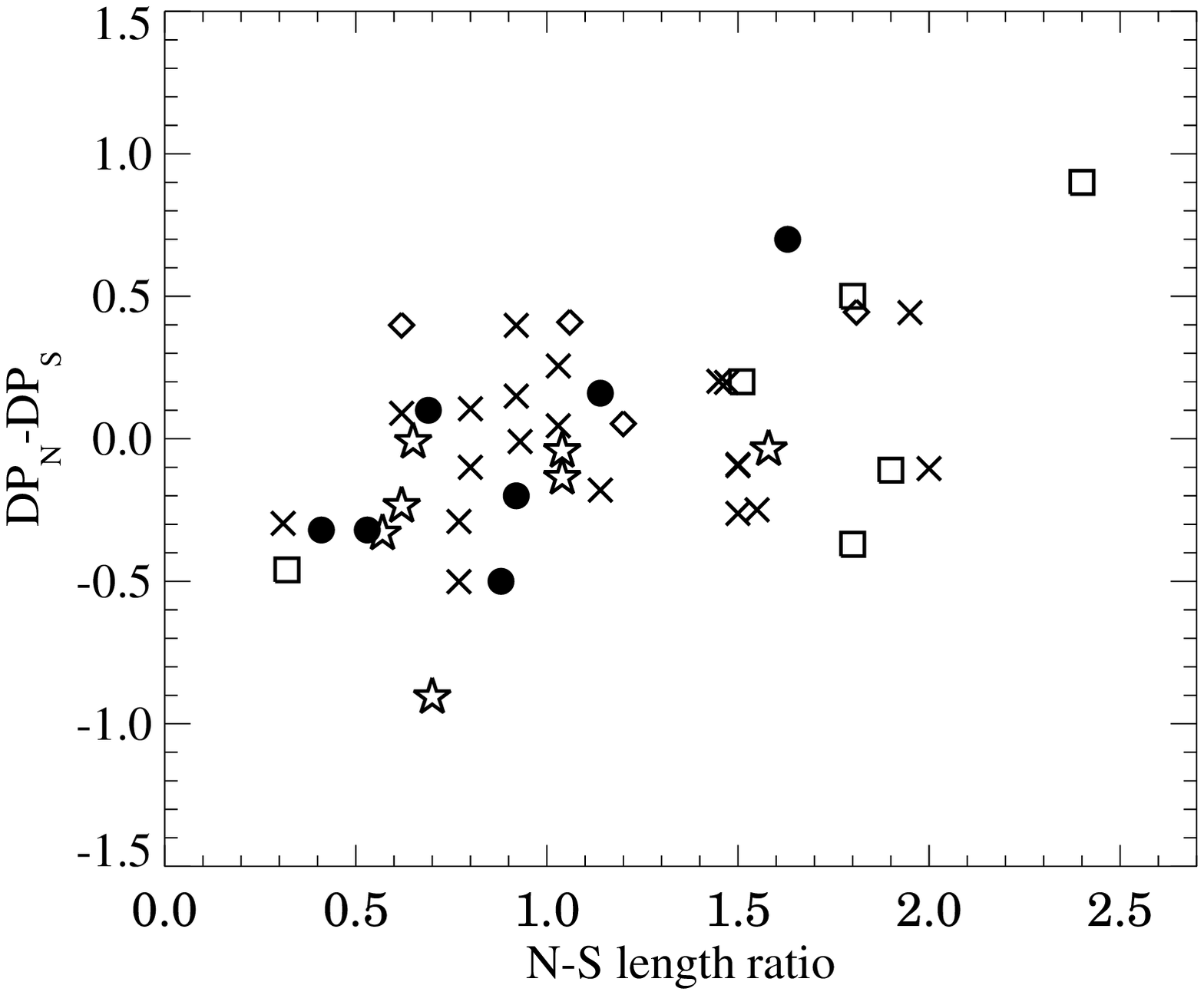}}
\caption{(Left) The North-to-South lobe difference in the spectral index
versus the depolarization parameter. 
(Right) The North-to-South lobe difference in depolarization versus 
the North-to-South arm-length ratio.
Filled circles denote our sample FRII radio galaxies, stars and crosses
denote radio galaxies from \citet{LiuPooleyA91} and \citet{Goodlet04},
respectively, while squares and diamonds are galaxies from
\citet{Pedelty89} and \citet{GarringtonConway91}, respectively.}
\label{figdpmalr}
\end{figure}

\citet{LiuPooleyB91} found evidence for a correlation between the 
depolarization measure and lobe spectral index. They found that the radio 
lobe with the flatter spectrum is associated with lower depolarization. 
Among the 13 sample FRIIs, eleven sources have lobe depolarization measure and 
spectral index estimates, while only seven have depolarization estimates
for both the lobes (see Table~\ref{tablobes}). Six out of these seven sources
show lower depolarization in the lobe with the flatter spectral index.
We examined the Liu-Pooley correlation with the combined dataset and found 
that the correlation is significant at the 99.99\% significance level
(Fig.~\ref{figdpmalr}). This is a significant improvement from
the original radio galaxy (excluding quasars) correlation which was observed 
at the $\sim80$\% significance level \citep[also see][]{Ishwara-Chandra01}.

\citet{LiuPooleyB91} have pointed out that the line-of-sight 
effects which derive from the orientation of the jet do not give a clear 
explanation for the $DP$-$\alpha$ correlation
\citep[see also][]{Dennett-Thorpe97}. They concluded that 
differences in the medium surrounding the two radio lobes influence both the 
spectrum and the depolarization. A denser medium around one radio lobe would
result in greater confinement of the lobe, thereby decreasing the expansion 
losses and possibly increasing the radiative
losses, resulting in a steeper spectral index and greater depolarization. 
\citet{McCarthy91} have indeed demonstrated that the emission-line gas is 
intrinsically asymmetric in powerful radio sources, and the differences in the 
density of the surrounding medium is sufficient to explain the observed 
arm-length ratios in radio galaxies \citep[however see][]{Best95}. 
We do find the presence of the ``alignment effect'' \citep{McCarthy93},
in 4 of our 13 FRII radio galaxies (see Notes in Appendix A).

\begin{figure}[h]
\centering{
\includegraphics[width=8.0cm]{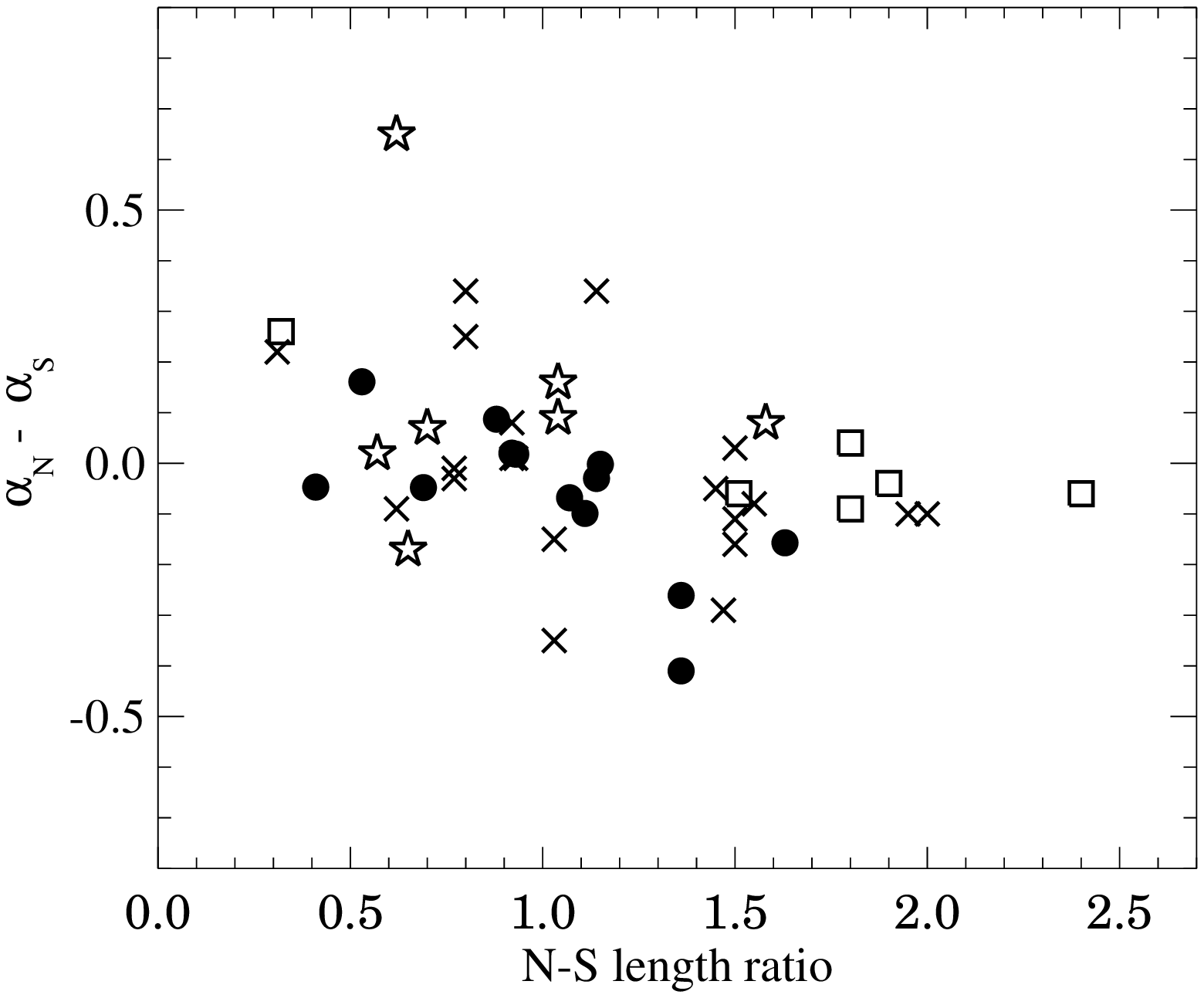}
\includegraphics[width=8.0cm]{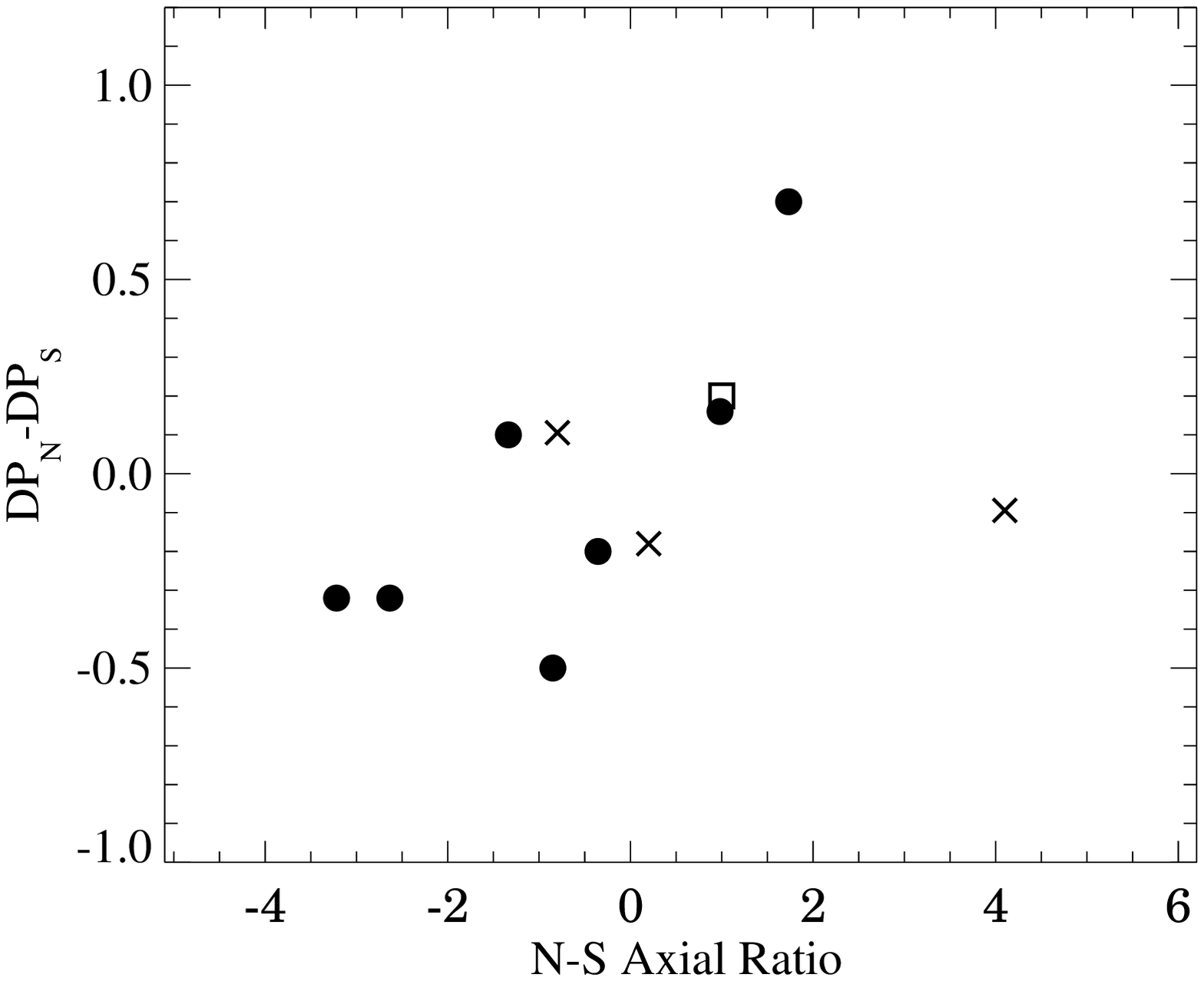}}
\caption{(Left) The North-to-South lobe difference in the spectral index versus
the North-to-South arm-length ratio. (Right) The North-to-South lobe
depolarization difference versus the axial ratio difference between the radio 
lobes. More data are clearly required to re-examine this correlation.
Filled circles denote our sample FRII radio galaxies, stars and crosses
denote radio galaxies from \citet{LiuPooleyA91} and \citet{Goodlet04},
respectively, while squares and diamonds are galaxies from
\citet{Pedelty89} and \citet{GarringtonConway91}, respectively.}
\label{figalph}
\end{figure}

With the intention of singling out factors that could contribute to the 
$DP-\alpha$ correlation, we examined the dependence of lobe depolarization
and spectral index with (possible) orientation and environmental indicators.
We find that the lobe depolarization difference is correlated with the North 
to South arm-length ratio (see Fig.~\ref{figdpmalr}), 
\citep[as found by][]{Pedelty89,Laing96}, in the sense that 
the shorter side of the source is more depolarized. 
This is in keeping with the idea of greater confinement around the shorter 
lobe by the source environment, which leads to greater depolarization
\citep{LiuPooleyA91}.
We also find a tentative correlation between the difference in the
lobe depolarization and the difference in axial ratios at a significance 
level greater than 99.8\% (see Fig.~\ref{figalph}). 
No trend is observed between the depolarization and
axial ratio taken for each lobe separately. More data are clearly 
required to re-examine these relations. The trend between lobe depolarization
difference and the axial ratio difference implies that shorter and/or
fatter lobes are more depolarized.
Again this is consistent with the picture of greater confinement around
the lobes by the source environment, which gives rise to fatter lobes and
depolarization. 

\begin{figure}[h]
\centering{
\includegraphics[width=8.0cm]{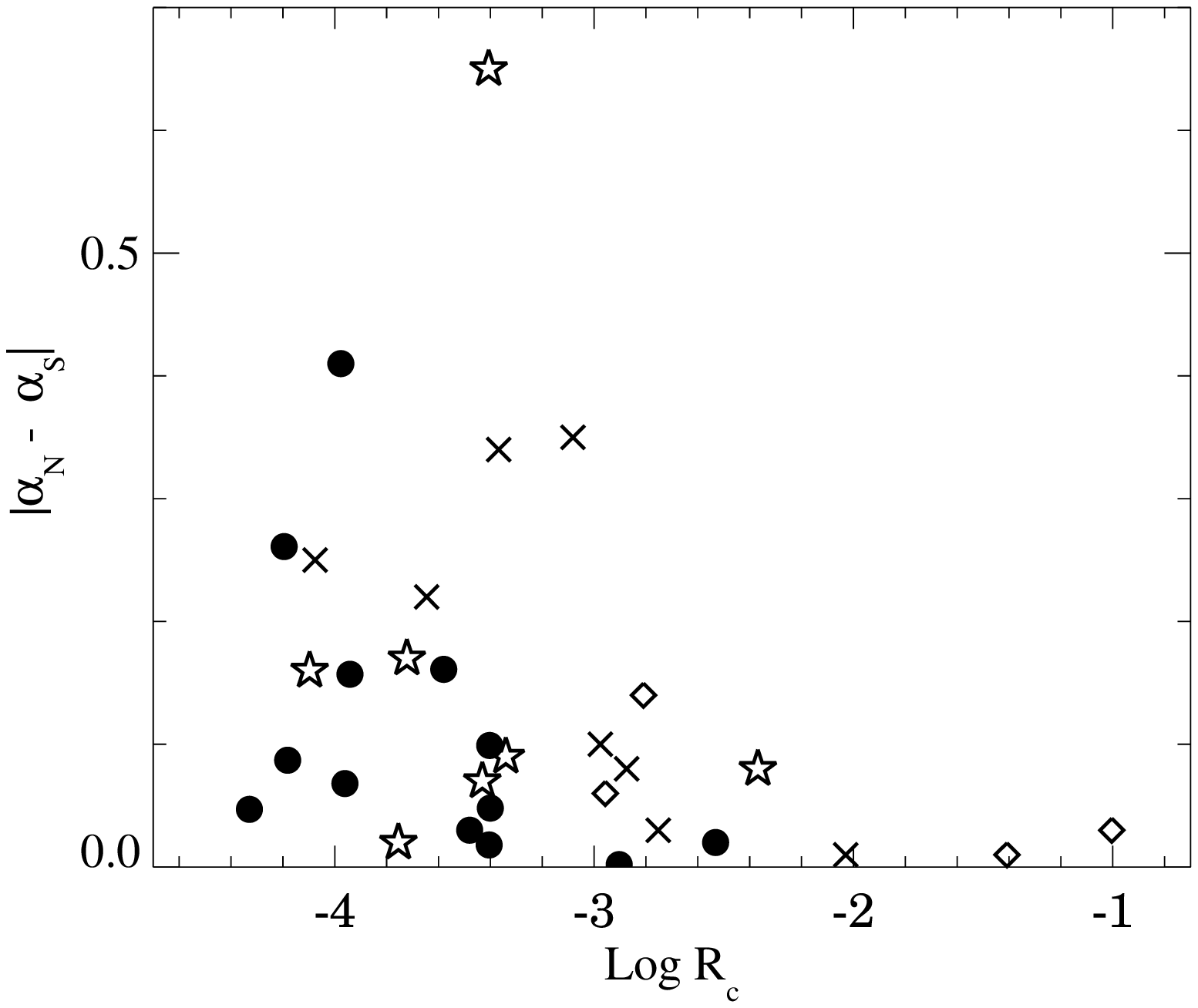}
\includegraphics[width=8.0cm]{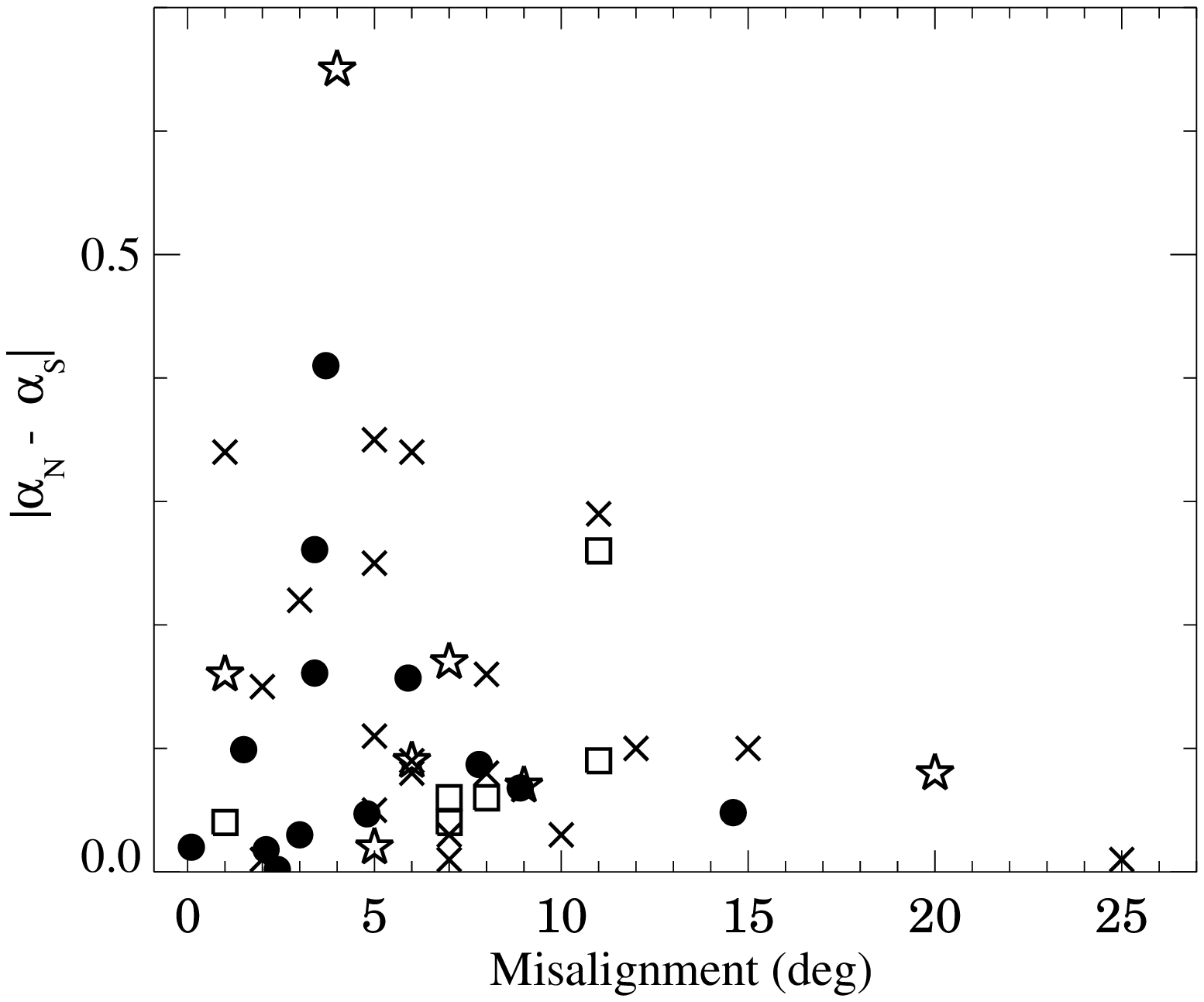}}
\caption{The absolute North-to-South lobe difference in the spectral index 
versus 
(Left) the radio core prominence and (Right) misalignment angle. 
Filled circles denote our sample FRII radio galaxies, stars and crosses
denote radio galaxies from \citet{LiuPooleyA91} and \citet{Goodlet04},
respectively, while squares and diamonds are galaxies from
\citet{Pedelty89} and \citet{GarringtonConway91}, respectively.}
\label{figalprc}
\end{figure}

We find that the absolute lobe-to-lobe difference in spectral index does not
correlate with the misalignment angle but shows a marginal correlation with 
the radio core prominence $-$ core-dominant sources seem to have smaller 
differences in their lobe spectral indices (see Fig.~\ref{figalprc} and 
Table~\ref{tabcorrel}). However, if relativistic beaming was indeed present 
in these radio galaxies, one would expect a correlation in the opposite sense. 
We confirm the significant correlation of the lobe spectral index difference 
with the North to South arm-length ratio, ie., the shorter lobe has the 
steeper spectral index (Fig.~\ref{figalph}). 

\begin{deluxetable}{llrrrrrrrr}
\tabletypesize{\scriptsize}
\tablecaption{Correlation Statistics}
\tablewidth{0pt}
\tablehead{
\colhead{Property 1}&\colhead{Property 2}&\colhead{Spearman}& 
\colhead{Spearman} & \colhead{Kendall} & \colhead{Kendall} \\
&&\colhead{Statistic}& \colhead{Prob.}& \colhead{Statistic}& 
\colhead{Prob.}}
\startdata
\multicolumn{6}{c}{Significant Correlations}\\
$r_h$ & $l$ & 0.537 & 1.6E-7  & 0.373  & 5.3E-7 \\
$\alpha_{lobe}$ & $z$ & 0.355 & 0.0002 & 0.243 & 0.0003 \\
$\alpha_{HS}$ & $z$ & 0.393 & 0.0007 & 0.299 & 0.0002 \\
DP$_N$-DP$_S$ & $\alpha_N-\alpha_S$ & --0.466 & 0.001 & --0.317 & 0.002 \\
$\alpha_N-\alpha_S$ & N-S Length ratio & --0.445 & 0.001 & --0.296 & 0.003 \\
AR & Q & --0.352 & 0.003 & --0.250 & 0.002\\
DP$_N$-DP$_S$ & N-S Length ratio & 0.448 & 0.002 & 0.315 & 0.002 \\
$\alpha_{HS}$ & $\alpha_{lobe}$ & 0.508 & 0.007 & 0.361 & 0.009 \\
\multicolumn{6}{c}{Marginal Correlations}\\
Q & $\zeta$ & 0.306 & 0.010 & 0.206 & 0.012 \\
$|RM|$ & $|b|^\dagger$ & --0.287  & 0.014  & --0.205  & 0.011  \\
DP$_N$-DP$_S$ & AR$_N$-AR$_S$ & 0.669 & 0.024 & 0.550 & 0.018 \\ 
DP & $z$ & --0.237 & 0.024 & --0.160 & 0.024 \\ 
$\zeta$ & $R_c$ & 0.367 & 0.038 & 0.234 & 0.059 \\
$|$$\alpha_N-\alpha_S|$ & $R_c$ & --0.452 & 0.009 & --0.308 & 0.013 \\
\multicolumn{6}{c}{No Correlations}\\
$\sigma_{RM}$ & $|b|^\ddagger$ & 0.205 & 0.100  & 0.141  & 0.095 \\
$\alpha_N-\alpha_S$ & $z$ & --0.162 & 0.275 & --0.116 & 0.249 \\
AR & $\zeta$ & --0.128 & 0.294 & --0.084 & 0.308 \\
DP & AR  & 0.167 & 0.414 & 0.132 & 0.341 \\
$|$DP$_N$-DP$_S|$ & $\zeta$ & 0.095 & 0.539 & 0.043 & 0.677 \\
Q & $R_c$ & --0.092 & 0.615 & --0.060 & 0.624 \\
$\zeta$ & $z$ & --0.092 & 0.454 & --0.065 & 0.431 \\
$|$DP$_N$-DP$_S|$ & $R_c$ & --0.122 & 0.550 & --0.080 & 0.565 \\
$|\alpha_N-\alpha_S|$ & $\zeta$ & --0.049 & 0.745 & --0.034 & 0.734 \\
DP$_N$-DP$_S$ & $z$ & --0.032 & 0.762 & --0.028 & 0.695 \\
DP & $\sigma_{RM}$ & --0.014 & 0.919 & --0.009 & 0.921 \\
Q  & $z$ &  0.002 & 0.981 & 0.012 & 0.882 \\
\enddata
\vspace{-0.4cm}

\tablecomments
{The correlations arranged according to decreasing probability.
Col.s 1 \& 2: The parameters examined for a correlation -
$|RM|$ = Absolute value of rotation measure, $|b|$ = Absolute Galactic
latitude, DP = Depolarization, Q = Arm-length ratio, 
$\zeta$ = Misalignment angle, AR = Axial ratio, $R_c$ = Radio core 
prominence, $r_h$ = Hotspot size, $l$ = Core-hotspot distance, 
$\alpha_{HS}$ = Hotspot spectral index between 1.4 and 5 GHz, 
$\alpha_{lobe}$ = Lobe spectral index between 1.4 and 5 GHz, $z$ = Redshift.
Col.s 3 \& 5: Spearman's (rho) and Kendalls's (tau) rank correlation
coefficient. 
Col.s 4 \& 6: Probability that the two properties in Cols. 1 \& 2 are not
correlated, using Spearman's (rho) and Kendalls's (tau) rank correlation
tests, respectively. A small value indicates a significant correlation.
$\dagger$ The correlation is significant at the 99.99\%
significance level, when the single high ${RM}$ value from \citet{Goodlet04} is 
excluded. $\ddagger$ There is no correlation (Spearman prob. = 0.196) when 
the two high $\sigma_{RM}$ values from \citet{Goodlet04} are excluded.}
\label{tabcorrel}
\end{deluxetable}

\subsection{Structural Asymmetries and Trends with Redshifts}
Although the 13 FRIIs have largely symmetric structures, trends with 
structural asymmetries emerge when the larger combined dataset
is considered \citep[see][for a review on structural asymmetries]
{Gopal-Krishna04}. 
We find a correlation between the arm-length ratio and 
misalignment angle at the 99.9\% significance level for the combined radio
galaxy sample (Fig.~\ref{figaxlr}) - sources whose lobes are asymmetric 
in length show greater misalignment between the two sides.
This effect has been previously observed by, for example,
\citet{Macklin81} and \citet{KapahiSaikia82}. 
This correlation could suggest that the environmental 
asymmetries that give rise to the arm-length ratio could also be 
contributing to the misalignment angles in these radio galaxies,
as also suggested by \citet{Macklin81}.

The amount of misalignment also places a limit on the systematic jet axis 
stability and/or the dentist drill effect. If a source has two or more 
hotspots on one side of the source, and these two hotspots plus the core 
(or these two hotspots and the hotspot on the other side of the source) 
aligned to within some angle, say one degree, then the outflow axis must not 
change by more than an amount $\theta$ in a time $\delta t$. $\theta$ is the 
accuracy with which the hotspots and the core line up and $\delta t$ is equal 
to or larger than the separation of the two hotspots on one side of the source 
divided by the speed of light. Thus if the separation of the two hotspots on 
one side is 10 kpc, and all of the hotspots line up to better than a degree, 
the wobble of the outflow axis must be less than about $10^{-14}$ rad~s$^{-1}$.
Wobbling could also produce substructure within the hotspots with a 
characteristic size that is proportional to the core-hotspot separation, and 
this could impact the overall hotspot size. As the misalignment angles 
are typically less than $10\degr$ in our 13 FRII radio galaxies, we infer 
that the wobble of the outflow axis in these sources is less that 
10$^{-13}$ rad~s$^{-1}$. This argument is based on the relative positions of 
the hotspots and the core, and does not account for possible light travel time effects. These effects will be small if the source is close to the plane of 
the sky.

\begin{figure}[h]
\centering{
\includegraphics[width=8.1cm]{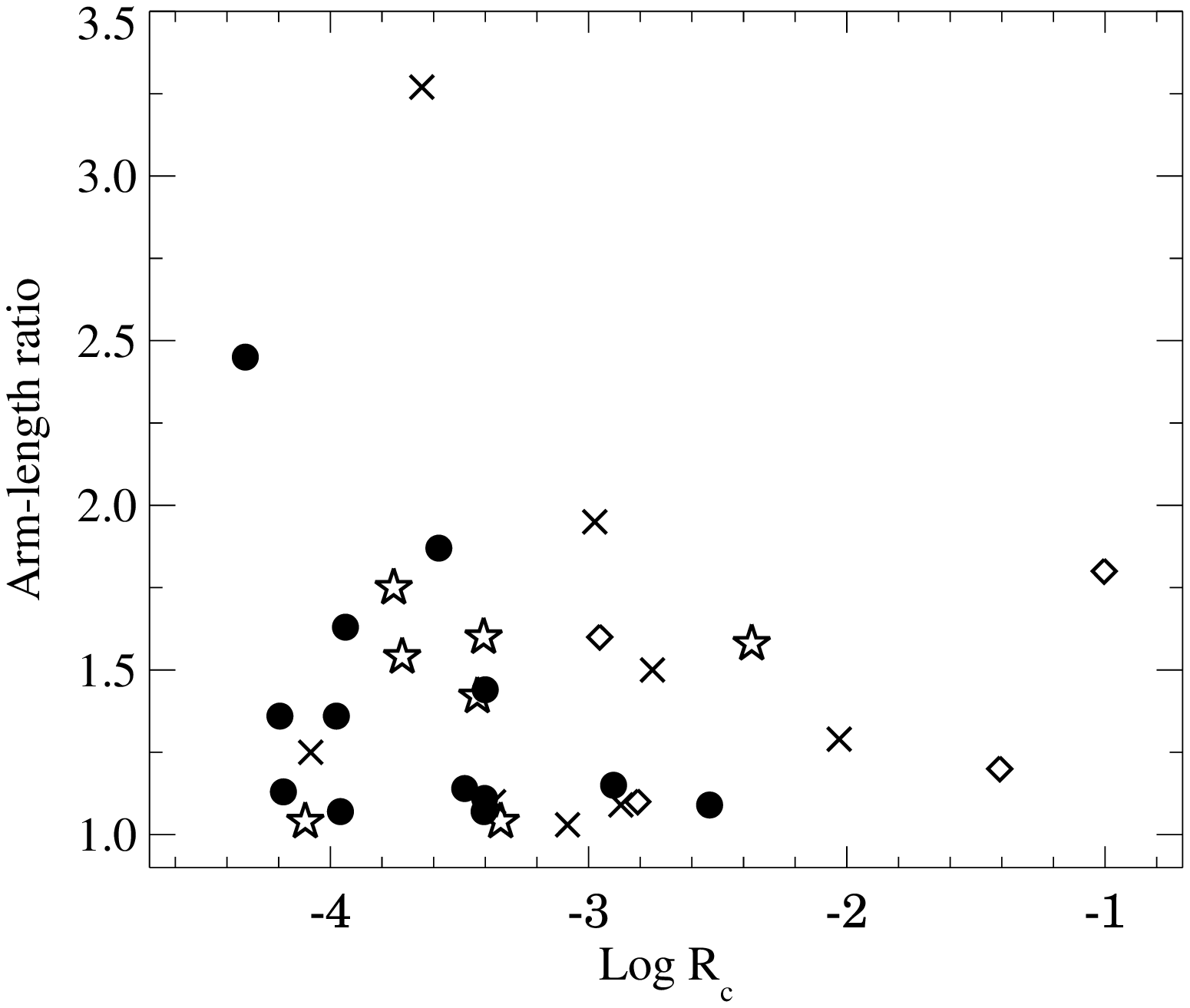}
\includegraphics[width=8.1cm]{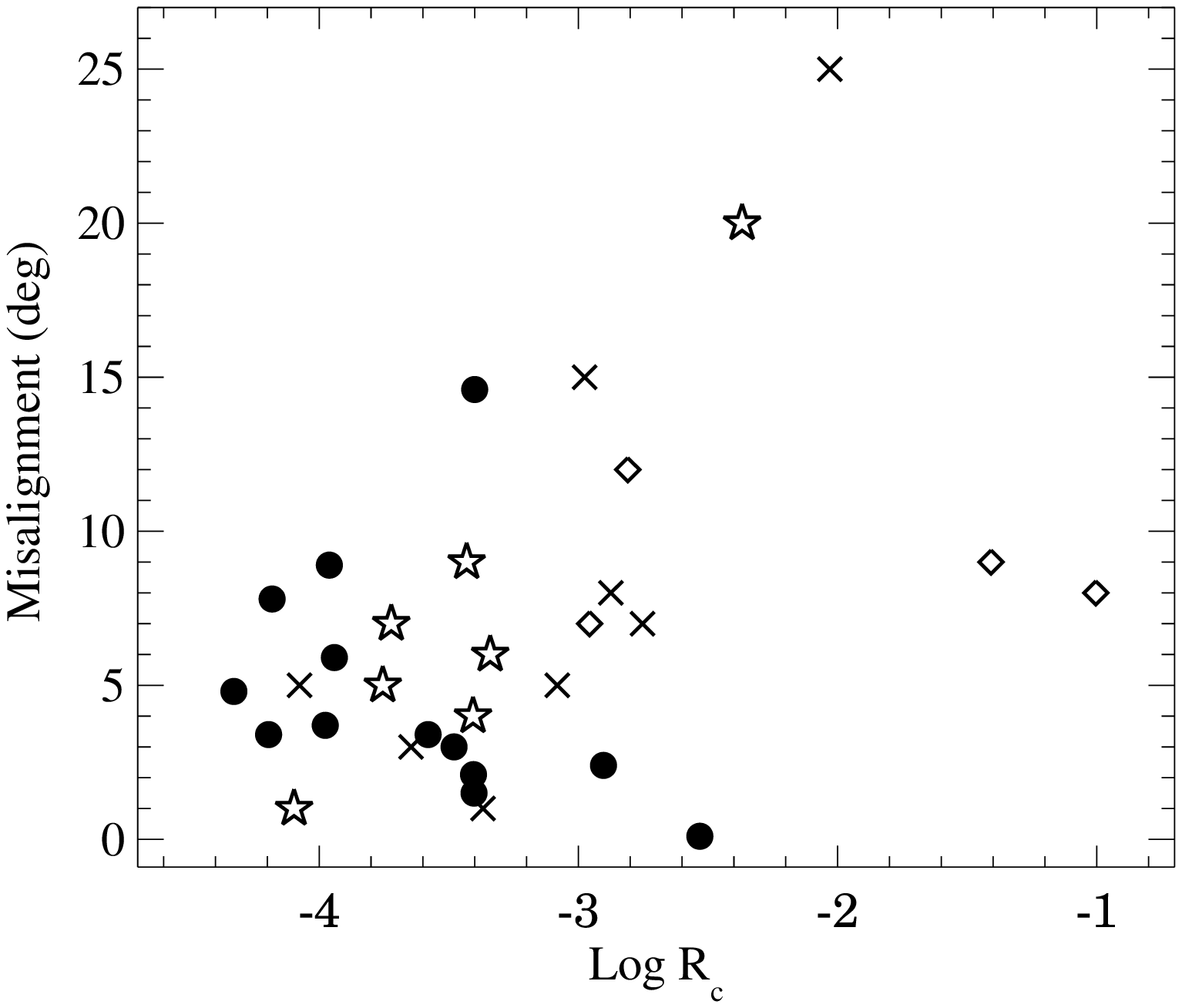}}
\caption{(Left) Arm-length ratio versus the radio core prominence parameter.
(Right) Misalignment angle versus the radio core prominence.
Filled circles denote our sample FRII radio galaxies, stars and crosses
denote radio galaxies from \citet{LiuPooleyA91} and \citet{Goodlet04},
respectively, diamonds are galaxies from \citet{GarringtonConway91}.}
\label{figrcmis}
\end{figure}

The radio core prominence is plotted with respect to the arm-length ratio and 
misalignment angle in Fig.~\ref{figrcmis}. 
We observe a weak correlation between core prominence $R_c$ and 
misalignment angle $\zeta$, 
but none between $R_c$ and arm-length ratio $Q$ (see Table~\ref{tabcorrel}). 
The radio core prominence-misalignment angle ($R_c-\zeta$) correlation has 
previously been observed by, for example,
\citet{KapahiSaikia82} and \citet{HoughReadhead89}. This correlation could be 
suggestive of the presence of relativistic beaming effects in the core 
due to orientation in the combined radio galaxy sample. On the other hand, 
there could be an environmental contribution to both the radio core 
prominence and misalignment angle. For
example, the interaction of the jet as it is launched, with the 
surrounding nuclear environment could
give rise to a brighter radio core along with changes in its
direction of propagation, resulting in greater misalignment angles.

\begin{figure}[h]
\centering{
\includegraphics[width=8.0cm]{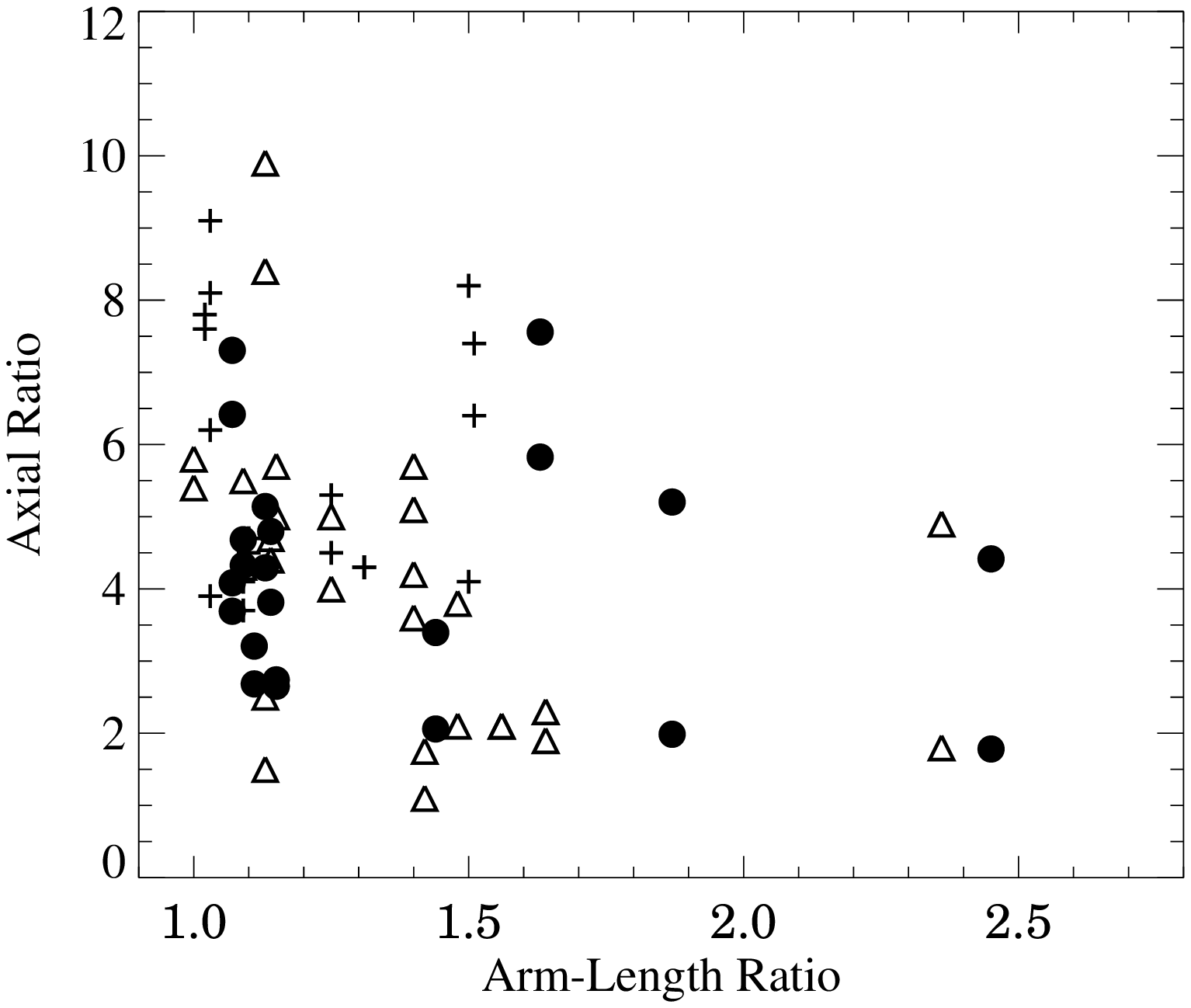}
\includegraphics[width=8.0cm]{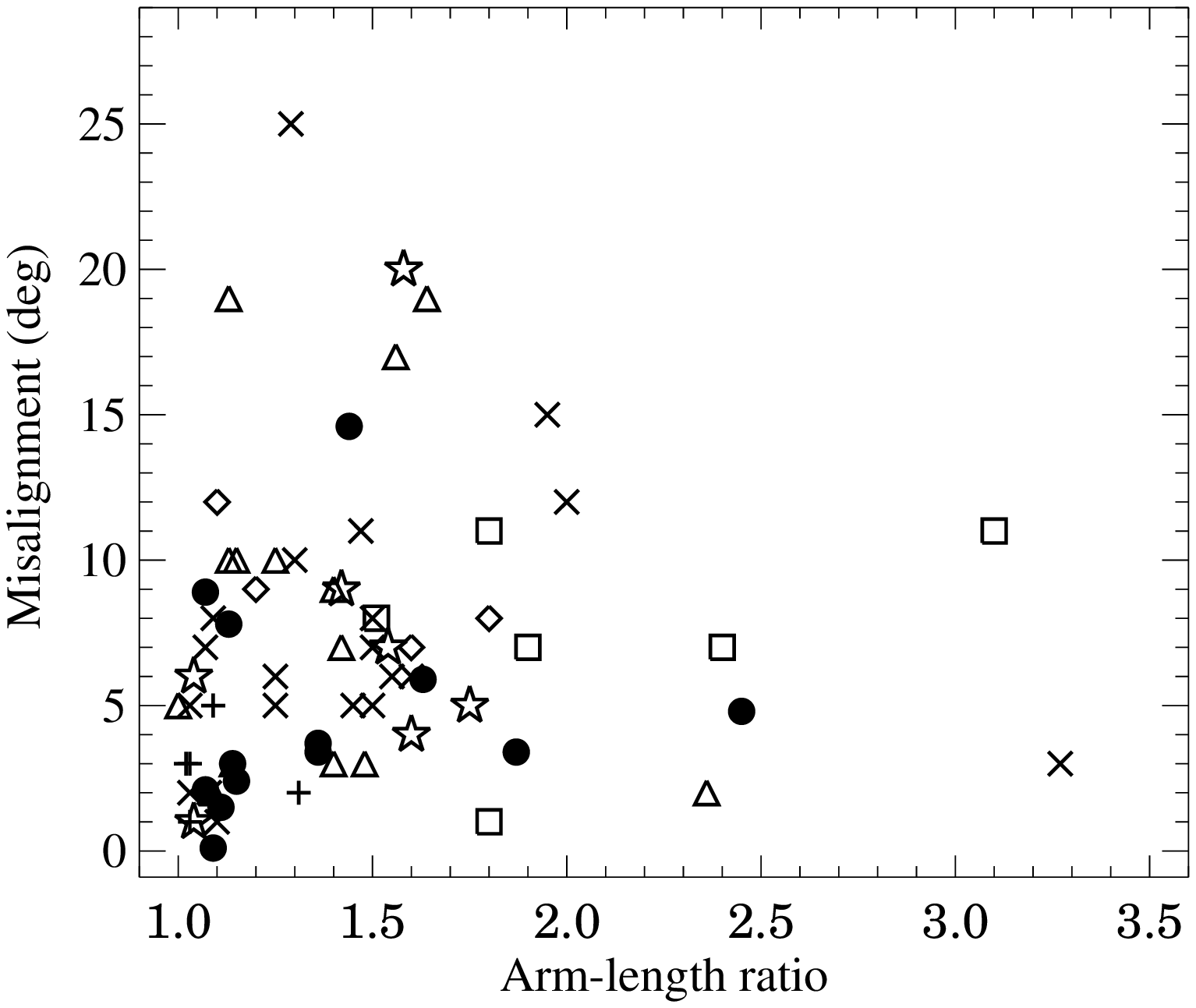}}
\caption{
(Left) Axial ratio for each radio lobe versus the arm-length ratio. 
(Right) Misalignment angle versus arm-length ratio.
The filled circles denote the radio galaxies from our sample,
stars are data from \citet{LiuPooleyA91}, squares from \citet{Pedelty89},
crosses from \citet{Goodlet04}, diamonds from \citet{GarringtonConway91},
triangles from \citet{LeahyWilliams84},
and plus~signs from \citet{LeahyMuxlow89}.}
\label{figaxlr}
\end{figure}
We find an anti-correlation between the arm-length ratios and axial ratios 
for the radio galaxies. In Fig.~\ref{figaxlr}, we have plotted the axial
ratios for each lobe separately, with respect to the arm-length ratio,
for the combined radio galaxy dataset. The (anti)-correlation implies that
sources which have a fatter lobe on one side, are also the most asymmetric 
in length. This effect is also clearly visible in the radio maps. 
If one radio lobe is shorter than the other, it could perhaps be due to
a difference in the surrounding medium which slows down the expansion
and confines one radio lobe preferentially. 
The overpressured radio plasma expands laterally when the jet meets 
resistance in 
the forward direction, thus producing fatter lobes.
A larger variation in outflow direction can also lead to larger misalignment,
fatter lobes and smaller axial ratios.
The axial ratios and misalignment angles are however not correlated
(see Table~\ref{tabcorrel}). The lack of a correlation does not diminish the
importance of the 
variation in outflow direction, because symmetric motions like the precession
of the entire blackhole-double-jet system, could result in zero misalignment
but still increase the axial ratios.

The anticorrelation between axial ratio and arm-length ratio and the 
correlation between depolarization and axial ratio (Figure~\ref{figalph}) 
leads to an interesting new result. While it has been known for a long
time that the shorter lobe is more depolarized in these radio galaxies, we
find that shorter lobe is also the fatter one. This explains the 
correlation between depolarization and axial ratio. In a nutshell, we
find that the shorter lobe is fatter, more depolarized and has
a steeper spectrum.

\begin{figure}[h]
\centering{
\includegraphics[width=8.0cm]{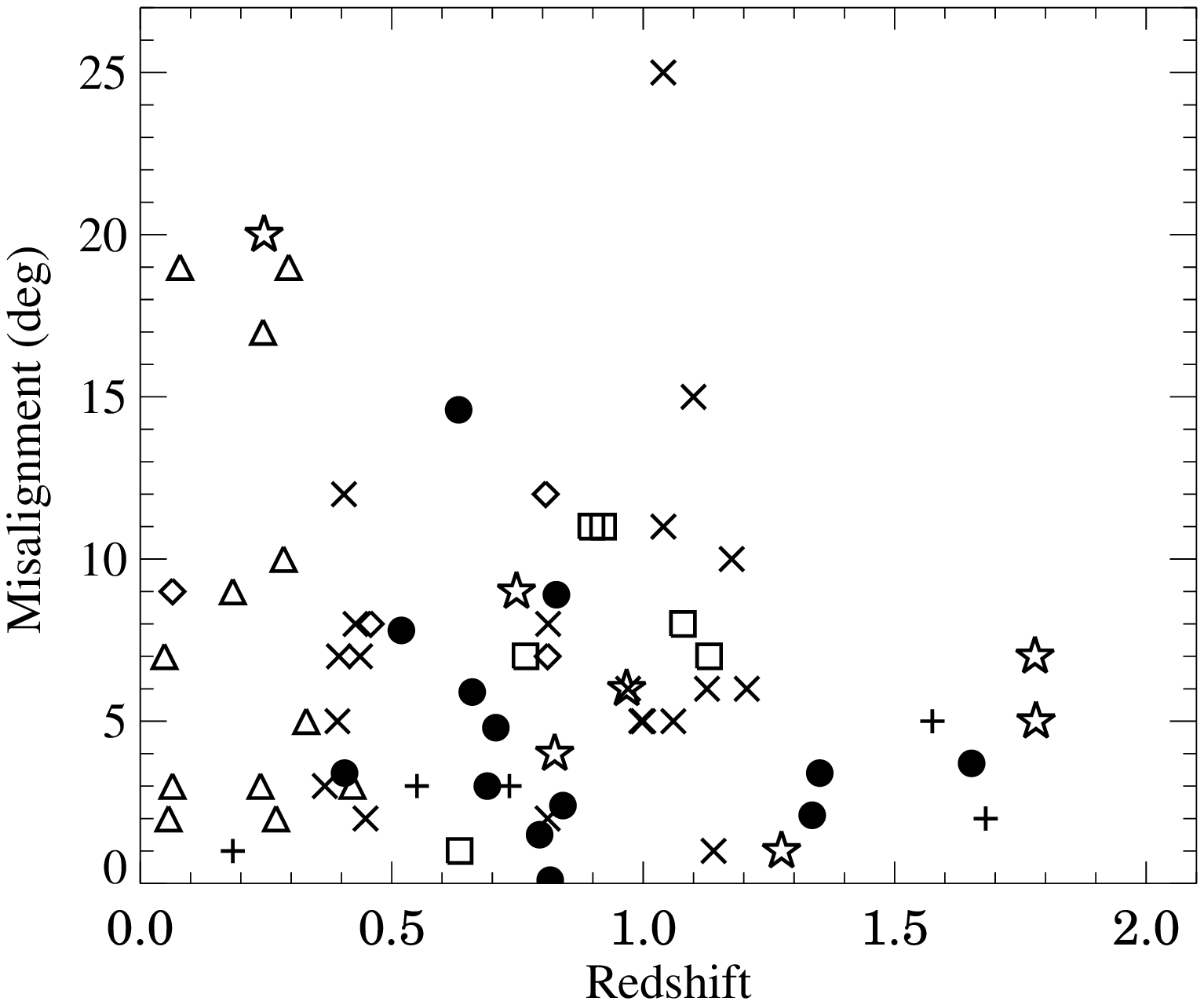}
\includegraphics[width=8.0cm]{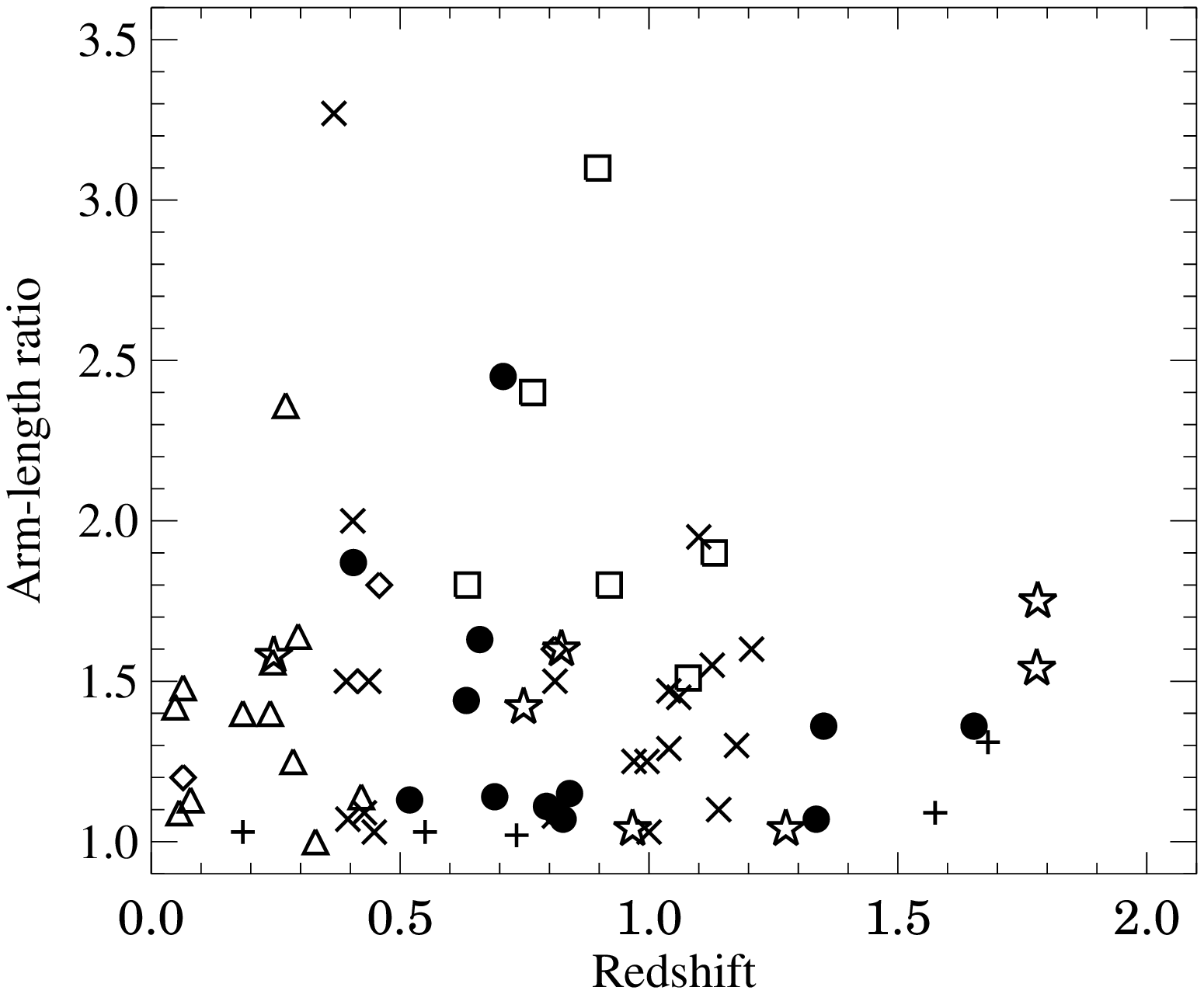}}
\caption{(Left) Misalignment angle and (Right) arm-length ratio versus
redshift.
The filled circles denote our 13 radio galaxies,
stars are galaxies from \citet{LiuPooleyA91}, squares from \citet{Pedelty89},
crosses from \citet{Goodlet04}, diamonds from \citet{GarringtonConway91},
triangles from \citet{LeahyWilliams84}, and plus~signs from 
\citet{LeahyMuxlow89}. No correlation is observed between the parameters.}
\label{figred}
\end{figure}
We do not find any correlation between misalignment angle and redshift, nor
between arm-length ratio and redshift (Fig.~\ref{figred}).
\citet{BarthelMiley88} had found a tentative correlation between misalignment
angle and redshift but \citet{KapahiKulkarni90} and \citet{Best95} did not
find such a correlation, consistent with our result.

\begin{figure}[h]
\centering{
\includegraphics[width=8.7cm]{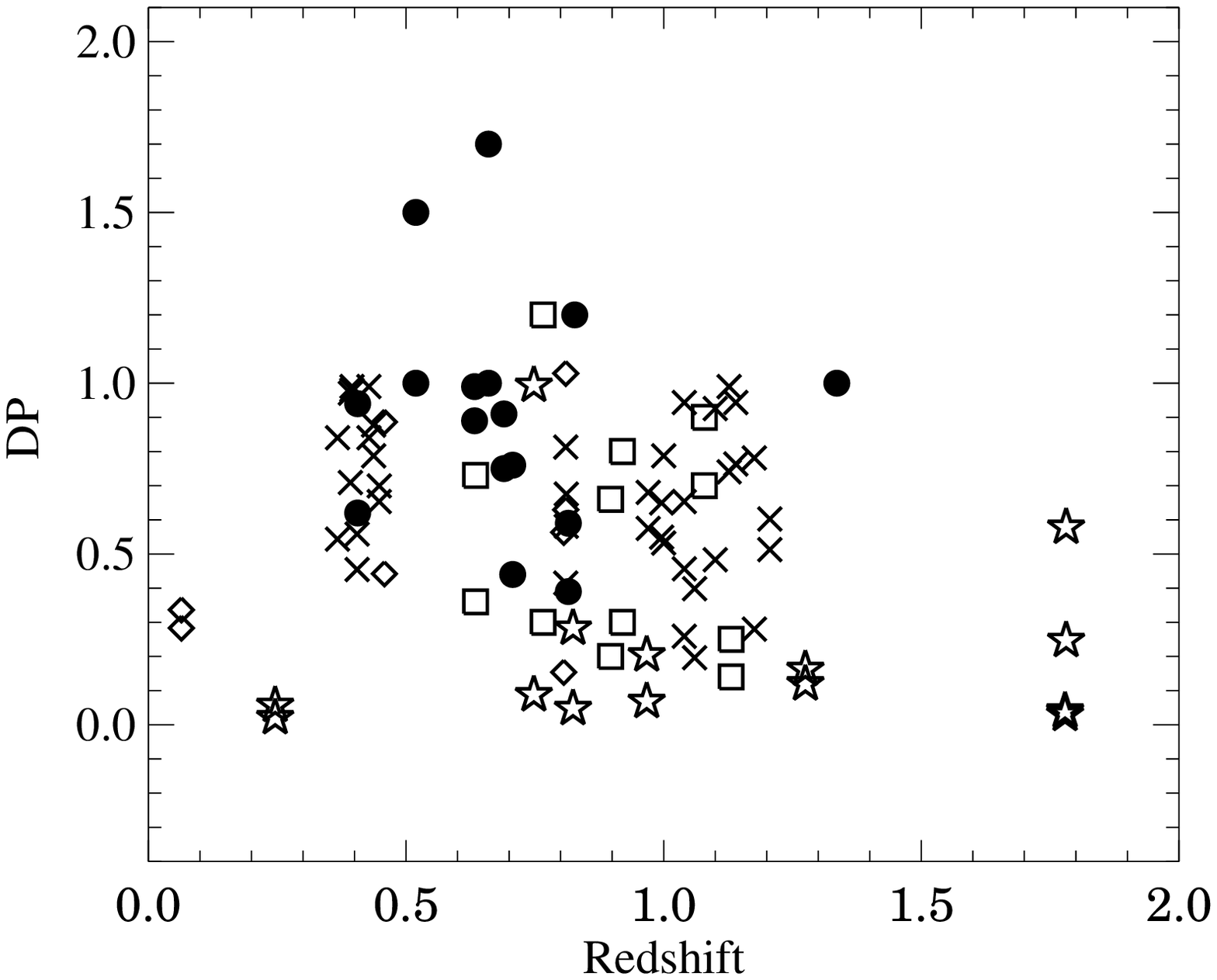}}
\caption{Depolarization for each radio lobe versus redshift.
The filled circles denote our 13 radio galaxies,
stars are galaxies from \citet{LiuPooleyA91}, squares from \citet{Pedelty89},
crosses from \citet{Goodlet04}, diamonds from \citet{GarringtonConway91}. }
\label{figDPalz}
\end{figure}

We find a correlation between the lobe depolarization $DP$ and redshift
(Fig~\ref{figDPalz}). \citet{Kronberg72} and \citet{Goodlet05} have also 
found that the lobes of high redshift galaxies are more depolarized that 
their local counterparts.
This would indicate that source environments vary with redshift, which
would be consistent with the increasing alignment of emission line gas 
with the radio source with increasing redshift 
\citep[e.g.,][]{McCarthy93,Best00,Inskip02,Privon07}.
However, \citet{Morris73} discovered that lobe depolarization showed a 
stronger dependence on the radio luminosity rather than on redshift. 
\citet{Goodlet05} have also found that lobe depolarization depends on both
radio luminosity and redshift. We do not find a correlation between
the depolarization difference between the lobes and redshift 
(see Table~\ref{tabcorrel}). \citet{Goodlet05} have reported the presence of 
this correlation, contrary to our finding. We note however that our
sample-size is twice as large as Goodlet's sample and spans a larger 
redshift range.

\section{SUMMARY AND CONCLUSIONS}
\label{conclusions}
We have observed 13 large and powerful FRII radio galaxies with multiple
arrays of the Very Large Array at the frequencies 330~MHz, 1.4, 5 and 8~GHz.
Extended radio bridges are detected in all but two FRII galaxies
(3C13 and 3C470). 
The rotation measures between 1.4 and 5 GHz are small, comparable to the 
expected Galactic contribution.
The spectral index maps between 1.4 and 5 GHz show a gradual steepening of the 
radio spectra away from the hotspots with some spectral index variations 
superimposed, in agreement with previous studies. 

The 13 FRII radio galaxies do not show the presence of bright jets and
the core-to-lobe flux density ratio, $i.e.,$ the radio core prominence
is typically small ($-4 \leqslant$ Log~$R_c$~$\leqslant -2$), indicating 
that these sources lie close to the plane of the sky.
This is also consistent with the small misalignment angles which are 
less than 8$\degr$ in all but one source.
These characteristics make our sample useful for a spectral aging analysis,
which will be presented in paper II.

We have supplemented our small sample with similar radio galaxy data 
gleaned from the 
literature and examined some well-known radio galaxy correlations. 
\begin{enumerate}
\item{We confirm that the hotspot size $r_h$ is correlated with the 
core-hotspot distance $l$ of the source and follows the relation 
$r_h\propto l^{0.7}$. This is consistent with the hotspot maintaining ram 
pressure balance as
the source propagates through a medium with declining ambient density.
This result is further consistent with a 
self-similar model of a jet propagating in a medium where the ambient 
density $\rho_a$ falls off with distance from source $d$ as 
$\rho_a\propto d^{-0.2}$. 
This could be due to the jets spanning hundreds of kiloparsecs in these 
sources, propagating through 
a roughly constant medium such as that which might be found in the 
core of a cluster or proto-cluster of galaxies.}

\item{The hotspot spectral index varies with redshift as 
$\alpha_{HS} \propto z^{0.4}$, consistent with previous studies on radio 
galaxies. A simple redshifting of the curved spectrum
along with an increase in inverse-Compton cooling due to scattering with
microwave background photons at larger redshifts, could be responsible for
this correlation. Alternatively, enhanced synchrotron losses in more powerful 
hotspots could be producing a spectral steepening at higher luminosity.
This correlation could also be a result of a higher fraction of radio galaxies
being located, as a function of redshift,
in environments with densities similar to nearby rich clusters.}

\item The hotspot spectral index is correlated with and flatter than
the lobe spectral index,
consistent with the assumptions of spectral aging models.

\item{We find that the trend between the lobe depolarization and 
orientation, $i.e.,$ the `Laing-Garrington effect' is weak in these radio
galaxies, suggesting that orientation/Doppler effects are not 
strong in them - these galaxies lie largely in the plane of the sky.}

\item{The correlation between lobe depolarization and lobe spectral index, 
$i.e.,$ the `Liu-Pooley effect', gains in statistical significance - radio 
lobes with a flatter spectrum exhibit lower depolarization.}

\item{The lobe depolarization difference
is correlated with the North-to-South arm-length ratio, in the sense that the
shorter side of the source is more depolarized. This strongly suggests that
lobe depolarization depends significantly on environmental asymmetries in
radio galaxies. The weak correlation between the lobe depolarization and the 
axial ratio difference is consistent with this inference.}

\item{In agreement with previous studies, we find that the lobe spectral index
and depolarization are correlated with redshift. While the lobe
spectral index $-$ redshift correlation follows from points 2 and 3 above, the
lobe depolarization $-$ redshift correlation could suggest a variation in the
source environments. However this effect cannot be differentiated from the
dependence of depolarization on radio luminosity.}

\item{Based on the small misalignment angles in our 13 FRII 
radio galaxies, we infer that the wobble of the outflow axis in these
sources is less that 10$^{-13}$ rad~s$^{-1}$.}

\item{We observe a weak correlation between the radio core prominence and 
misalignment angle but none between core prominence and
arm-length ratio. This is consistent with both core prominence and misalignment 
angle being orientation indicators, and the arm-length ratio being more
sensitive towards asymmetries in the environment.
The radio core prominence and misalignment can however also be correlated
if the inner jet interacts with the ambient medium, resulting in brighter radio
cores and larger misalignment angles.}

\item{The arm-length ratio seems to be significantly correlated with the
misalignment angle between the two sides of the radio source but 
anti-correlated with the axial ratio. This is again suggestive of
environmental asymmetries close to the radio sources. Such asymmetries can 
cause a variation in the outflow direction which can result in larger 
misalignments. Variation in the jet direction can also result in
fatter radio lobes and lower axial ratios.}

\item{We find that the shorter lobe is fatter, more depolarized and has a 
steeper spectrum.}
\end{enumerate}

To summarize, the FRII radio galaxy attributes can be influenced by 
many factors such as their local environment and asymmetries in the gas 
in the interstellar and intergalactic medium, and relativistic beaming
effects. Changes in the outflow direction and variations in the beam power 
with time, may also be playing a role. Radio core prominence and misalignment 
angles can be used as statistical indicators of orientation while the 
arm-length and axial ratios can serve to highlight environmental asymmetries, 
in large radio galaxy samples. However the environmental contribution
to the orientation-indicators, and the presence of variable motion of the 
outflow axis cannot be ignored.

\clearpage
\begin{figure}
\epsscale{1.0}
\plottwo{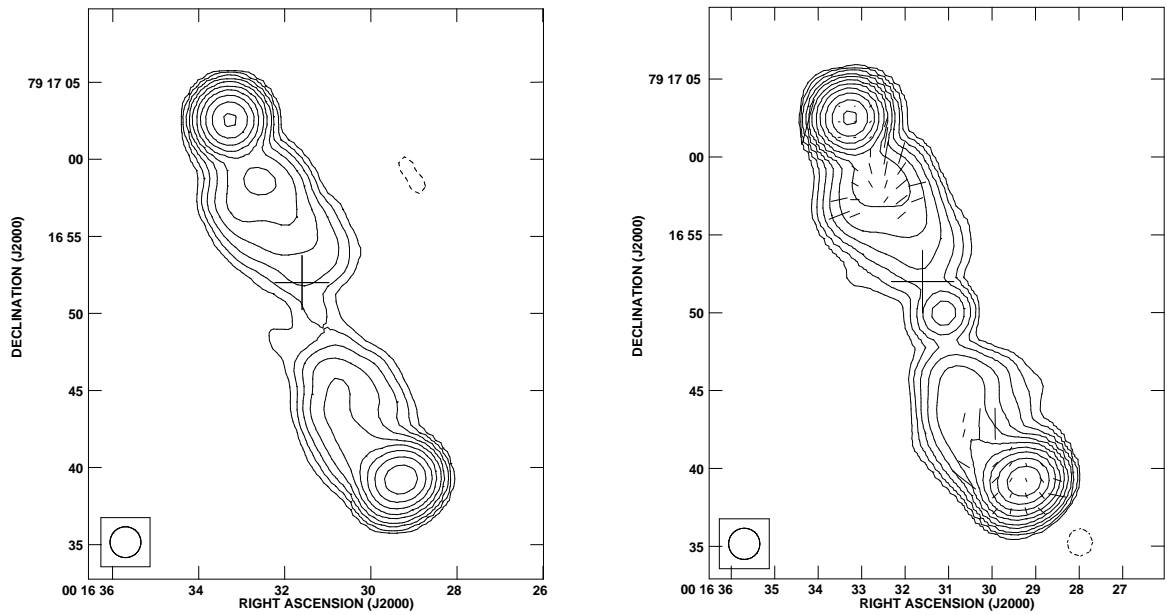}{f13b.ps}
\caption{Total intensity image of 3C6.1 at (left) 1.66~GHz and (right) 4.86~GHz,
at a resolution of 2$\arcsec$ with fractional polarization vectors 
superimposed. The peak surface brightness and contour levels increasing in 
steps of 2 are 
(left) 1.2 Jy/beam, (--0.17,0.17,...,90)\% of peak brightness,  
(right) 419.9 mJy/beam, (--0.085,0.085,...,90)\% of peak brightness, 
1$\arcsec$ vector = 17\% polarization. }
\label{fig3c6.1}
\end{figure}

\begin{figure}
\epsscale{1.0}
\plotone{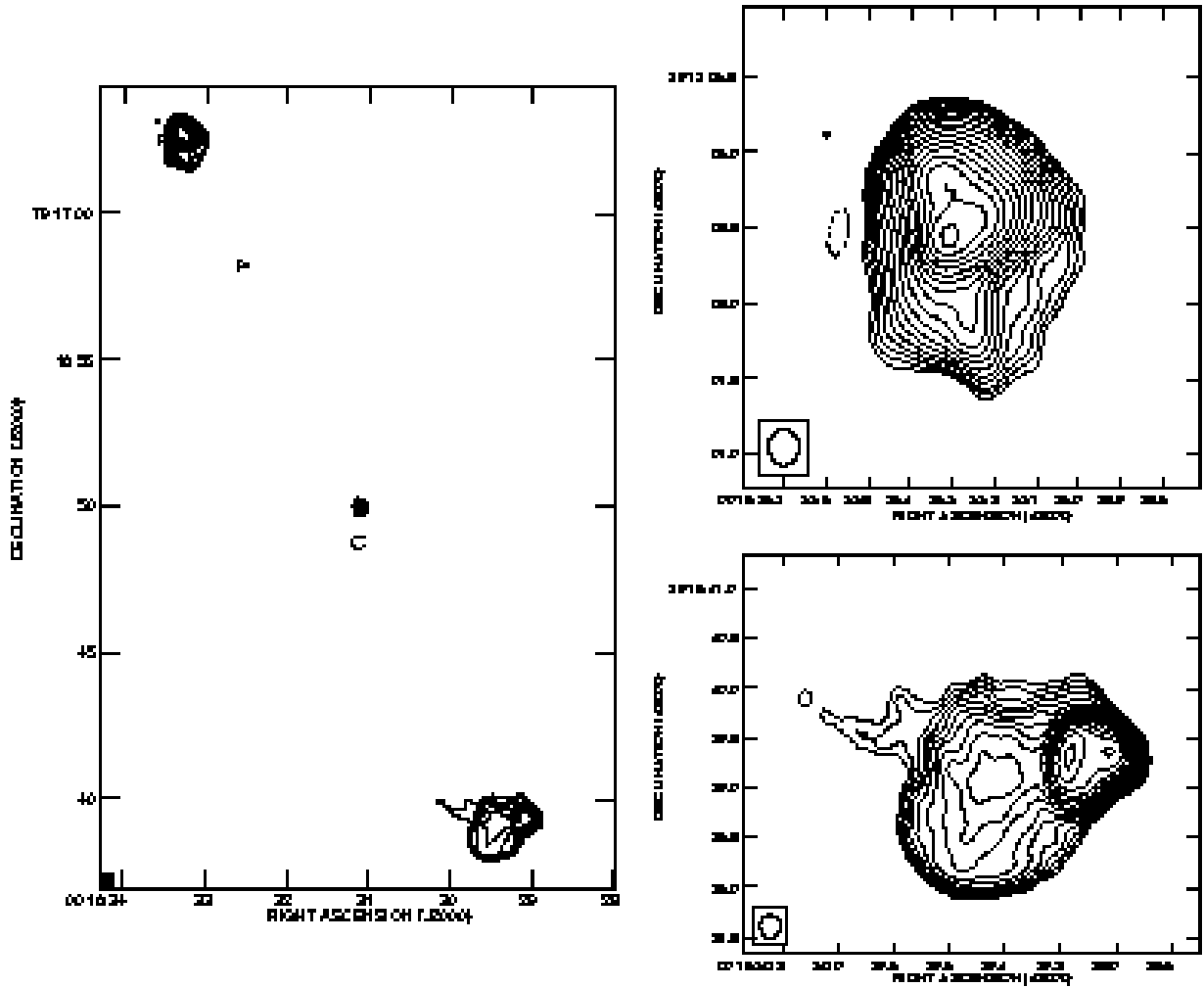}
\caption{Total intensity image of 3C6.1 at 8.4~GHz. The peak surface brightness
and contour levels increasing in steps of 2
($\sqrt{2}$ for the closeup views of the hotspot regions) are
49.2 mJy/beam and (--0.35,0.35,...,90)\% of peak brightness,
respectively.}
\end{figure}

\clearpage
\begin{figure}
\epsscale{1.0}
\plottwo{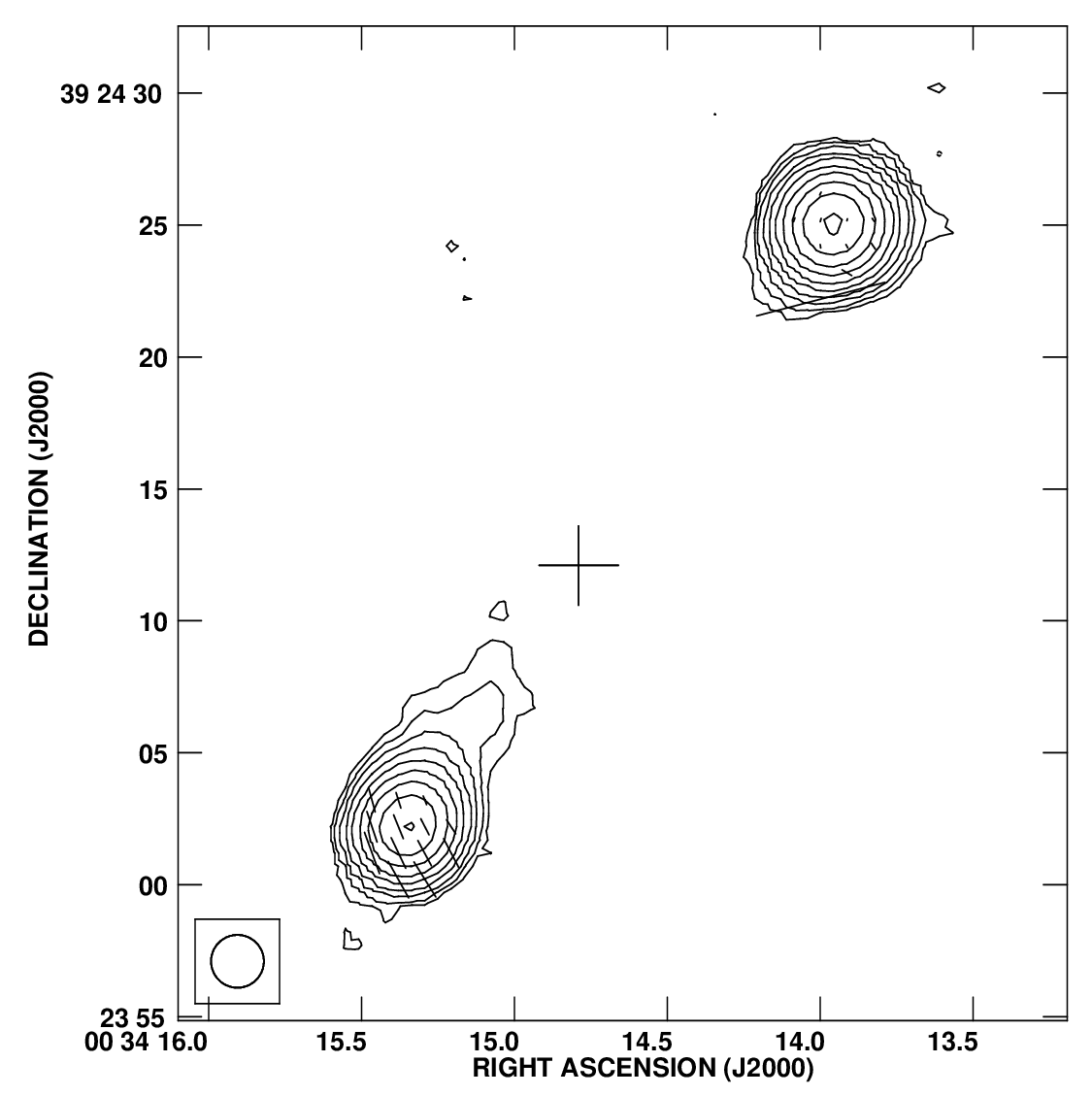}{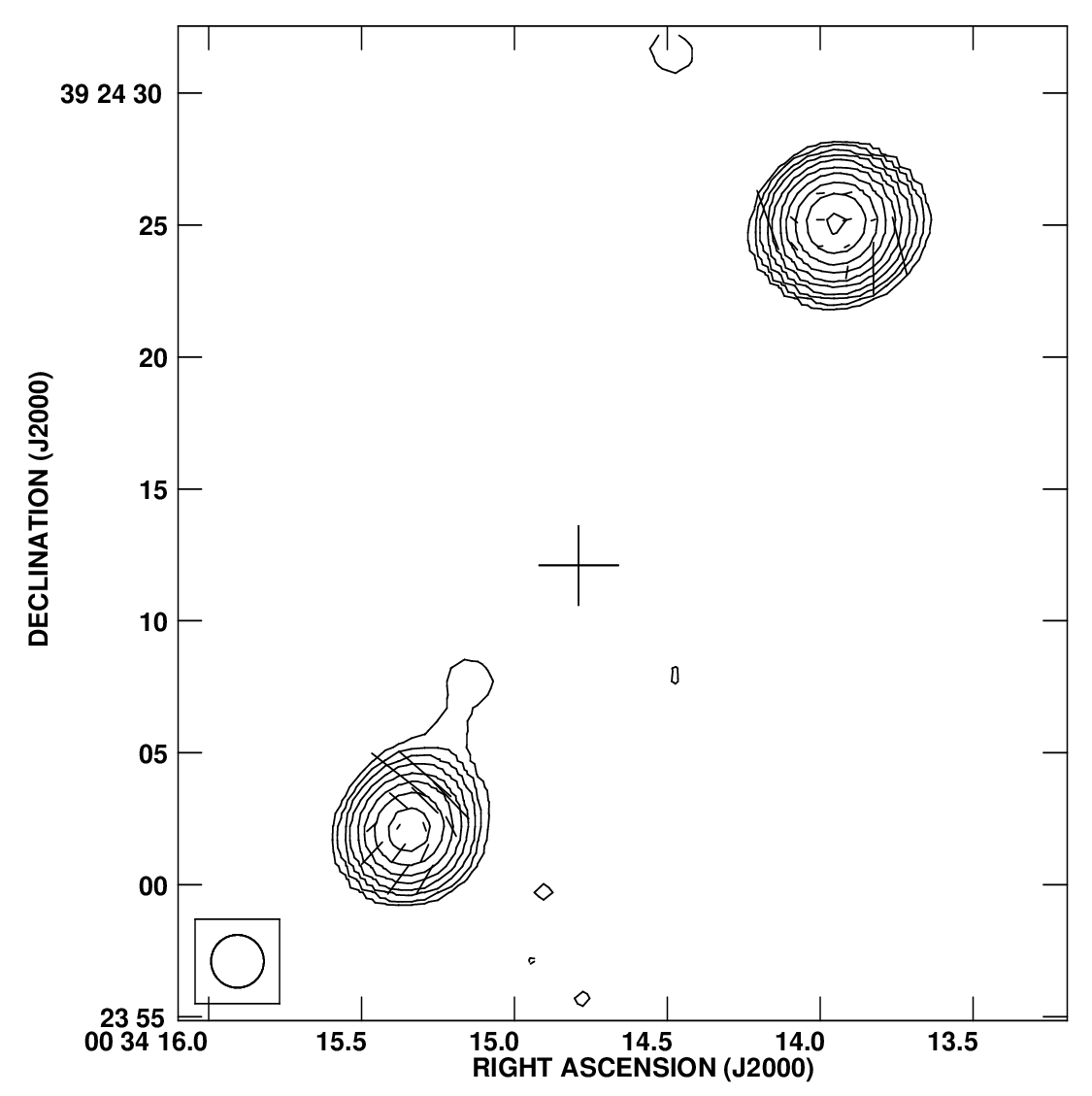}
\caption{Total intensity image of 3C13 at (left) 1.46~GHz and (right) 4.86~GHz,
at a resolution of 2$\arcsec$ with fractional polarization vectors 
superimposed. The peak surface brightness and contour levels increasing in 
steps of 2 are 
(left) 1.06 Jy/beam, (--0.17,0.17,...,90)\% of peak brightness,  
1$\arcsec$ vector = 7\% polarization,
(right) 270.0 mJy/beam, (--0.17,0.17,...,90)\% of peak brightness, 
1$\arcsec$ vector = 7\% polarization. }
\label{fig3c13}
\end{figure}

\begin{figure}
\epsscale{0.5}
\plotone{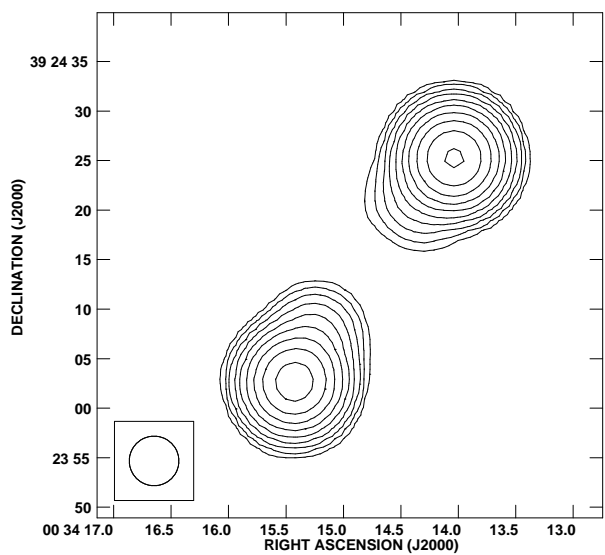}
\caption{Total intensity image of 3C13 at 327.6~MHz. 
The peak surface brightness 
and contour levels increasing in steps of 2 are 4.3 Jy/beam, 
(--0.17,0.17,...,90)\% of peak brightness.}
\end{figure}

\begin{figure}
\epsscale{1.0}
\plotone{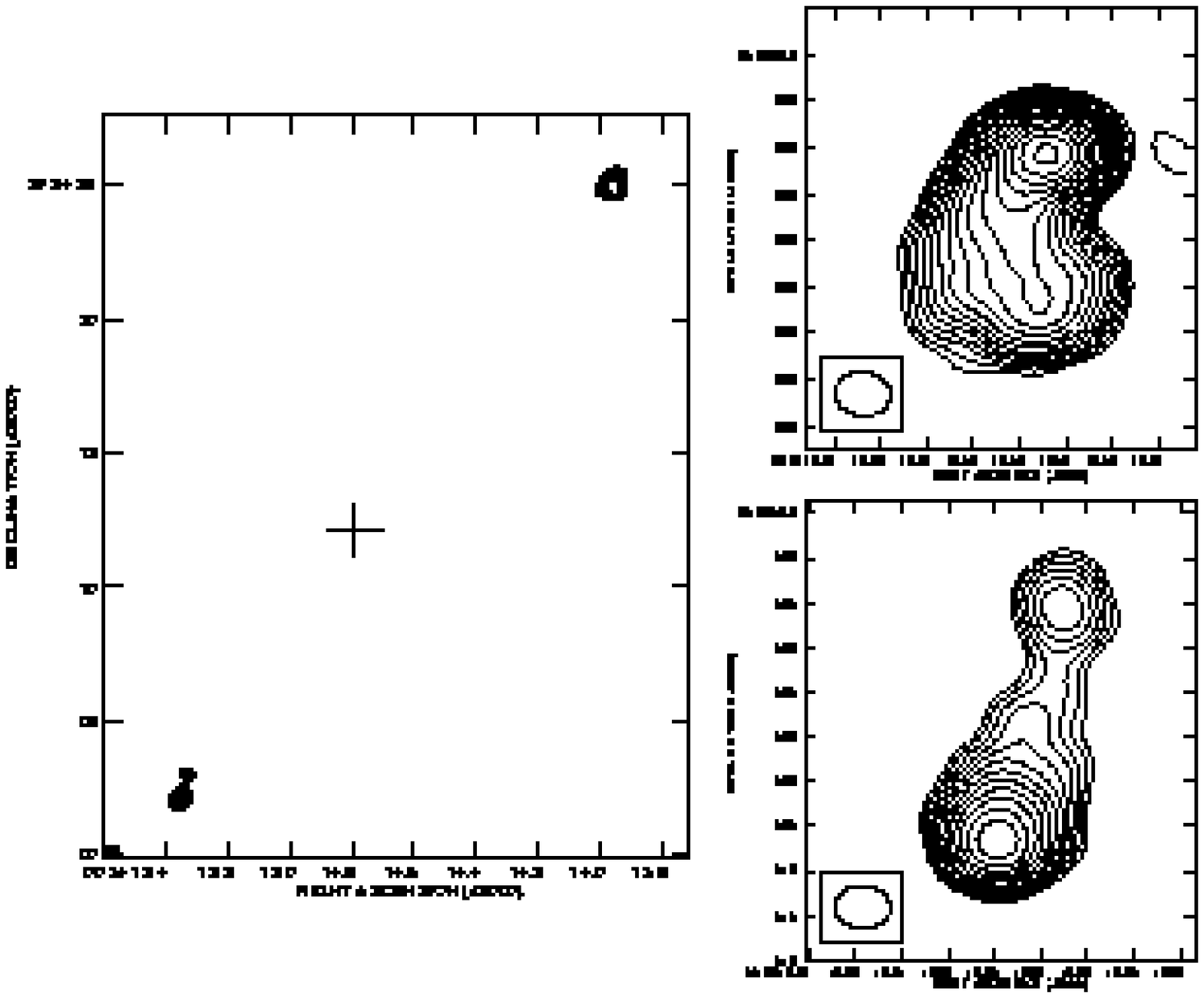}
\caption{Total intensity image of 3C13 at 8.4~GHz. The peak surface brightness
and contour levels increasing in steps of 2
($\sqrt{2}$ for the closeup views of the hotspot regions) are 60.3 mJy/beam, 
(--0.7,0.7,...,90)\% of peak brightness. }
\end{figure}

\clearpage
\begin{figure}
\epsscale{1.0}
\plottwo{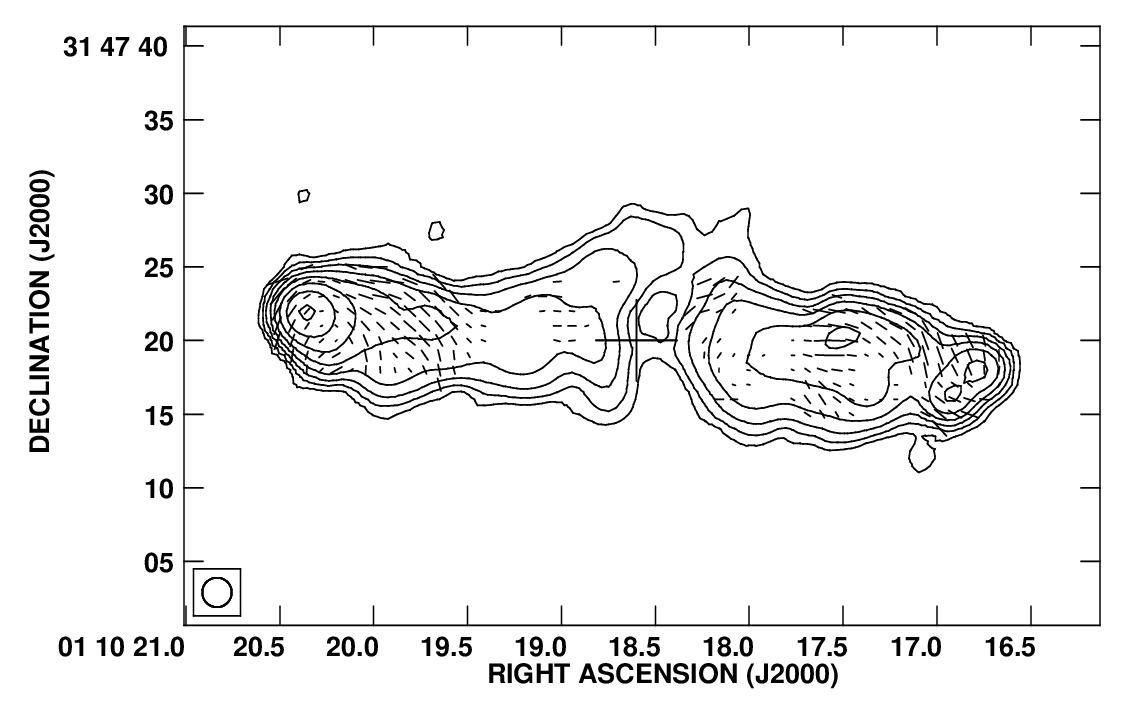}{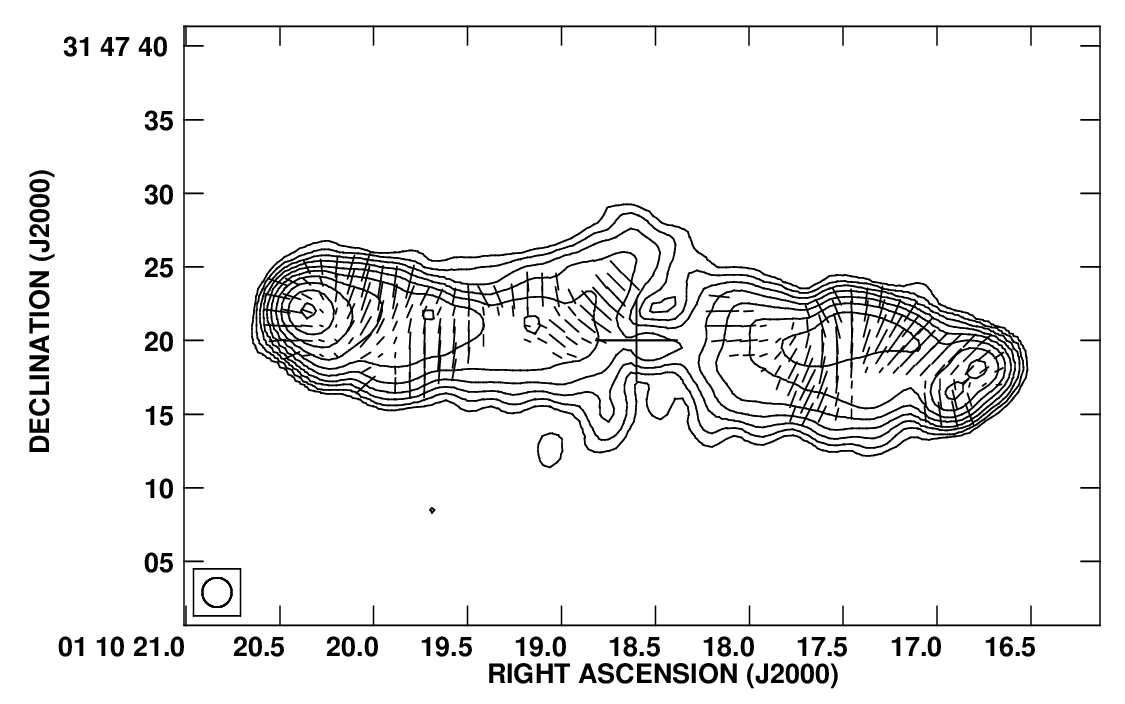}
\caption{Total intensity image of 3C34 at (left) 1.66~GHz and (right) 4.86~GHz,
at a resolution of 2$\arcsec$ with fractional polarization vectors 
superimposed. The peak surface brightness and contour levels increasing in 
steps of 2 are 
(left) 135.5 mJy/beam, (--0.7,0.7,...,90)\% of peak brightness,  
1$\arcsec$ vector = 20\% polarization, 
(right) 57.8 mJy/beam, (--0.17,0.17,...,90)\% of peak brightness, 
1$\arcsec$ vector = 17\% polarization. }
\label{fig3c34}
\end{figure}

\begin{figure}
\epsscale{0.5}
\plotone{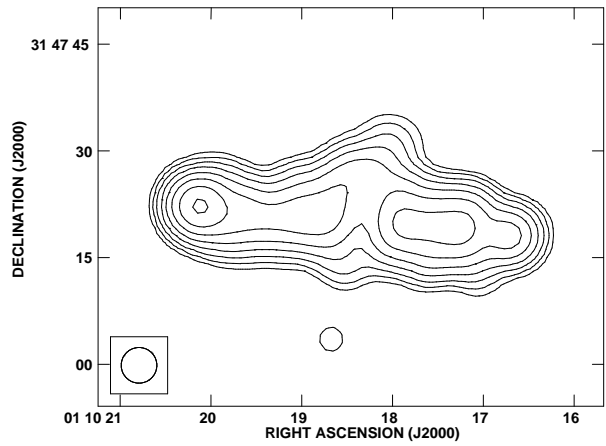}
\caption{Total intensity image of 3C34 at 327.6~MHz. 
The peak surface brightness 
and contour levels increasing in steps of 2 are 
913.8 mJy/beam, (--0.7,0.7,...,90)\% of peak brightness}
\label{fig3c34px}
\end{figure}

\begin{figure}
\epsscale{1.0}
\plotone{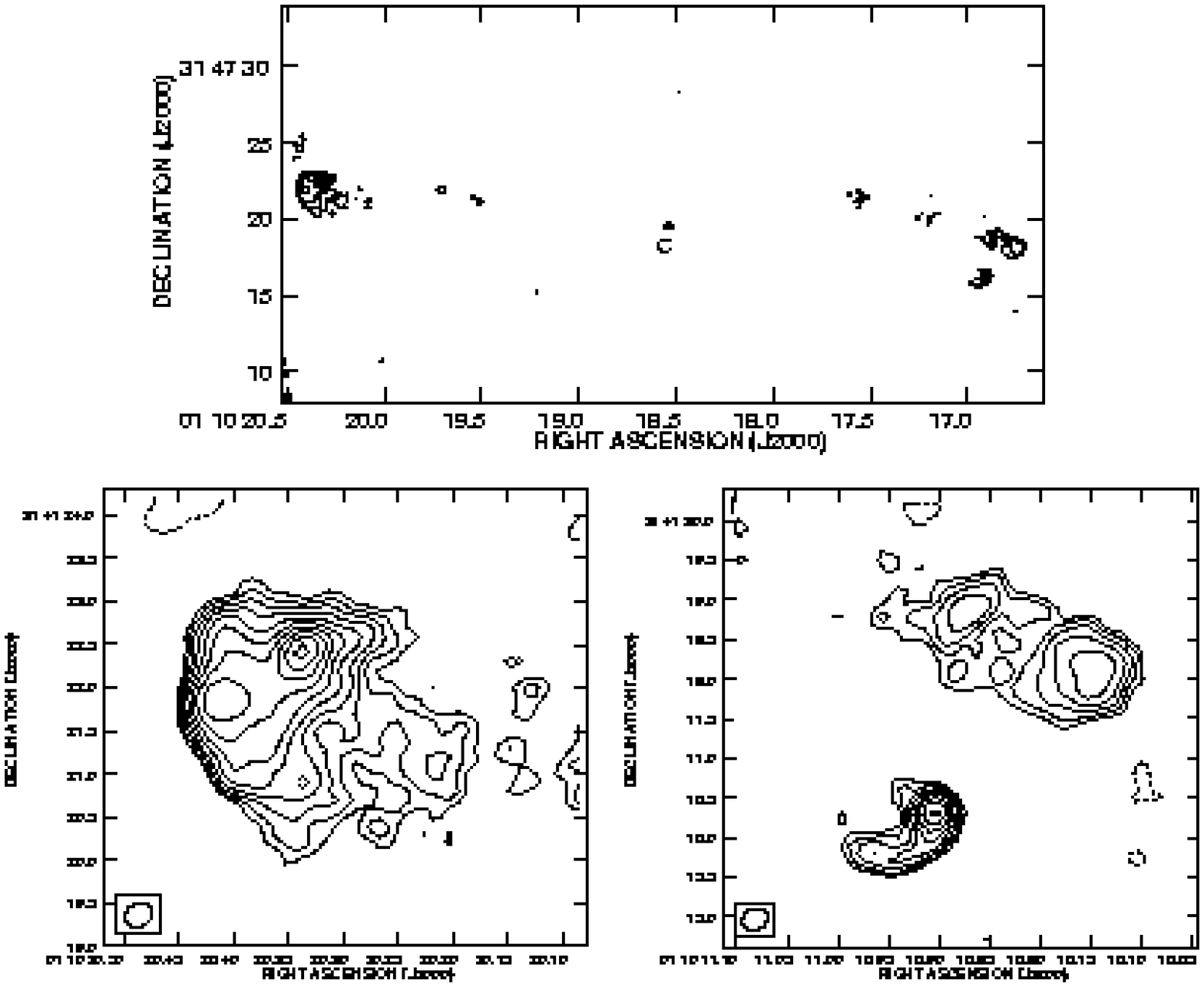}
\caption{Total intensity image of 3C34 at 8.4~GHz. The peak surface
brightness and contour levels increasing in steps of 2 
($\sqrt{2}$ for the closeup views of the hotspot regions)
are 3.4 mJy/beam, (--5.6,5.6,...,90)\% of peak brightness. The
lowest contour level in the hotspot blowup is 4\% of the peak brightness.}
\end{figure}

\clearpage

\begin{figure}
\epsscale{1.0}
\plottwo{f21a.ps}{f21b.ps}
\caption{Total intensity image of 3C41 at (left) 1.34~GHz and (right) 4.86~GHz,
at a resolution of 2$\arcsec$ with fractional polarization vectors 
superimposed. The peak surface brightness and contour levels increasing in 
steps of 2 are 
(left) 977.0 mJy/beam, (--0.17,0.17,...,90)\% of peak brightness,  
(right) 502.6 mJy/beam, (--0.085,0.085,...,90)\% of peak brightness, 
1$\arcsec$ vector = 10\% polarization. }
\label{fig3c41}
\end{figure}

\begin{figure}
\epsscale{1.0}
\plotone{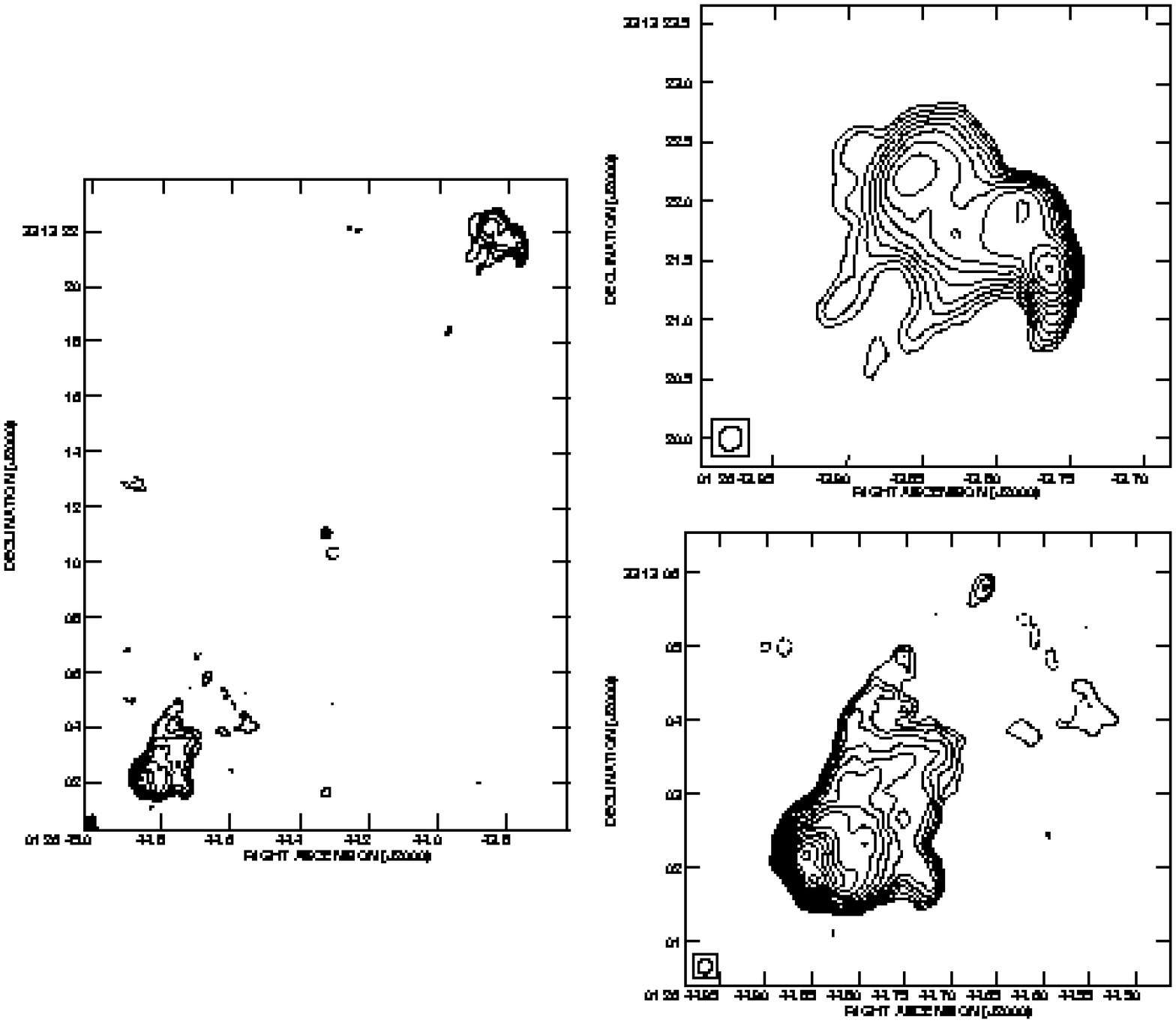}
\caption{Total intensity image of 3C41 at 8.4~GHz. The peak surface brightness
and contour levels increasing in steps of 2 
($\sqrt{2}$ for the closeup views of the hotspot regions)
are 38.2 mJy/beam and (--0.7,0.7,...,90)\% of peak brightness, respectively.
The lowest contour level in the hotspot blowup is 0.5\% of the peak 
brightness.}
\end{figure}

\clearpage

\begin{figure}
\epsscale{1.0}
\plottwo{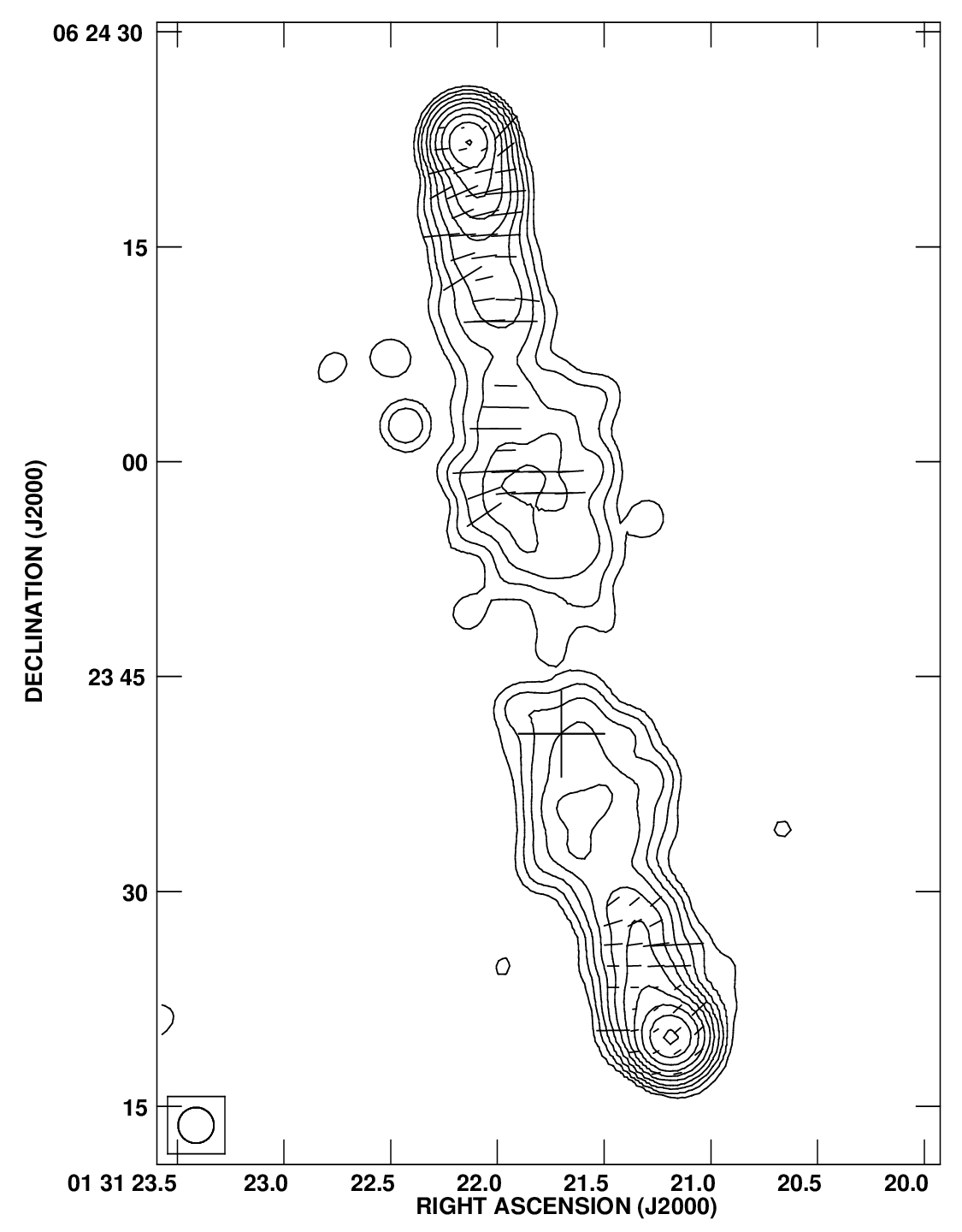}{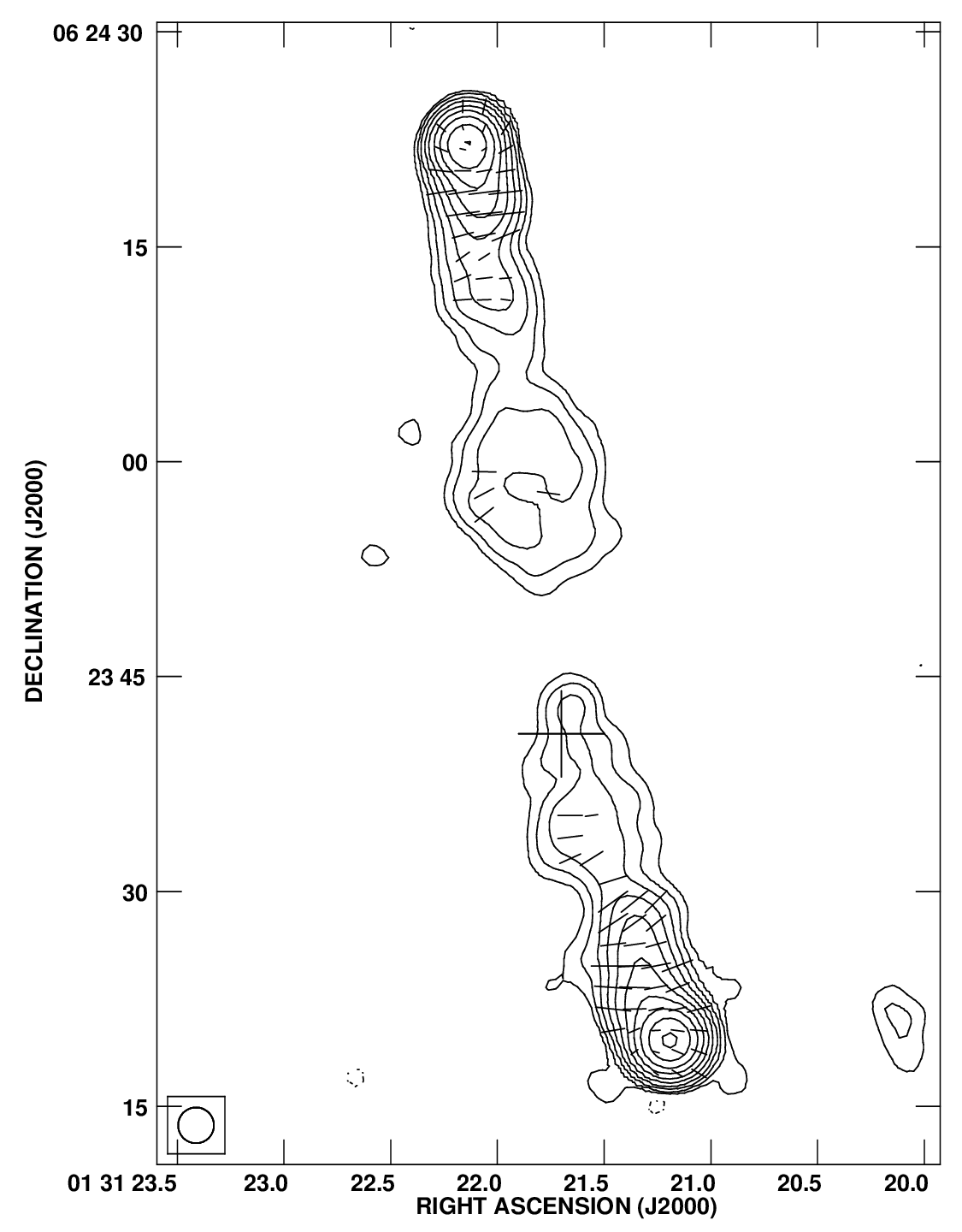}
\caption{Total intensity image of 3C44 at (left) 1.66~GHz and (right) 4.86~GHz,
at a resolution of 2.5$\arcsec$ with fractional polarization vectors 
superimposed. The peak surface brightness and contour levels increasing in 
steps of 2 are 
(left) 432.1 mJy/beam, (--0.17,0.17,...,90)\% of peak brightness,  
1$\arcsec$ vector = 13\% polarization,
(right) 136.1 mJy/beam, (--0.17,0.17,...,90)\% of peak brightness, 
1$\arcsec$ vector = 10\% polarization. }
\label{fig3c44}
\end{figure}

\begin{figure}
\epsscale{0.5}
\plotone{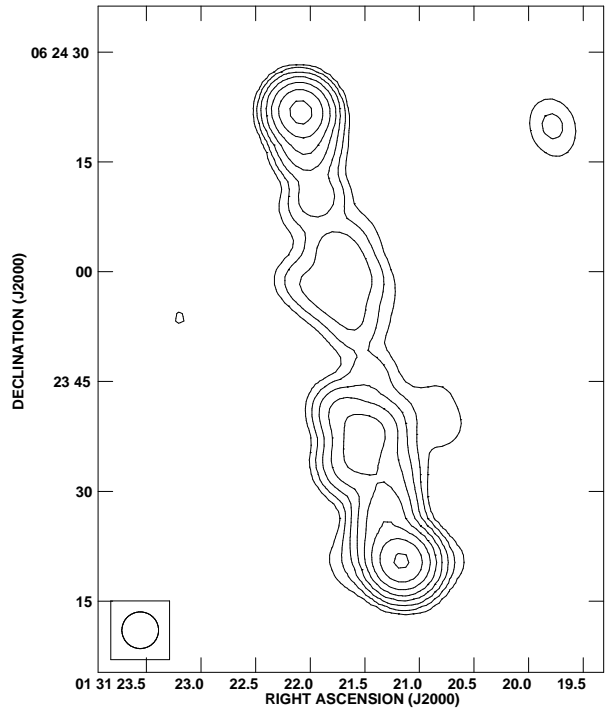}
\caption{Total intensity image of 3C44 at 327.6~MHz. The peak surface brightness 
and contour levels increasing in steps of 2 are 
1.6 Jy/beam, (--0.7,0.7,...,90)\% of peak brightness.}
\end{figure}

\begin{figure}
\epsscale{1.0}
\plotone{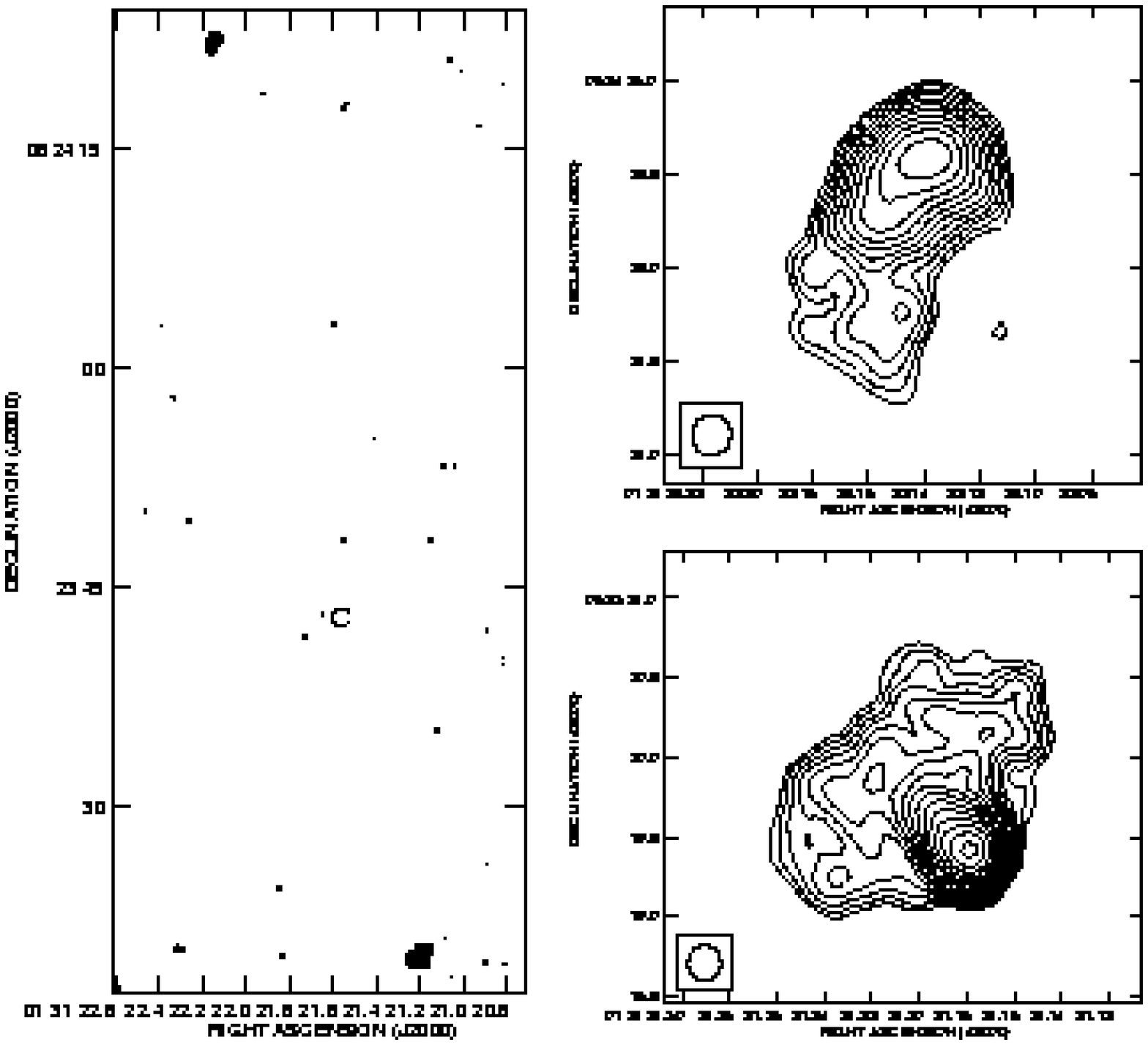}
\caption{Total intensity image of 3C44 at 8.4~GHz. The peak surface brightness
and contour levels increasing in steps of 2 
($\sqrt{2}$ for the closeup views of the hotspot regions)
are 24.3 mJy/beam, (--0.7,0.7,...,90)\% of peak brightness.
The lowest contour level in the hotspot blowup is 0.5\% of the peak 
brightness.}
\end{figure}

\clearpage

\begin{figure}
\epsscale{1.0}
\plottwo{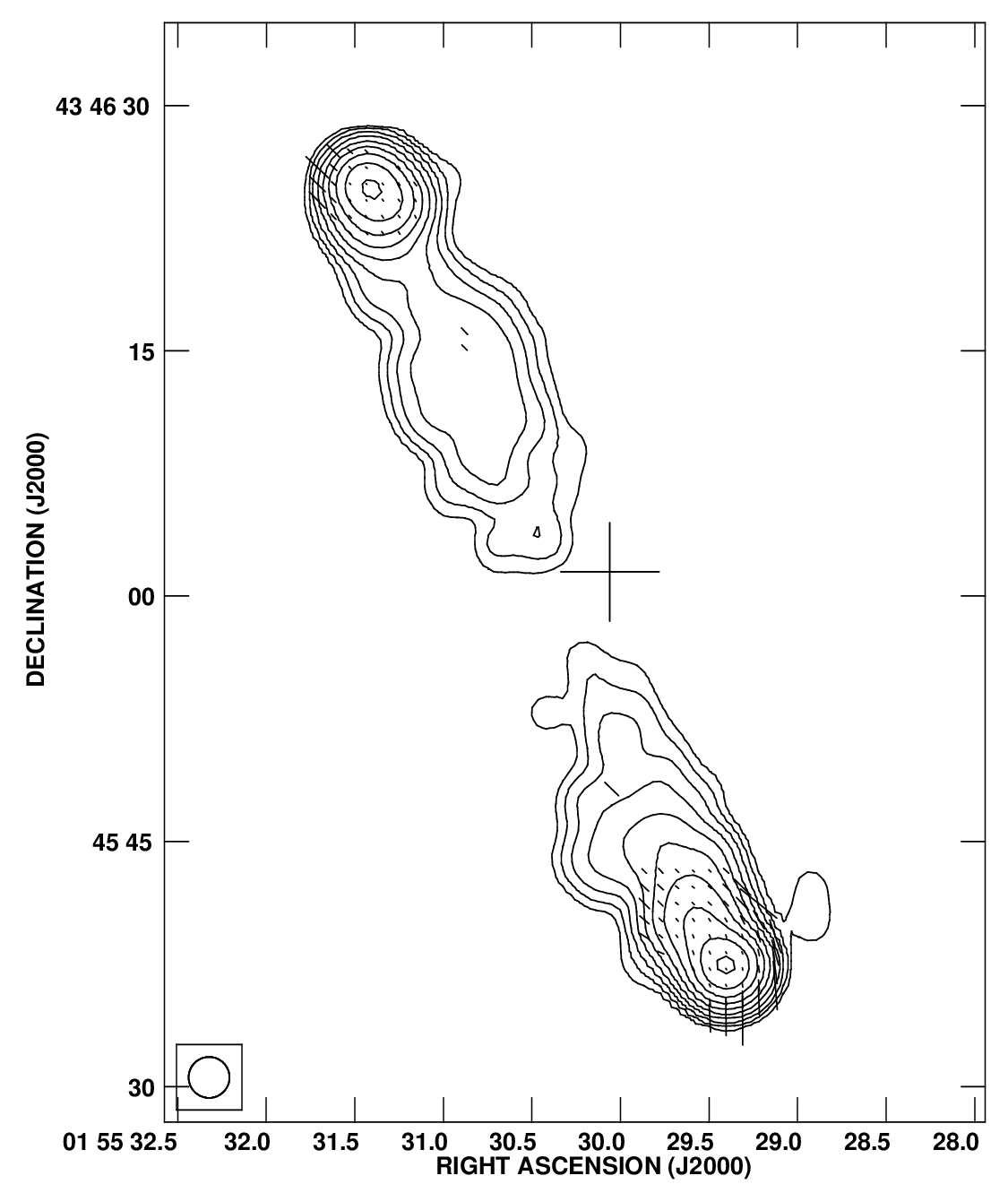}{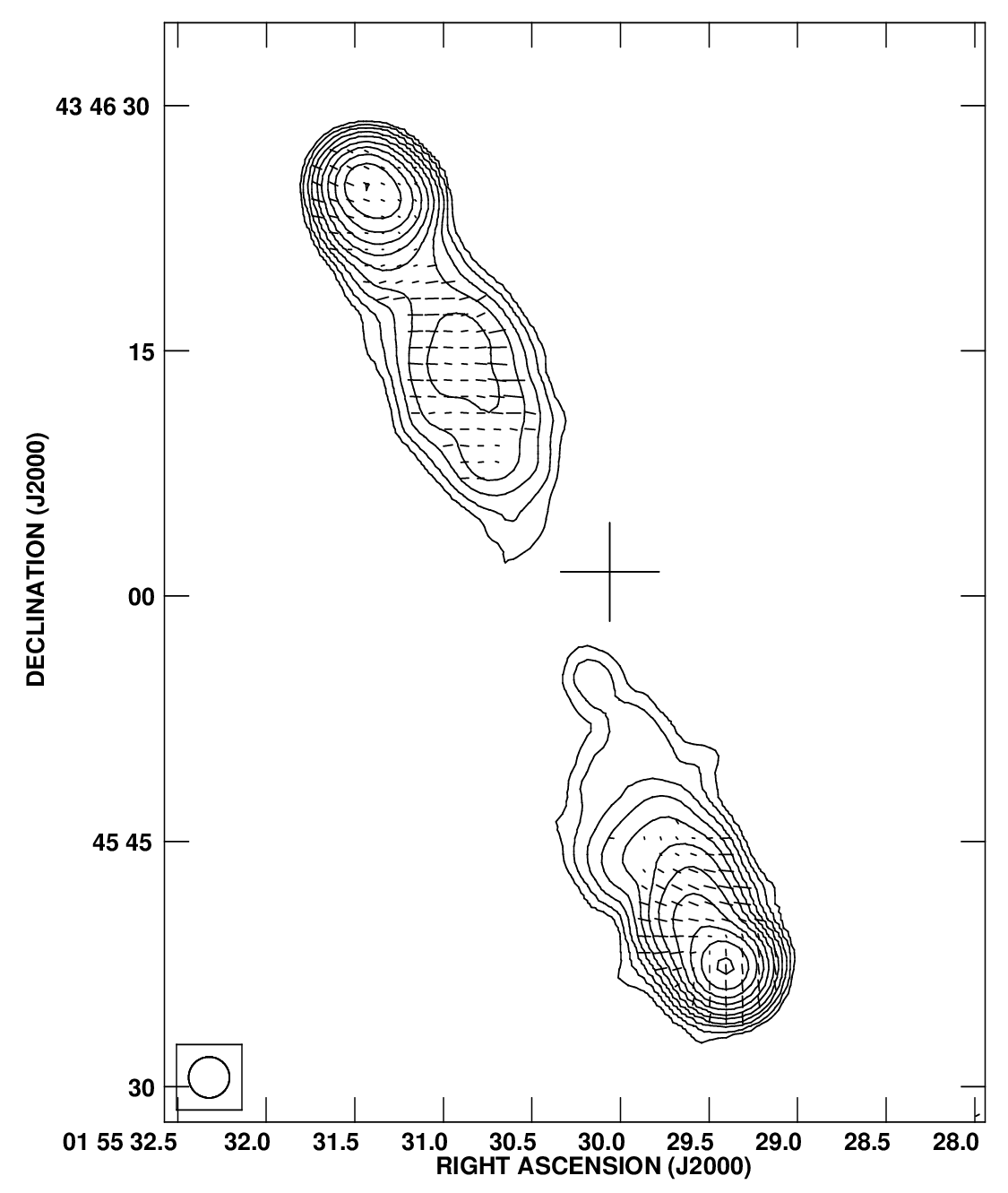}
\caption{Total intensity image of 3C54 at (left) 1.34~GHz and (right) 4.86~GHz,
at a resolution of 2.5$\arcsec$ with fractional polarization vectors 
superimposed. The peak surface brightness and contour levels increasing in 
steps of 2 are 
(left) 581.9 mJy/beam, (--0.17,0.17,...,90)\% of peak brightness,  
1$\arcsec$ vector = 25\% polarization,
(right) 227.8 mJy/beam, (--0.085,0.085,...,90)\% of peak brightness, 
1$\arcsec$ vector = 25\% polarization. }
\label{fig3c54}
\end{figure}

\begin{figure}
\epsscale{0.5}
\plotone{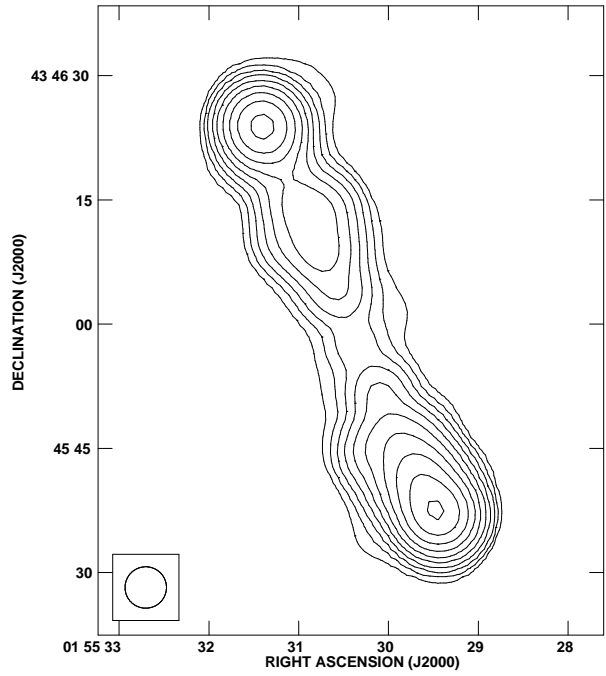}
\caption{Total intensity image of 3C54 at 327.6~MHz. The peak surface brightness 
and contour levels increasing in steps of 2 are 
2.1 Jy/beam, (--0.17,0.17,...,90)\% of peak brightness}
\end{figure}

\begin{figure}
\epsscale{1.0}
\plotone{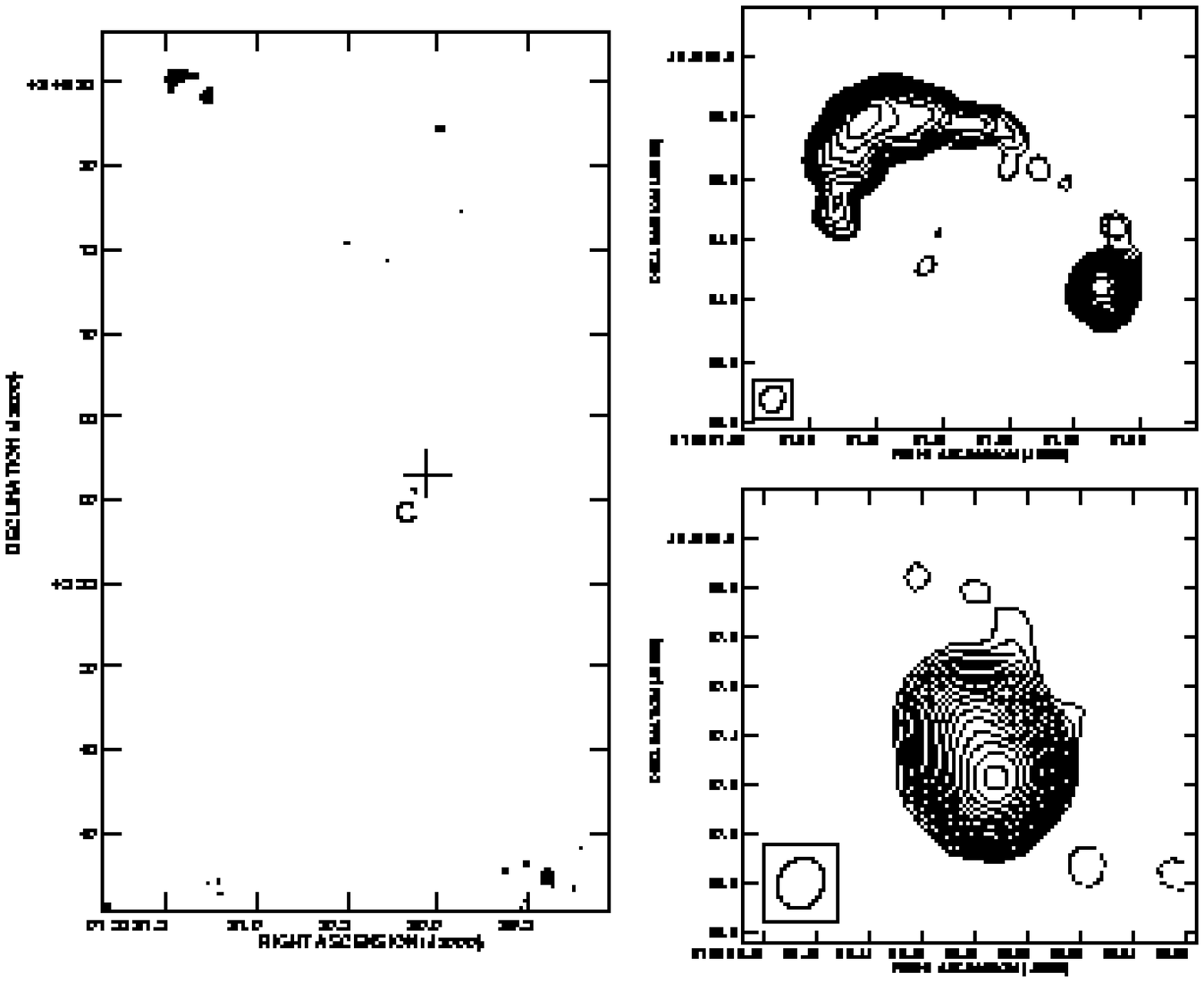}
\caption{Total intensity image of 3C54 at 8.4~GHz. The peak surface brightness
and contour levels increasing in steps of 2 
($\sqrt{2}$ for the closeup views of the hotspot regions)
are 69.6 mJy/beam, (--0.35,0.35,...,90)\% of peak brightness.
The lowest contour level in the hotspot blowup is 0.25\% of the peak 
brightness.}
\end{figure}

\clearpage

\begin{figure}
\epsscale{1.0}
\plottwo{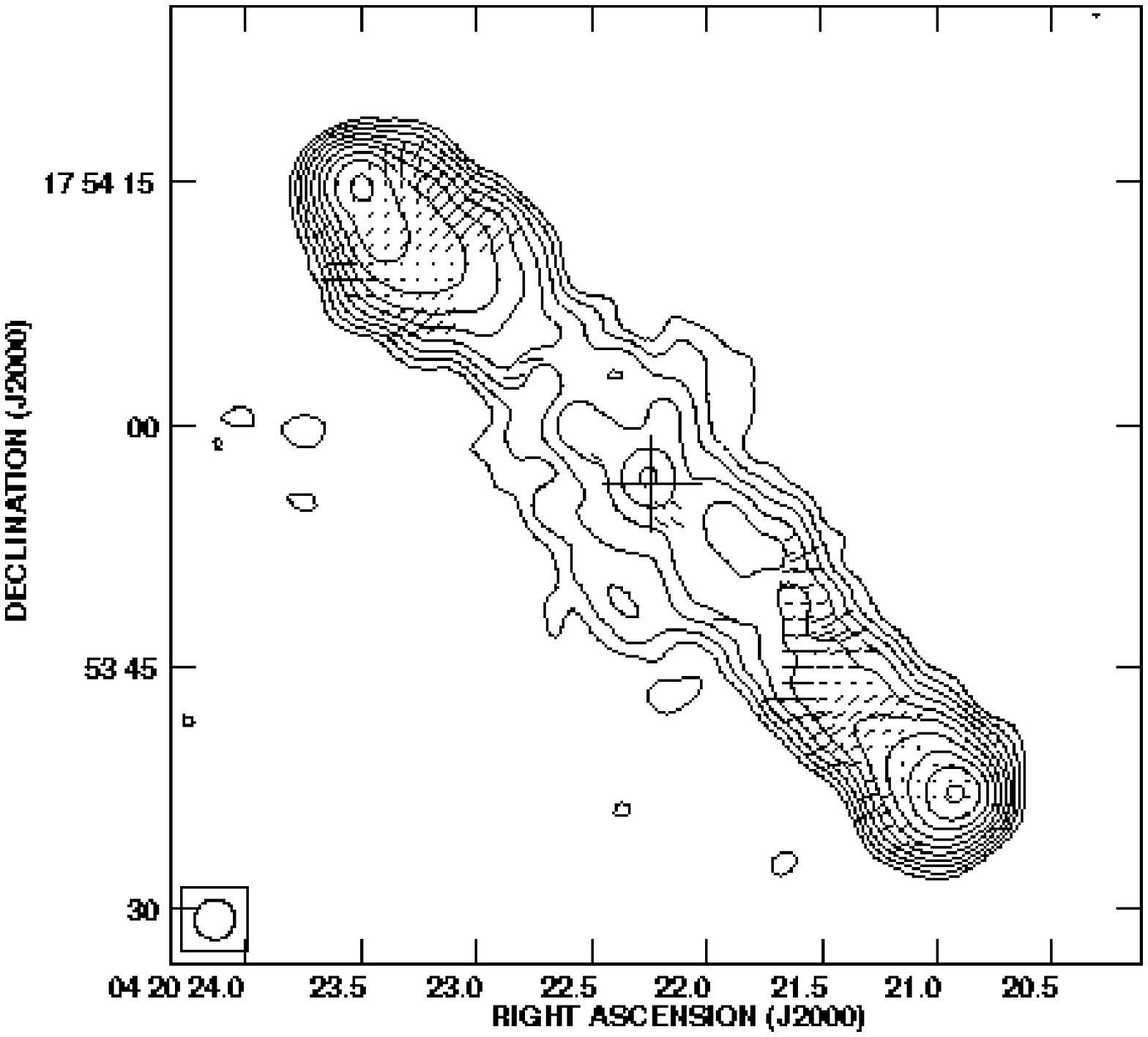}{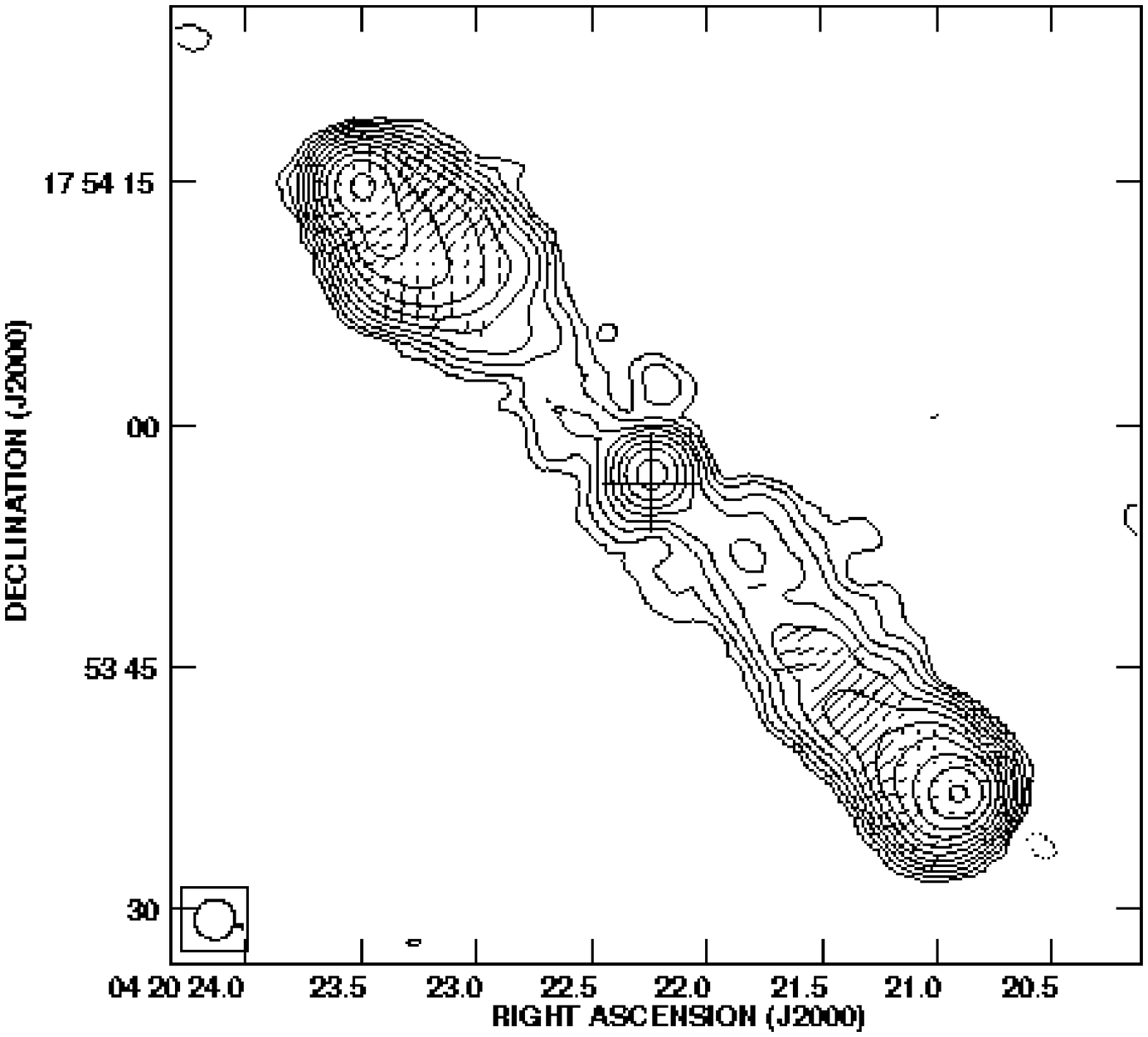}
\caption{Total intensity image of 3C114 at (left) 1.66~GHz and (right) 4.86~GHz,
at a resolution of 2.5$\arcsec$ with fractional polarization vectors 
superimposed. The peak surface brightness and contour levels increasing in 
steps of 2 are 
(left) 203.7 mJy/beam, (--0.17,0.17,...,90)\% of peak brightness,  
1$\arcsec$ vector = 17\% polarization,
(right) 69.5 mJy/beam, (--0.085,0.085,...,90)\% of peak brightness, 
1$\arcsec$ vector = 17\% polarization. }
\label{fig3c114}
\end{figure}

\begin{figure}
\epsscale{0.5}
\plotone{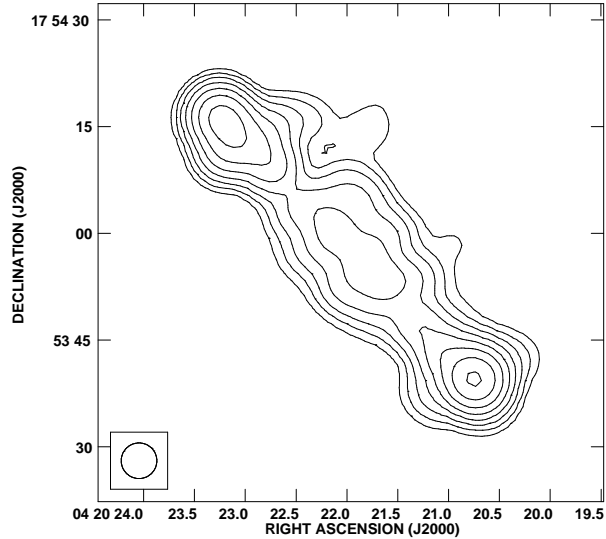}
\caption{Total intensity image of 3C114 at 327.6~MHz. The peak surface brightness 
and contour levels increasing in steps of 2 are 
898.5 mJy/beam, (--0.7,0.7,...,90)\% of peak brightness}
\end{figure}

\begin{figure}
\epsscale{1.0}
\plotone{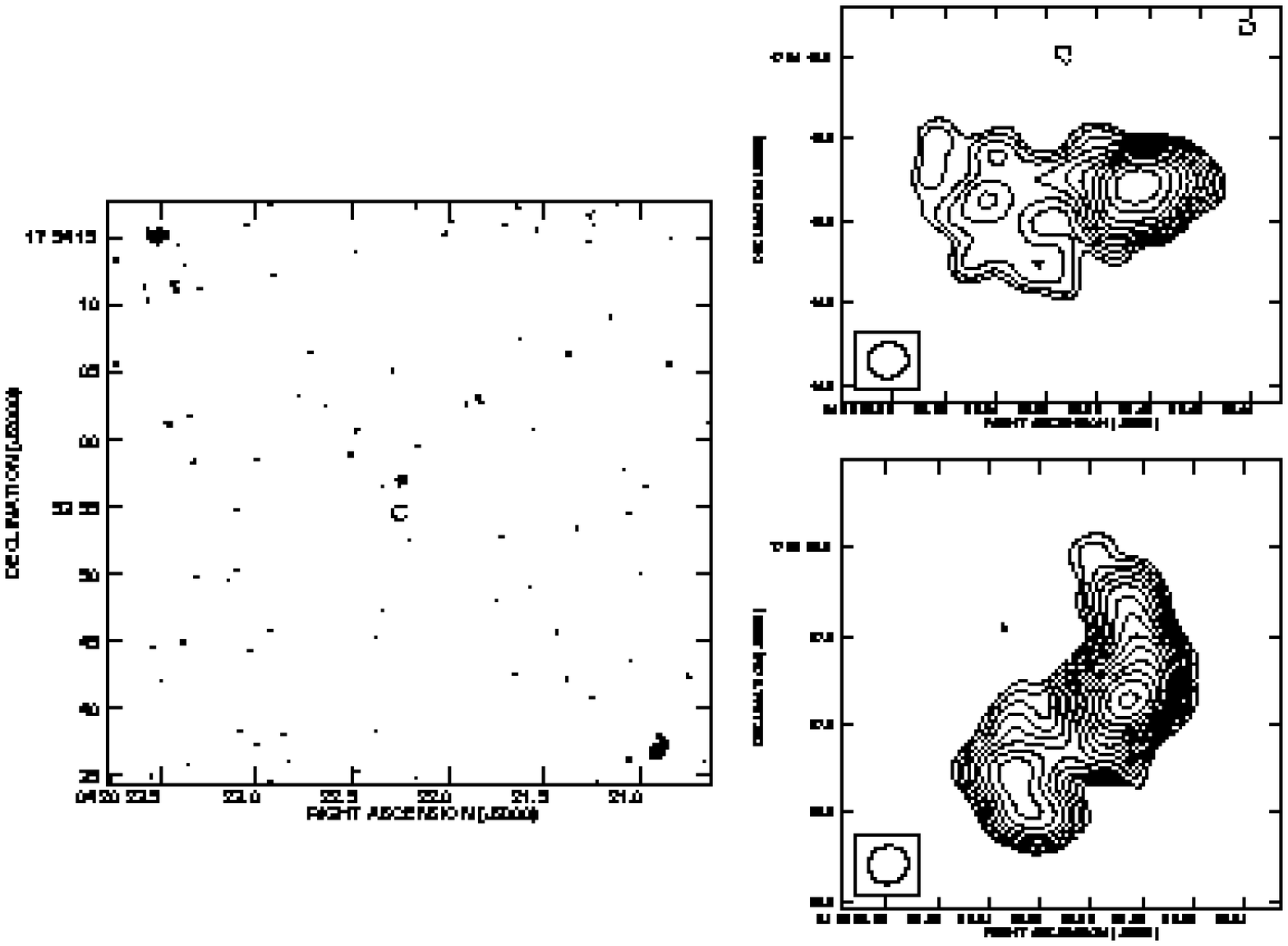}
\caption{Total intensity image of 3C114 at 8.4~GHz. The peak surface brightness
and contour levels increasing in steps of 2 
($\sqrt{2}$ for the closeup views of the hotspot regions)
are 7.3 mJy/beam, (--1.4,1.4,...,90)\% of peak brightness.
The lowest contour level in the hotspot blowup is 1\% of the peak 
brightness.}
\end{figure}

\clearpage

\begin{figure}
\epsscale{1.0}
\plottwo{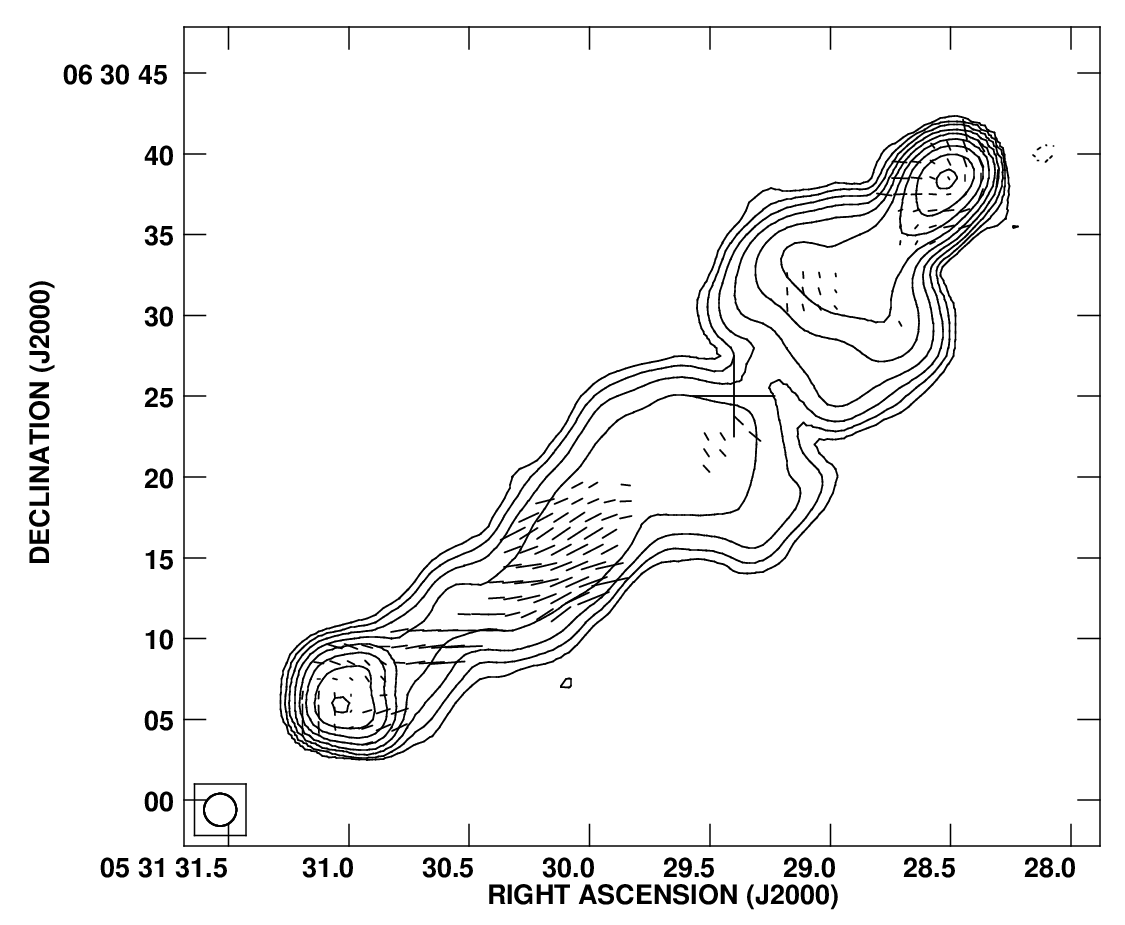}{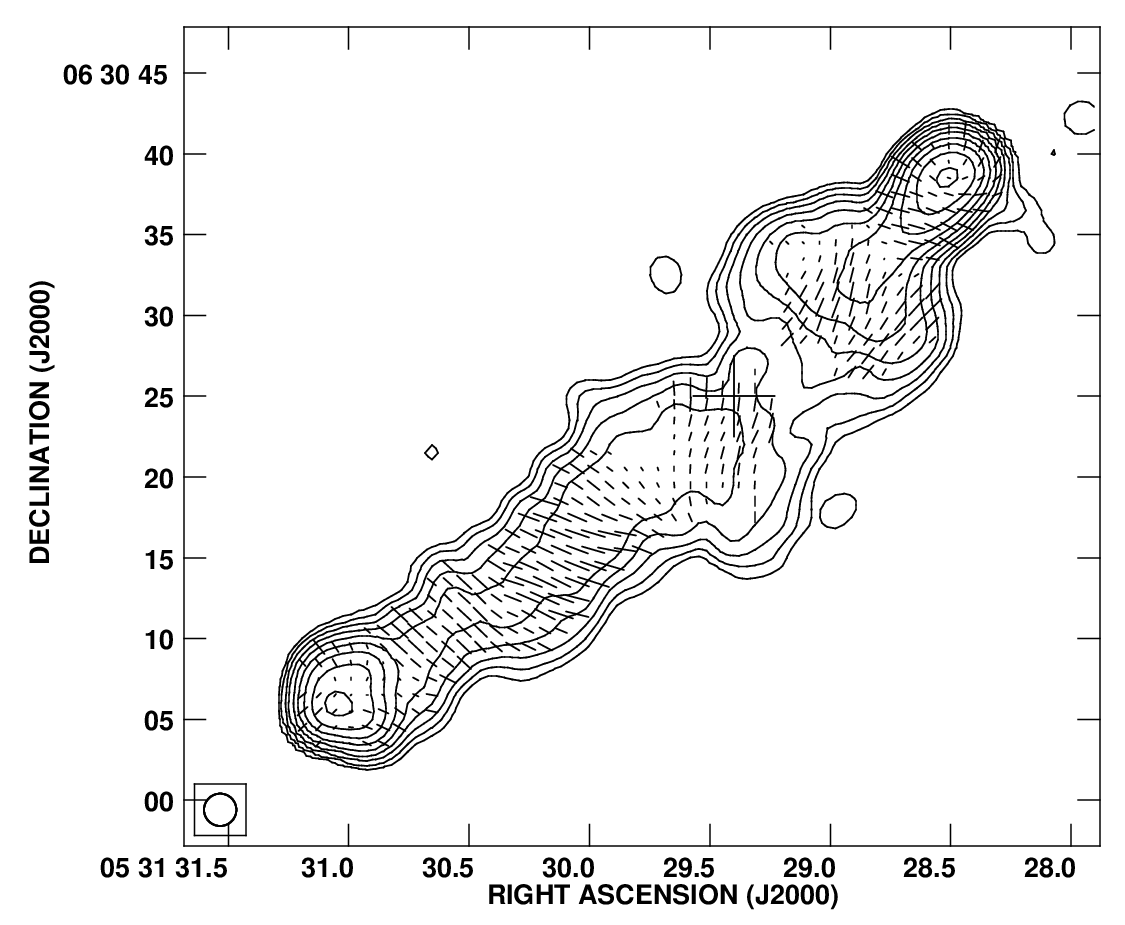}
\caption{Total intensity image of 3C142.1 at (left) 1.34~GHz and (right) 4.86~GHz,
at a resolution of 2$\arcsec$ with fractional polarization vectors 
superimposed. The peak surface brightness and contour levels increasing in 
steps of 2 are 
(left) 378.8 mJy/beam, (--0.35,0.35,...,90)\% of peak brightness,  
1$\arcsec$ vector = 20\% polarization,
(right) 123.0 mJy/beam, (--0.17,0.17,...,90)\% of peak brightness, 
1$\arcsec$ vector = 25\% polarization. }
\label{fig3c142.1}
\end{figure}

\begin{figure}
\epsscale{0.5}
\plotone{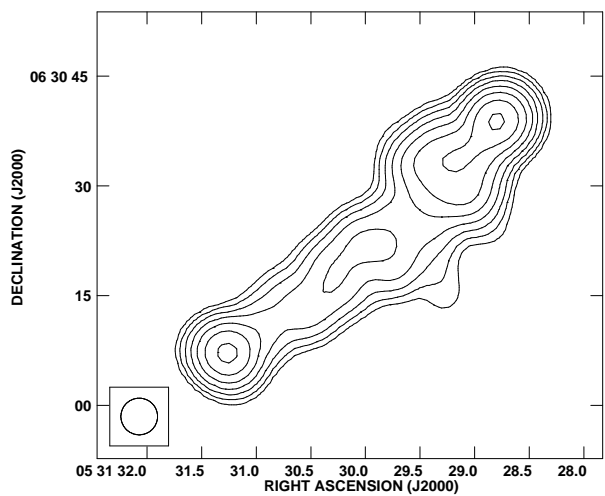}
\caption{Total intensity image of 3C142.1 at 327.6~MHz. The peak surface 
brightness and contour levels increasing in steps of 2 are 
2.6 Jy/beam, (--0.7,0.7,...,90)\% of peak brightness}
\end{figure}

\begin{figure}
\epsscale{1.0}
\plotone{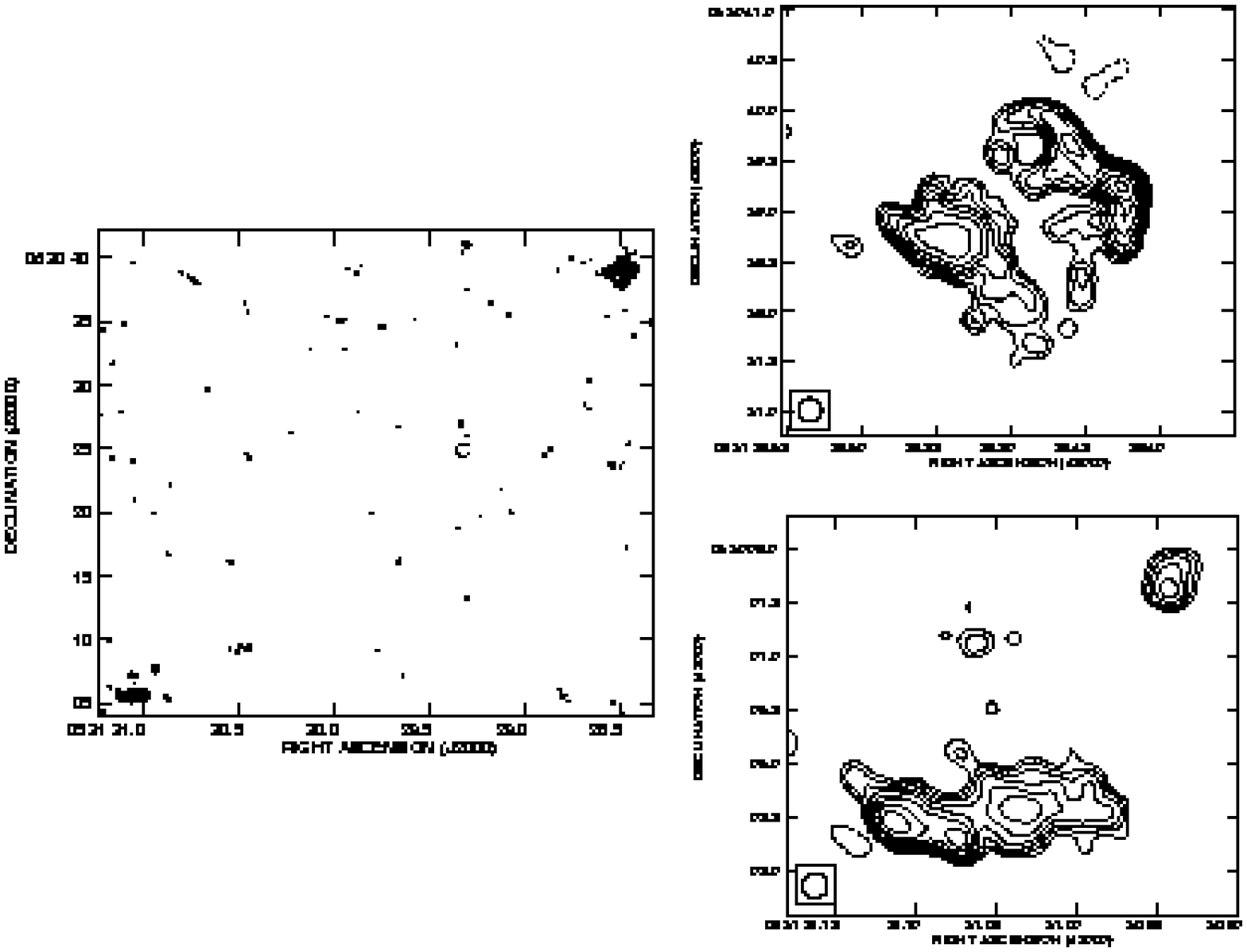}
\caption{Total intensity image of 3C142.1 at 8.4~GHz. The peak surface brightness
and contour levels increasing in steps of 2 
($\sqrt{2}$ for the closeup views of the hotspot regions)
are 2.1 mJy/beam, (--5.6,5.6,...,90)\% of peak brightness.}
\end{figure}

\clearpage

\begin{figure}
\plottwo{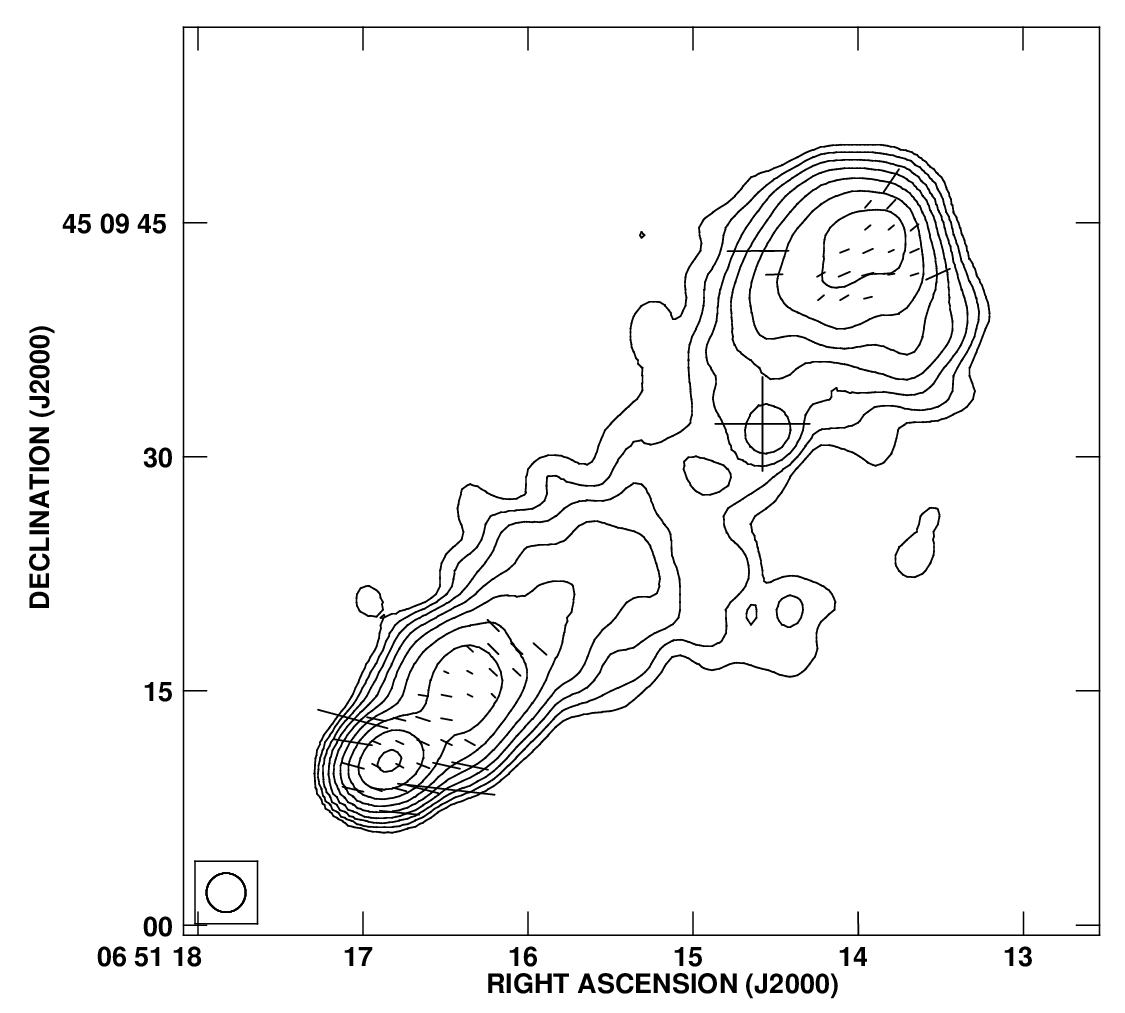}{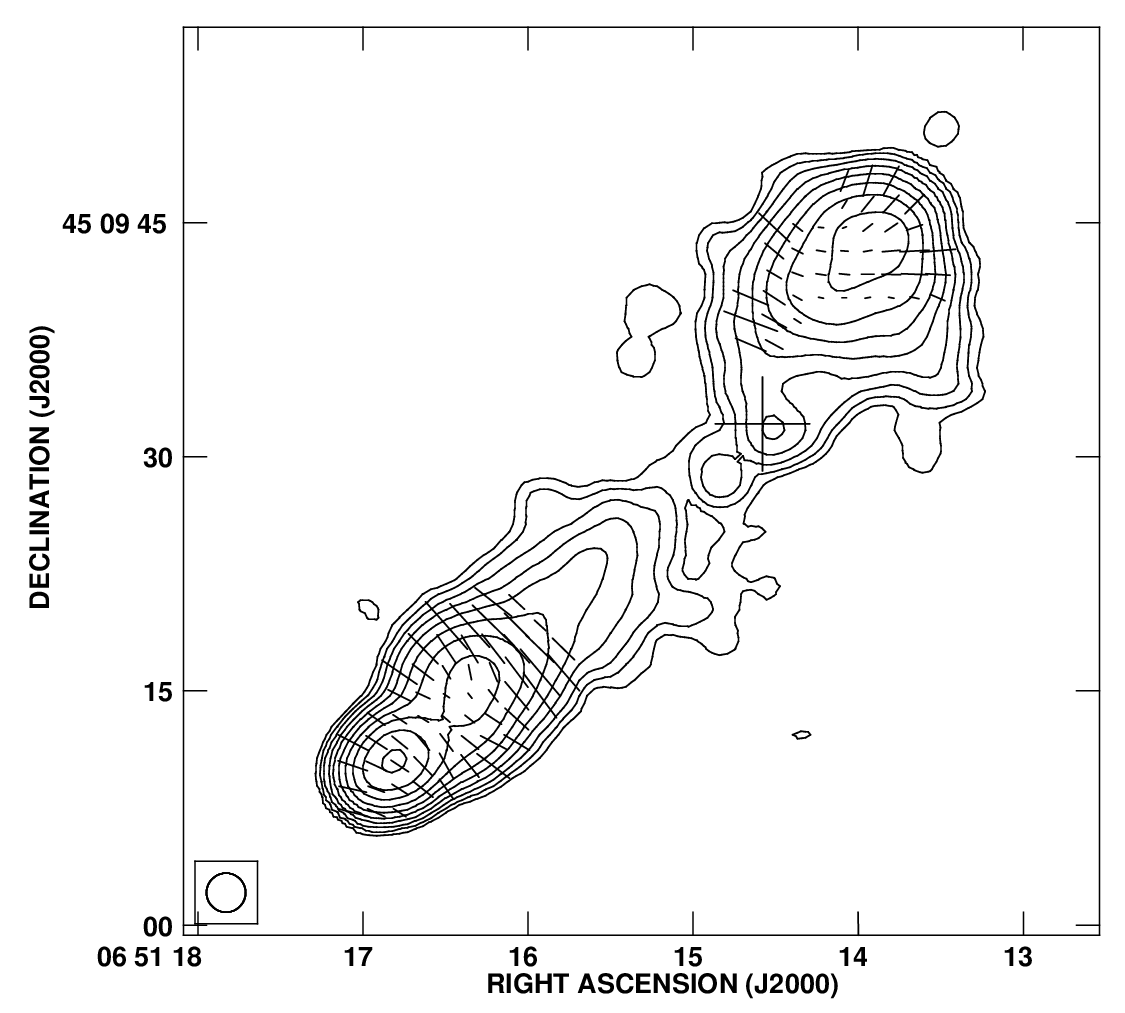}
\caption{Total intensity image of 3C169.1 at (left) 1.66~GHz and (right) 4.86~GHz,
at a resolution of 2.5$\arcsec$ with fractional polarization vectors 
superimposed. The peak surface brightness and contour levels increasing in 
steps of 2 are 
(left) 153.2 mJy/beam, (--0.35,0.35,...,90)\% of peak brightness,  
1$\arcsec$ vector = 10\% polarization,
(right) 56.7 mJy/beam, (--0.17,0.17,...,90)\% of peak brightness, 
1$\arcsec$ vector = 10\% polarization. }
\label{fig3c169.1}
\end{figure}

\begin{figure}
\epsscale{0.5}
\plotone{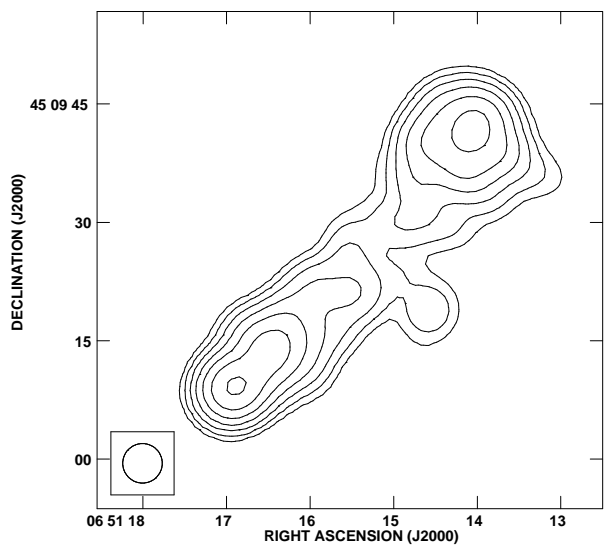}
\caption{Total intensity image of 3C169.1 at 327.6~MHz.
The peak surface brightness 
and contour levels increasing in steps of 2 are 
744.1 Jy/beam, (--1.4,1.4,...,90)\% of peak brightness.}
\end{figure}

\begin{figure}
\epsscale{1.0}
\plotone{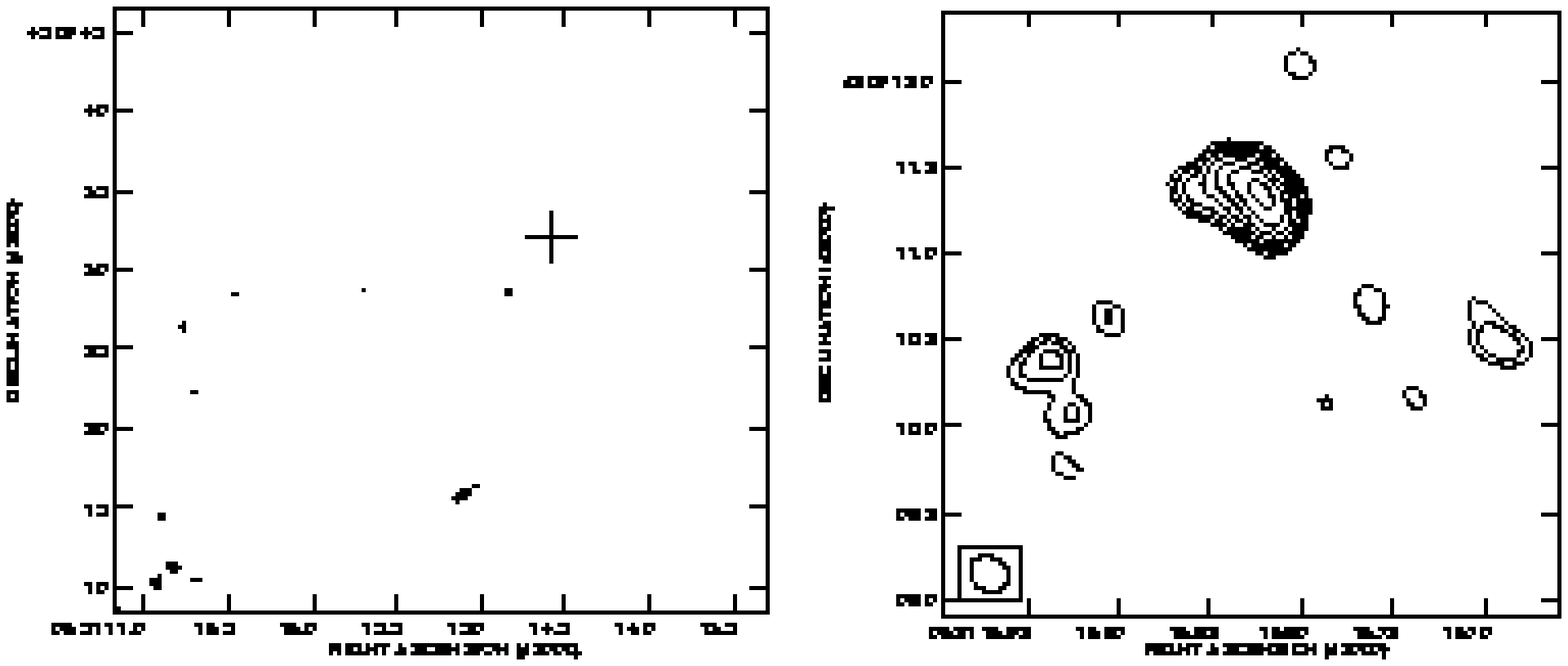}
\caption{Total intensity image of 3C169.1 at 8.4~GHz.
The peak surface brightness and contour levels increasing in steps of 2 
($\sqrt{2}$ for the closeup views of the hotspot regions)
are 1.5 mJy/beam, (--16,16,32,64)\% of peak brightness.
The lowest contour level in the hotspot blowup is 11.2\% of the peak 
brightness.}
\end{figure}

\clearpage

\begin{figure}
\plottwo{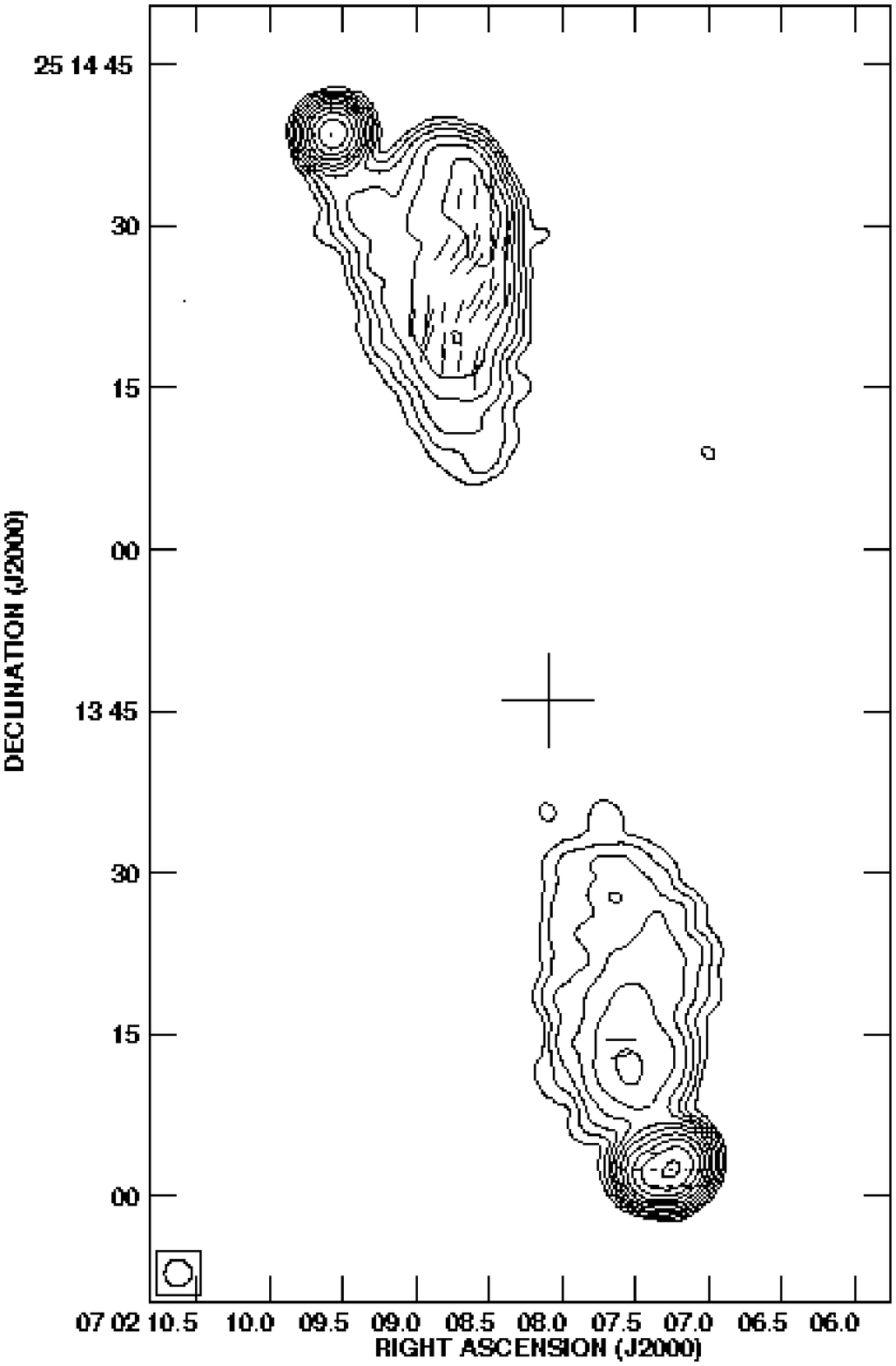}{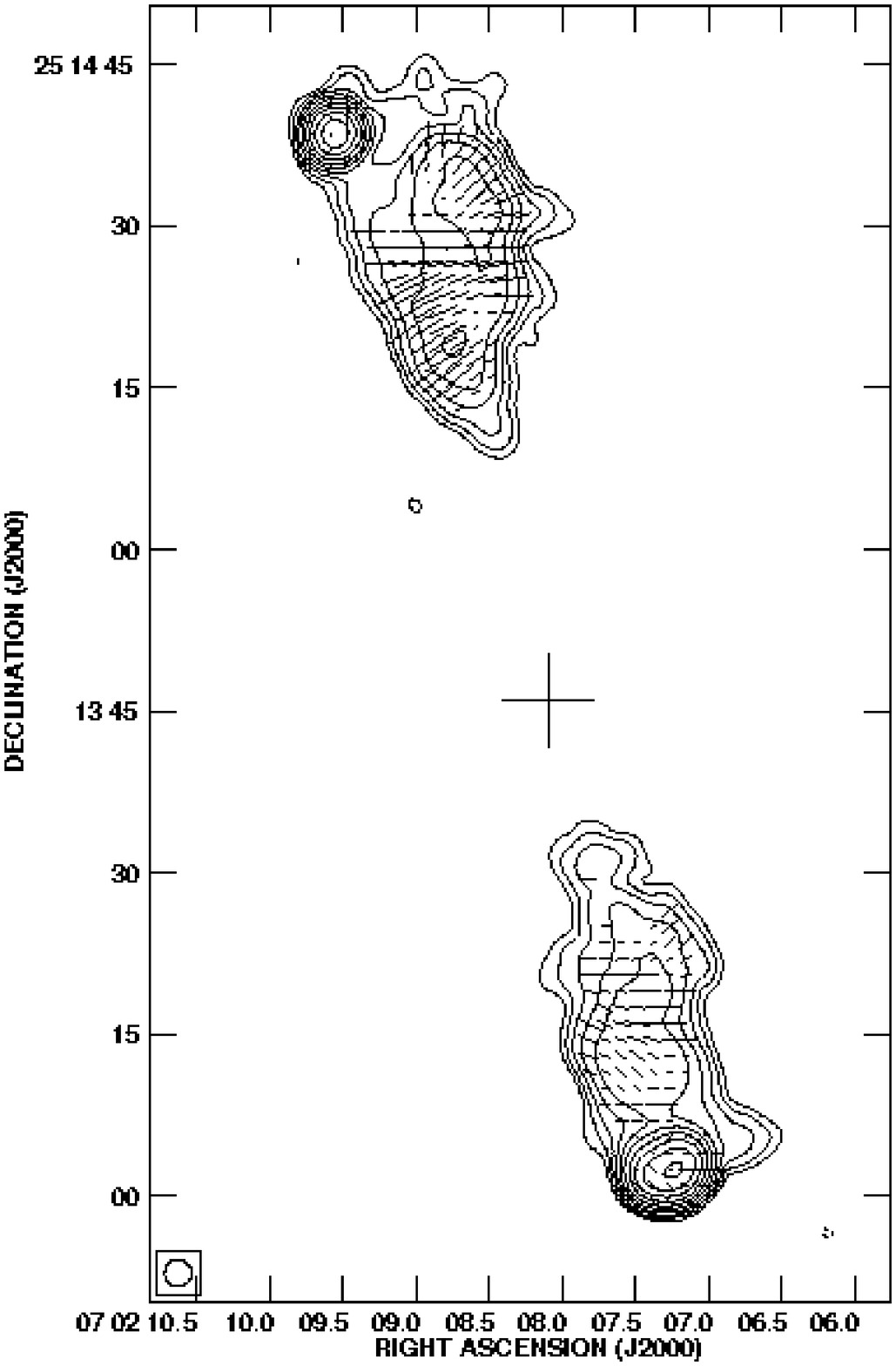}
\caption{Total intensity image of 3C172 at (left) 1.66~GHz and (right) 4.86~GHz,
at a resolution of 2.5$\arcsec$ with fractional polarization vectors 
superimposed. The peak surface brightness and contour levels increasing in 
steps of 2 are 
(left) 478.1 mJy/beam, (--0.17,0.17,...,90)\% of peak brightness,  
1$\arcsec$ vector = 10\% polarization,
(right) 153.7 mJy/beam, (--0.17,0.17,...,90)\% of peak brightness, 
1$\arcsec$ vector = 11\% polarization. }
\label{fig3c172}
\end{figure}

\begin{figure}
\epsscale{0.5}
\plotone{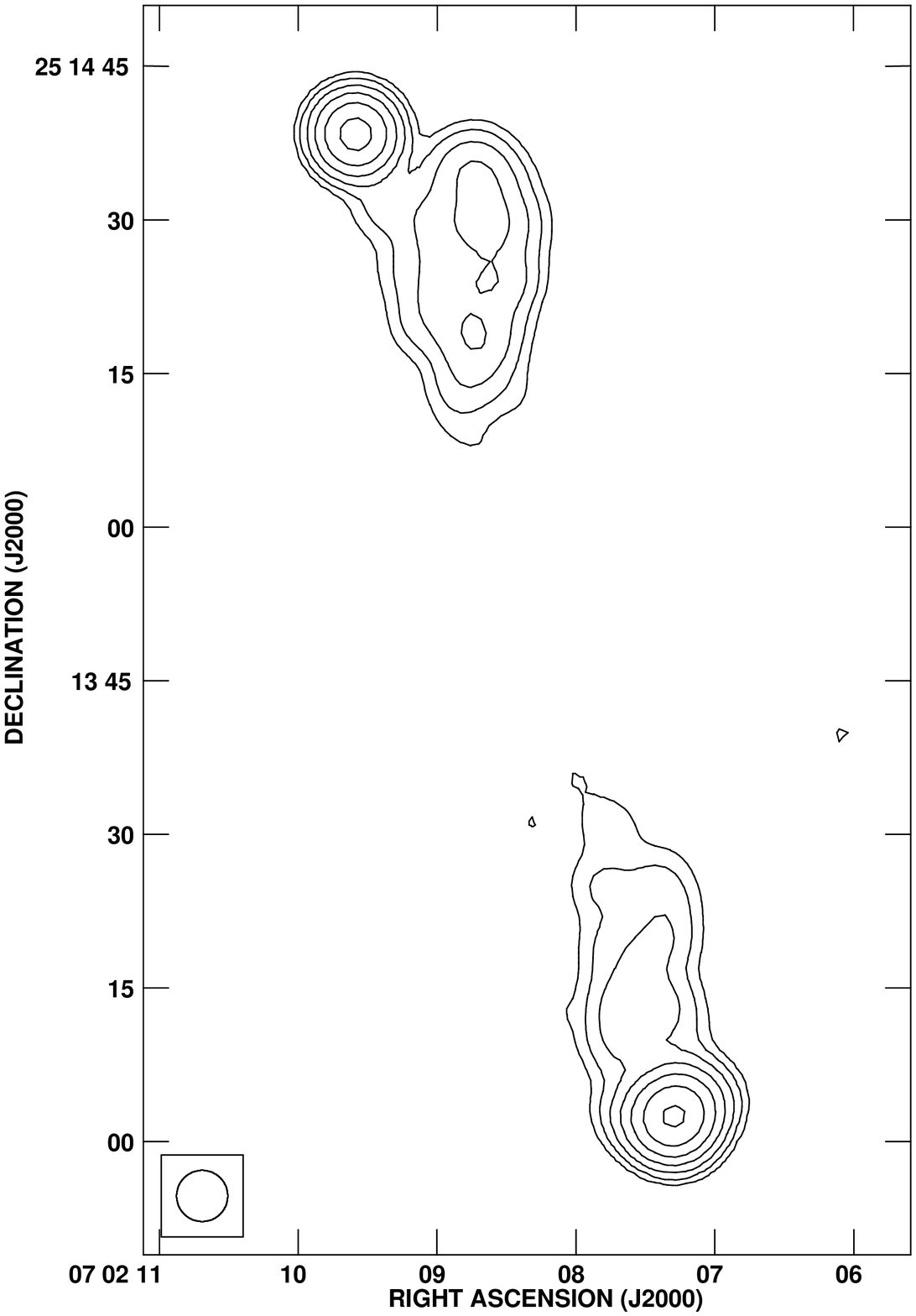}
\caption{Total intensity image of 3C172 at 327.6~MHz. The peak surface brightness 
and contour levels increasing in steps of 2 are 
2.3 Jy/beam, (--1.4,1.4,...,90)\% of peak brightness.}
\end{figure}

\begin{figure}
\epsscale{1.0}
\plotone{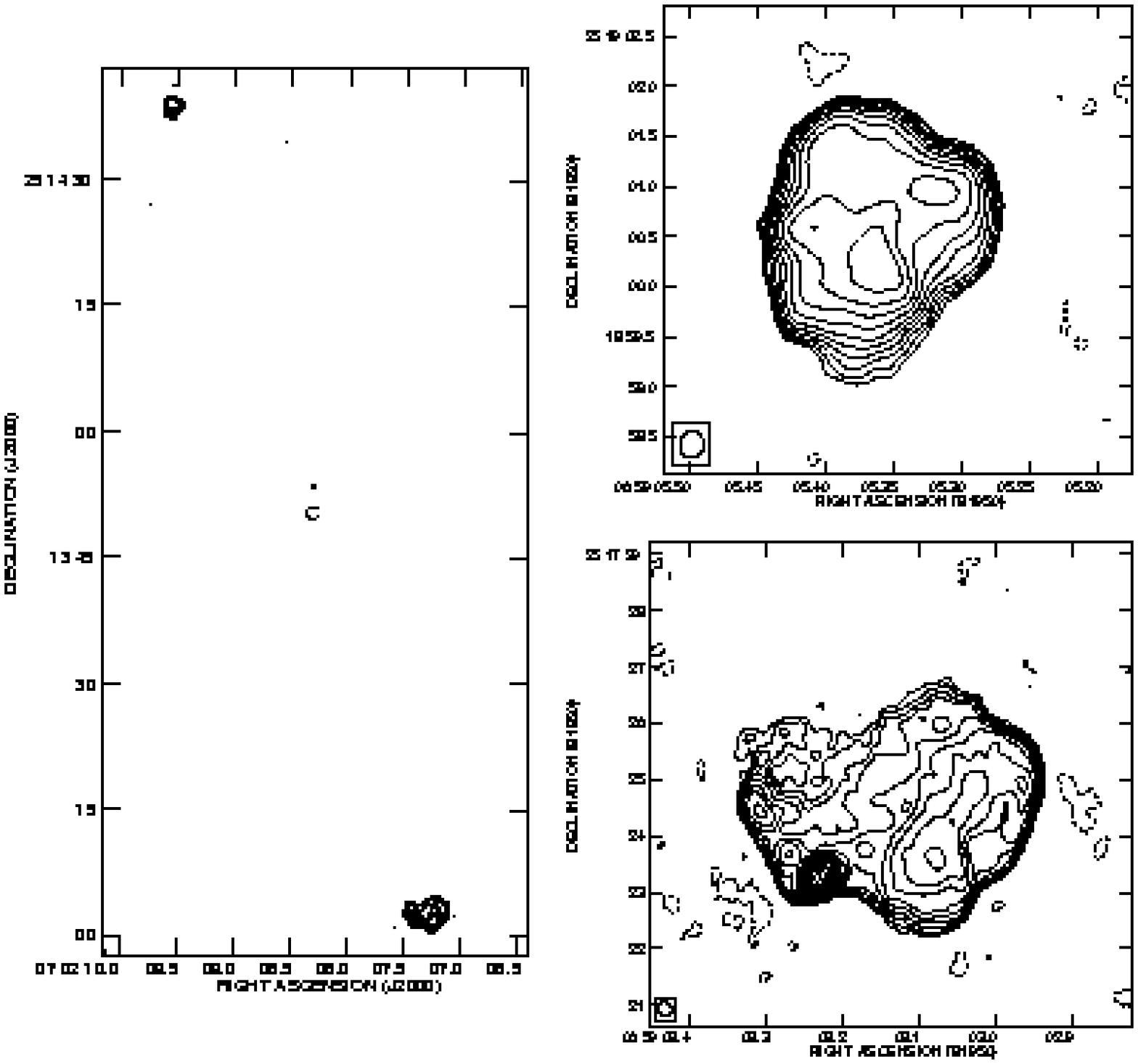}
\caption{Total intensity image of 3C172 at 8.4~GHz. The peak surface brightness
and contour levels increasing in steps of 2 
($\sqrt{2}$ for the closeup views of the hotspot regions)
are 11.9 mJy/beam, (--1.4,1.4,...,90)\% of peak brightness.
The lowest contour level in the hotspot blowup is 1\% of the peak
brightness.}
\end{figure}

\clearpage
\begin{figure}
\epsscale{1.0}
\plottwo{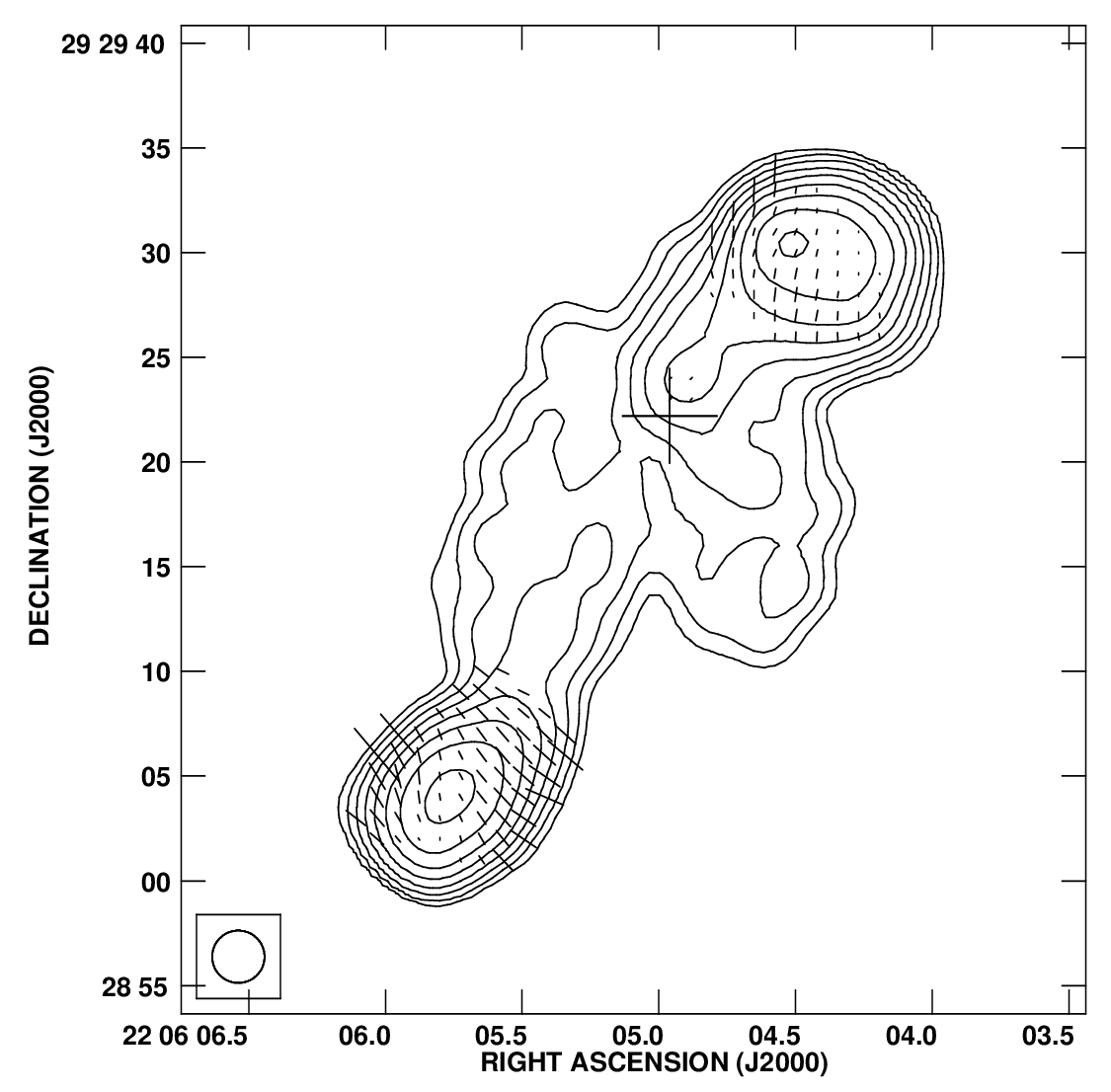}{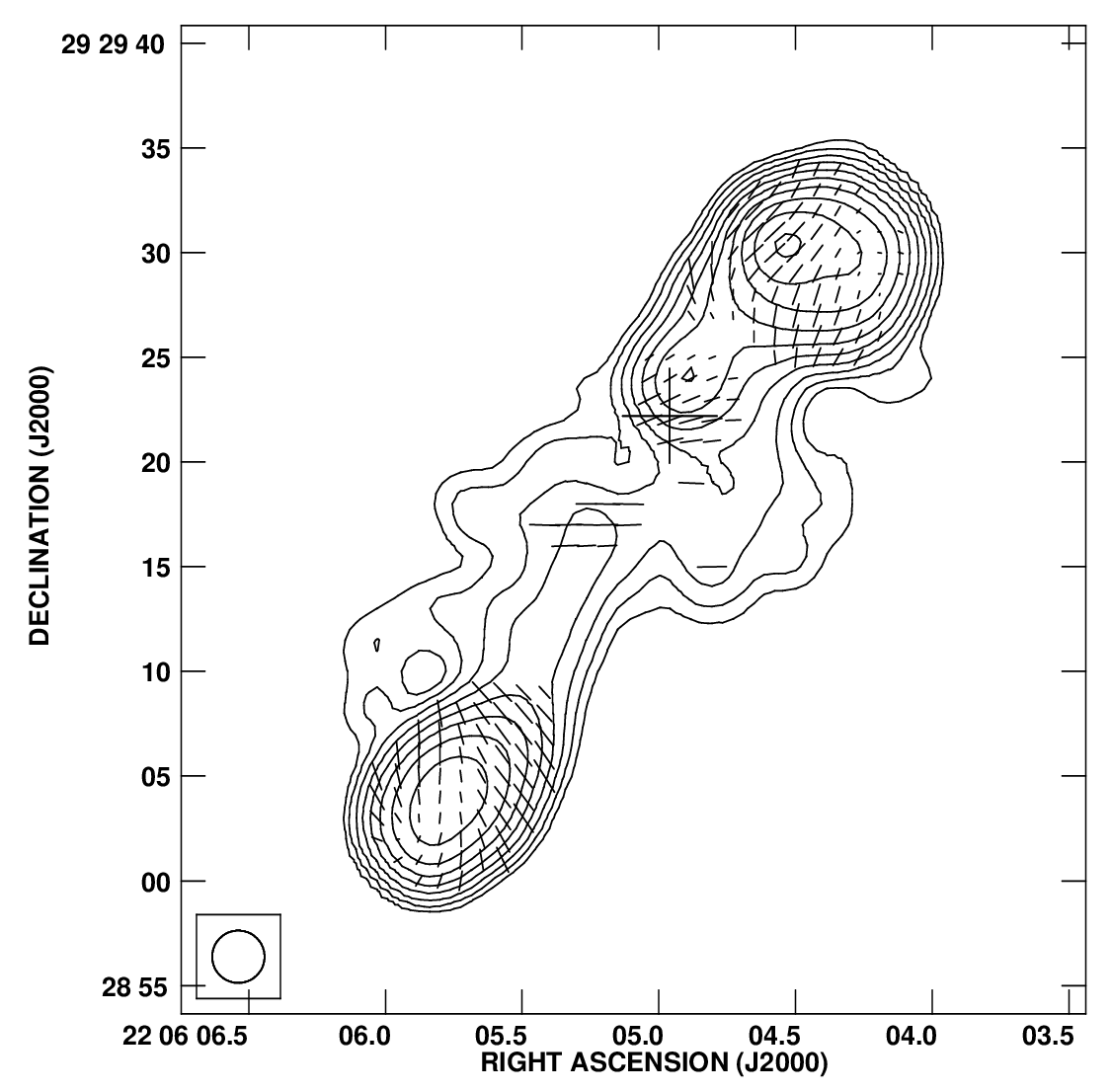}
\caption{Total intensity image of 3C441 at (left) 1.34~GHz and (right) 4.86~GHz,
at a resolution of 2.5$\arcsec$ with fractional polarization vectors 
superimposed. The peak surface brightness and contour levels increasing in 
steps of 2 are 
(left) 420.1 mJy/beam, (--0.35,0.35,...,90)\% of peak brightness,  
1$\arcsec$ vector = 20\% polarization,
(right) 184.8 mJy/beam, (--0.17,0.17,...,90)\% of peak brightness, 
1$\arcsec$ vector = 20\% polarization. }
\label{fig3c441}
\end{figure}

\begin{figure}
\epsscale{0.5}
\plotone{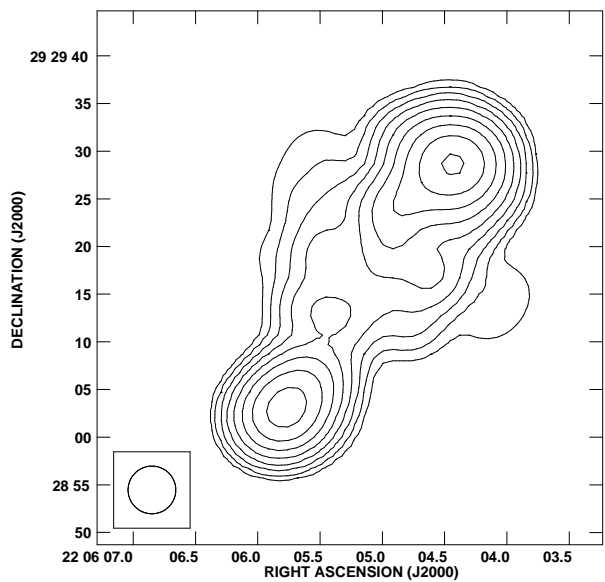}
\caption{Total intensity image of 3C441 at 327.6~MHz. The peak surface brightness 
and contour levels increasing in steps of 2 are 
2.8 Jy/beam, (--0.35,0.35,...,90)\% of peak brightness.}
\end{figure}

\begin{figure}
\epsscale{1.0}
\plotone{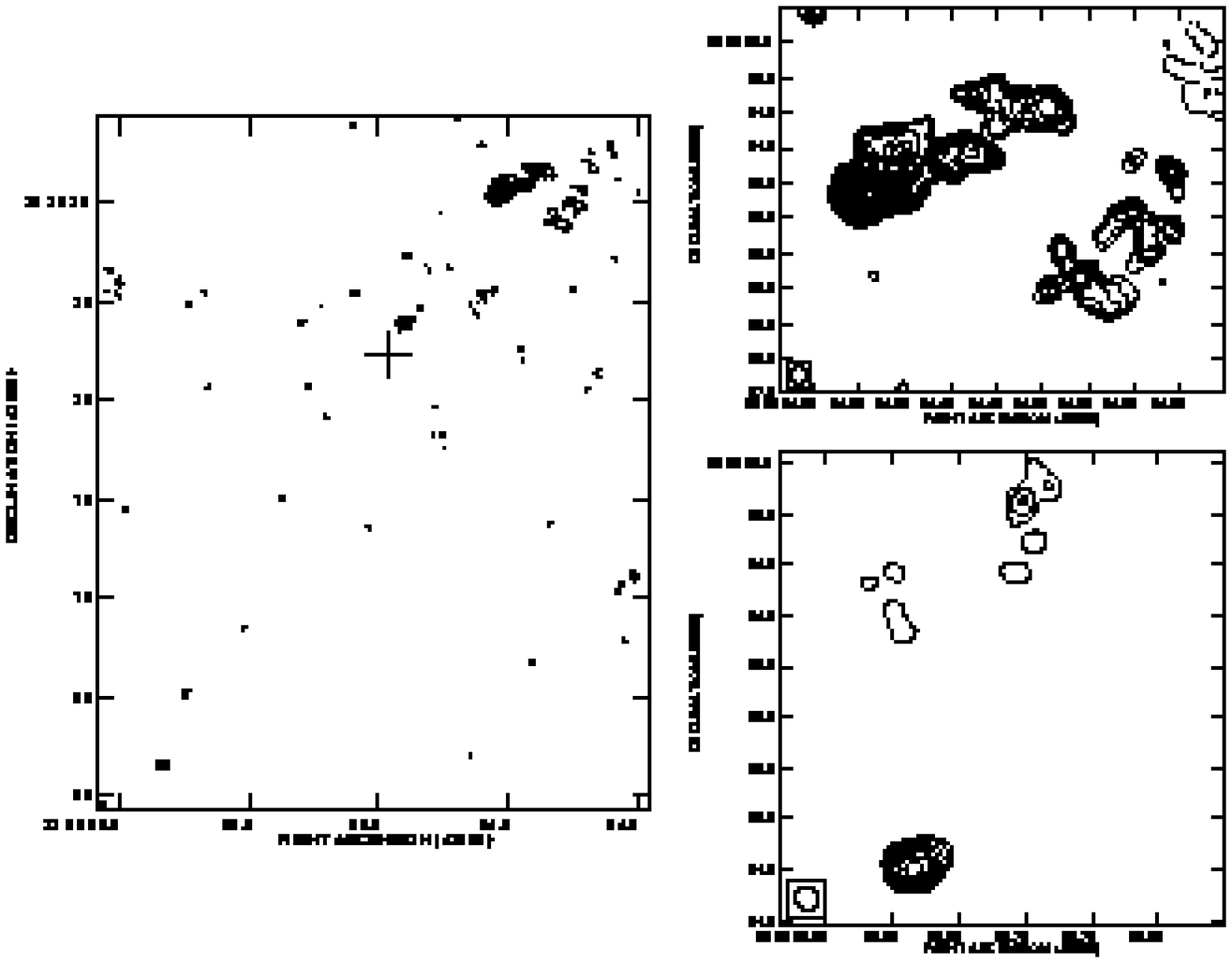}
\caption{Total intensity image of 3C441 at 8.4~GHz. The peak surface brightness
and contour levels increasing in steps of 2 
($\sqrt{2}$ for the closeup views of the hotspot regions)
are 33.9 mJy/beam, (--0.7,0.7,...,90)\% of peak brightness.
The lowest contour level in the hotspot blowup is 0.5\% of the peak
brightness.}
\end{figure}

\clearpage

\begin{figure}
\epsscale{1.0}
\plottwo{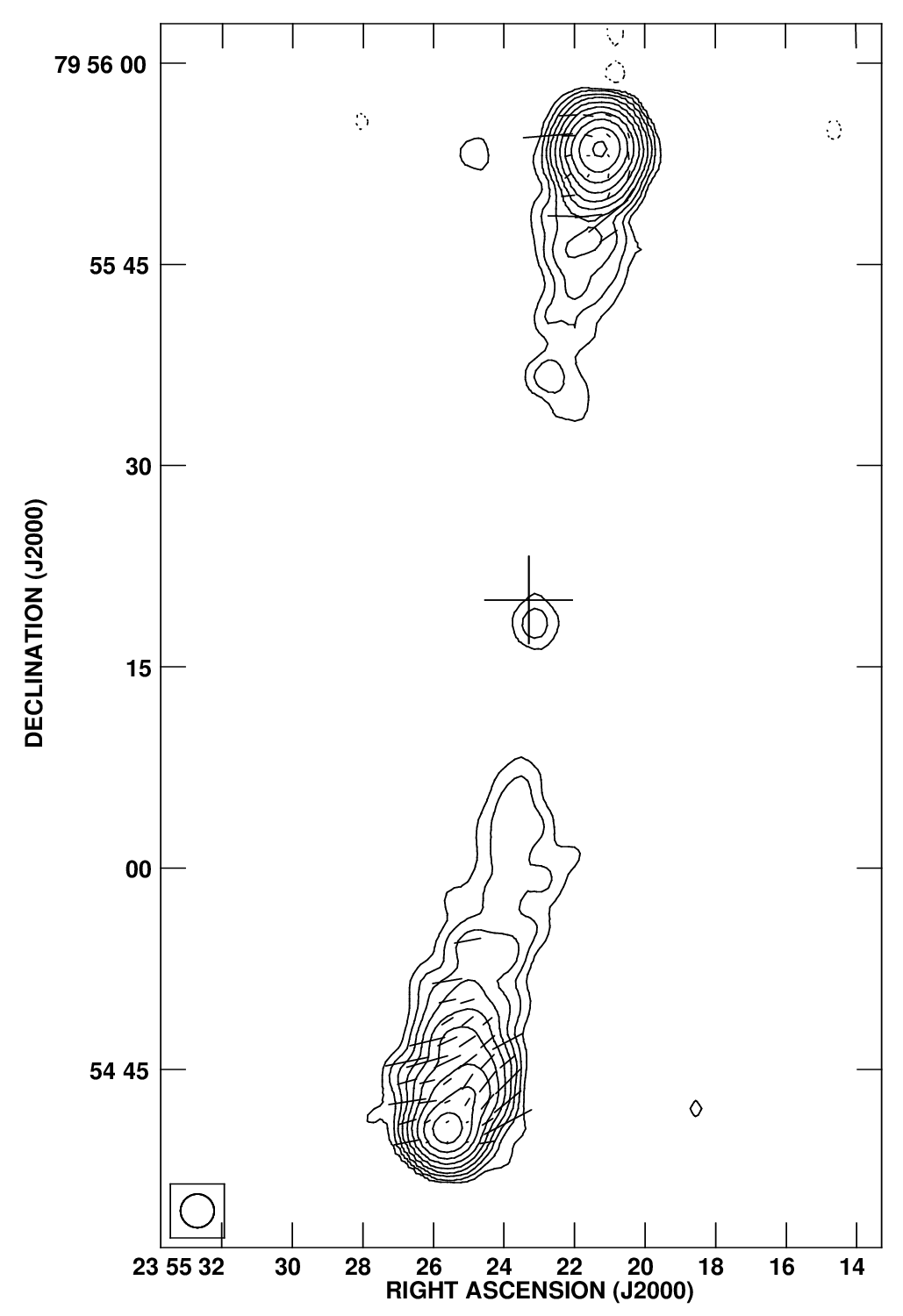}{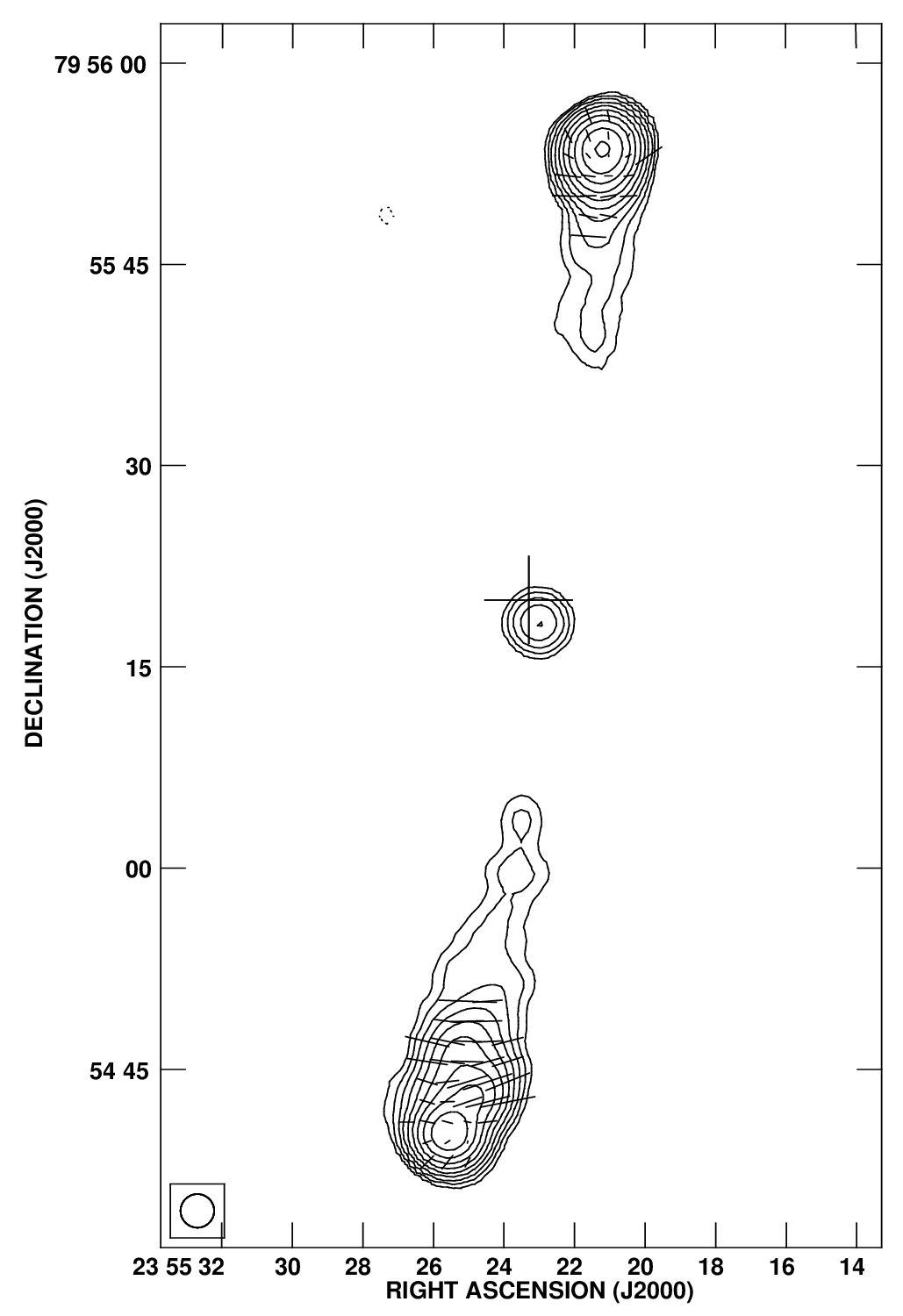}
\caption{Total intensity image of 3C469.1 at (left) 1.34~GHz and (right) 4.86~GHz,
at a resolution of 2.5$\arcsec$ with fractional polarization vectors 
superimposed. The peak surface brightness and contour levels increasing in 
steps of 2 are 
(left) 559.2 mJy/beam, (--0.17,0.17,...,90)\% of peak brightness,  
1$\arcsec$ vector = 12.5\% polarization,
(right) 131.7 mJy/beam, (--0.17,0.17,...,90)\% of peak brightness, 
1$\arcsec$ vector = 7\% polarization. }
\label{fig3c469.1}
\end{figure}

\begin{figure}
\epsscale{0.5}
\plotone{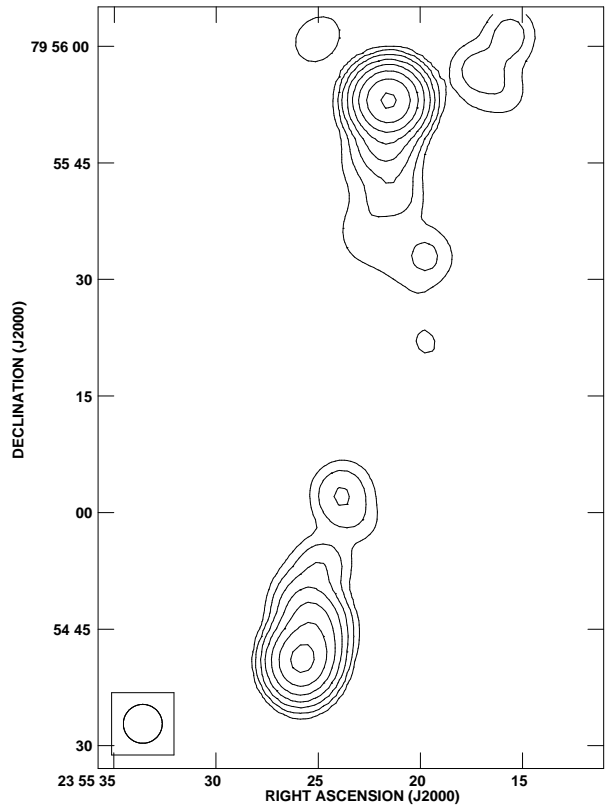}
\caption{Total intensity image of 3C469.1 at 327.6~MHz. The peak surface 
brightness and contour levels increasing in steps of 2 are 
2.9 Jy/beam, (--0.7,0.7,...,90)\% of peak brightness.}
\end{figure}

\begin{figure}
\epsscale{1.0}
\plotone{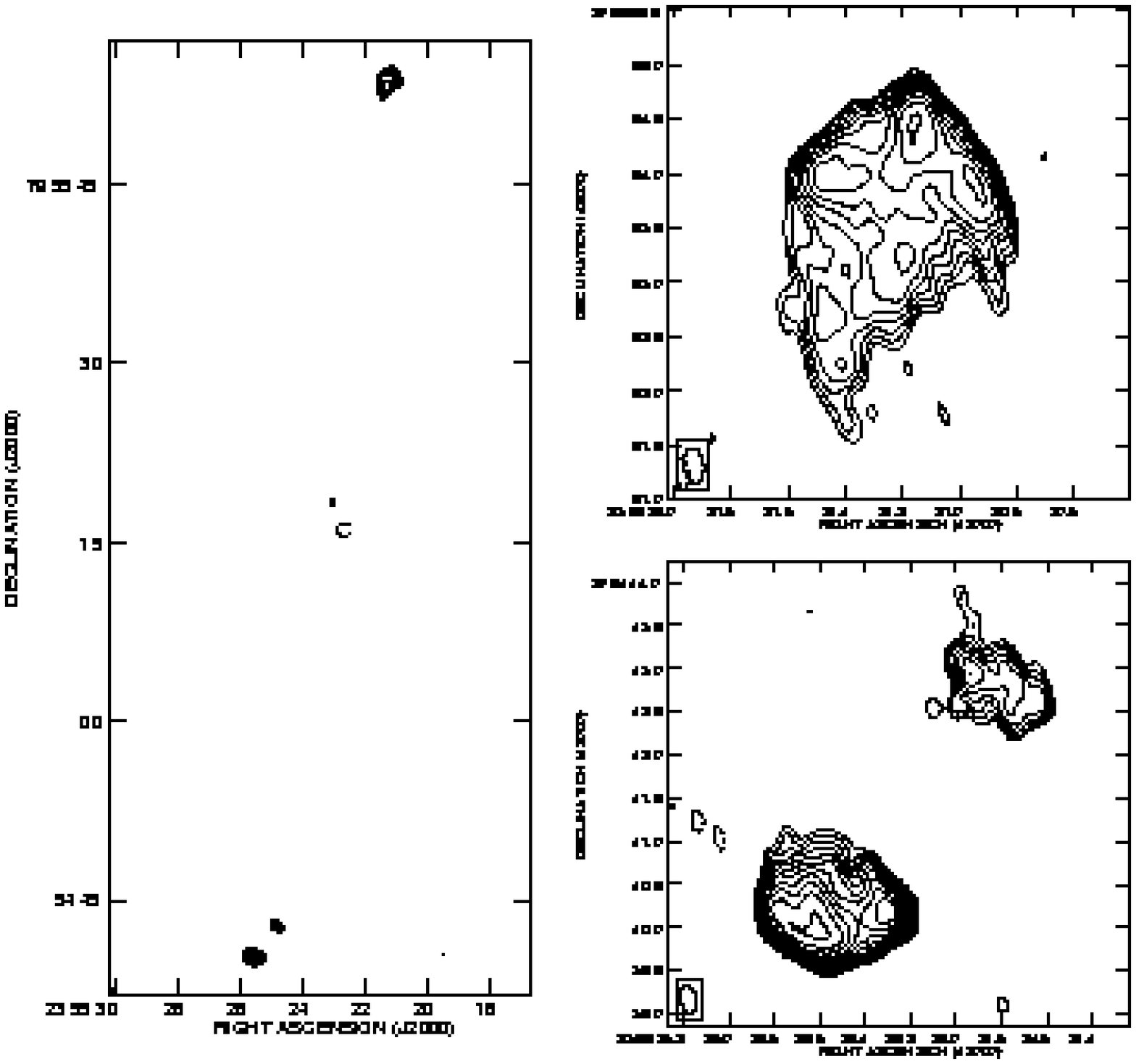}
\caption{Total intensity image of 3C469.1 at 8.4~GHz. The peak surface brightness 
and contour levels increasing in steps of 2 
($\sqrt{2}$ for the closeup views of the hotspot regions)
are 4.0 mJy/beam, (--5.6,5.6,...,90)\% of peak brightness.
The lowest contour level in the hotspot blowup is 2.8\% of the peak
brightness.}
\end{figure}

\clearpage

\begin{figure}
\epsscale{1.0}
\plottwo{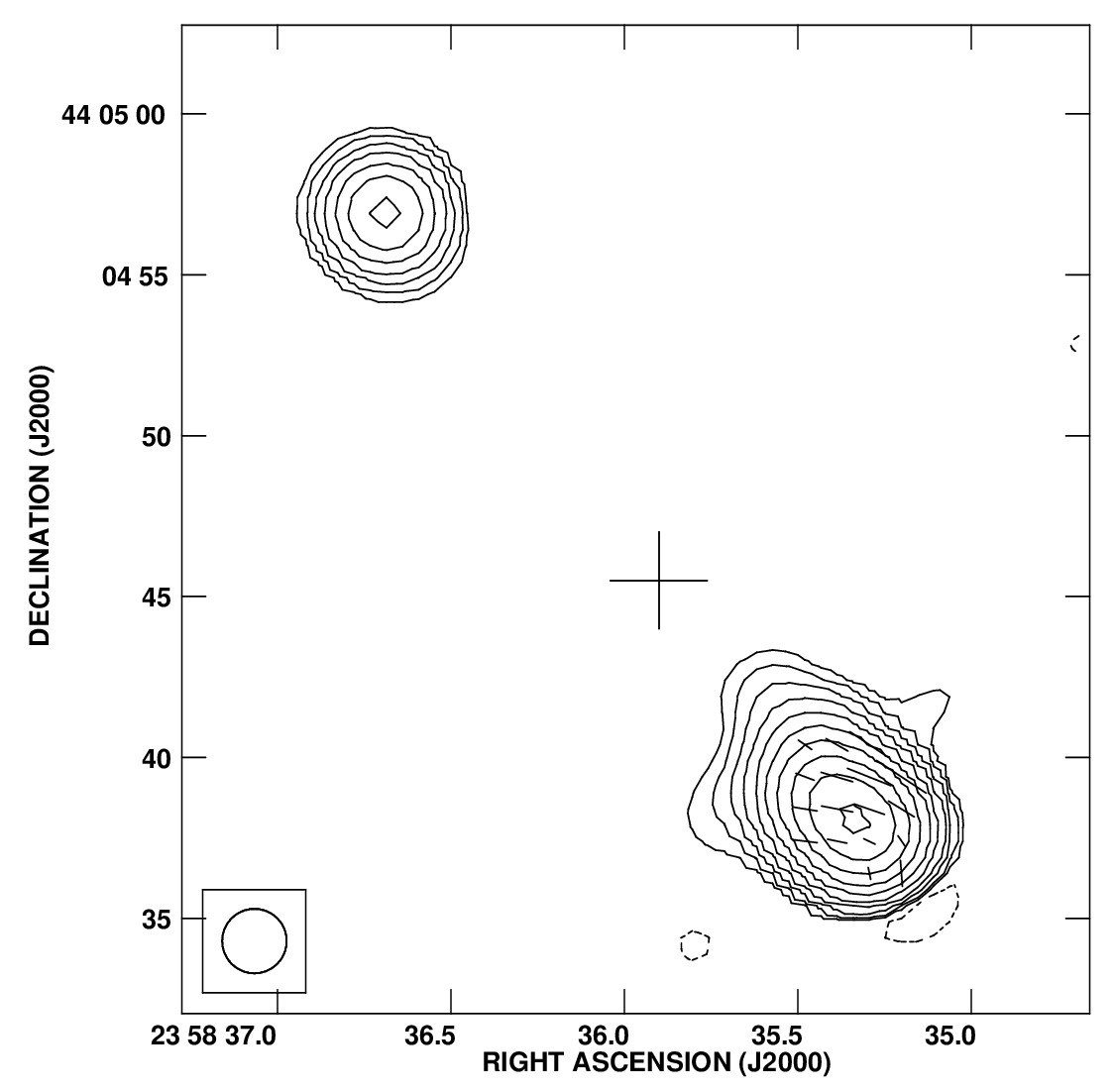}{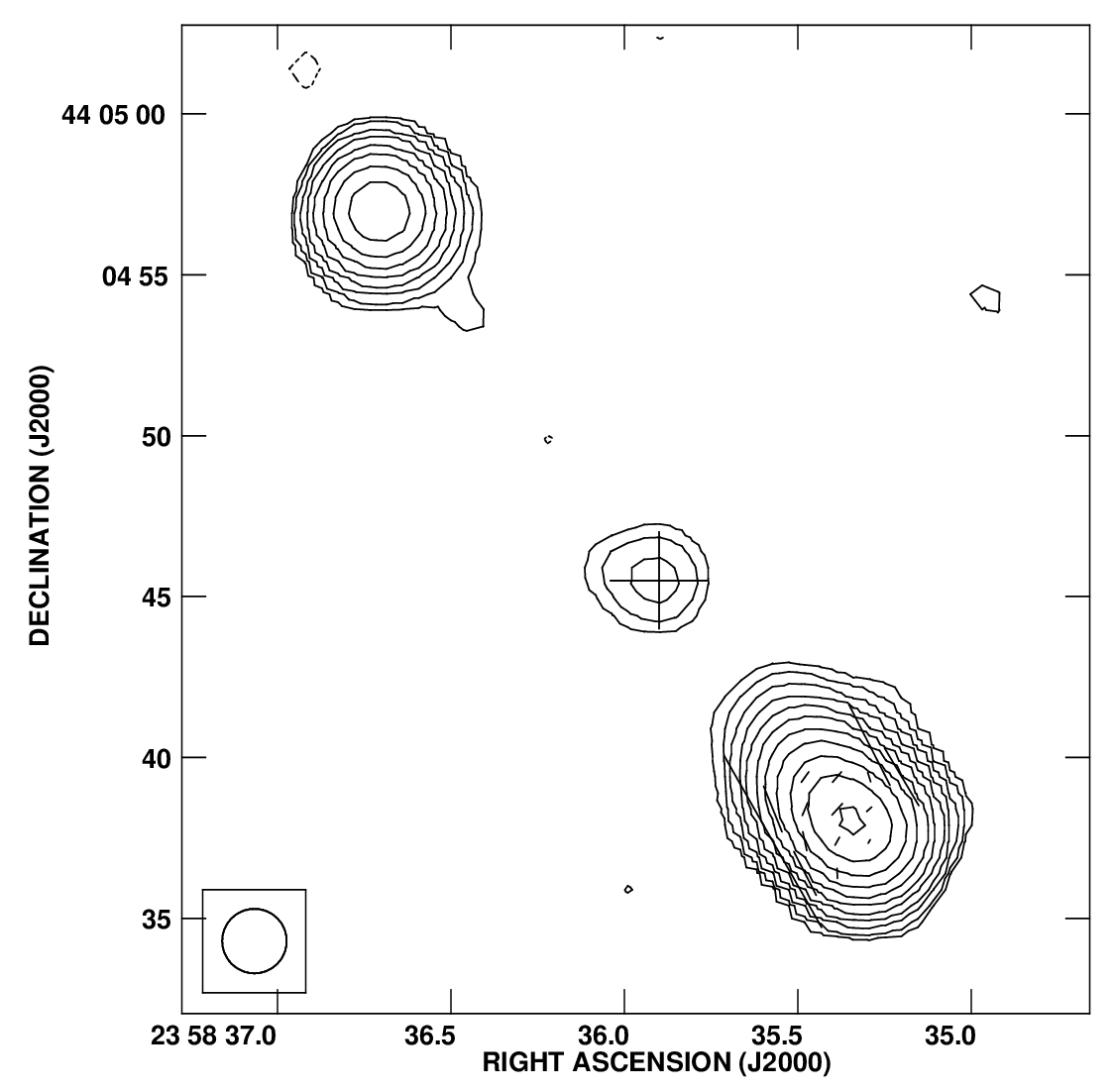}
\caption{Total intensity image of 3C470 at (left) 1.34~GHz and (right) 4.86~GHz,
at a resolution of 2$\arcsec$ with fractional polarization vectors 
superimposed. The peak surface brightness and contour levels increasing in 
steps of 2 are 
(left) 1.2 Jy/beam, (--0.17,0.17,...,90)\% of peak brightness,  
1$\arcsec$ vector = 7.7\% polarization,
(right) 318.8 mJy/beam, (--0.085,0.085,...,90)\% of peak brightness, 
1$\arcsec$ vector = 14\% polarization. }
\end{figure}

\begin{figure}
\epsscale{0.5}
\plotone{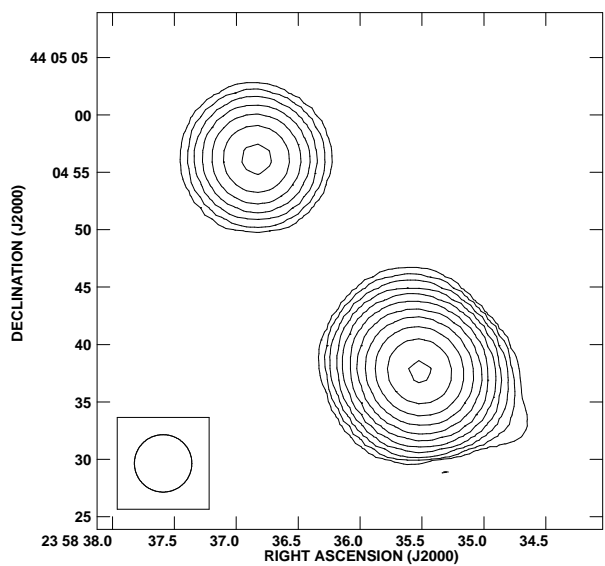}
\caption{Total intensity image of 3C470 at 327.6~MHz. The peak surface 
brightness and contour levels increasing in steps of 2 are 
5.9 Jy/beam, (--0.085,0.085,...,90)\% of peak brightness.}
\end{figure}

\begin{figure}
\epsscale{1.0}
\plotone{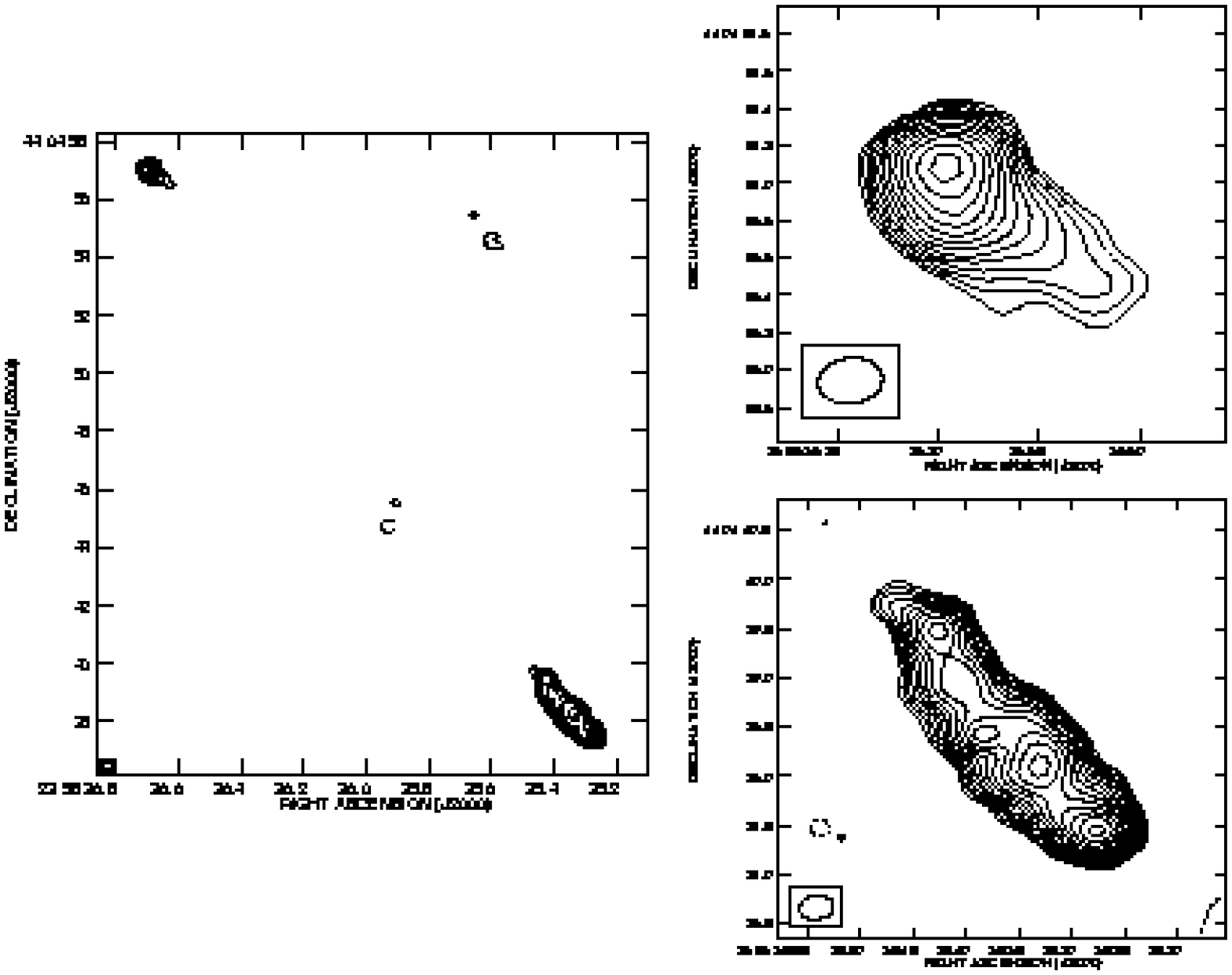}
\caption{Total intensity image of 3C470 at 8.4~GHz. The peak surface brightness 
and contour levels increasing in steps of 2 
($\sqrt{2}$ for the closeup views of the hotspot regions)
are 39.9 mJy/beam, (--1.4,1.4,...,90)\% of peak brightness.
The lowest contour level in the hotspot blowup is 0.7\% of the peak
brightness.}
\end{figure}
\label{figs3c470}
\clearpage

\begin{figure}
\plottwo{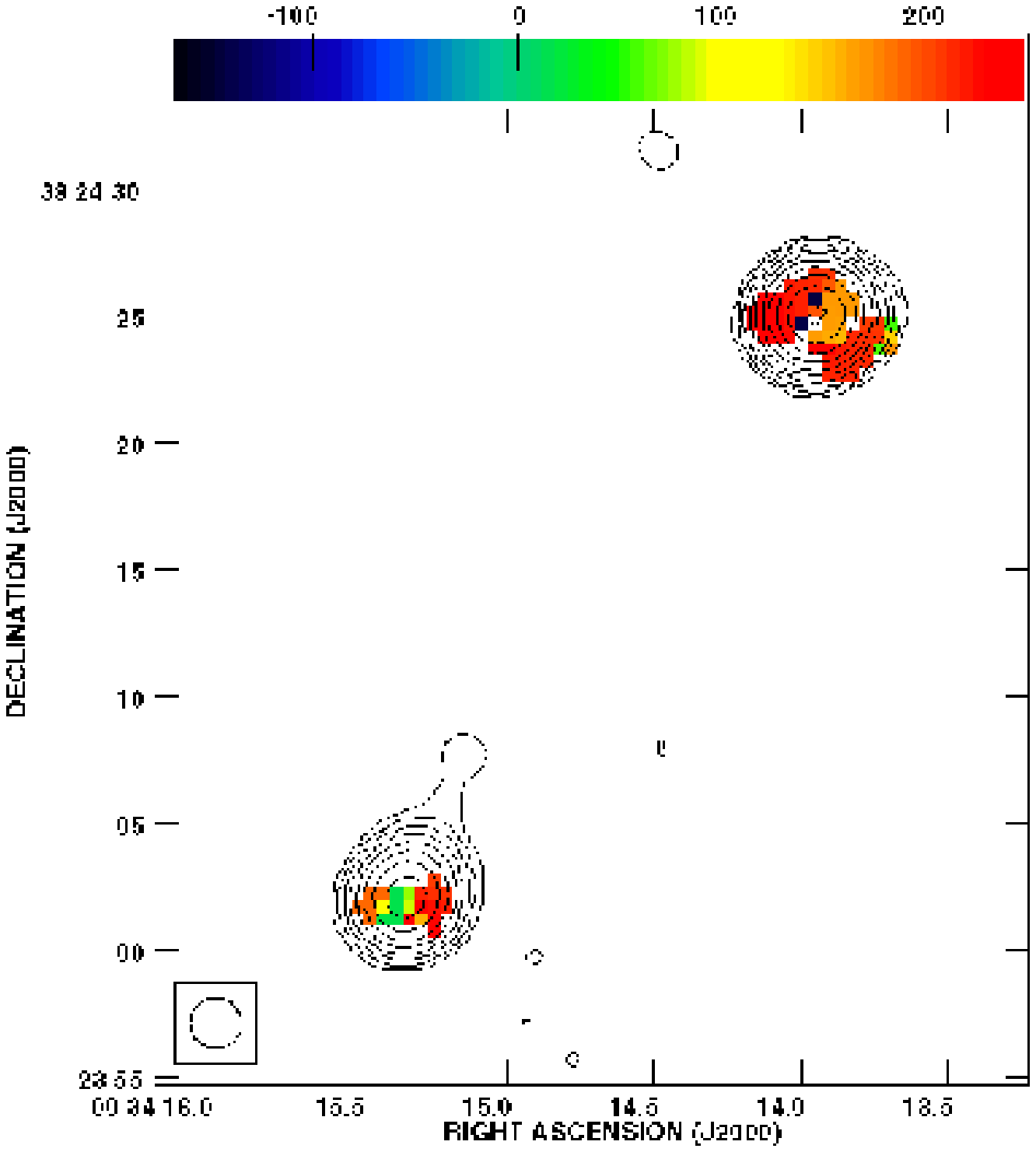}{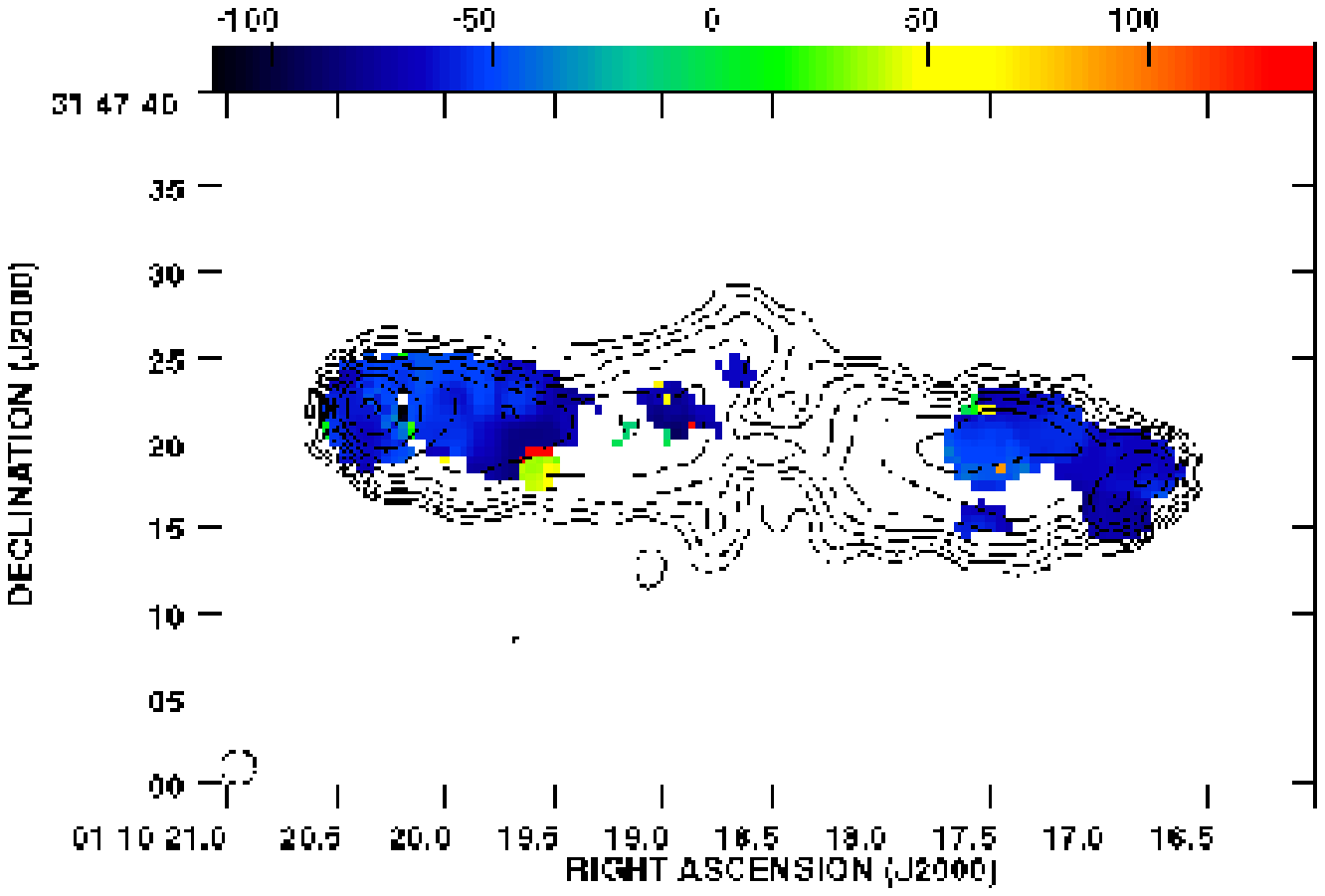}
\caption{Rotation measure image in colour superimposed by 5~GHz radio 
contours for (left) 3C13 and (right) 3C34. The RM was obtained from
polarization maps at 1.4 and 5 GHz. The colour-scale units are in
rad~m$^2$.}
\label{fig3c13rm_01}
\end{figure}

\begin{figure}
\plottwo{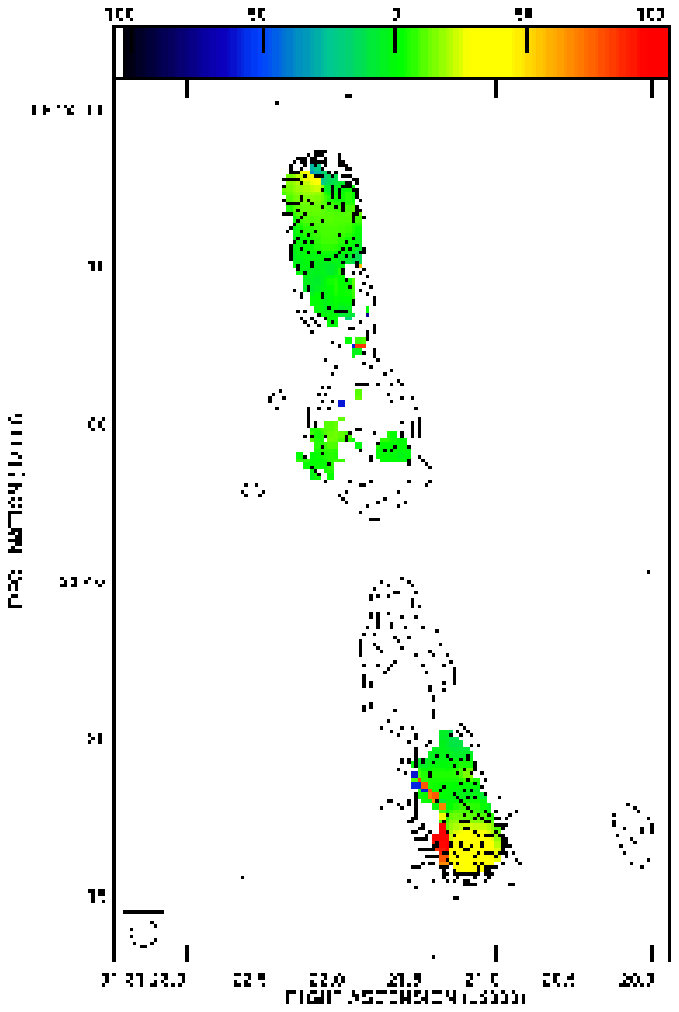}{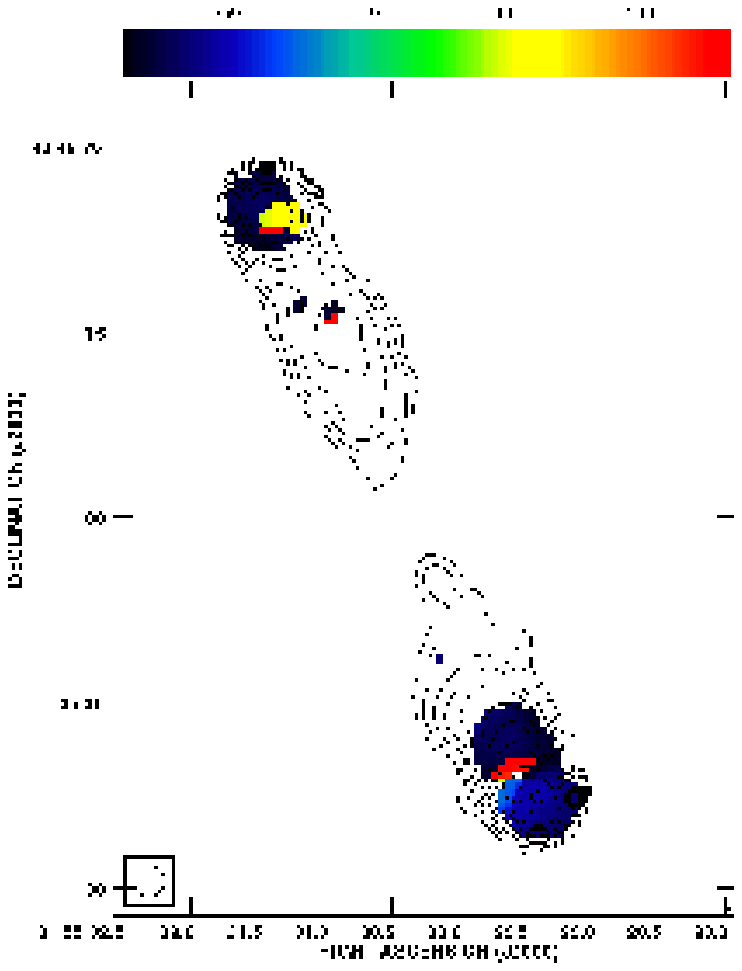}
\caption{Rotation measure image in colour superimposed by 5~GHz radio 
contours for (left) 3C44 and (right) 3C54. The RM was obtained from
polarization maps at 1.4 and 5 GHz. The colour-scale units are in
rad~m$^2$.}
\label{fig3c54rm_01}
\end{figure}

\clearpage

\begin{figure}
\plottwo{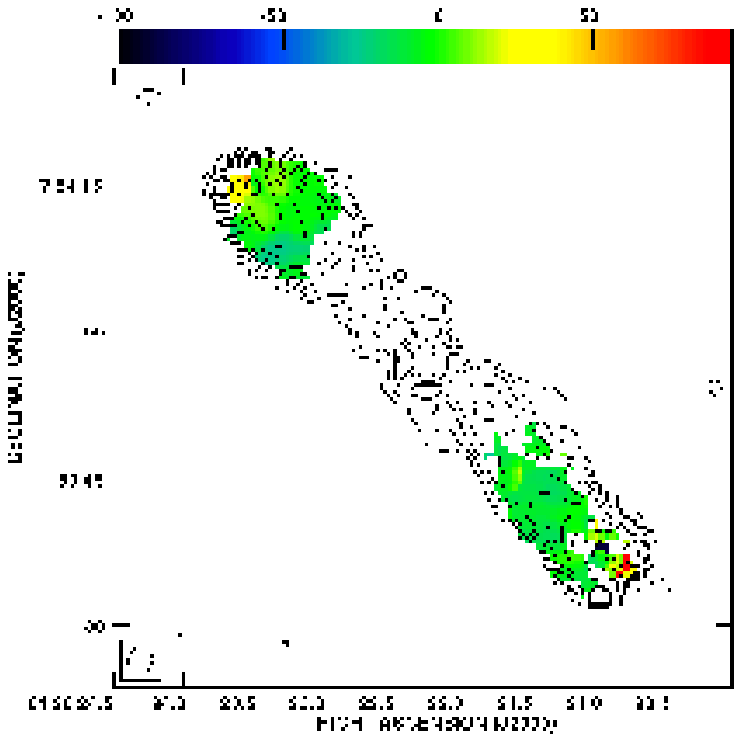}{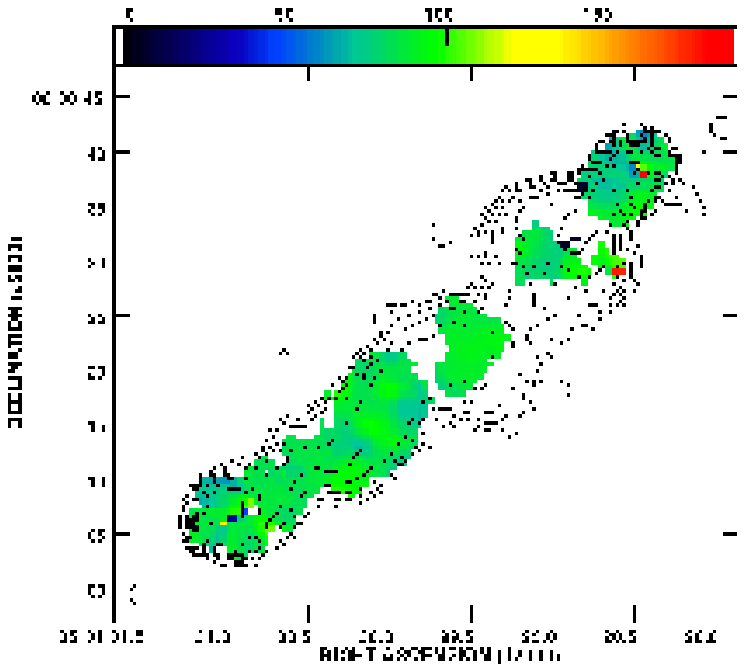}
\caption{Rotation measure image in colour superimposed by 5~GHz radio 
contours for (left) 3C114 and (right) 3C142.1. The RM was obtained from
polarization maps at 1.4 and 5 GHz. The colour-scale units are in
rad~m$^2$.}
\label{fig3c13rm_02}
\end{figure}

\begin{figure}
\plottwo{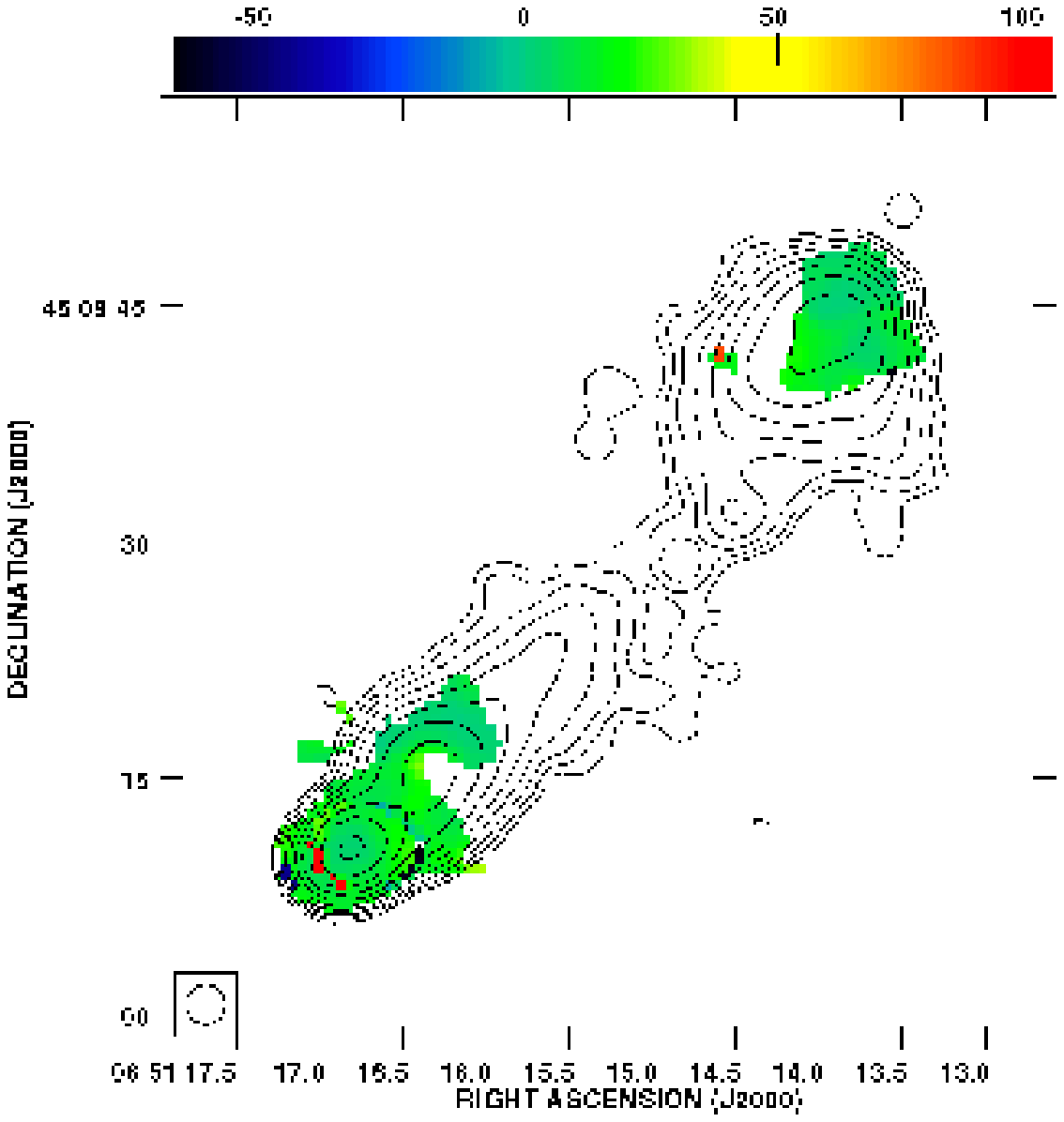}{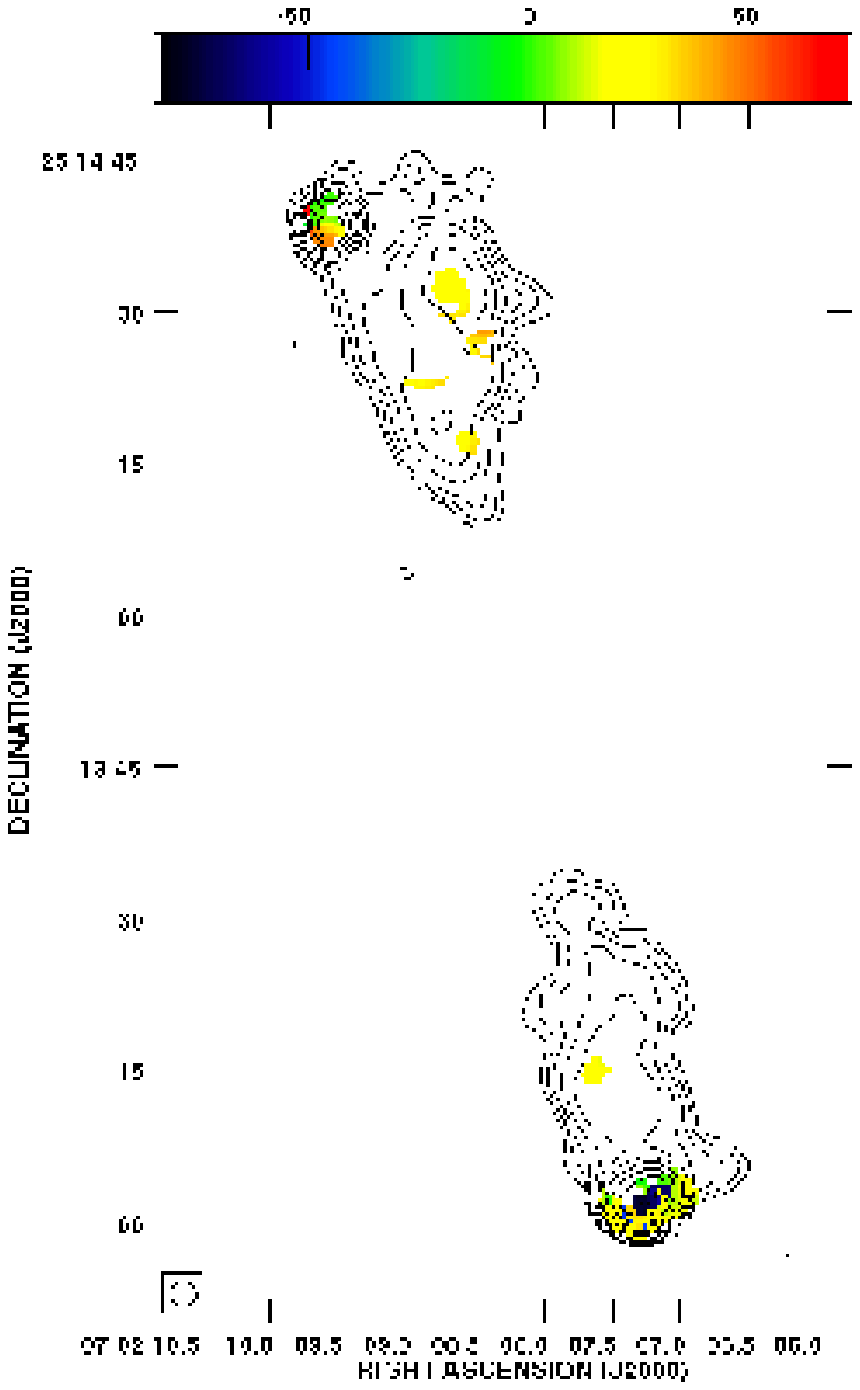}
\caption{Rotation measure image in colour superimposed by 5~GHz radio 
contours for (left) 3C169.1 and (right) 3C172. The RM was obtained from
polarization maps at 1.4 and 5 GHz. The colour-scale units are in
rad~m$^2$.}
\label{fig3c54rm_02}
\end{figure}

\clearpage

\begin{figure}
\plottwo{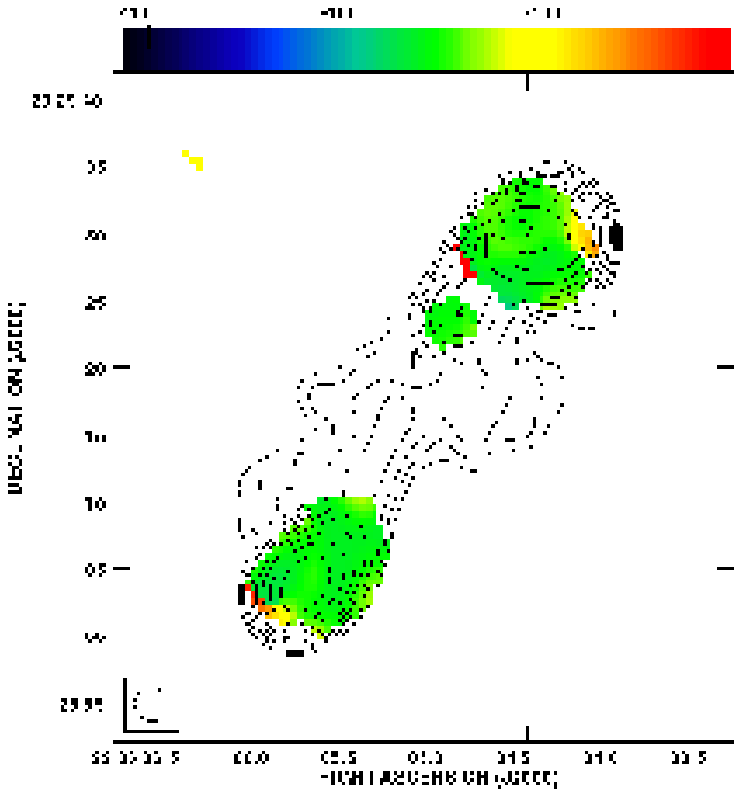}{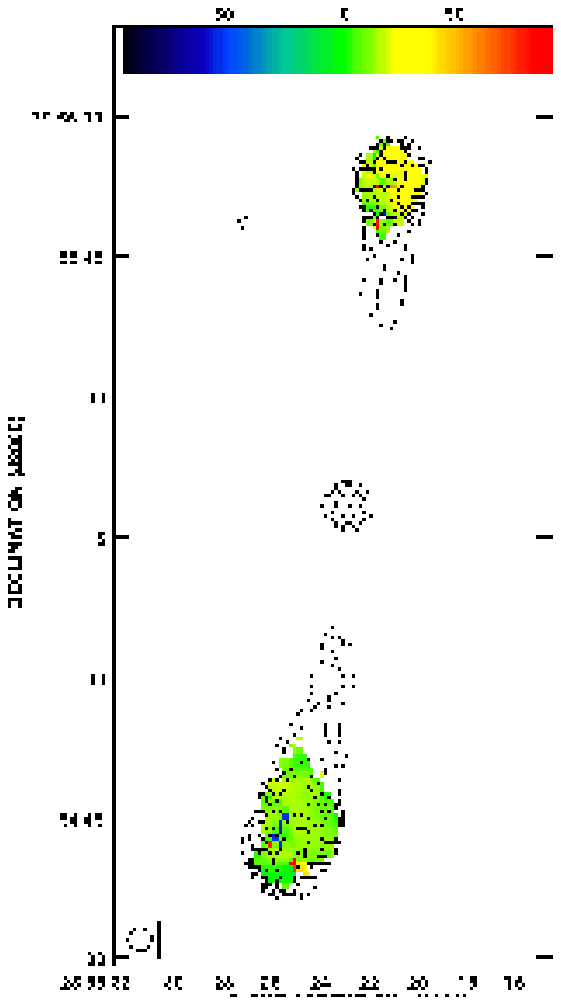}
\caption{Rotation measure image in colour superimposed by 5~GHz radio 
contours for (left) 3C441 and (right) 3C469.1. The RM was obtained from
polarization maps at 1.4 and 5 GHz. The colour-scale units are in
rad~m$^2$.}
\label{fig3c13rm_03}
\end{figure}

\begin{figure}
\epsscale{0.4}
\plotone{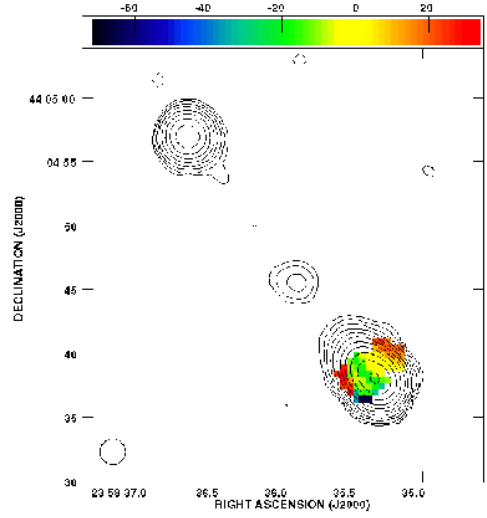}
\caption{Rotation measure image in colour superimposed by 5~GHz radio 
contours for 3C470. The RM was obtained from
polarization maps at 1.4 and 5 GHz. The colour-scale units are in
rad~m$^2$.}
\label{fig3c54rm_03}
\end{figure}

\clearpage

\begin{figure}
\epsscale{1.0}
\plottwo{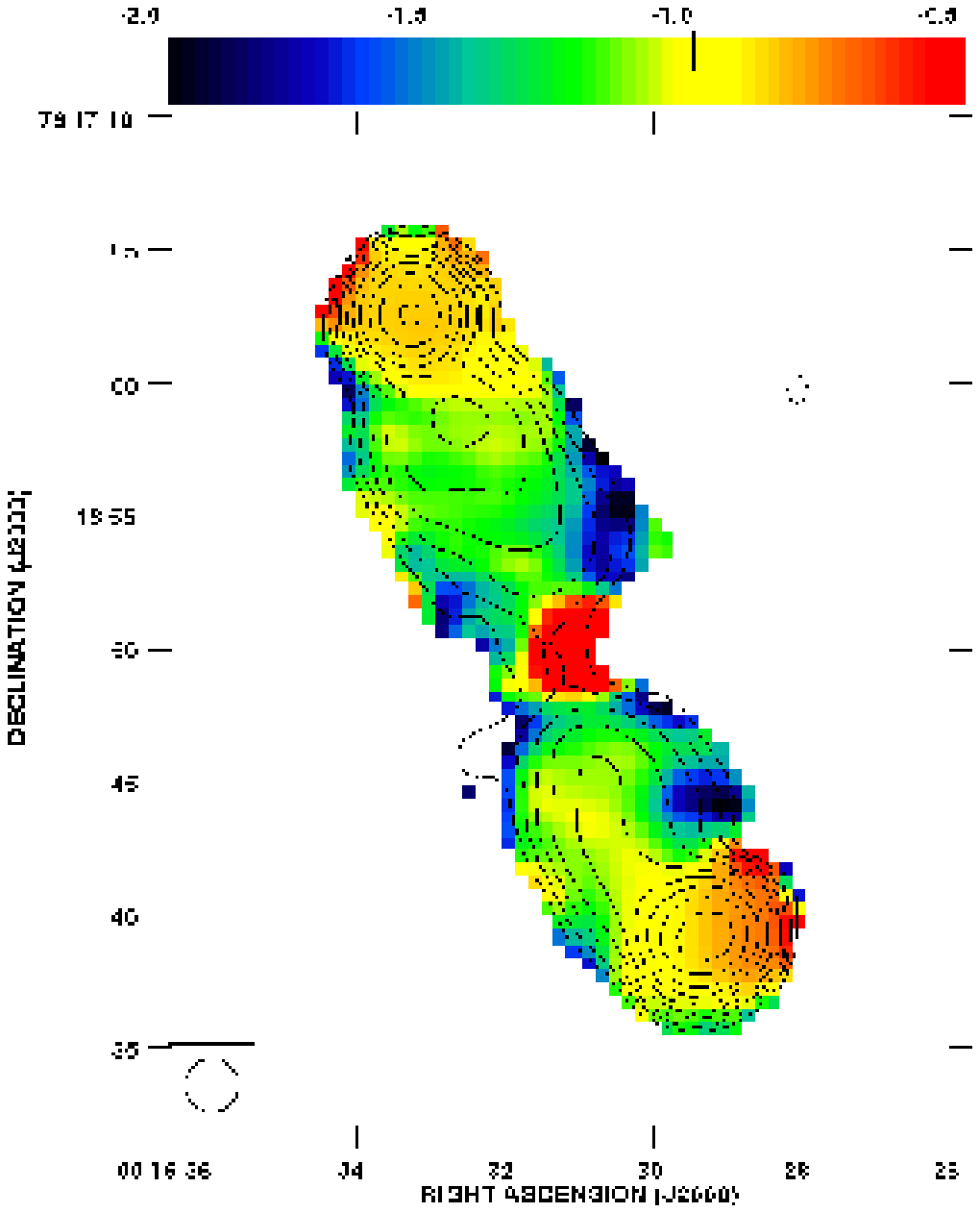}{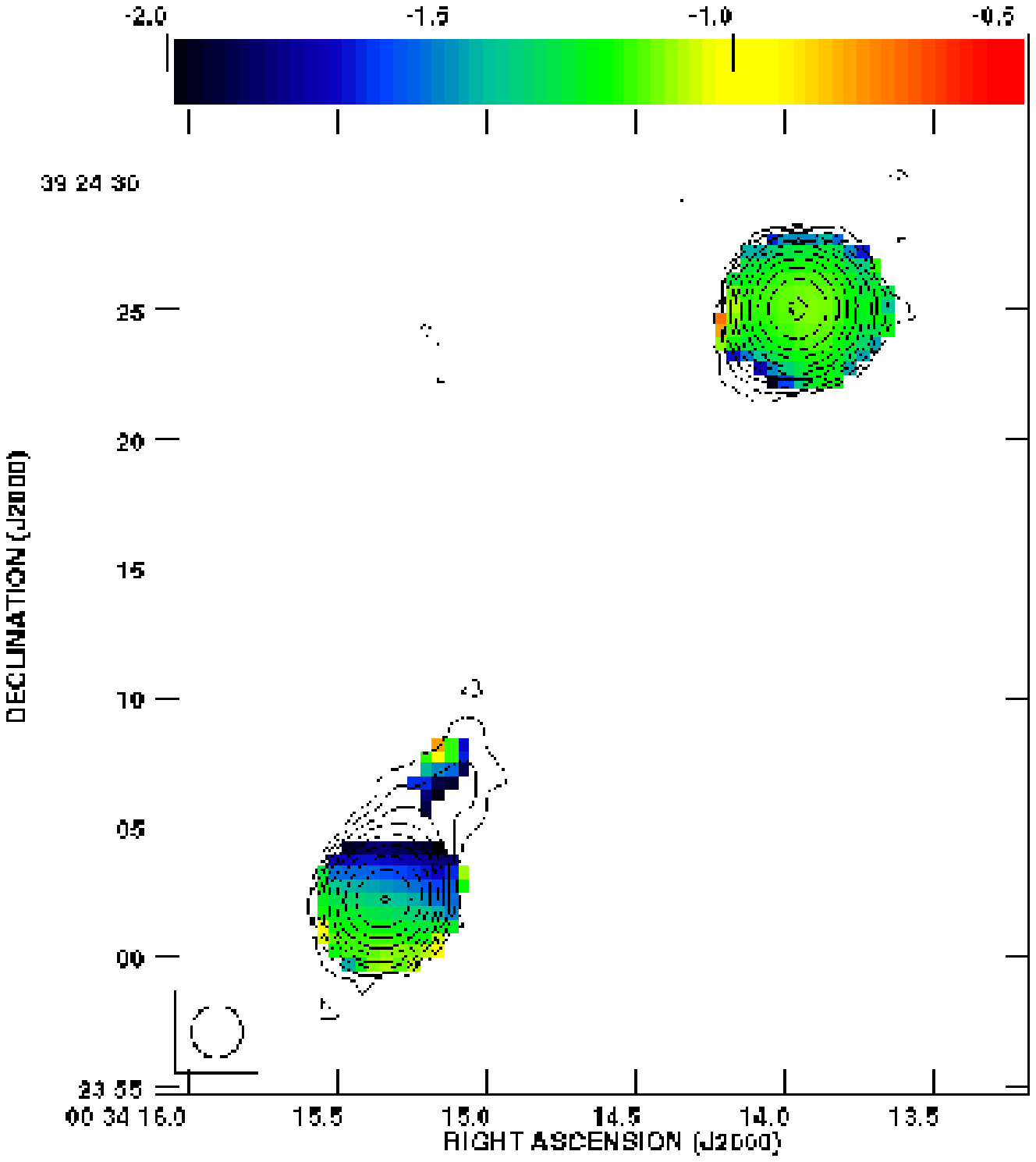}
\caption{Spectral index image in colour superimposed by 1.4~GHz radio
contours at a resolution of $\sim$2$\arcsec$ for 
(left) 3C6.1 and (right) 3C13. The spectral index image was obtained with
1.4 and 5 GHz data.}
\label{fig3c13sp_01}
\end{figure}

\begin{figure}
\epsscale{1.0}
\plottwo{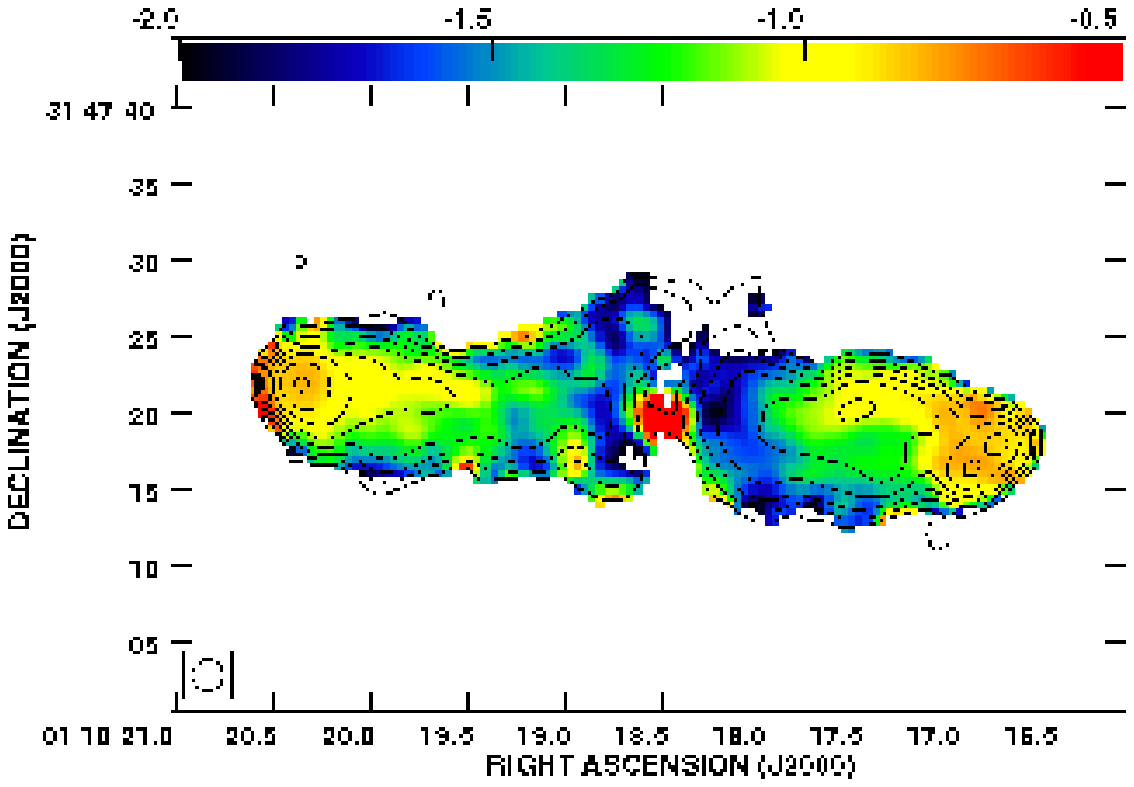}{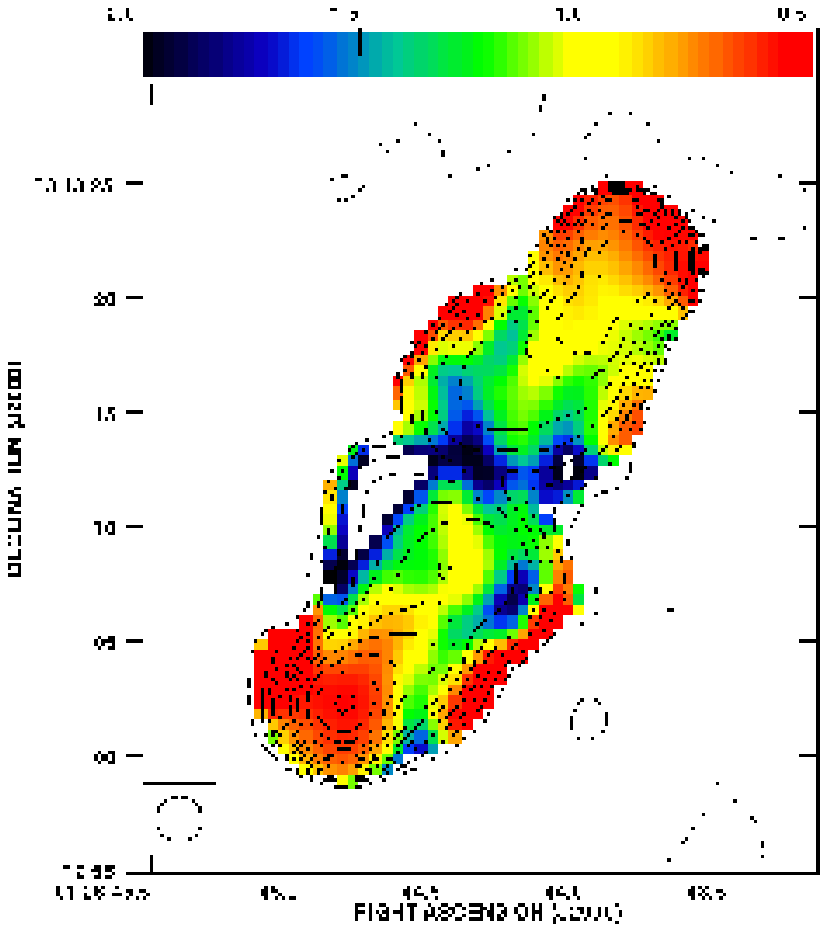}
\caption{Spectral index image in colour superimposed by 1.4~GHz radio
contours at a resolution of $\sim$2$\arcsec$ for
(left) 3C34 and (right) 3C41. The spectral index image was obtained with
1.4 and 5 GHz data.}
\label{fig3c34sp}
\end{figure}

\clearpage

\begin{figure}
\epsscale{1.0}
\plottwo{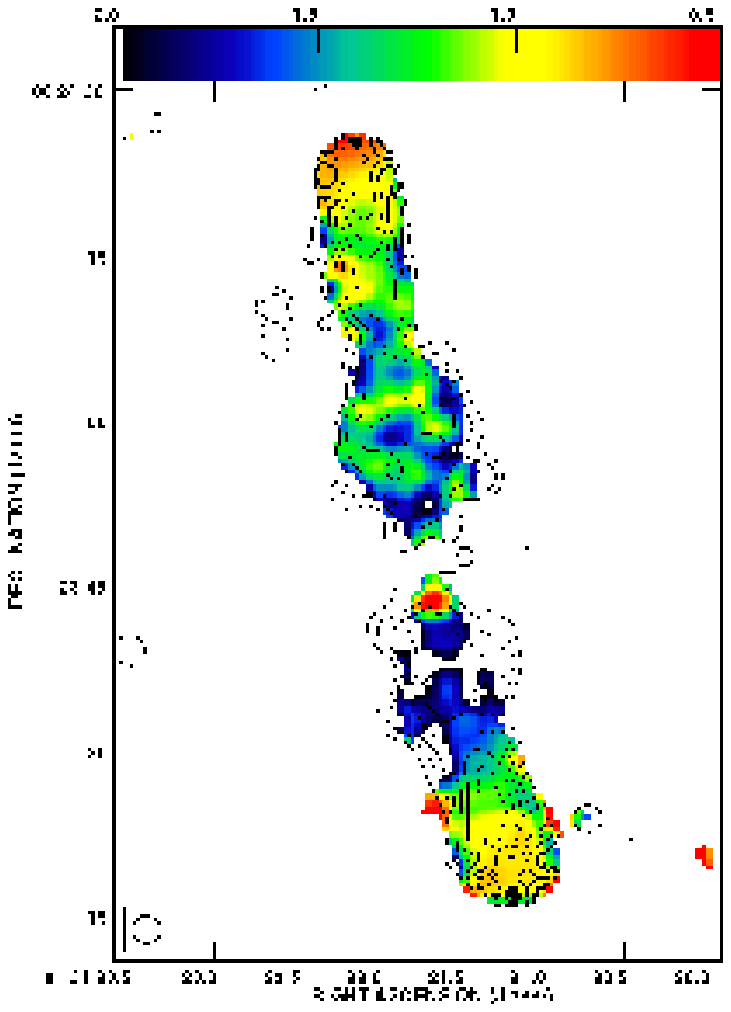}{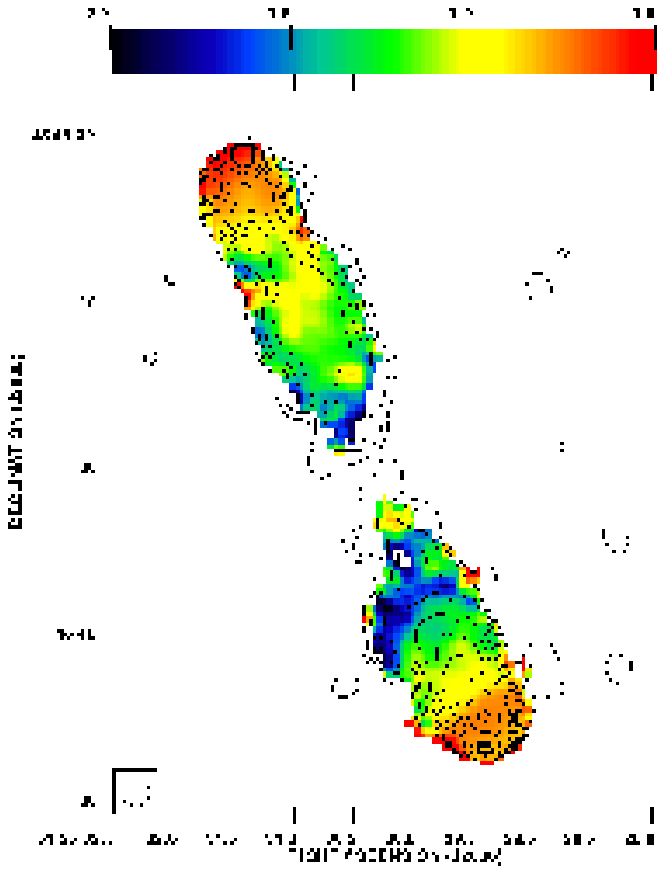}
\caption{Spectral index image in colour superimposed by 1.4~GHz radio
contours at a resolution of $\sim$2$\arcsec$ for 
(left) 3C44 and (right) 3C54. The spectral index image was obtained with
1.4 and 5 GHz data.}
\label{fig3c44sp}
\end{figure}

\begin{figure}
\epsscale{1.0}
\plottwo{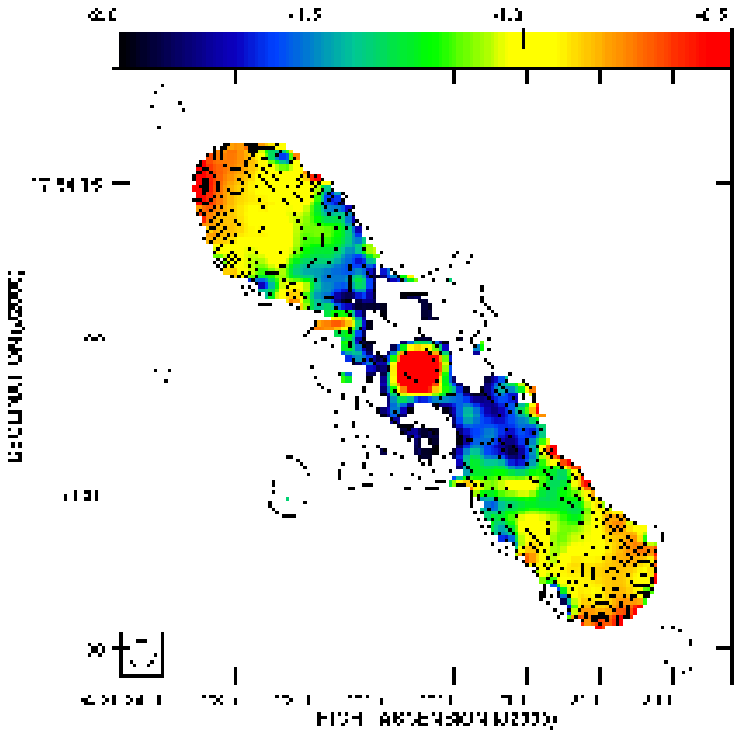}{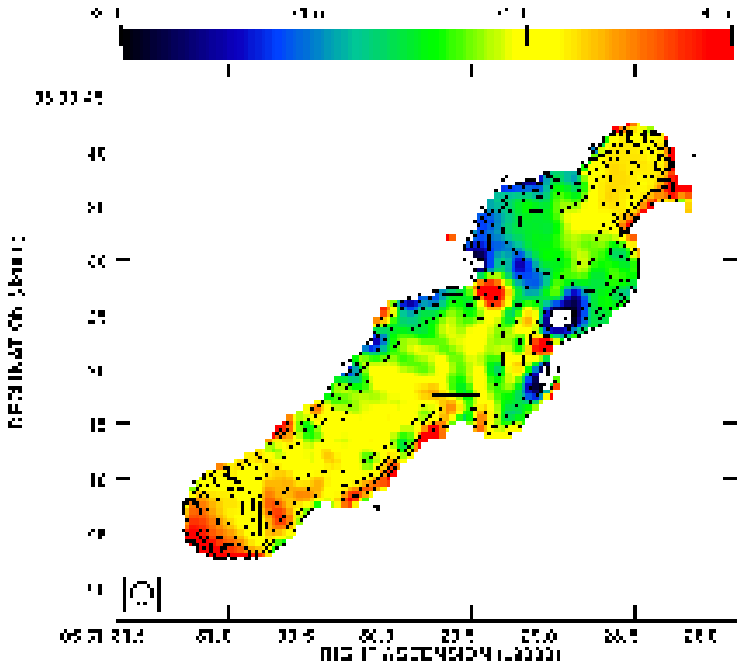}
\caption{Spectral index image in colour superimposed by 1.4~GHz radio
contours at a resolution of $\sim$2$\arcsec$ for 
(left) 3C114 and (right) 3C142.1. The spectral index image was obtained with
1.4 and 5 GHz data.}
\label{fig3c114sp}
\end{figure}

\clearpage

\begin{figure}
\epsscale{1.0}
\plottwo{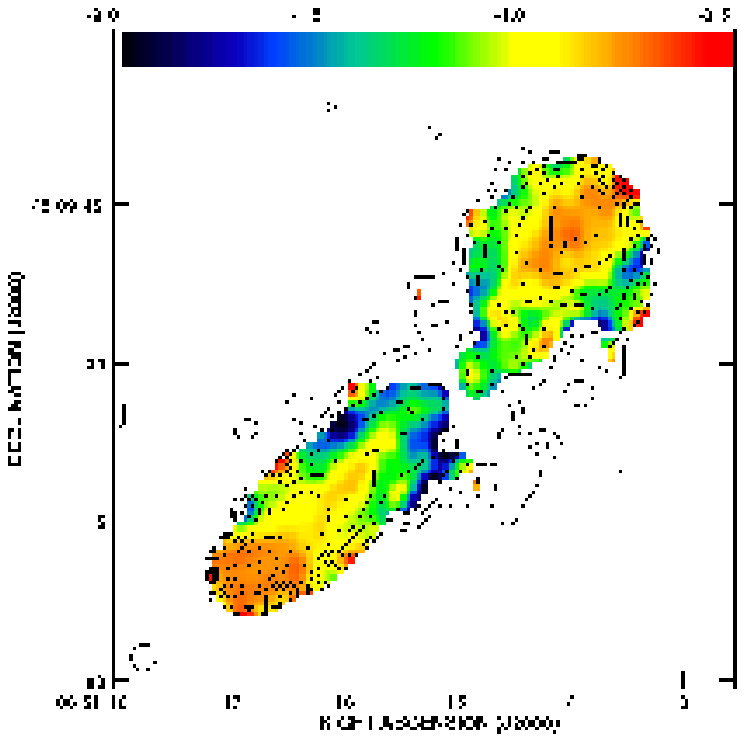}{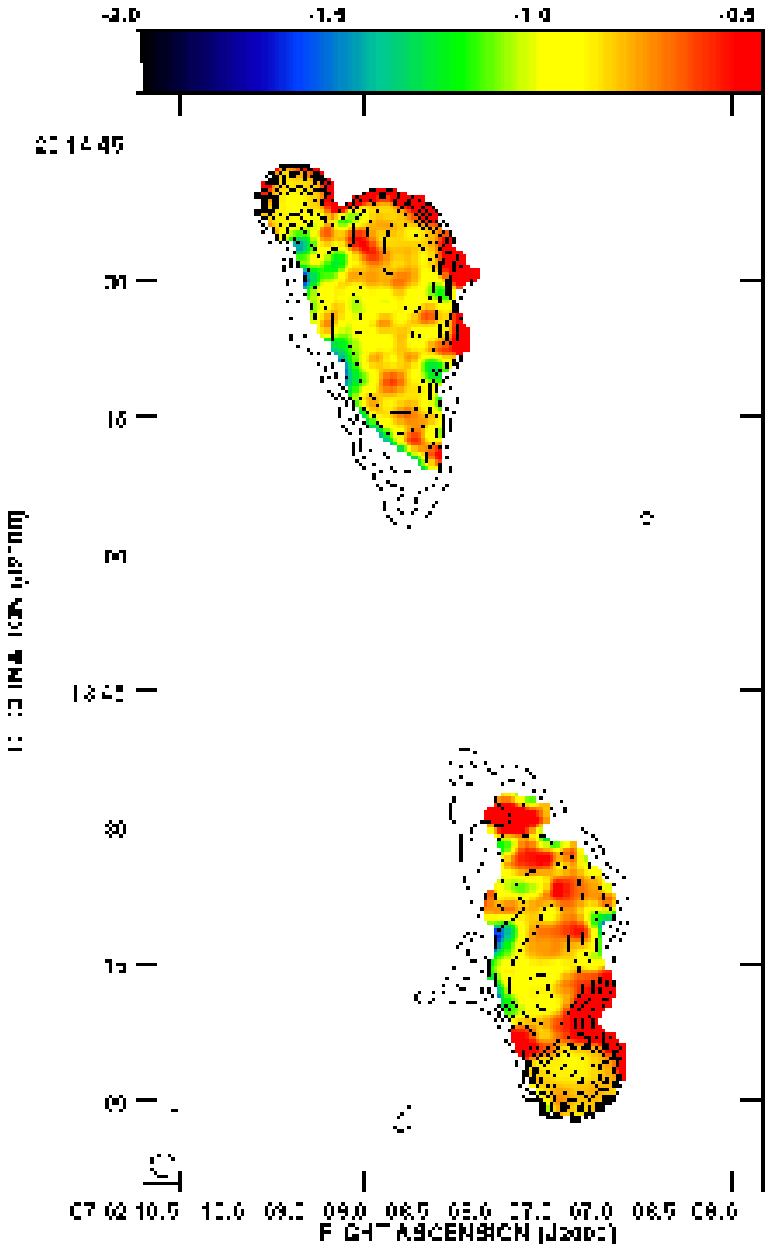}
\caption{Spectral index image in colour superimposed by 1.4~GHz radio
contours at a resolution of $\sim$2$\arcsec$ for 
(left) 3C169.1 and (right) 3C172. The spectral index image was obtained with
1.4 and 5 GHz data.}
\label{fig3c169sp}
\end{figure}

\begin{figure}
\epsscale{1.0}
\plottwo{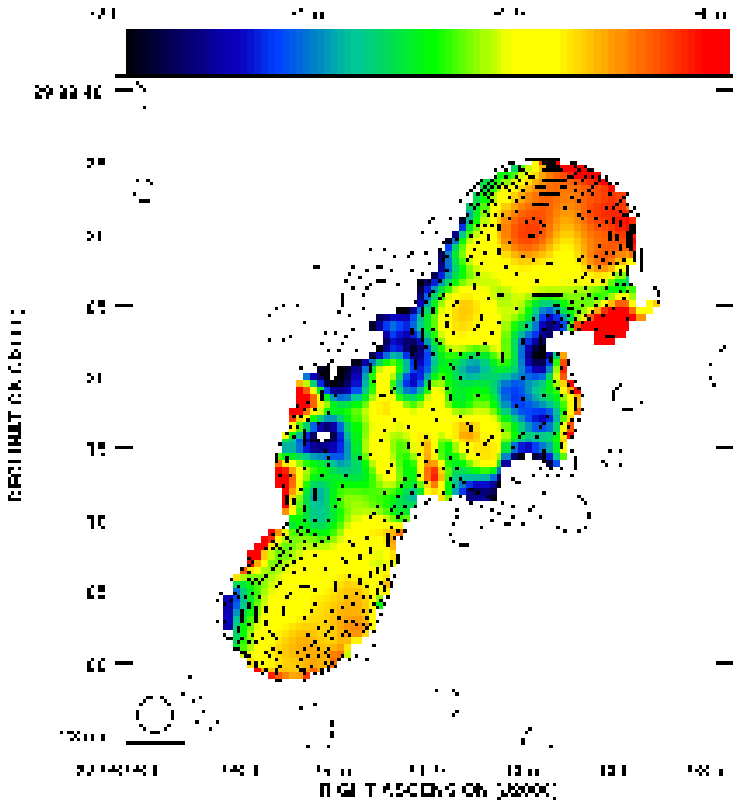}{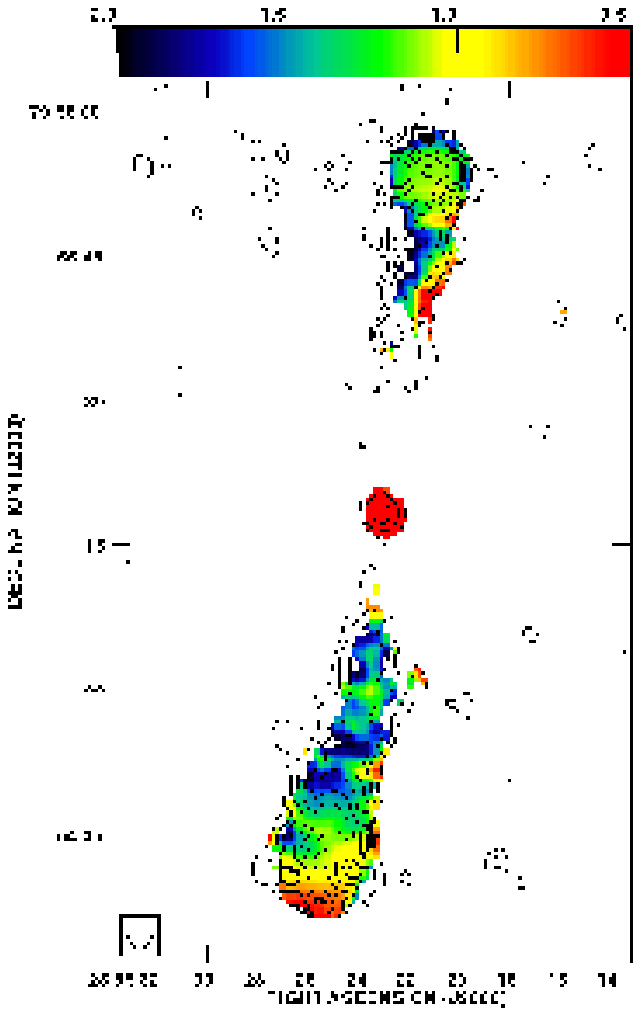}
\caption{Spectral index image in colour superimposed by 1.4~GHz radio
contours at a resolution of $\sim$2$\arcsec$ for 
(left) 3C441 and (right) 3C469.1. The spectral index image was obtained with
1.4 and 5 GHz data.}
\label{fig3c441sp}
\end{figure}

\clearpage

\begin{figure}
\epsscale{0.4}
\plotone{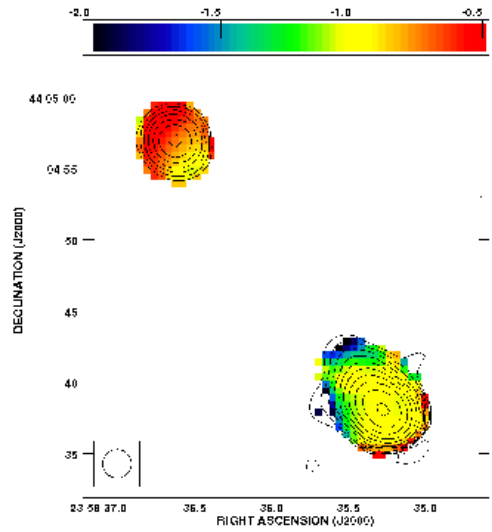}
\caption{Spectral index image in colour superimposed by 1.4~GHz radio
contours at a resolution of $\sim$2$\arcsec$ for 
3C470. The spectral index image was obtained with
1.4 and 5 GHz data.}
\label{fig3c13sp_02}
\end{figure}

\begin{figure}
\epsscale{1.0}
\plottwo{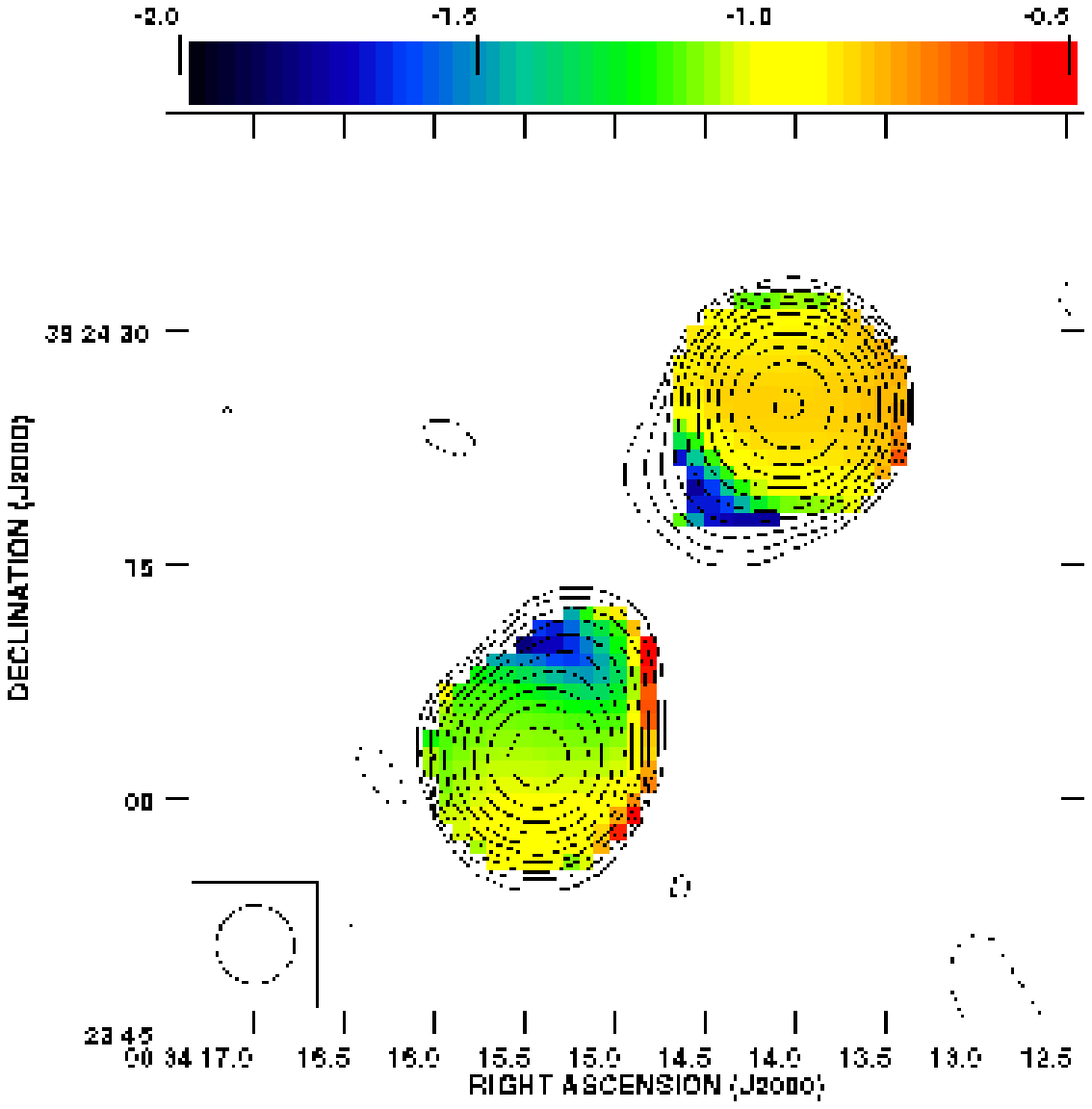}{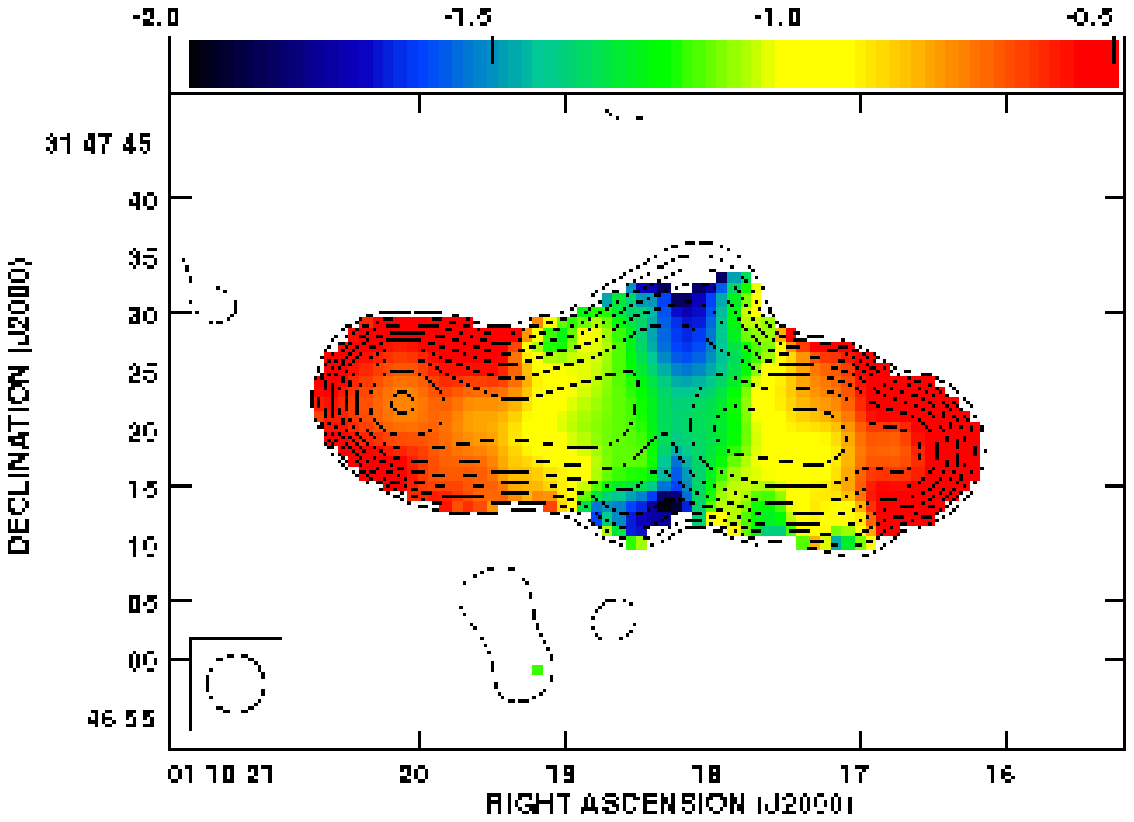}
\caption{Spectral index image in colour superimposed by 1.4~GHz radio
contours at a resolution of 5$\arcsec$ for
(left) 3C13 and (right) 3C34. The spectral index image was obtained
with 327.6~MHz and 1.4 GHz data.}
\label{fig3c13spix}
\end{figure}

\clearpage

\begin{figure}
\epsscale{1.0}
\plottwo{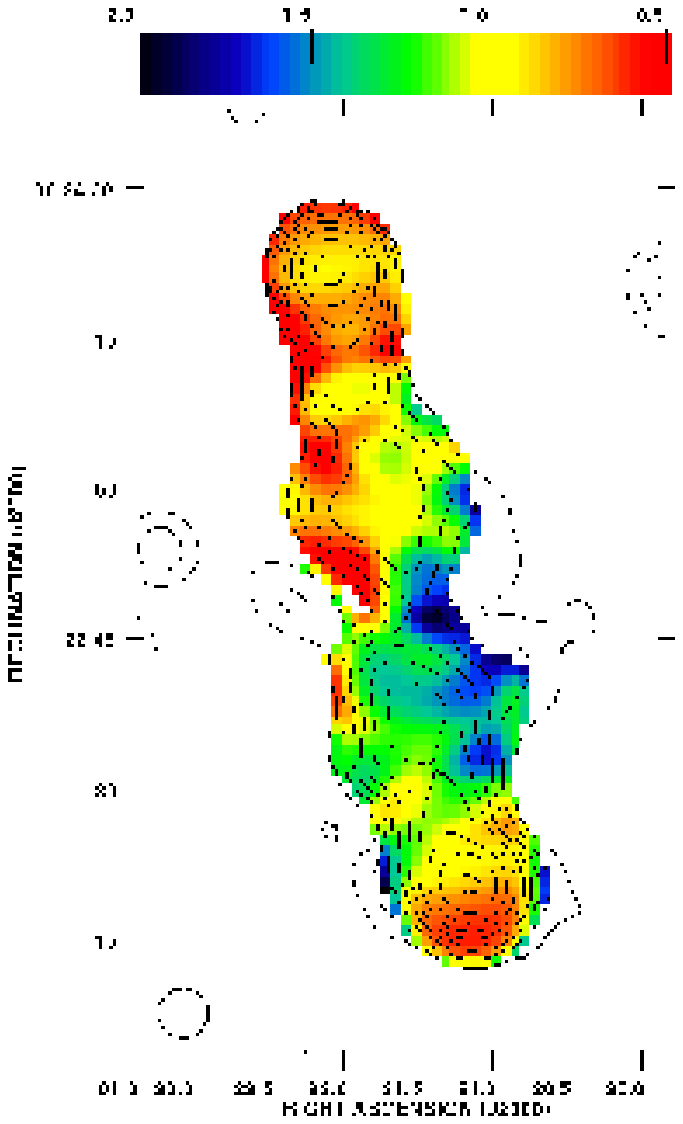}{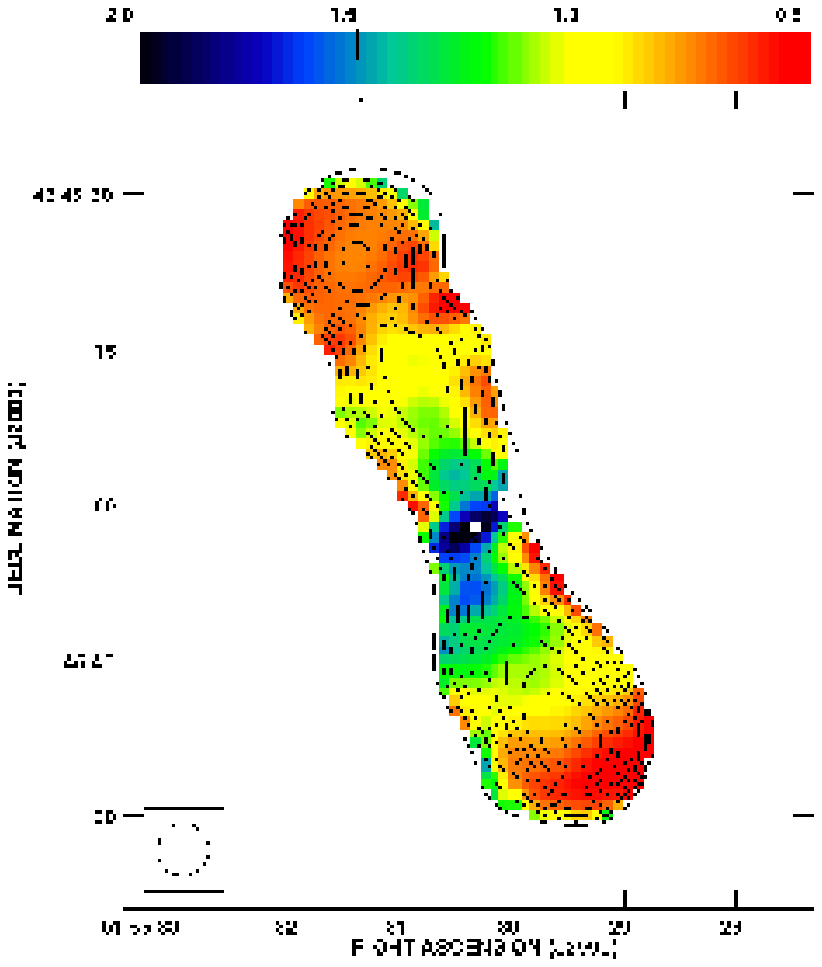}
\caption{Spectral index image in colour superimposed by 1.4~GHz radio
contours at a resolution of 5$\arcsec$ for
(left) 3C44 and (right) 3C54. The spectral index image was obtained
with 327.6~MHz and 1.4 GHz data.}
\label{fig3c44spix}
\end{figure}

\begin{figure}
\epsscale{1.0}
\plottwo{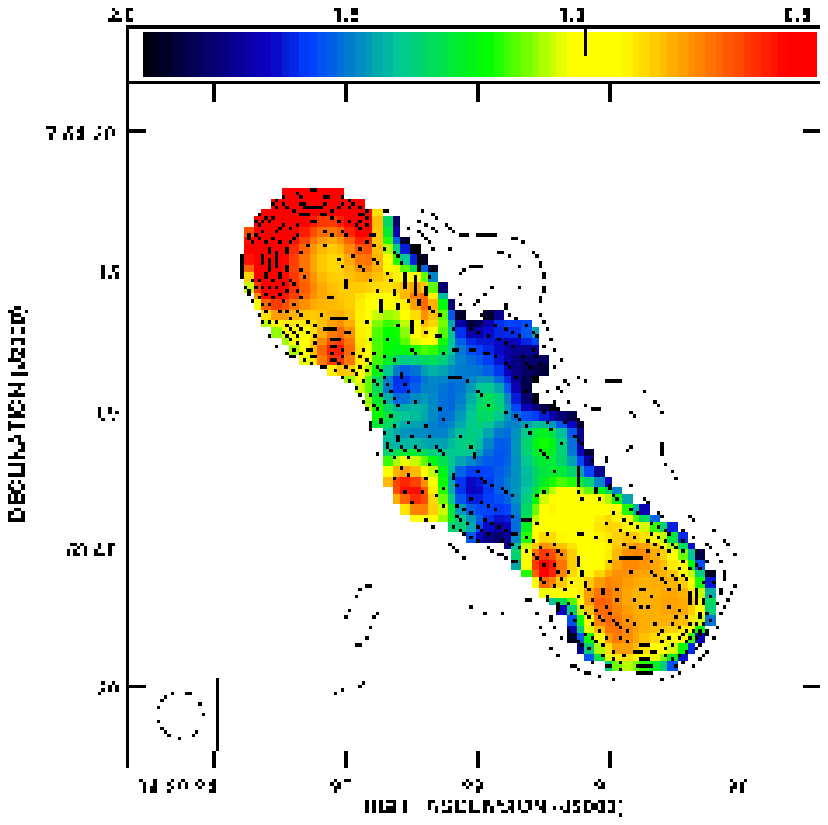}{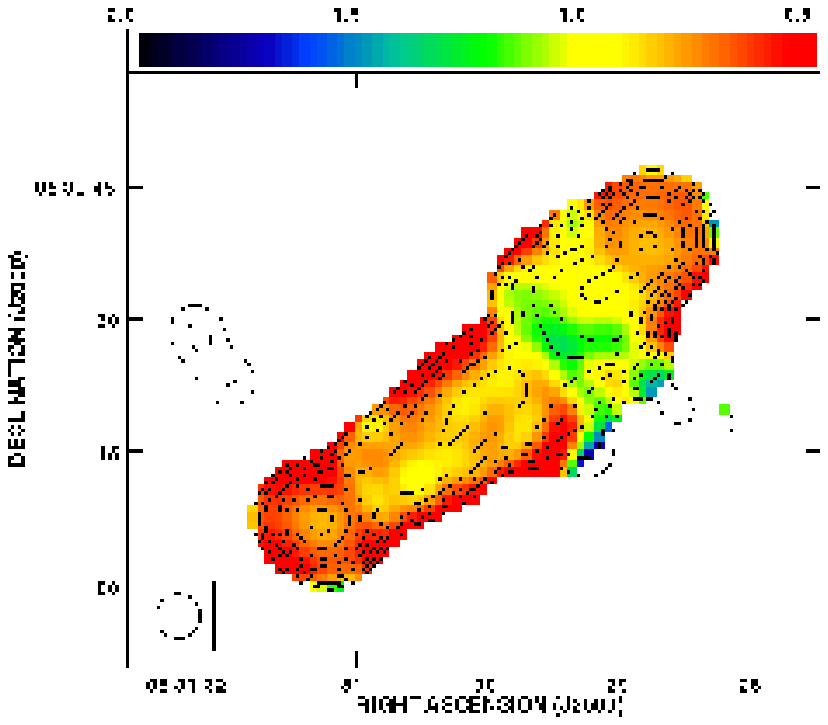}
\caption{Spectral index image in colour superimposed by 1.4~GHz radio
contours at a resolution of 5$\arcsec$ for
(left) 3C114 and (right) 3C142.1. The spectral index image was obtained
with 327.6~MHz and 1.4 GHz data.}
\label{fig3c114spix}
\end{figure}

\clearpage

\begin{figure}
\epsscale{1.0}
\plottwo{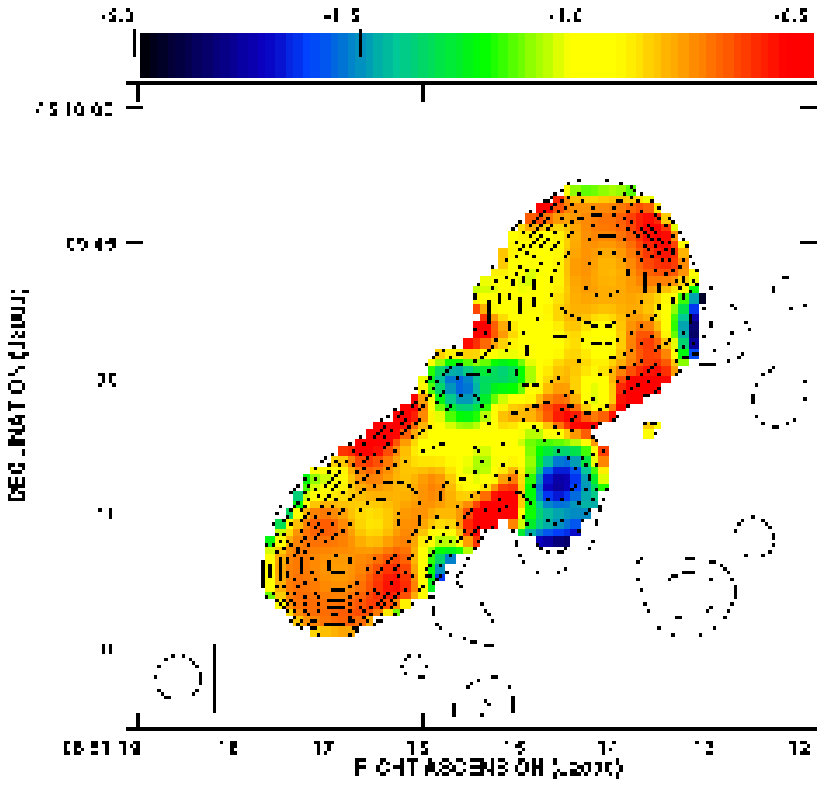}{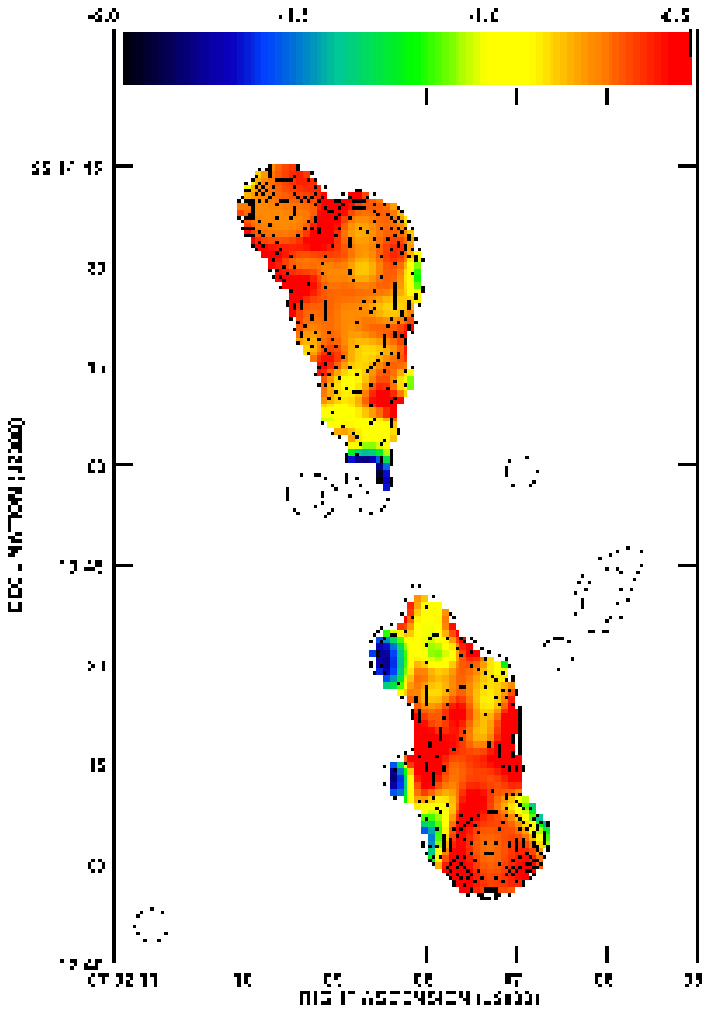}
\caption{Spectral index image in colour superimposed by 1.4~GHz radio
contours at a resolution of 5$\arcsec$ for
(left) 3C169.1 and (right) 3C172. The spectral index image was obtained
with 327.6~MHz and 1.4 GHz data.}
\label{fig3c169spix}
\end{figure}

\begin{figure}
\epsscale{1.0}
\plottwo{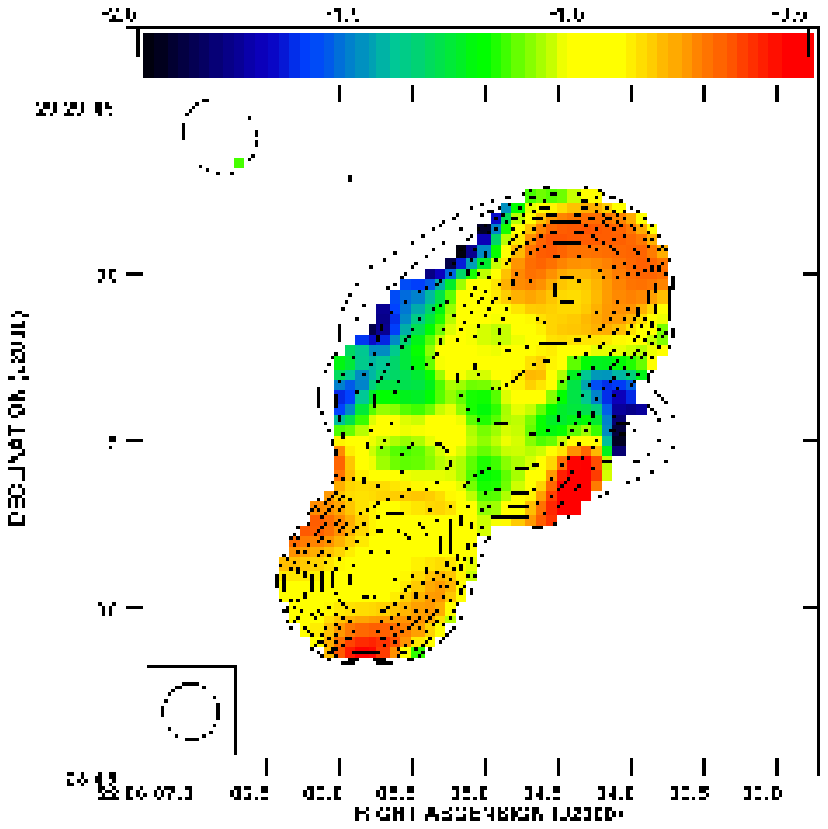}{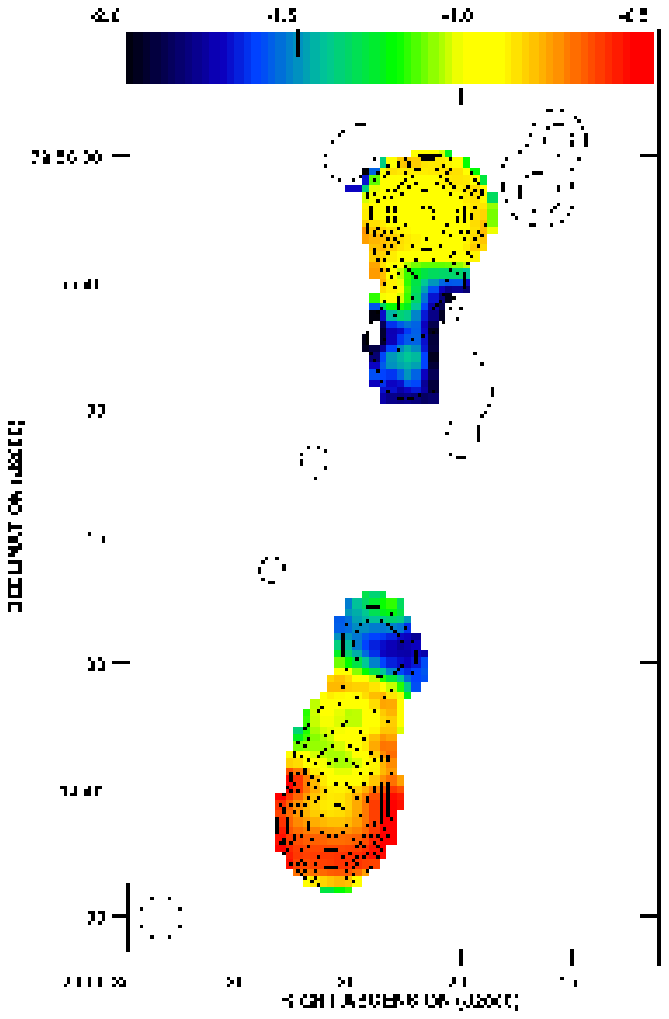}
\caption{Spectral index image in colour superimposed by 1.4~GHz radio
contours at a resolution of 5$\arcsec$ for
(left) 3C441 and (right) 3C469.1. The spectral index image was obtained
with 327.6~MHz and 1.4 GHz data.}
\label{fig3c441spix}
\end{figure}

\begin{figure}
\epsscale{0.4}
\plotone{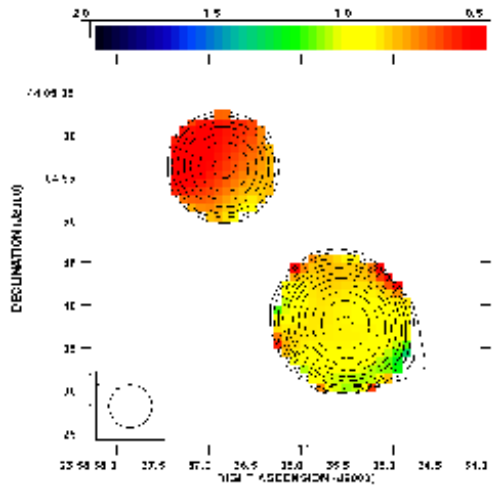}
\caption{Spectral index image in colour superimposed by 1.4~GHz radio
contours at a resolution of 5$\arcsec$ for 3C470. The spectral index image was obtained with 327.6~MHz and 1.4 GHz data.}
\label{fig3c470sp}
\end{figure}

\begin{figure}
\centering{
\includegraphics[width=8.7cm]{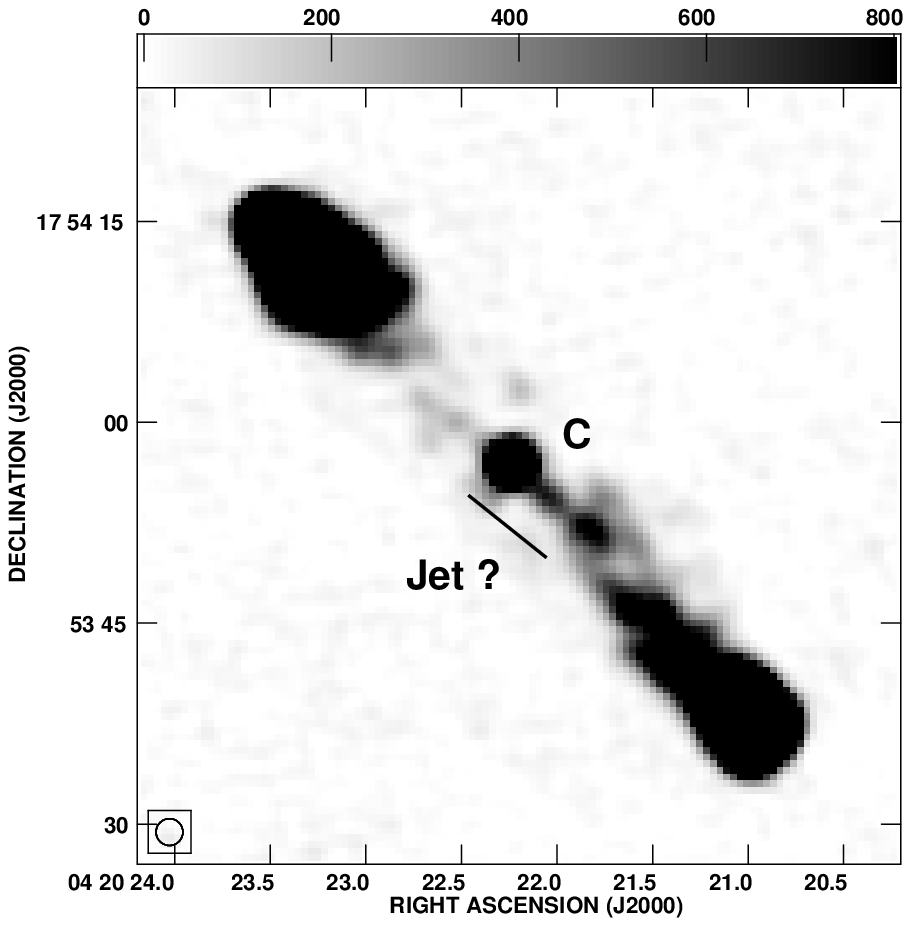}
\includegraphics[width=7.0cm]{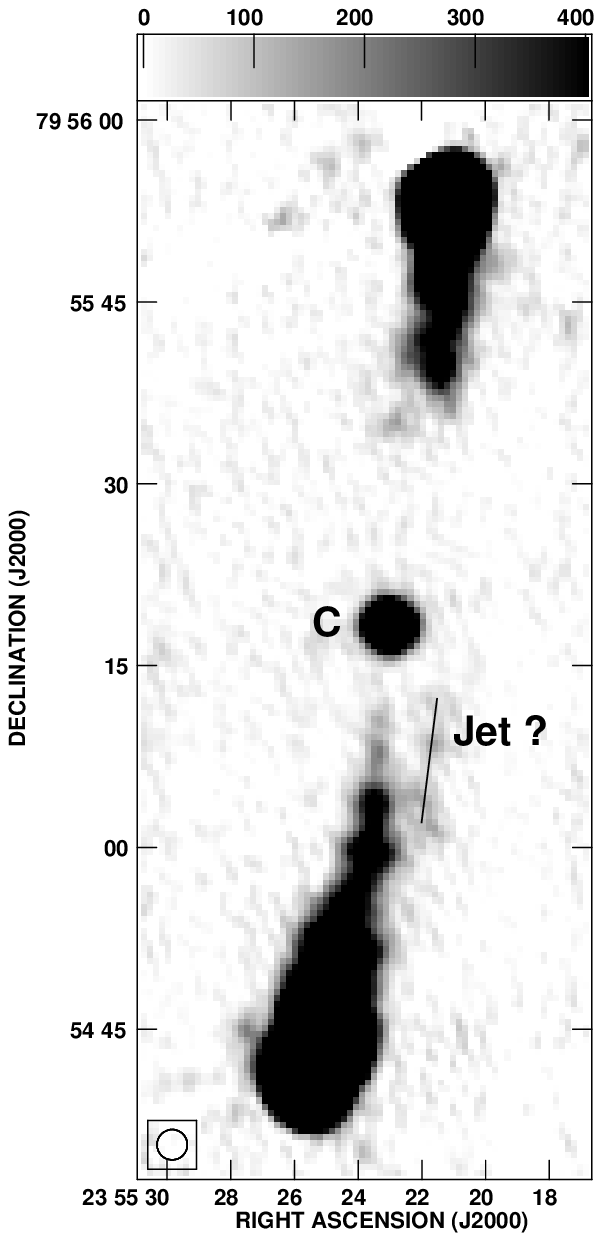}}
\caption{The 5 GHz grey-scale image of (left) 3C114 and (right) 3C469.1, 
showing the jet-like feature.}
\label{figgrey3c44}
\end{figure}
\clearpage

\acknowledgements
We would like to express our thanks to the referee for a careful assessment
of our work, which has significantly improved this paper.
We thank Joel C. Carvalho for stimulating discussions on this work. 
This work was supported in part by the U. S. National Science Foundation
under grant AST-0507465 (R. A. D.).
The National Radio Astronomy Observatory is a facility of the National
Science Foundation operated under cooperative agreement by Associated
Universities, Inc.
This research has made use of the NASA/IPAC Extragalactic Database (NED)
which is operated by the Jet Propulsion Laboratory, California Institute
of Technology, under contract with the National Aeronautics and Space
Administration.

{\it Facility:} \facility{VLA (A, B, C and D-array configurations)}

\appendix
\section{Notes on Individual Sources}

{\bf 3C6.1:}
The Hubble Space Telescope ($HST$) observations with WFPC2/F702W by
\citet{McCarthy97} show the optical galaxy to be compact with the lower
isophotes extending at PA$\sim30\degr$, roughly aligned with the radio
axis \citep{Pooley74}.
The overall radio morphology could be classified as type LW1.
The spectral index $\alpha^{1.4}_{5.0}$ shows a general steepening away from
the hotspot and towards the core.

{\bf 3C13:}
The $HST$ observations of this source show the optical galaxy to be
extended over nearly
4$\arcsec$ with two tail-like structures to the south and east of the
nucleus \citep{McCarthy97}. The extended emission in both the nucleus and on
the 1$\arcsec$ scale is aligned with the radio axis at a PA of 145$\degr$.
\citet{Best97,Best98} find that the infrared emission from the host
galaxy is also strongly aligned with the radio axis.
Our radio image shows clear hotspots but no radio core in this high-redshift
source (Fig.~\ref{fig3c13}). More sensitive observations are required to
be able to image the full radio bridge.

{\bf 3C34:}
This source lies in a compact cluster of galaxies \citep{McCarthy95}. The
$HST$ observations of \citet{McCarthy97} find that the
central galaxy is diffuse with a low surface brightness.
\citet{BestLongair97} have proposed that the aligned optical emission
observed in the $HST$ images of 3C34 is associated with a region of
massive star formation, induced by the passage of the radio jet through a
galaxy within the surrounding cluster. In Fig.~\ref{fig3c34} we see that
3C34 shows twin hotspots on the eastern side. The polarization image shows
extensive structure and changes in the orientation of the magnetic field.
The 5~GHz greyscale image shows a faint jet on both sides of the core
with the jet on the western side being brighter closer to the core. 

{\bf 3C41:}
The $HST$ observations of \citet{McCarthy97,Best97} show that the optical
galaxy is compact and symmetrical. The K-band image of the galaxy indicates
that it is misaligned by about 25$\degr$ with the radio axis
\citep{Best97,Best98}. We observe extended radio bridge region in this
source, but the radio core is observed only in the 8 GHz image
(see Fig.~\ref{fig3c41}). The high resolution image at 8 GHz
also shows a complex hotspot region. The radio morphology shows a LW3-type
distortion, although the southern lobe seems to
turn back towards the core close to it.
The spectral index map shows a flatter spectral
index extending through the centre of the northern lobe, perhaps suggesting
the presence of the radio jet (Fig~\ref{fig3c34sp}).

{\bf 3C44: }
This galaxy lies in a cluster \citep{Spinrad86,McCarthy95}. The $HST$
observations show that the galaxy is either composed of two components
oriented north-south or, more likely, is bisected by a dust lane running
east-west \citep{McCarthy97}. Figure~\ref{fig3c44} shows that this source
exhibits a LW1-type radio morphology. 
We detect a weak radio core at the position, 
RA=01h,31m,21.6459s, 
Dec=06$\degr$,23$\arcmin$,43.1048$\arcsec$.
There is a bright radio source on
the eastern side about 50$\arcsec$ from the core in 3C44.
The spectral index shows a general steepening away from the hotspots
(Fig.~\ref{fig3c44sp}).
3C44 shows a hint of a jet-like structure extending towards the southern
hotspot in the 5 GHz image (Fig.~\ref{figgrey3c44}). 

{\bf 3C54:}
The $HST$ observations reveal a compact nucleus in this galaxy with an
extension to the southwest in PA 220$\degr$ \citep{McCarthy97}. There is a
compact object near the end of the 2$\arcsec$ extension from the nucleus.
It is unclear if the extension is a bridge, tidal tail, or jet. There is
another companion object a few arcseconds to the southeast. In
Fig.\ref{fig3c54} we see that this source shows a LW1-type lobe morphology.
No radio core is observed. The northern hotspot shows two components in the
5 GHz image. 3C54 shows a jet-like structure
extending towards the southern hotspot in the 5 GHz image.

{\bf 3C114:}
The low surface brightness optical galaxy of 3C114 is poorly detected in
the $HST$ observations \citep{McCarthy97}. A faint compact nucleus with
several clumps within the central few arcseconds is observed in this source.
The radio bridge region and core are clearly detected in our
observations (Fig.\ref{fig3c114}).
3C114 shows a jet-like structure extending towards the southern hotspot in
the 5 GHz image. However, the jet-like feature is more prominent on the
northern side in the 1.4 GHz image. 

{\bf 3C142.1:}
Extended [OII] emission in an elongated east-west structure $\sim5\arcsec$
across has been observed for 3C142.1 by \citet{Hes96}.
The radio source lies close to a PA$\sim40\degr$, not aligned with the
emission-line region. We see clear extended radio lobes in 3C142.1.
(Fig.~\ref{fig3c142.1}).
The southern lobe shows a nearly constant spectral index across it
(Fig.~\ref{fig3c114sp}).
This is an interesting source with a bridge
structure that is different from most sources. It does not show the
classic surface brightness decline from the hotspot toward the core, has
a roughly constant spectral index, and a complex hotspot structure.

{\bf 3C169.1:}
\citet{McCarthy97} have observed two galaxies in this system, both with
extended structure. The host galaxy of 3C169.1 shows a clear extension to the
east with a number of faint clumps within the central arcsecond.
The two components lie along the radio axis, and the companion object has an
extension pointing towards the radio galaxy. In Fig.~\ref{fig3c169.1} we see
that the northern hotspot is much fainter than the southern hotspot.
The northern hotspot is not detected in the 8~GHz image.

{\bf 3C172:}
The $HST$ image of 3C712 shows that it lies in a complex system -- four
objects are detected in the optical image \citep{McCarthy97}. There are
two galaxies lying to the southwest that appear to be associated with 3C172,
and they lie along the axis of the radio source. We find that this source
has an interesting radio structure (see Fig.~\ref{fig3c172}). The bridge
appears to pull away to the west side of the source. The polarization image
suggests that the magnetic field has been carried to the side of the radio
bridge along with the relativistic plasma. However, the 8~GHz image
indicates that the two brightest hotspots and the weak radio core lie
roughly along a line. 3C172 shows a jet-like feature closer to the northern
hotspot in the 5~GHz image. 

{\bf 3C441:}
The $HST$ observations of 3C441 by \citet{McCarthy97} and \citet{Best97} show
that the host galaxy is compact and lies in a cluster. The optical galaxy
shows a slight extension to the northwest, along the radio axis.
The $HST$ image shows a diffuse object located 12$\arcsec$ north and
8$\arcsec$ west of the core galaxy. We find that this source has a complex
radio morphology, especially in the northern radio lobe and hotspot
(Fig.~\ref{fig3c441}). The maps suggest that the jet has shifted position
moving from the west to the present position indicated by the brightest
hotspot. The bridge seems to exhibit a type LW3 lobe morphology.
The 5 GHz greyscale image shows a jet-like feature extending towards the
northern hotspot. 

{\bf 3C469.1:}
The $HST$ observations of \citet{McCarthy97} find that the host galaxy
of 3C469.1 shows a double structure at PA 95$\degr$, not aligned with the
radio axis. We detect a bright radio core in this source but not
sufficient radio bridge emission (Fig.~\ref{fig3c469.1}).
The source is at a redshift of $z$=1.336 and more sensitive observations
are needed to observe the full bridge region in the source
3C469.1 shows a jet-like feature extending towards the southern hotspot in
the 5 GHz image. 

{\bf 3C470:}
\citet{Best98} find that the host galaxy of 3C470 is slightly extended
in both the optical and the K-band images, with the elongation being
highly misaligned (by 80$\degr$) with the radio axis. We find that the
radio bridge in 3C470 is partly visible only on the southern side
(Fig.~\ref{fig3c469.1}). This is the highest redshift object in our
sample ($z$=1.653). Therefore more sensitive observations are required to
observe the full bridge region. We detect a radio core at the position,
RA=23h,58m,35.9063s, 
Dec=44$\degr$,04$\arcmin$,45.5180$\arcsec$.
3C470 shows a slight extension in the radio core towards the southern
hotspot in the 5 GHz image.


\begin{thebibliography}{}

\bibitem[\protect\citeauthoryear{{Alexander}}{{Alexander}}{1987}]{Alexander87}
{Alexander} P.,  1987, \mnras, 225, 27

\bibitem[\protect\citeauthoryear{{Alexander} \& {Leahy}}{{Alexander} \&
  {Leahy}}{1987}]{AlexanderLeahy87}
{Alexander} P.,  {Leahy} J.~P.,  1987, \mnras, 225, 1

\bibitem[\protect\citeauthoryear{{Athreya} \& {Kapahi}}{{Athreya} \&
  {Kapahi}}{1999}]{Athreya99}
{Athreya} R.~M.,  {Kapahi} V.~K.,  1999, in {R{\"o}ttgering} H.~J.~A.,  {Best}
  P.~N.,   {Lehnert} M.~D.,  eds, The Most Distant Radio Galaxies {Radio
  Galaxies from the MRC/1Jy sample}.
p.~453

\bibitem[\protect\citeauthoryear{{Baars}, {Genzel}, {Pauliny-Toth} \&
  {Witzel}}{{Baars} et~al.}{1977}]{Baars77}
  {Baars} J.~W.~M.,  {Genzel} R.,  {Pauliny-Toth} I.~I.~K.,    {Witzel} A.,
    1977, \aap, 61, 99

\bibitem[\protect\citeauthoryear{{Baldwin} \& {Scott}}{{Baldwin} \&
  {Scott}}{1973}]{BaldwinScott73}
{Baldwin} J.~E.,  {Scott} P.~F.,  1973, \mnras, 165, 259

\bibitem[\protect\citeauthoryear{{Barthel} \& {Miley}}{{Barthel} \&
  {Miley}}{1988}]{BarthelMiley88}
{Barthel} P.~D.,  {Miley} G.~K.,  1988, \nat, 333, 319

\bibitem[\protect\citeauthoryear{{Begelman}, {Blandford} \& {Rees}}{{Begelman}
  et~al.}{1984}]{Begelman84}
{Begelman} M.~C.,  {Blandford} R.~D.,    {Rees} M.~J.,  1984, Reviews of Modern
  Physics, 56, 255

\bibitem[\protect\citeauthoryear{{Begelman} \& {Cioffi}}{{Begelman} \&
  {Cioffi}}{1989}]{BegelmanCioffi89}
{Begelman} M.~C.,  {Cioffi} D.~F.,  1989, \apjl, 345, L21

\bibitem[\protect\citeauthoryear{{Bennett}}{{Bennett}}{1962}]{Bennett62}
{Bennett} A.~S.,  1962, \memras, 68, 163

\bibitem[\protect\citeauthoryear{{Best}, {Bailer}, {Longair} \& {Riley}}{{Best}
  et~al.}{1995}]{Best95}
{Best} P.~N.,  {Bailer} D.~M.,  {Longair} M.~S.,    {Riley} J.~M.,  1995,
  \mnras, 275, 1171

\bibitem[\protect\citeauthoryear{{Best}, {Longair} \& {Roettgering}}{{Best}
  et~al.}{1998}]{Best98}
{Best} P.~N.,  {Longair} M.~S.,    {Roettgering} H.~J.~A.,  1998, \mnras, 295,
  549

\bibitem[\protect\citeauthoryear{{Best}, {Longair} \& {Roettgering}}{{Best}
  et~al.}{1997}]{Best97}
{Best} P.~N.,  {Longair} M.~S.,    {Roettgering} J.~H.~A.,  1997, \mnras, 292,
  758

\bibitem[\protect\citeauthoryear{{Best}, {Longair} \& {Rottgering}}{{Best}
  et~al.}{1997}]{BestLongair97}
{Best} P.~N.,  {Longair} M.~S.,    {Rottgering} H.~J.~A.,  1997, \mnras, 286,
  785

\bibitem[\protect\citeauthoryear{{Best}, {R{\"o}ttgering} \& {Longair}}{{Best}
  et~al.}{2000}]{Best00}
{Best} P.~N.,  {R{\"o}ttgering} H.~J.~A.,    {Longair} M.~S.,  2000, \mnras,
  311, 23

\bibitem[\protect\citeauthoryear{{Blundell}, {Rawlings} \&
  {Willott}}{{Blundell} et~al.}{1999}]{Blundell99}
{Blundell} K.~M.,  {Rawlings} S.,    {Willott} C.~J.,  1999, \aj, 117, 677

\bibitem[\protect\citeauthoryear{{Bogers}, {Hes}, {Barthel} \&
  {Zensus}}{{Bogers} et~al.}{1994}]{Bogers94}
{Bogers} W.~J.,  {Hes} R.,  {Barthel} P.~D.,    {Zensus} J.~A.,  1994, \aaps,
  105, 91

\bibitem[\protect\citeauthoryear{{Bridle}, {Hough}, {Lonsdale}, {Burns} \&
  {Laing}}{{Bridle} et~al.}{1994}]{Bridle94}
{Bridle} A.~H.,  {Hough} D.~H.,  {Lonsdale} C.~J.,  {Burns} J.~O.,    {Laing}
  R.~A.,  1994, \aj, 108, 766

\bibitem[\protect\citeauthoryear{{Carilli}, {Perley}, {Dreher} \&
  {Leahy}}{{Carilli} et~al.}{1991}]{Carilli91}
{Carilli} C.~L.,  {Perley} R.~A.,  {Dreher} J.~W.,    {Leahy} J.~P.,  1991,
  \apj, 383, 554

\bibitem[\protect\citeauthoryear{{Carvalho} \& {O'Dea}}{{Carvalho} \&
  {O'Dea}}{2002a}]{Carvalho02a}
{Carvalho} J.~C.,  {O'Dea} C.~P.,  2002a, \apjs, 141, 337

\bibitem[\protect\citeauthoryear{{Carvalho} \& {O'Dea}}{{Carvalho} \&
  {O'Dea}}{2002b}]{Carvalho02b}
{Carvalho} J.~C.,  {O'Dea} C.~P.,  2002b, \apjs, 141, 371

\bibitem[\protect\citeauthoryear{{Chambers}, {Miley} \& {van
  Breugel}}{{Chambers} et~al.}{1990}]{Chambers90}
{Chambers} K.~C.,  {Miley} G.~K.,    {van Breugel} W.~J.~M.,  1990, \apj, 363,
  21

\bibitem[\protect\citeauthoryear{{Daly}}{{Daly}}{1990}]{Daly90}
{Daly} R.~A.,  1990, \apj, 355, 416

\bibitem[\protect\citeauthoryear{{De Breuck}, {van Breugel}, {R{\"o}ttgering}
  \& {Miley}}{{De Breuck} et~al.}{2000}]{DeBreuck00}
{De Breuck} C.,  {van Breugel} W.,  {R{\"o}ttgering} H.~J.~A.,    {Miley} G.,
  2000, \aaps, 143, 303

\bibitem[\protect\citeauthoryear{{Dennett-Thorpe}, {Bridle}, {Laing} \&
  {Scheuer}}{{Dennett-Thorpe} et~al.}{1999}]{Dennett-Thorpe99}
{Dennett-Thorpe} J.,  {Bridle} A.~H.,  {Laing} R.~A.,    {Scheuer} P.~A.~G.,
  1999, \mnras, 304, 271

\bibitem[\protect\citeauthoryear{{Dennett-Thorpe}, {Bridle}, {Scheuer}, {Laing}
  \& {Leahy}}{{Dennett-Thorpe} et~al.}{1997}]{Dennett-Thorpe97}
{Dennett-Thorpe} J.,  {Bridle} A.~H.,  {Scheuer} P.~A.~G.,  {Laing} R.~A.,
  {Leahy} J.~P.,  1997, \mnras, 289, 753

\bibitem[\protect\citeauthoryear{{Fanaroff} \& {Riley}}{{Fanaroff} \&
  {Riley}}{1974}]{FanaroffRiley74}
{Fanaroff} B.~L.,  {Riley} J.~M.,  1974, \mnras, 167, 31P

\bibitem[\protect\citeauthoryear{{Fernini}, {Burns} \& {Perley}}{{Fernini}
  et~al.}{1997}]{Fernini97}
{Fernini} I.,  {Burns} J.~O.,    {Perley} R.~A.,  1997, \aj, 114, 2292

\bibitem[\protect\citeauthoryear{{Garrington}, {Conway} \&
  {Leahy}}{{Garrington} et~al.}{1991}]{GarringtonConway91}
{Garrington} S.~T.,  {Conway} R.~G.,    {Leahy} J.~P.,  1991, \mnras, 250, 171

\bibitem[\protect\citeauthoryear{{Garrington}, {Leahy}, {Conway} \&
  {Laing}}{{Garrington} et~al.}{1988}]{Garrington88}
{Garrington} S.~T.,  {Leahy} J.~P.,  {Conway} R.~G.,    {Laing} R.~A.,  1988,
  \nat, 331, 147

\bibitem[\protect\citeauthoryear{{Gilbert}, {Riley}, {Hardcastle}, {Croston},
  {Pooley} \& {Alexander}}{{Gilbert} et~al.}{2004}]{Gilbert04}
{Gilbert} G.~M.,  {Riley} J.~M.,  {Hardcastle} M.~J.,  {Croston} J.~H.,
  {Pooley} G.~G.,    {Alexander} P.,  2004, \mnras, 351, 845

\bibitem[\protect\citeauthoryear{{Goodlet} \& {Kaiser}}{{Goodlet} \&
  {Kaiser}}{2005}]{Goodlet05}
{Goodlet} J.~A.,  {Kaiser} C.~R.,  2005, \mnras, 359, 1456

\bibitem[\protect\citeauthoryear{{Goodlet}, {Kaiser}, {Best} \&
  {Dennett-Thorpe}}{{Goodlet} et~al.}{2004}]{Goodlet04}
{Goodlet} J.~A.,  {Kaiser} C.~R.,  {Best} P.~N.,    {Dennett-Thorpe} J.,  2004,
  \mnras, 347, 508

\bibitem[\protect\citeauthoryear{{Gopal-Krishna}}{{Gopal-Krishna}}{1988}]{Gopa%
l-Krishna88}
{Gopal-Krishna} 1988, \aap, 192, 37

\bibitem[\protect\citeauthoryear{{Gopal-Krishna} \& {Wiita}}{{Gopal-Krishna} \&
  {Wiita}}{2004}]{Gopal-Krishna04}
{Gopal-Krishna} {Wiita} P.~J.,  2004, ArXiv Astrophysics e-prints

\bibitem[\protect\citeauthoryear{{Gower}, {Scott} \& {Wills}}{{Gower}
  et~al.}{1967}]{Gower67}
  {Gower} J.~F.~R.,  {Scott} P.~F.,    {Wills} D.,  1967, \memras, 71, 49

\bibitem[\protect\citeauthoryear{{Hardcastle}, {Alexander}, {Pooley} \&
  {Riley}}{{Hardcastle} et~al.}{1998}]{Hardcastle98}
{Hardcastle} M.~J.,  {Alexander} P.,  {Pooley} G.~G.,    {Riley} J.~M.,  1998,
  \mnras, 296, 445

\bibitem[\protect\citeauthoryear{{Hes}, {Barthel} \& {Fosbury}}{{Hes}
  et~al.}{1996}]{Hes96}
{Hes} R.,  {Barthel} P.~D.,    {Fosbury} R.~A.~E.,  1996, \aap, 313, 423

\bibitem[\protect\citeauthoryear{{Hough} \& {Readhead}}{{Hough} \&
  {Readhead}}{1989}]{HoughReadhead89}
{Hough} D.~H.,  {Readhead} A.~C.~S.,  1989, \aj, 98, 1208

\bibitem[\protect\citeauthoryear{{Inskip}, {Best}, {Rawlings}, {Longair},
  {Cotter}, {R{\"o}ttgering} \& {Eales}}{{Inskip} et~al.}{2002}]{Inskip02}
{Inskip} K.~J.,  {Best} P.~N.,  {Rawlings} S.,  {Longair} M.~S.,  {Cotter} G.,
  {R{\"o}ttgering} H.~J.~A.,    {Eales} S.,  2002, \mnras, 337, 1381

\bibitem[\protect\citeauthoryear{{Ishwara-Chandra} \&
  {Saikia}}{{Ishwara-Chandra} \& {Saikia}}{2000}]{Ishwara-Chandra00}
{Ishwara-Chandra} C.~H.,  {Saikia} D.~J.,  2000, \mnras, 317, 658

\bibitem[\protect\citeauthoryear{{Ishwara-Chandra}, {Saikia}, {McCarthy} \&
  {van Breugel}}{{Ishwara-Chandra} et~al.}{2001}]{Ishwara-Chandra01}
{Ishwara-Chandra} C.~H.,  {Saikia} D.~J.,  {McCarthy} P.~J.,    {van Breugel}
  W.~J.~M.,  2001, \mnras, 323, 460

\bibitem[\protect\citeauthoryear{{Jackson} \& {Rawlings}}{{Jackson} \&
  {Rawlings}}{1997}]{JacksonRawlings97}
{Jackson} N.,  {Rawlings} S.,  1997, \mnras, 286, 241

\bibitem[\protect\citeauthoryear{{Johnson}, {Leahy} \& {Garrington}}{{Johnson}
  et~al.}{1995}]{Johnson95}
{Johnson} R.~A.,  {Leahy} J.~P.,    {Garrington} S.~T.,  1995, \mnras, 273, 877

\bibitem[\protect\citeauthoryear{{Kaiser} \& {Alexander}}{{Kaiser} \&
  {Alexander}}{1997}]{KaiserAlexander97}
{Kaiser} C.~R.,  {Alexander} P.,  1997, \mnras, 286, 215

\bibitem[\protect\citeauthoryear{{Kapahi} \& {Kulkarni}}{{Kapahi} \&
  {Kulkarni}}{1990}]{KapahiKulkarni90}
{Kapahi} V.~K.,  {Kulkarni} V.~K.,  1990, \aj, 99, 1397

\bibitem[\protect\citeauthoryear{{Kapahi} \& {Saikia}}{{Kapahi} \&
  {Saikia}}{1982}]{KapahiSaikia82}
{Kapahi} V.~K.,  {Saikia} D.~J.,  1982, Journal of Astrophysics and Astronomy
  (ISSN 0250-6335), vol.~3, D ec.~1982, p.~465-483., 3, 465

\bibitem[\protect\citeauthoryear{{Klamer}, {Ekers}, {Bryant}, {Hunstead},
  {Sadler} \& {De Breuck}}{{Klamer} et~al.}{2006}]{Klamer06}
{Klamer} I.~J.,  {Ekers} R.~D.,  {Bryant} J.~J.,  {Hunstead} R.~W.,  {Sadler}
  E.~M.,    {De Breuck} C.,  2006, \mnras, 371, 852

\bibitem[\protect\citeauthoryear{{Krolik} \& {Chen}}{{Krolik} \&
  {Chen}}{1991}]{KrolikChen91}
{Krolik} J.~H.,  {Chen} W.,  1991, \aj, 102, 1659

\bibitem[\protect\citeauthoryear{{Kronberg}, {Conway} \& {Gilbert}}{{Kronberg}
  et~al.}{1972}]{Kronberg72}
{Kronberg} P.~P.,  {Conway} R.~G.,    {Gilbert} J.~A.,  1972, \mnras, 156, 275

\bibitem[\protect\citeauthoryear{{Kuehr}, {Witzel}, {Pauliny-Toth} \&
  {Nauber}}{{Kuehr} et~al.}{1996}]{Kuehr96}
{Kuehr} H.,  {Witzel} A.,  {Pauliny-Toth} I.~I.~K.,    {Nauber} U.,  1996,
  VizieR Online Data Catalog, 8005, 0

\bibitem[\protect\citeauthoryear{{Lacy}, {Hill}, {Kaiser} \& {Rawlings}}{{Lacy}
  et~al.}{1993}]{Lacy93}
{Lacy} M.,  {Hill} G.~J.,  {Kaiser} M.~E.,    {Rawlings} S.,  1993, \mnras,
  263, 707

\bibitem[\protect\citeauthoryear{{Laing}}{{Laing}}{1989}]{Laing89}
{Laing} R.,  1989, Lecture Notes in Physics, Berlin Springer Verlag, 327, 27

\bibitem[\protect\citeauthoryear{{Laing}}{{Laing}}{1981}]{Laing81}
{Laing} R.~A.,  1981, \mnras, 195, 261

\bibitem[\protect\citeauthoryear{{Laing}}{{Laing}}{1988}]{Laing88}
{Laing} R.~A.,  1988, \nat, 331, 149

\bibitem[\protect\citeauthoryear{{Laing}}{{Laing}}{1993}]{Laing93}
{Laing} R.~A.,  1993, in {Davis} R.~J.,  {Booth} R.~S.,  eds, Sub-arcsecond
  Radio Astronomy {Relativistic flow in low luminosity radio jets}.
p.~346

\bibitem[\protect\citeauthoryear{{Laing}}{{Laing}}{1996}]{Laing96}
{Laing} R.~A.,  1996, in {Ekers} R.~D.,  {Fanti} C.,   {Padrielli} L.,  eds,
  IAU Symp. 175: Extragalactic Radio Sources {Large Scale Structure: Jets on
  kiloparsec Scales (Review)}.
p.~147

\bibitem[\protect\citeauthoryear{{Laing} \& {Peacock}}{{Laing} \&
  {Peacock}}{1980}]{LaingPeacock80}
{Laing} R.~A.,  {Peacock} J.~A.,  1980, \mnras, 190, 903

\bibitem[\protect\citeauthoryear{{Laing}, {Riley} \& {Longair}}{{Laing}
  et~al.}{1983}]{LaingRiley83}
  {Laing} R.~A.,  {Riley} J.~M.,    {Longair} M.~S.,  1983, \mnras, 204, 151

\bibitem[\protect\citeauthoryear{{Law-Green}, {Leahy}, {Alexander},
  {Allington-Smith}, {van Breugel}, {Eales}, {Rawlings} \&
  {Spinrad}}{{Law-Green} et~al.}{1995}]{LawGreen95}
{Law-Green} J.~D.~B.,  {Leahy} J.~P.,  {Alexander} P.,  {Allington-Smith}
  J.~R.,  {van Breugel} W.~J.~M.,  {Eales} S.~A.,  {Rawlings} S.~G.,
  {Spinrad} H.,  1995, \mnras, 274, 939

\bibitem[\protect\citeauthoryear{{Leahy}}{{Leahy}}{1987}]{Leahy87}
{Leahy} J.~P.,  1987, \mnras, 226, 433

\bibitem[\protect\citeauthoryear{{Leahy}, {Muxlow} \& {Stephens}}{{Leahy}
  et~al.}{1989}]{LeahyMuxlow89}
{Leahy} J.~P.,  {Muxlow} T.~W.~B.,    {Stephens} P.~W.,  1989, \mnras, 239, 401

\bibitem[\protect\citeauthoryear{{Leahy} \& {Williams}}{{Leahy} \&
  {Williams}}{1984}]{LeahyWilliams84}
{Leahy} J.~P.,  {Williams} A.~G.,  1984, \mnras, 210, 929

\bibitem[\protect\citeauthoryear{{Liu} \& {Pooley}}{{Liu} \&
  {Pooley}}{1991a}]{LiuPooleyB91}
{Liu} R.,  {Pooley} G.,  1991a, \mnras, 249, 343

\bibitem[\protect\citeauthoryear{{Liu} \& {Pooley}}{{Liu} \&
  {Pooley}}{1991b}]{LiuPooleyA91}
{Liu} R.,  {Pooley} G.,  1991b, \mnras, 253, 669

\bibitem[\protect\citeauthoryear{{Liu}, {Pooley} \& {Riley}}{{Liu}
  et~al.}{1992}]{Liu92}
{Liu} R.,  {Pooley} G.,    {Riley} J.~M.,  1992, \mnras, 257, 545

\bibitem[\protect\citeauthoryear{{Longair}}{{Longair}}{1975}]{Longair75}
{Longair} M.~S.,  1975, \mnras, 173, 309

\bibitem[\protect\citeauthoryear{{MacDonald}, {Kenderdine} \&
  {Neville}}{{MacDonald} et~al.}{1968}]{MacDonald68}
{MacDonald} G.~H.,  {Kenderdine} S.,    {Neville} A.~C.,  1968, \mnras, 138,
  259

\bibitem[\protect\citeauthoryear{{Macklin}}{{Macklin}}{1981}]{Macklin81}
{Macklin} J.~T.,  1981, \mnras, 196, 967

\bibitem[\protect\citeauthoryear{{McCarthy}}{{McCarthy}}{1993}]{McCarthy93}
{McCarthy} P.~J.,  1993, \araa, 31, 639

\bibitem[\protect\citeauthoryear{{McCarthy}, {Miley}, {de Koff}, {Baum},
  {Sparks}, {Golombek}, {Biretta} \& {Macchetto}}{{McCarthy}
  et~al.}{1997}]{McCarthy97}
{McCarthy} P.~J.,  {Miley} G.~K.,  {de Koff} S.,  {Baum} S.~A.,  {Sparks}
  W.~B.,  {Golombek} D.,  {Biretta} J.,    {Macchetto} F.,  1997, \apjs, 112,
  415

\bibitem[\protect\citeauthoryear{{McCarthy}, {Spinrad} \& {van
  Breugel}}{{McCarthy} et~al.}{1995}]{McCarthy95}
{McCarthy} P.~J.,  {Spinrad} H.,    {van Breugel} W.,  1995, \apjs, 99, 27

\bibitem[\protect\citeauthoryear{{McCarthy} \& {van Breugel}}{{McCarthy} \&
  {van Breugel}}{1989}]{McCarthy89}
{McCarthy} P.~J.,  {van Breugel} W.,  1989, in {Merus} E.~J.~A.,  {Fosbury}
  R.~A.~E.,  eds, Extranuclear Activity in Galaxies {Emission Line Properties
  of High Redshift Radio Galaxies}.
p.~55

\bibitem[\protect\citeauthoryear{{McCarthy}, {van Breugel} \&
  {Kapahi}}{{McCarthy} et~al.}{1991}]{McCarthy91}
{McCarthy} P.~J.,  {van Breugel} W.,    {Kapahi} V.~K.,  1991, \apj, 371, 478

\bibitem[\protect\citeauthoryear{{Morris} \& {Tabara}}{{Morris} \&
  {Tabara}}{1973}]{Morris73}
{Morris} D.,  {Tabara} H.,  1973, \pasj, 25, 295

\bibitem[\protect\citeauthoryear{{Mullin}, {Hardcastle} \& {Riley}}{{Mullin}
  et~al.}{2006}]{Mullin06}
{Mullin} L.~M.,  {Hardcastle} M.~J.,    {Riley} J.~M.,  2006, \mnras, 372, 113

\bibitem[\protect\citeauthoryear{{Myers} \& {Spangler}}{{Myers} \&
  {Spangler}}{1985}]{MyersSpangler85}
{Myers} S.~T.,  {Spangler} S.~R.,  1985, \apj, 291, 52

\bibitem[\protect\citeauthoryear{{Napier}, {Thompson} \& {Ekers}}{{Napier}
  et~al.}{1983}]{Napier83}
{Napier} P.~J.,  {Thompson} A.~R.,    {Ekers} R.~D.,  1983, IEEE Proceedings,
  71, 1295

\bibitem[\protect\citeauthoryear{{Neff}, {Roberts} \& {Hutchings}}{{Neff}
  et~al.}{1995}]{Neff95}
{Neff} S.~G.,  {Roberts} L.,    {Hutchings} J.~B.,  1995, \apjs, 99, 349

\bibitem[\protect\citeauthoryear{{Orr} \& {Browne}}{{Orr} \&
  {Browne}}{1982}]{OrrBrowne82}
{Orr} M.~J.~L.,  {Browne} I.~W.~A.,  1982, \mnras, 200, 1067

\bibitem[\protect\citeauthoryear{{Pedelty}, {Rudnick}, {McCarthy} \&
  {Spinrad}}{{Pedelty} et~al.}{1989}]{Pedelty89}
{Pedelty} J.~A.,  {Rudnick} L.,  {McCarthy} P.~J.,    {Spinrad} H.,  1989, \aj,
  97, 647

\bibitem[\protect\citeauthoryear{{Perley} \& {Taylor}}{{Perley} \&
  {Taylor}}{1991}]{PerleyTaylor91}
{Perley} R.~A.,  {Taylor} G.~B.,  1991, \aj, 101, 1623

\bibitem[\protect\citeauthoryear{{Perucho} \& {Mart{\'{\i}}}}{{Perucho} \&
  {Mart{\'{\i}}}}{2003}]{Perucho03}
{Perucho} M.,  {Mart{\'{\i}}} J.~M.,  2003, Publications of the Astronomical
  Society of Australia, 20, 94

\bibitem[\protect\citeauthoryear{{Pooley} \& {Henbest}}{{Pooley} \&
  {Henbest}}{1974}]{Pooley74}
{Pooley} G.~G.,  {Henbest} S.~N.,  1974, \mnras, 169, 477

\bibitem[\protect\citeauthoryear{{Privon}, {O'Dea}, {Baum}, {Axon}, {Kharb},
  {Buchanan}, {Sparks} \& {Chiaberge}}{{Privon} et~al.}{2007}]{Privon07}
{Privon} G.~C.,  {O'Dea} C.~P.,  {Baum} S.~A.,  {Axon} D.~J.,  {Kharb} P.,
  {Buchanan} C.~L.,  {Sparks} W.,    {Chiaberge} M.,  2007, submitted to ApJ
  Suppl. Ser.

\bibitem[\protect\citeauthoryear{{Rees} \& {Setti}}{{Rees} \&
  {Setti}}{1968}]{ReesSetti68}
{Rees} M.~J.,  {Setti} G.,  1968, \nat, 219, 127

\bibitem[\protect\citeauthoryear{{Riley} \& {Pooley}}{{Riley} \&
  {Pooley}}{1975}]{RileyPooley75}
{Riley} J.~M.,  {Pooley} G.~G.,  1975, \memras, 80, 105

\bibitem[\protect\citeauthoryear{{Roger}, {Costain} \& {Bridle}}{{Roger}
  et~al.}{1973}]{Roger73}
  {Roger} R.~S.,  {Costain} C.~H.,    {Bridle} A.~H.,  1973, \aj, 78, 1030

\bibitem[\protect\citeauthoryear{{Scheuer}}{{Scheuer}}{1982}]{Scheuer82}
{Scheuer} P.~A.~G.,  1982, in {Heeschen} D.~S.,  {Wade} C.~M.,  eds, IAU Symp.
  97: Extragalactic Radio Sources {Morphology and power of radio sources}.
pp 163--165

\bibitem[\protect\citeauthoryear{{Schilizzi}, {Kapahi} \& {Neff}}{{Schilizzi}
  et~al.}{1982}]{Schilizzi82}
{Schilizzi} R.~T.,  {Kapahi} V.~K.,    {Neff} S.~G.,  1982, Journal of
  Astrophysics and Astronomy, 3, 173

\bibitem[\protect\citeauthoryear{{Schwab}}{{Schwab}}{1980}]{Schwab80}
{Schwab} F.~R.,  1980, SPIE, 231, 18

\bibitem[\protect\citeauthoryear{{Simard-Normandin}, {Kronberg} \&
  {Button}}{{Simard-Normandin} et~al.}{1981}]{SimardNormandin81}
{Simard-Normandin} M.,  {Kronberg} P.~P.,    {Button} S.,  1981, \apjs, 45, 97

\bibitem[\protect\citeauthoryear{{Simonetti}, {Cordes} \&
  {Spangler}}{{Simonetti} et~al.}{1984}]{Simonetti84}
{Simonetti} J.~H.,  {Cordes} J.~M.,    {Spangler} S.~R.,  1984, \apj, 284, 126

\bibitem[\protect\citeauthoryear{{Spinrad}}{{Spinrad}}{1986}]{Spinrad86}
{Spinrad} H.,  1986, \pasp, 98, 269

\bibitem[\protect\citeauthoryear{{Strom} \& {Conway}}{{Strom} \&
  {Conway}}{1985}]{Strom85}
{Strom} R.~G.,  {Conway} R.~G.,  1985, \aaps, 61, 547

\bibitem[\protect\citeauthoryear{{Strom}, {Riley}, {Spinrad}, {van Breugel},
  {Djorgovski}, {Liebert} \& {McCarthy}}{{Strom} et~al.}{1990}]{Strom90}
{Strom} R.~G.,  {Riley} J.~M.,  {Spinrad} H.,  {van Breugel} W.~J.~M.,
  {Djorgovski} S.,  {Liebert} J.,    {McCarthy} P.~J.,  1990, \aap, 227, 19

\bibitem[\protect\citeauthoryear{{Tribble}}{{Tribble}}{1991}]{Tribble91}
{Tribble} P.~C.,  1991, \mnras, 250, 726

\bibitem[\protect\citeauthoryear{{Wardle} \& {Kronberg}}{{Wardle} \&
  {Kronberg}}{1974}]{Wardle74}
{Wardle} J.~F.~C.,  {Kronberg} P.~P.,  1974, \apj, 194, 249

\bibitem[\protect\citeauthoryear{{Wellman}, {Daly} \& {Wan}}{{Wellman}
  et~al.}{1997}]{WellmanDaly97}
{Wellman} G.~F.,  {Daly} R.~A.,    {Wan} L.,  1997, \apj, 480, 96

\bibitem[\protect\citeauthoryear{{Williams} \& {Gull}}{{Williams} \&
  {Gull}}{1985}]{WilliamsGull85}
{Williams} A.~G.,  {Gull} S.~F.,  1985, \nat, 313, 34

\end{thebibliography}

\end{document}